\documentclass[aps,pra,twocolumn,superscriptaddress]{revtex4}
\usepackage{graphics}
\usepackage[thmmarks]{ntheorem}
\usepackage{graphicx}
\usepackage{amsmath}
\usepackage{amsfonts}
\usepackage{amssymb}
\usepackage{float}
\usepackage{longtable}
\usepackage{epsfig}
\usepackage{latexsym}
\usepackage{theorem}
\usepackage{bbm}
\usepackage{bm}
\usepackage{psfrag}
\usepackage{color}
\theoremseparator{:} \theoremheaderfont{\normalfont}
\theorembodyfont{\normalfont}
\newtheorem{theorem}{\emph{Theorem}}
\newtheorem{lemma}[theorem]{\emph{Lemma}}
\newtheorem{proposition}[theorem]{\emph{Proposition}}

\newtheorem{remark}[theorem]{\emph{Remark}}
\newcommand{\SP}[1]{{{\textcolor{black}{#1}}}}

\begin{document}
\title{Fundamental Limits of Repeaterless Quantum Communications}
\author{Stefano Pirandola}
\affiliation{Computer Science and York Centre for Quantum Technologies, University of York,
York YO10 5GH, United Kingdom}
\author{Riccardo Laurenza}
\affiliation{Computer Science and York Centre for Quantum Technologies, University of York,
York YO10 5GH, United Kingdom}
\author{Carlo Ottaviani}
\affiliation{Computer Science and York Centre for Quantum Technologies, University of York,
York YO10 5GH, United Kingdom}
\author{Leonardo Banchi}
\affiliation{Department of Physics and Astronomy, University College London, Gower Street,
London WC1E 6BT, United Kingdom}

\begin{abstract}
Quantum communications promises reliable transmission of quantum information,
efficient distribution of entanglement and generation of completely secure
keys. For all these tasks, we need to determine the optimal point-to-point
rates that are achievable by two remote parties at the ends of a quantum
channel, without restrictions on their local operations and classical
communication, which can be unlimited and two-way. These two-way assisted
capacities represent the ultimate rates that are reachable without quantum
repeaters. By constructing an upperbound based on the relative entropy of
entanglement and devising a dimension-independent technique dubbed
\textquotedblleft teleportation stretching\textquotedblright, we establish
these capacities for many fundamental channels, namely bosonic lossy channels,
quantum-limited amplifiers, dephasing and erasure channels in arbitrary
dimension. In particular, we determine the fundamental rate-loss trade-off
affecting any protocol of quantum key distribution.\ Our findings set the
ultimate limits of point-to-point quantum communications and provide the most
precise and general benchmarks for quantum repeaters.
\end{abstract}
\maketitle

Quantum information~\cite{NielsenChuangm,RMP,HolevoBOOK} is moving towards
practical implementations, gradually evolving into next-generation quantum
technologies. The development of completely secure quantum
communications~\cite{BB84,GisinQKD,Scaranim} is an appealing alternative to
the present and insecure classical infrastructure. The idea of a quantum
Internet~\cite{Kimble2008} is attracting huge efforts from different fields,
with hybrid solutions being considered very
promising~\cite{HybridINTERNET,telereview,Ulrikreview}. But quantum
information is more fragile than its classical counterpart. Due to inevitable
interactions with the environment, the ideal performances of quantum protocols
may rapidly degrade. Unfortunately, this limitation affects any point-to-point
quantum communication and requires the use of quantum
repeaters~\cite{Briegel98}. An open problem is therefore to determine the
ultimate point-to-point limits that we can reach without these devices, so
that we fully understand when they are actually needed and what benchmarks
they need to surpass.

Our work investigates this basic problem and establishes the optimal rates
that are achievable by point-to-point (i.e., repeaterless) quantum
communications in the most relevant settings. In fact, we consider two remote
parties connected by a quantum channel, who may exploit unlimited two-way
classical communication (CC) and adaptive local operations (LOs), briefly
called adaptive LOCCs. In this general scenario, we determine the maximum
achievable rates for transmitting quantum information (two-way quantum
capacity $Q_{2}$), distributing entanglement (two-way entanglement
distribution capacity $D_{2}$) and generating secret keys (secret key capacity
$K$), through the most fundamental quantum channels.

The two-way assisted capacities are benchmarks for quantum repeaters because
they are derived by removing any restriction from the point-to-point protocols
between the remote parties, who may perform the most general strategies
allowed by quantum mechanics in the absence of pre-shared entanglement.
Clearly these ultimate limits cannot be achieved by imposing restrictions on
the number of channel uses or enforcing energy constraints. Furthermore, we
note that current protocols of quantum key distribution (QKD) already exploit
relatively large data blocks and high-energy Gaussian
modulations~\cite{Fred,CVMDIQKD}.

To achieve our results we suitably combine the relative entropy of
entanglement (REE)~\cite{VedFORMm,Pleniom,RMPrelent} with
teleportation~\cite{tele93,Samtele,telereview} in a novel reduction method
which completely simplifies quantum protocols based on adaptive LOCCs. The
first step is to show that two-way capacities cannot exceed a bound based on
the REE. The second step is the application of a technique, dubbed
\textquotedblleft teleportation stretching\textquotedblright, which is valid
at any dimension. This allows us to reduce any adaptive protocol to a block
form, so that the REE bound becomes a single-letter quantity. In this way, we
upperbound the two-way capacities of bosonic Gaussian channels~\cite{RMP},
Pauli channels, erasure channels and amplitude damping
channels~\cite{NielsenChuangm}.

Most importantly, by showing coincidence with suitable lower bounds, we can
prove strikingly simple formulas for the two-way quantum capacity $Q_{2}$
($=D_{2}$) and the secret-key capacity $K$ of several fundamental channels. In
fact, for the erasure channel we show that $K=1-p$ where $p$ is the erasure
probability; for the dephasing channel we show that $Q_{2}=K=1-H_{2}(p)$,
where $H_{2}$ is the binary Shannon entropy and $p$ is the dephasing
probability (we also extend these results to any finite dimension). Then, for
a quantum-limited amplifier, we show that $Q_{2}=K=-\log_{2}(1-g^{-1})$ where
$g$ is the gain. Finally, for the lossy channel, we show that $Q_{2}%
=K=-\log_{2}(1-\eta)$ where $\eta$ is the transmissivity.

Note that only the $Q_{2}$ of the erasure channel was previously
known~\cite{ErasureChannelm}. It took about $20$ years to find the other
two-way capacities, which should give an idea of the novelty of our reduction
method. Furthermore, the secret-key capacity\ of the lossy channel determines
the maximum rate achievable in QKD. At high loss $\eta\simeq0$ we find the
optimal rate-loss scaling of $K\simeq1.44\eta$ secret bits per channel use.
This result completely characterizes the fundamental rate-loss law which
restricts secure quantum optical communications, closing a long-standing
investigation with a series of preliminary results achieved in the last years,
such as the lower bound of ref.~\cite{ReverseCAP} based on the reverse
coherent information~\cite{RevCohINFO} and the non-tight upper
bound~\cite{TGW} based on the squashed entanglement~\cite{Matthias2} (see also
other results~\cite{DWrates2,Devetak,Fred,Lutken}). Finally, our approach
using the REE and teleportation stretching closes the investigation, setting
the ultimate and precise limit that only quantum repeaters can surpass.

\section*{Results}

Our machinery builds on several tools but the key features are the REE\ bound
for the two-way capacities, and teleportation stretching, that reduces this
bound to a single-letter quantity. The dimension of the Hilbert space is
arbitrary and may be infinite, i.e., we consider both discrete variable (DV)
and continuous variable (CV) systems~\cite{RMP}. Here we first formulate our
reduction method. Then we derive the results for CV and DV channels.

\bigskip

\begin{figure}[ptbh]
\vspace{-2.3cm}
\par
\begin{center}
\includegraphics[width=0.52\textwidth]{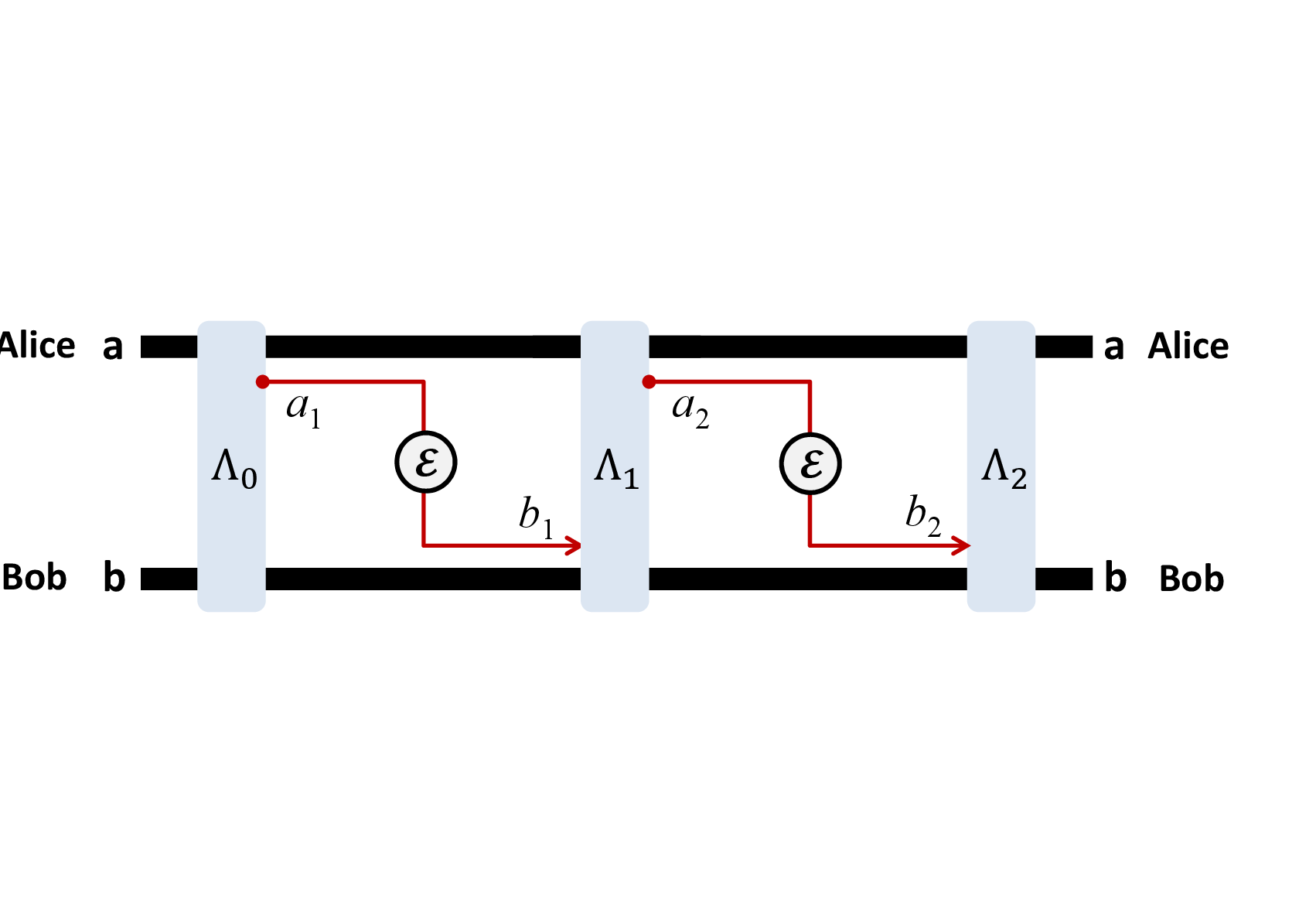} \vspace{-3.1cm}
\end{center}
\caption{Adaptive quantum protocol. The first step is the preparation of the
initial separable state $\rho_{\mathbf{ab}}^{0}$ of $\mathbf{a}$ and
$\mathbf{b}$ by some adaptive LOCC $\Lambda_{0}$. After the preparation of the
local registers, there is the first transmission: Alice picks a system from
her local register $a_{1}\in\mathbf{a}$, so that the register is updated as
$\mathbf{a}\rightarrow\mathbf{a}a_{1}$; system $a_{1}$ is sent through the
channel $\mathcal{E}$, with Bob getting the output $b_{1}$; Bob includes the
output in his local register, which is updated as $b_{1}\mathbf{b}%
\rightarrow\mathbf{b}$; finally, Alice and Bob apply another adaptive LOCC
$\Lambda_{1}$ to their registers $\mathbf{a}$ and $\mathbf{b}$. In the second
transmission, Alice picks and sends another system $a_{2}\in\mathbf{a}$
through channel $\mathcal{E}$ with output $b_{2}$ for Bob. The parties apply a
further adaptive LOCC $\Lambda_{2}$\ to their registers and so on. This
procedure is repeated $n$ times, with output state $\rho_{\mathbf{ab}}^{n}$
for the Alice's and Bob's local registers.}%
\label{protocolAD}%
\end{figure}

\textbf{Adaptive protocols and two-way capacities}. Suppose that Alice and Bob
are separated by a quantum channel $\mathcal{E}$ and want to implement the
most general quantum protocol assisted by adaptive LOCCs. This protocol may be
stated for an arbitrary quantum task and then specified for the transmission
of quantum information, distribution of entanglement or secret correlations.
Assume that Alice and Bob have countable sets of systems, $\mathbf{a}$ and
$\mathbf{b}$, respectively. These are local registers which are updated before
and after each transmission. The steps of an arbitrary adaptive protocol are
described in Fig.~\ref{protocolAD}.

After $n$ transmissions, Alice and Bob share an output state $\rho
_{\mathbf{ab}}^{n}:=\rho_{\mathbf{ab}}(\mathcal{E}^{\otimes n})$
depending on the sequence of adaptive LOCCs
$\mathcal{L}=\{\Lambda_{0},\cdots,\Lambda _{n}\}$. By definition,
this adaptive protocol has a rate equal to $R_{n}$ if the output
$\rho_{\mathbf{ab}}^{n}$ is sufficiently close to a target state
$\phi_{n}$ with $nR_{n}$ bits, i.e., we may write $\left\Vert \rho
_{\mathbf{ab}}^{n}-\phi_{n}\right\Vert \leq\varepsilon$ in trace
norm. The rate of the protocol is an average quantity, which means
that the sequence $\mathcal{L}$ is assumed to be averaged over
local measurements, so that it becomes trace-preserving. Thus, by
taking the asymptotic limit in $n$\ and optimizing over
$\mathcal{L}$, we define the generic two-way capacity
of the channel as%
\begin{equation}
\mathcal{C}(\mathcal{E}):=\sup_{\mathcal{L}}\lim_{n}R_{n}. \label{assisted}%
\end{equation}

In particular, if the aim of the protocol is entanglement distribution, then
the target state $\phi_{n}$ is a maximally entangled state and $\mathcal{C}%
(\mathcal{E})=D_{2}(\mathcal{E})$. Because an ebit can teleport a qubit and a
qubit can distribute an ebit, $D_{2}(\mathcal{E})$ coincides with the two-way
quantum capacity $Q_{2}(\mathcal{E})$. If the goal is to implement QKD, then
the target state $\phi_{n}$ is a private state~\cite{KD1} and $\mathcal{C}%
(\mathcal{E})=K(\mathcal{E})$. Here the secret-key capacity satisfies
$K(\mathcal{E})\geq D_{2}(\mathcal{E})$, because ebits are specific types of
secret bits and LOCCs are equivalent to LOs and public
communication~\cite{KD1}. Thus, the generic two-way capacity $\mathcal{C}$ can
be any of $D_{2}$, $Q_{2}$, or $K$, and these capacities must satisfy
$D_{2}=Q_{2}\leq K$. Also note that we may consider the two-way private
capacity $P_{2}(\mathcal{E})$, which is the maximum rate at which classical
messages can be securely transmitted~\cite{Devetak}. Because of the unlimited
two-way CCs and the one-time pad, we have $P_{2}(\mathcal{E})=K(\mathcal{E})$,
so that this equivalence is implicitly assumed hereafter.

\bigskip

\textbf{General bounds for two-way capacities.} Let us design suitable bounds
for $\mathcal{C}(\mathcal{E})$. From below we know that we may use the
coherent~\cite{QC1,QC2} or reverse coherent~\cite{RevCohINFO,ReverseCAP}
information. Take a maximally entangled state of two systems $A$ and $B$,
i.e., an Einstein-Podolsky-Rosen (EPR) state $\Phi_{AB}$. Propagating the
$B$-part through the channel defines its Choi matrix $\rho_{\mathcal{E}%
}:=\mathcal{I}\otimes\mathcal{E}(\Phi_{AB})$. This allows us to introduce the
coherent information of the channel $I_{\mathrm{C}}(\mathcal{E})$ and its
reverse counterpart $I_{\mathrm{RC}}(\mathcal{E})$, defined as
$I_{\mathrm{C(RC)}}(\mathcal{E}):=S[\mathrm{Tr}_{A(B)}(\rho_{\mathcal{E}%
})]-S(\rho_{\mathcal{E}})$, where $S(\cdot)$ is the von Neumann entropy. These
quantities represent lower bounds for the entanglement that is distillable
from the Choi matrix $\rho_{\mathcal{E}}$ via one-way CCs, denoted as
$D_{1}(\rho_{\mathcal{E}})$. In other words, we can write the hashing
inequality~\cite{DWrates2}
\begin{equation}
\max\{I_{\mathrm{C}}(\mathcal{E}),I_{\mathrm{RC}}(\mathcal{E})\}\leq
D_{1}(\rho_{\mathcal{E}})\leq\mathcal{C}(\mathcal{E}). \label{hashing}%
\end{equation}

For bosonic systems, the ideal EPR\ state has infinite energy, so that the
Choi matrix of a bosonic channel is energy-unbounded (see Methods for notions
on bosonic systems). In this case we consider a sequence of two-mode squeezed
vacuum (TMSV) states~\cite{RMP} $\Phi^{\mu}$ with variance $\mu=\bar{n}+1/2$,
where $\bar{n}$ is the mean number of thermal photons in each mode. This
sequence defines the bosonic EPR state as $\Phi:=\lim_{\mu}\Phi^{\mu}$. At the
output of the channel, we have the sequence of quasi-Choi matrices%
\begin{equation}
\rho_{\mathcal{E}}^{\mu}:=\mathcal{I}\otimes\mathcal{E}(\Phi^{\mu}),
\end{equation}
defining the asymptotic Choi matrix $\rho_{\mathcal{E}}:=\lim_{\mu}%
\rho_{\mathcal{E}}^{\mu}$. As a result, the coherent information quantities
must be computed as limits on $\rho_{\mathcal{E}}^{\mu}$ and the hashing
inequality needs to be suitably extended (see Supplementary Note~2, which
exploits the truncation tools of Supplementary Note~1).

In this work the crucial tool is the upper bound. Recall that, for any
bipartite state $\rho$, the REE is defined as $E_{\mathrm{R}}(\rho
)=\inf_{\sigma_{s}}S(\rho||\sigma_{s})$, where $\sigma_{s}$ is an arbitrary
separable state and $S(\rho||\sigma_{s}):=\mathrm{Tr}\left[  \rho(\log_{2}%
\rho-\log_{2}\sigma_{s})\right]  $ is the relative entropy~\cite{RMPrelent}.
Hereafter we extend this definition to include asymptotic (energy-unbounded)
states. For an asymptotic state $\sigma:=\lim_{\mu}\sigma^{\mu}$ defined by a
sequence of states $\sigma^{\mu}$, we define its REE as
\begin{equation}
E_{\mathrm{R}}(\sigma):=\inf_{\sigma_{s}^{\mu}}\underset{\mu\rightarrow
+\infty}{\lim\inf}S(\sigma^{\mu}||\sigma_{s}^{\mu}), \label{REE_weaker}%
\end{equation}
\SP{where $\sigma_{s}^{\mu}$ is an arbitrary sequence of separable states.} 
In general, we also consider the regularized REE
\begin{equation}
E_{\mathrm{R}}^{\infty}(\sigma):=\lim_{n}n^{-1}E_{\mathrm{R}}(\sigma^{\otimes
n})\leq E_{\mathrm{R}}(\sigma),
\end{equation}
where $\sigma^{\otimes n}:=\lim_{\mu}\sigma^{\mu\otimes n}$ for an asymptotic
state $\sigma$.

Thus, the REE\ of a Choi matrix $E_{\mathrm{R}}\left(  \rho_{\mathcal{E}%
}\right)  $ is correctly defined for channels of any dimension, both finite
and infinite. We may also define the channel's REE as
\begin{equation}
E_{\mathrm{R}}(\mathcal{E}):=\sup_{\rho}E_{\mathrm{R}}\left[  \mathcal{I}%
\otimes\mathcal{E}(\rho)\right]  \geq E_{\mathrm{R}}\left(  \rho_{\mathcal{E}%
}\right)  , \label{channelREEmain}%
\end{equation}
where the supremum includes asymptotic states for bosonic channels. In the
following, we prove that these single-letter quantities, $E_{\mathrm{R}%
}(\mathcal{E})$ and $E_{\mathrm{R}}\left(  \rho_{\mathcal{E}}\right)  $, bound
the two-way capacity $\mathcal{C}(\mathcal{E})$ of basic channels. The first
step is the following general result.

\bigskip

\begin{theorem}
[\textit{general weak converse}]\label{TheoMAIN}At any dimension, finite or
infinite, the generic two-way capacity of a quantum channel $\mathcal{E}$ is
upper bounded by the REE bound%
\begin{equation}
\mathcal{C}(\mathcal{E})\leq E_{\mathrm{R}}^{\bigstar}(\mathcal{E}%
):=\sup_{\mathcal{L}}\lim_{n}\frac{E_{\mathrm{R}}(\rho_{\mathbf{ab}}^{n})}{n}.
\label{Theo1}%
\end{equation}

\end{theorem}


In Supplementary Note~3 we provide various equivalent proofs. The simplest one
assumes an exponential growth of the shield system\ in the target private
state~\cite{KD1} as proven by ref.~\cite{MatthiasMAIN} and trivially adapted
to CVs. Another proof is completely independent from the shield system. Once
established the bound $E_{\mathrm{R}}^{\bigstar}(\mathcal{E})$, our next step
is to simplify it by applying the technique of teleportation stretching, which
is in turn based on a suitable simulation of quantum channels.

\bigskip

\textbf{Simulation of quantum channels.} The idea of simulating
channels by teleportation was first
developed~\cite{B2main,HoroTELm} for Pauli
channels~\cite{SougatoBowenm}, and further studied in finite
dimension~\cite{Albeveriom,Alex,Leungm} after the introduction of
generalized teleportation protocols~\cite{WernerTELEmain}. Then,
ref.~\cite{NisetMAIN} moved the first steps in the simulation of
Gaussian channels via the CV teleportation
protocol~\cite{Samtele,Samtele2}. Another type of simulation is a
deterministic version~\cite{qSIM} of a programmable quantum gate
array~\cite{Processorsm}. Developed for DV systems, this is based
on joint quantum operations, therefore failing to catch the LOCC
structure of quantum communication. Here, not only we fully extend
the teleportation-simulation to CV\ systems but we also design the
most general channel simulation in a communication scenario; this
is based on arbitrary LOCCs and may involve systems of any
dimension, finite or infinite (see Supplementary Note~8 for
comparisons and advances).

\begin{figure*}[ptbh]
\begin{center}
\vspace{-3.3cm} \vspace{-0.8cm}
\includegraphics[width=1.02\textwidth]{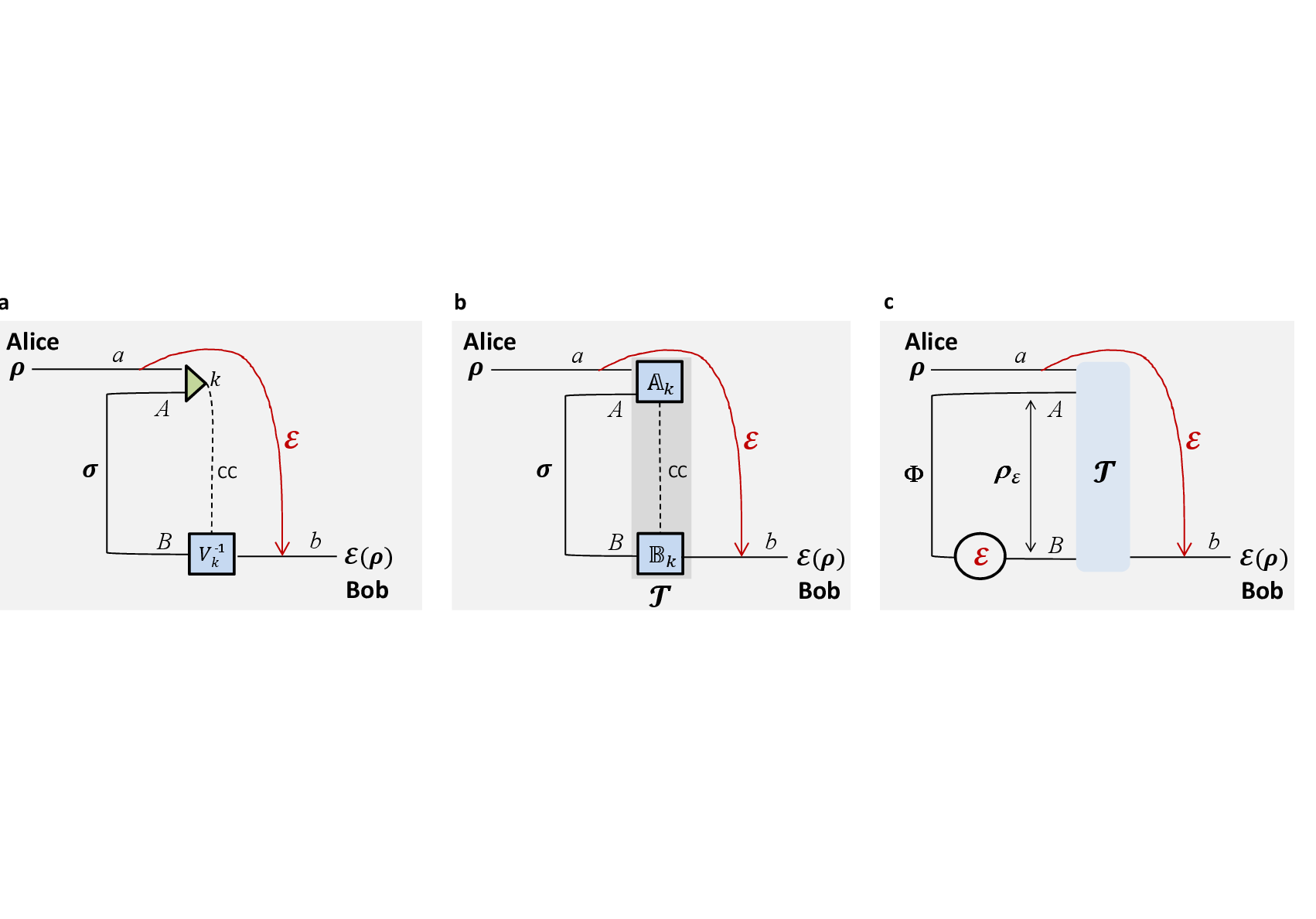} \vspace{-4.8cm}
\end{center}
\caption{From teleportation- to LOCC-simulation of quantum channels.
(\textbf{a}) Consider the generalized teleportation of an input state $\rho$
of a $d$-dimensional system $a$ by using a resource state $\sigma$ of two
systems, $A$ and $B$, with corresponding dimensions $d$ and $d^{\prime}$
(finite or infinite). Systems $a$ and $A$ are subject to a Bell detection
(triangle) with random outcome $k$. This outcome is associated with a
projection onto a maximally entangled state up to an associated teleportation
unitary $U_{k}$ which is a Pauli operator for $d<+\infty$ and a
phase-displacement for $d=+\infty$ (see Methods for the basics of quantum
teleportation and the characterization of the teleportation unitaries). The
classical outcome $k$ is communicated to Bob, who applies a correction unitary
$V_{k}^{-1}$ to his system $B$ with output $b$. In general, $V_{k}$ does not
necessarily belong to the set $\{U_{k}\}$. On average, this teleportation LOCC
defines a teleportation channel $\mathcal{E}$\ from $a$ to $b$. It is clear
that this construction also teleports part $a$ of an input state involving
ancillary systems.\ (\textbf{b}) In general we may replace the teleportation
LOCC (Bell detection and unitary corrections) with an arbitrary LOCC
$\mathcal{T}$: Alice performs a quantum operation $\mathbb{A}_{k}$ on her
systems $a$ and $A$, communicates the classical variable $k$\ to Bob, who then
applies another quantum operation $\mathbb{B}_{k}$\ on his system $B$. By
averaging over the variable $k$, so that\ $\mathcal{T}$\ is certainly
trace-preserving, we achieve the simulation $\mathcal{E}(\rho)=\mathcal{T}%
(\rho\otimes\sigma)$ for any input state $\rho$. We say that a channel
$\mathcal{E}$\ is \textquotedblleft$\sigma$-stretchable\textquotedblright\ if
it can be simulated by a resource state $\sigma$ for some LOCC $\mathcal{T}$.
Note that Alice's and Bob's LOs $\mathbb{A}_{k}$ and $\mathbb{B}_{k}$ are
arbitrary quantum operations; they may involve other local ancillas and also
have extra labels (due to additional local measurements), in which case
$\mathcal{T}$ is assumed to be averaged over all these labels. (\textbf{c}%
)~The most important case is when channel $\mathcal{E}$ can be simulated by a
trace-preserving LOCC $\mathcal{T}$ applied to its Choi matrix $\rho
_{\mathcal{E}}:=\mathcal{I}\otimes$ $\mathcal{E}(\Phi)$, with $\Phi$ being an
EPR state. In this case, we say that the channel is \textquotedblleft
Choi-stretchable\textquotedblright. These definitions are suitably extended to
bosonic channels.}%
\label{stretchingMAIN}%
\end{figure*}

As explained in Fig.~\ref{stretchingMAIN}a, performing a teleportation LOCC
(i.e., Bell detection and unitary corrections) over a mixed state $\sigma$ is
a way to simulate a (certain type of) quantum channel $\mathcal{E}$ from Alice
to Bob. However, more generally, the channel simulation can be realized using
an arbitrary trace-preserving LOCC $\mathcal{T}$ and an arbitrary resource
state $\sigma$ (see Fig.~\ref{stretchingMAIN}b). Thus, at any dimension, we
say that a channel $\mathcal{E}$ is \textquotedblleft$\sigma$%
-stretchable\textquotedblright\ or \textquotedblleft stretchable into $\sigma
$\textquotedblright\ if there is a trace-preserving LOCC $\mathcal{T}$ such
that%
\begin{equation}
\mathcal{E}(\rho)=\mathcal{T}(\rho\otimes\sigma). \label{simulationCHOI}%
\end{equation}

In general, we can simulate the same channel $\mathcal{E}$\ with different
choices of $\mathcal{T}$ and $\sigma$. In fact, any channel is stretchable
into some state $\sigma$: A trivial choice is decomposing $\mathcal{E}%
=\mathcal{I}\circ\mathcal{E}$, inserting $\mathcal{E}$ in Alice's LO and
simulating $\mathcal{I}$ with teleportation over the ideal EPR state
$\sigma=\Phi$. Therefore, among all simulations, one needs to identify the
best resource state that optimizes the functional under study. In our work,
the best results are achieved when the state $\sigma$ can be chosen as the
Choi matrix of the channel. This is not a property of any channel but defines
a class. Thus, we define \textquotedblleft Choi-stretchable\textquotedblright%
\ a channel that can be LOCC-simulated over its Choi matrix, so that we may
write Eq.~(\ref{simulationCHOI}) with $\sigma=\rho_{\mathcal{E}}$ (see also
Fig.~\ref{stretchingMAIN}c).

In infinite dimension, the LOCC simulation may involve limits $\mathcal{T}%
:=\lim_{\mu}\mathcal{T}^{\mu}$ and $\sigma:=\lim_{\mu}\sigma^{\mu}$ of
sequences $\mathcal{T}^{\mu}$ and $\sigma^{\mu}$. For any finite $\mu$, the
simulation $(\mathcal{T}^{\mu},\sigma^{\mu})$ provides some teleportation
channel $\mathcal{E}^{\mu}$. Now, suppose that an asymptotic channel
$\mathcal{E}$ is defined as a pointwise limit of the sequence $\mathcal{E}%
^{\mu}$, i.e., we have $||\mathcal{I}\otimes\mathcal{E}(\rho)-\mathcal{I}%
\otimes\mathcal{E}^{\mu}(\rho)||\overset{\mu}{\rightarrow}0$ for any bipartite
state $\rho$. Then, we say that $\mathcal{E}$ is stretchable with asymptotic
simulation $(\mathcal{T},\sigma)$. This is important for bosonic channels, for
which Choi-based simulations can only be asymptotic and based on sequences
$\rho_{\mathcal{E}}^{\mu}$.

\bigskip

\textbf{Teleportation covariance.}~We now discuss a property\ which easily
identifies Choi-stretchable channels. Call $\mathbb{U}_{d}$\ the random
unitaries which are generated by the Bell detection in a teleportation
process. For a qudit, $\mathbb{U}_{d}$\ is composed of generalized Pauli
operators, i.e., the generators of the Weyl-Heisenberg group. For a CV system,
the set $\mathbb{U}_{\infty}$ is composed of displacement
operators~\cite{telereview}, spanning the infinite dimensional version of the
previous group. In arbitrary dimension (finite or infinite), we say that a
quantum channel is \textquotedblleft teleportation-covariant\textquotedblright%
\ if, for any teleportation unitary $U\in\mathbb{U}_{d}$, we may write%
\begin{equation}
\mathcal{E}(U\rho U^{\dagger})=V\mathcal{E}(\rho)V^{\dagger},
\label{tele-covariant}%
\end{equation}
for some another unitary $V$ (not necessarily in $\mathbb{U}_{d}$).

The key property of a teleportation-covariant channel is that the input
teleportation unitaries can be pushed out of the channel, where they become
other correctable unitaries. Because of this property, the transmission of a
system through the channel can be simulated by a generalized teleportation
protocol over its Choi matrix. This is the content of the following proposition.

\bigskip

\begin{proposition}
[\textit{tele-covariance}]\label{propPRELI}At any dimension, a
teleportation-covariant channel $\mathcal{E}$\ is Choi-stretchable. The
simulation is a teleportation LOCC over its Choi matrix $\rho_{\mathcal{E}}$,
which is asymptotic for a bosonic channel.
\end{proposition}

\bigskip

\begin{figure*}[ptbh]
\begin{center}
\vspace{-0.7cm} \vspace{-1cm}
\includegraphics[width=0.75\textwidth] {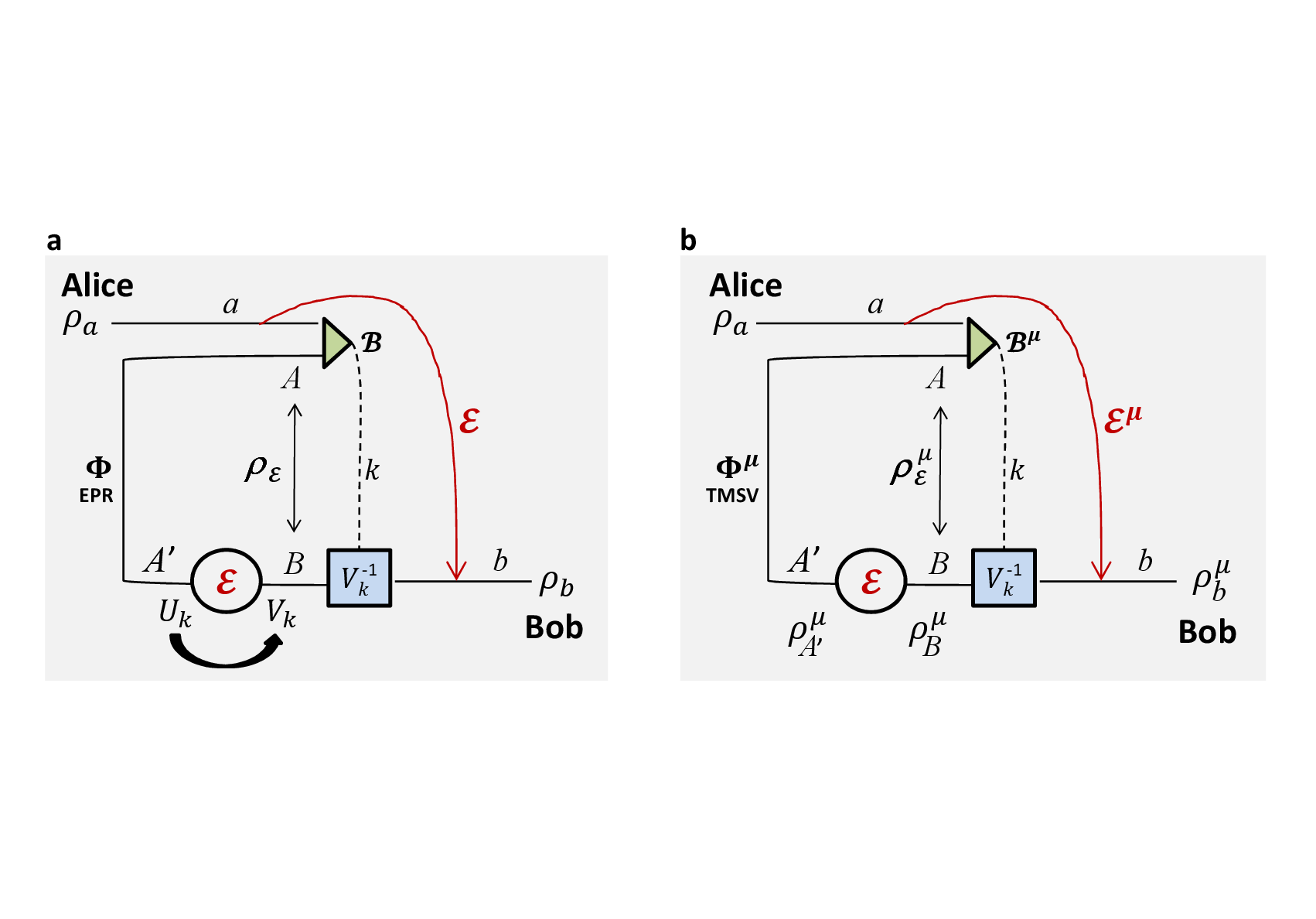} \vspace{-2.2cm}
\end{center}
\par
\vspace{-0.2cm}\caption{Teleportation-covariant channels are Choi-stretchable.
(\textbf{a})~Consider the teleportation of an input state $\rho_{a}$ by using
an EPR state $\Phi_{AA^{\prime}}$ of systems $A$ and $A^{\prime}$. The Bell
detection $\mathcal{B}$ on systems $a$ and $A$ teleports the input state onto
$A^{\prime}$, up to a random teleportation unitary, i.e., $\rho_{A^{\prime}%
}=U_{k}\rho_{a}U_{k}^{\dagger}$. Because $\mathcal{E}$ is
teleportation-covariant, $U_{k}$ is mapped into an output unitary $V_{k}$ and
we may write $\rho_{B}=\mathcal{E}(\rho_{A^{\prime}})=\mathcal{E}(U_{k}%
\rho_{a}U_{k}^{\dagger})=V_{k}\mathcal{E}(\rho_{a})V_{k}^{\dagger}$.
Therefore, Bob just needs to receive the outcome $k$ and apply $V_{k}^{-1}$,
so that $\rho_{b}=V_{k}^{-1}\rho_{B}(V_{k}^{-1})^{\dagger}=\mathcal{E}%
(\rho_{a})$. Globally, the process describes the simulation of channel
$\mathcal{E}$\ by means of a generalized teleportation protocol over the Choi
matrix $\rho_{\mathcal{E}}$. (\textbf{b})~The procedure is also valid for CV
systems. If the input $a$ is a bosonic mode, we need to consider finite-energy
versions for the EPR state $\Phi$ and the Bell detection $\mathcal{B}$, i.e.,
we use a TMSV state $\Phi^{\mu}$ and a corresponding quasi-projection
$\mathcal{B}^{\mu}$\ onto displaced TMSV states. At finite energy $\mu$, the
teleportation process from $a$ to $A^{\prime}$ is imperfect with some output
$\rho_{A^{\prime}}^{\mu}\neq\rho_{A^{\prime}}=U_{k}\rho_{a}U_{k}^{\dagger}$.
However, for any $\varepsilon>0$\ and input state $\rho_{a}$, there is a
sufficiently large value of $\mu$ such that $||\rho_{A^{\prime}}^{\mu}%
-\rho_{A^{\prime}}||\leq\varepsilon$~\cite{Samtele,Samtele2}. Consider the
transmitted state $\rho_{B}^{\mu}=\mathcal{E}(\rho_{A^{\prime}}^{\mu})$.
Because the trace distance decreases under channels, we have $||\rho_{B}^{\mu
}-\rho_{B}||\leq||\rho_{A^{\prime}}^{\mu}-\rho_{A^{\prime}}||\leq\varepsilon$.
After the application of the correction unitary $V_{k}^{-1}$, we have the
output state $\rho_{b}^{\mu}$ which satisfies $||\rho_{b}^{\mu}-\mathcal{E}%
(\rho_{a})||\leq\varepsilon$. Taking the asymptotic limit of large $\mu$, we
achieve $||\rho_{b}^{\mu}-\mathcal{E}(\rho_{a})||\rightarrow0$ for any input
$\rho_{a}$, therefore achieving the perfect asymptotic simulation of the
channel. The asymptotic teleportation-LOCC is therefore $(\mathcal{B}%
,\rho_{\mathcal{E}}):=\lim_{\mu}(\mathcal{B}^{\mu},\rho_{\mathcal{E}}^{\mu})$
where $\rho_{\mathcal{E}}^{\mu}:=\mathcal{I}\otimes\mathcal{E}(\Phi^{\mu})$.
The result is trivially extended to the presence of ancillas.}%
\label{stretchpic}%
\end{figure*}

The simple proof is explained in Fig.~\ref{stretchpic}. The class of
teleportation-covariant channels is wide and includes bosonic Gaussian
channels, Pauli and erasure channels at any dimension (see Methods for a more
detailed classification). All these fundamental channels are therefore
Choi-stretchable. There are channels which are not (or not known to be)
Choi-stretchable but still have decompositions $\mathcal{E=E}^{\prime\prime
}\circ\mathcal{\tilde{E}}\circ\mathcal{E}^{\prime}$ where the middle part
$\mathcal{\tilde{E}}$ is Choi-stretchable. In this case, $\mathcal{E}^{\prime
}$ and $\mathcal{E}^{\prime\prime}$ can be made part of Alice's and Bob's LOs,
so that channel $\mathcal{E}$ can be stretched into the state $\sigma
=\rho_{\mathcal{\tilde{E}}}$. An example is the amplitude damping channel as
we will see afterwards.

\bigskip

\begin{figure*}[pth]
\vspace{-1.1cm}
\par
\begin{center}
\includegraphics[width=0.76\textwidth]{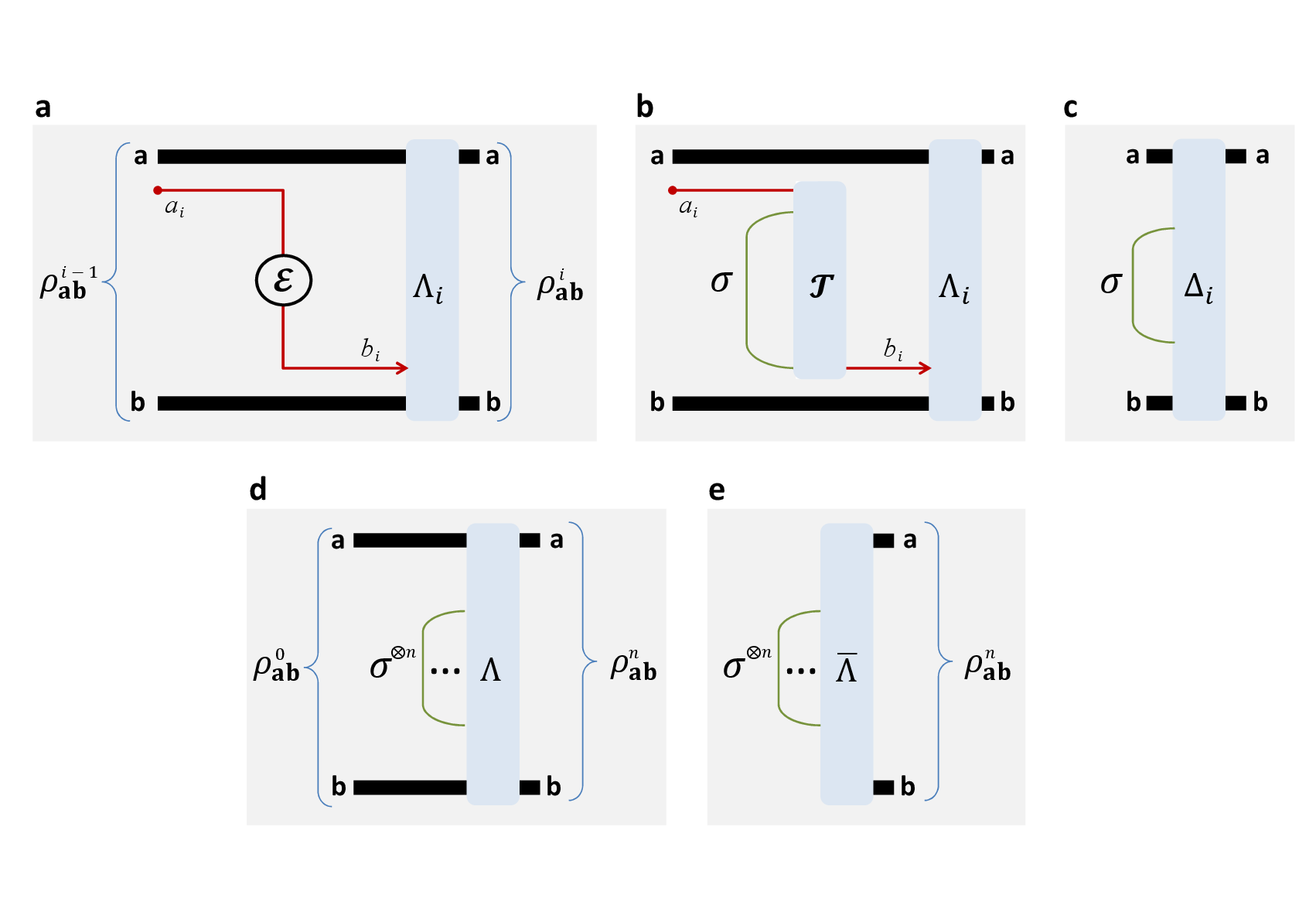} \vspace{-1.2cm}
\end{center}
\caption{Teleportation stretching of an adaptive quantum protocol.
\textbf{(a)}~Consider the $i$th transmission through channel $\mathcal{E}$,
where the input $(i-1)$th\ register state is given by $\rho_{\mathbf{ab}%
}^{i-1}:=\rho_{\mathbf{a}a_{i}\mathbf{b}}$. After transmission through
$\mathcal{E}$ and the adaptive LOCC $\Lambda_{i}$, the register state is
updated to $\rho_{\mathbf{ab}}^{i}=\Lambda_{i}\circ(\mathcal{I}_{\mathbf{a}%
}\otimes\mathcal{E}\otimes\mathcal{I}_{\mathbf{b}})(\rho_{\mathbf{a}%
a_{i}\mathbf{b}})$. \textbf{(b)}~Let us simulate the channel $\mathcal{E}$ by
a LOCC $\mathcal{T}$ and a resource state $\sigma$. \textbf{(c)}~The
simulation LOCC $\mathcal{T}$ can be combined with the adaptive LOCC
$\Lambda_{i}$ into a single \textquotedblleft extended\textquotedblright\ LOCC
$\Delta_{i}$ while the resource state $\sigma$ can be stretched back in time
and out of the adaptive operations. We may therefore write $\rho_{\mathbf{ab}%
}^{i}=\Delta_{i}(\rho_{\mathbf{ab}}^{i-1}\otimes\sigma)$. \textbf{(d)} We
iterate the previous steps for all transmissions, so as to stretch $n$ copies
$\sigma^{\otimes n}$ and collapse all the extended LOCCs $\Delta_{n}\circ$
$\ldots\circ\Delta_{1}$ into a single LOCC $\Lambda$. In other words, we may
write $\rho_{\mathbf{ab}}^{n}=$ $\Lambda(\rho_{\mathbf{ab}}^{0}\otimes
\sigma^{\otimes n})$. \textbf{(e)}~Finally, we include the preparation of the
separable state $\rho_{\mathbf{ab}}^{0}$ into $\Lambda$ and we also average
over all local measurements present in $\Lambda$, so that we may write the
output state as $\rho_{\mathbf{ab}}^{n}=$ $\bar{\Lambda}(\sigma^{\otimes n})$
for a trace-preserving LOCC $\bar{\Lambda}$. The procedure is asymptotic in
the presence of asymptotic channel simulations (bosonic channels).}%
\label{reduction}%
\end{figure*}

\textbf{Teleportation stretching of adaptive protocols.} We are now ready to
describe the reduction of arbitrary adaptive protocols. The procedure is
schematically shown in Fig.~\ref{reduction}. We start by considering the $i$th
transmission through the channel $\mathcal{E}$, so that Alice and Bob's
register state is updated from $\rho_{\mathbf{ab}}^{i-1}$ to $\rho
_{\mathbf{ab}}^{i}$. By using a\ simulation $(\mathcal{T},\sigma)$, we show
the input-output formula%
\begin{equation}
\rho_{\mathbf{ab}}^{i}=\Delta_{i}\left(  \rho_{\mathbf{ab}}^{i-1}\otimes
\sigma\right)  ,
\end{equation}
for some \textquotedblleft extended\textquotedblright\ LOCC $\Delta_{i}$ (see
Fig.~\ref{reduction}c). By iterating the previous formula $n$ times, we may
write the output state $\rho_{\mathbf{ab}}^{n}=\Lambda\left(  \rho
_{\mathbf{ab}}^{0}\otimes\sigma^{\otimes n}\right)  $ for $\Lambda:=\Delta
_{n}\circ\ldots\circ\Delta_{1}$ (as in Fig.~\ref{reduction}d). Because the
initial state $\rho_{\mathbf{ab}}^{0}$ is separable, its preparation can be
included in $\Lambda$ and we may directly write $\rho_{\mathbf{ab}}%
^{n}=\Lambda\left(  \sigma^{\otimes n}\right)  $. Finally, we average over all
local measurements present in $\Lambda$, so that $\rho_{\mathbf{ab}}^{n}%
=\bar{\Lambda}\left(  \sigma^{\otimes n}\right)  $ for a trace-preserving LOCC
$\bar{\Lambda}$ (see Fig.~\ref{reduction}e). More precisely, for any sequence
of outcomes $\mathbf{u}$\ with probability $p(\mathbf{u})$, there is
conditional LOCC $\Lambda_{\mathbf{u}}$ with output $\rho_{\mathbf{ab}}%
^{n}(\mathbf{u})=p(\mathbf{u})^{-1}\Lambda_{\mathbf{u}}\left(  \sigma^{\otimes
n}\right)  $. Thus, the mean output state $\rho_{\mathbf{ab}}^{n}$ is
generated by $\bar{\Lambda}=\sum_{\mathbf{u}}\Lambda_{\mathbf{u}}$ (see
Methods for more technical details on this LOCC\ averaging).

Note that the simulation of a bosonic channel $\mathcal{E}$ is typically
asymptotic, with infinite-energy limits $\mathcal{T}:=\lim_{\mu}%
\mathcal{T}^{\mu}$ and $\sigma:=\lim_{\mu}\sigma^{\mu}$. In this case, we
repeat the procedure for some $\mu$, with output $\rho_{\mathbf{ab}}^{n,\mu
}:=\bar{\Lambda}_{\mu}(\sigma^{\mu\otimes n})$, where $\bar{\Lambda}_{\mu}%
$\ is derived assuming the finite-energy LOCCs\ $\mathcal{T}^{\mu}$. Then, we
take the limit for large $\mu$, so that $\rho_{\mathbf{ab}}^{n,\mu}$ converges
to $\rho_{\mathbf{ab}}^{n}$ in trace norm (see Methods for details on
teleportation stretching with bosonic channels). Thus, at any dimension, we
have proven the following result.

\bigskip

\begin{lemma}
[\textit{Stretching}]\label{stretch_LEMMA}Consider arbitrary $n$ transmissions
through a channel $\mathcal{E}$ which is stretchable into a resource state
$\sigma$. The output of an adaptive protocol can be decomposed into the block
form%
\begin{equation}
\rho_{\mathbf{ab}}^{n}=\bar{\Lambda}(\sigma^{\otimes n})~, \label{Stretch111}%
\end{equation}
for some trace-preserving LOCC $\bar{\Lambda}$. If the channel $\mathcal{E}$
is Choi-stretchable, then we may write
\begin{equation}
\rho_{\mathbf{ab}}^{n}=\bar{\Lambda}(\rho_{\mathcal{E}}^{\otimes n})~.
\end{equation}
In particular, $\bar{\Lambda}(\sigma^{\otimes n}):=\lim_{\mu}\bar{\Lambda
}_{\mu}(\sigma^{\mu\otimes n})$\ for an asymptotic channel simulation
$(\mathcal{T},\sigma):=\lim_{\mu}(\mathcal{T}^{\mu},\sigma^{\mu})$.
\end{lemma}

\bigskip

According to this Lemma, teleportation stretching reduces an\textit{ }adaptive
protocol performing an arbitrary task (quantum communication, entanglement
distribution or key generation) into an equivalent block protocol, whose
output state $\rho_{\mathbf{ab}}^{n}$ is the same but suitably decomposed as
in Eq.~(\ref{Stretch111}) for any number $n$ of channel uses. In particular,
for Choi-stretchable channels, the output is decomposed into a tensor-product
of Choi matrices. An essential feature which makes the technique applicable to
many contexts is the fact that the adaptive-to-block reduction maintains task
and output of the original protocol so that, e.g., adaptive key generation is
reduced to block key generation and not entanglement distillation.

\bigskip

\begin{remark}
Some aspects of our method might be traced back to a precursory but very
specific argument discussed in Section~V of ref.~\cite{B2main}, where
protocols of quantum communication (through Pauli channels) were transformed
into protocols of entanglement distillation (the idea was developed for 1-way
CCs, with an implicit extension to 2-way CCs). However, while this argument
may be seen as precursory, it is certainly not developed at the level of
generality of the present work where the adaptive-to-block reduction is
explicitly proven for any type of protocol and any channel at any dimension
(see Supplementary Notes~9 and~10 for remarks on previous literature).
\end{remark}

\bigskip

\textbf{REE as a single-letter converse bound.} The combination of
Theorem~\ref{TheoMAIN} and Lemma~\ref{stretch_LEMMA} provides the insight of
our entire reduction method. In fact, let us compute the REE\ of the output
state $\rho_{\mathbf{ab}}^{n}$, decomposed as in Eq.~(\ref{Stretch111}). Using
the monotonicity of the REE under trace-preserving LOCCs, we derive%
\begin{equation}
E_{\mathrm{R}}(\rho_{\mathbf{ab}}^{n})\leq E_{\mathrm{R}}(\sigma^{\otimes n}),
\label{toREP}%
\end{equation}
where the complicated $\bar{\Lambda}$ is fully discarded. Then, by replacing
Eq.~(\ref{toREP}) into Eq.~(\ref{Theo1}), we can ignore the supremum in the
definition of $E_{\mathrm{R}}^{\bigstar}(\mathcal{E})$ and get the simple
bound
\begin{equation}
E_{\mathrm{R}}^{\bigstar}(\mathcal{E})\leq E_{\mathrm{R}}^{\infty}(\sigma)\leq
E_{\mathrm{R}}(\sigma). \label{UBREE}%
\end{equation}
Thus, we can state the following main result.

\bigskip

\begin{theorem}
[one-shot REE bound]\label{addTHEO}Let us stretch an arbitrary quantum channel
$\mathcal{E}$ into some resource state $\sigma$, according to
Eq.~(\ref{simulationCHOI}).
Then, we may write%
\begin{equation}
\mathcal{C}(\mathcal{E})\leq E_{\mathrm{R}}^{\infty}(\sigma)\leq
E_{\mathrm{R}}(\sigma). \label{th11}%
\end{equation}
In particular, if $\mathcal{E}$ is Choi-stretchable, we have%
\begin{equation}
\mathcal{C}(\mathcal{E})\leq E_{\mathrm{R}}^{\infty}(\rho_{\mathcal{E}})\leq
E_{\mathrm{R}}(\rho_{\mathcal{E}})=E_{\mathrm{R}}(\mathcal{E}). \label{th22}%
\end{equation}

\end{theorem}

\bigskip

\begin{figure}[ptbh]
\vspace{-0.9cm}
\par
\begin{center}
\vspace{-0.0cm} \includegraphics[width=0.55\textwidth]
{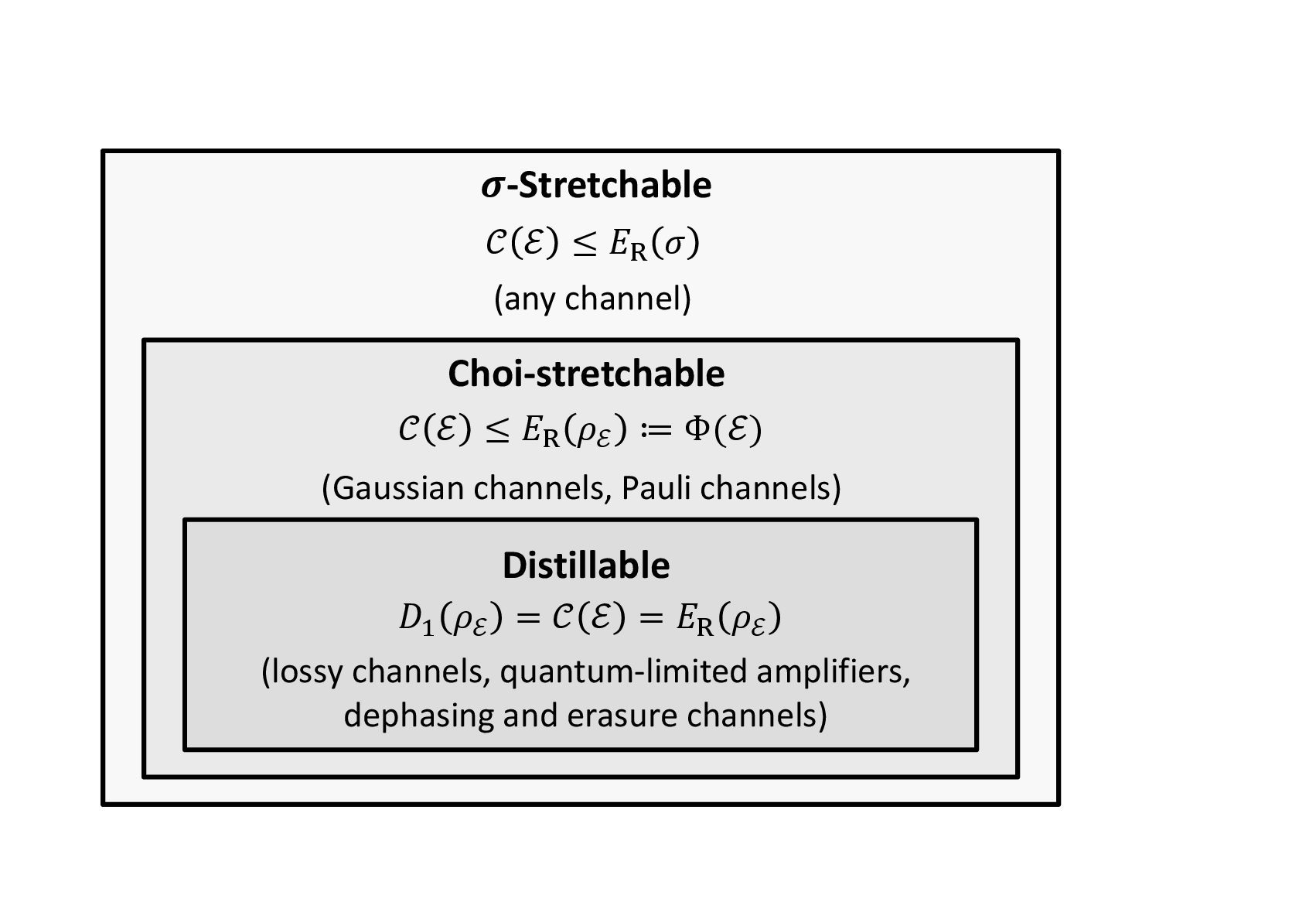} \vspace{-0.5cm} \vspace{-1cm}
\end{center}
\caption{Classification of channels in DVs and CVs.}%
\label{channelCLASS}%
\end{figure}

See Methods for a detailed proof, with explicit derivations for bosonic
channels. We have therefore reached our goal and found single-letter bounds.
In particular, note that $E_{\mathrm{R}}\left(  \rho_{\mathcal{E}}\right)  $
measures the entanglement distributed by a single EPR\ state, so that we may
call it the \textquotedblleft entanglement flux\textquotedblright\ of the
channel $\Phi(\mathcal{E}):=E_{\mathrm{R}}\left(  \rho_{\mathcal{E}}\right)
$.\ Remarkably, there is a sub-class of Choi-stretchable channels for which
$E_{\mathrm{R}}\left(  \rho_{\mathcal{E}}\right)  $ coincides with the lower
bound $D_{1}(\rho_{\mathcal{E}})$ in Eq.~(\ref{hashing}). We call these
\textquotedblleft distillable channels\textquotedblright. We establish all
their two-way capacities as $\mathcal{C}(\mathcal{E})=E_{\mathrm{R}}\left(
\rho_{\mathcal{E}}\right)  $. They include lossy channels, quantum-limited
amplifiers, dephasing and erasure channels. See also Fig.~\ref{channelCLASS}.

\bigskip

\textbf{Immediate generalizations.} Consider a fading channel, described by an
ensemble $\{p_{i},\mathcal{E}_{i}\}$, where channel $\mathcal{E}_{i}$ occurs
with probability $p_{i}$. Let us stretch $\mathcal{E}_{i}$ into a resource
state $\sigma_{i}$. For large $n$,
we may decompose the output of an adaptive protocol as $\rho_{\mathbf{ab}}%
^{n}=\bar{\Lambda}(\otimes_{i}\sigma_{i}^{\otimes np_{i}})$, so that the
two-way capacity of this channel is bounded by
\begin{equation}
\mathcal{C}(\{p_{i},\mathcal{E}_{i}\})\leq\sum_{i}p_{i}E_{\mathrm{R}}%
(\sigma_{i})~. \label{streFAD}%
\end{equation}

Then consider adaptive protocols of two-way quantum communication, where the
parties have forward ($\mathcal{E}$) and backward ($\mathcal{E}^{\prime}$)
channels. The capacity $\mathcal{C}(\mathcal{E},\mathcal{E}^{\prime})$
maximizes the number of target bits per channel use. Stretching $(\mathcal{E}%
,\mathcal{E}^{\prime})$ into a pair of states $(\sigma,\sigma^{\prime})$, we
find $\mathcal{C}(\mathcal{E},\mathcal{E}^{\prime})\leq\max\{E_{\mathrm{R}%
}(\sigma),E_{\mathrm{R}}(\sigma^{\prime})\}$.\ For Choi-stretchable channels,
this means $\mathcal{C}(\mathcal{E},\mathcal{E}^{\prime})\leq\max
\{\Phi(\mathcal{E}),\Phi(\mathcal{E}^{\prime})\}$, which reduces to
$\mathcal{C}(\mathcal{E},\mathcal{E}^{\prime})=\max\{\mathcal{C}%
(\mathcal{E}),\mathcal{C}(\mathcal{E}^{\prime})\}$ if they are distillable. In
the latter case, the optimal strategy is\ using the channel with the maximum
capacity (see Methods).

Yet another scenario is the multiband channel $\mathcal{E}_{\text{mb}}$, where
Alice exploits $m$ independent channels or \textquotedblleft
bands\textquotedblright\ $\{\mathcal{E}_{i}\}$, so that the capacity
$\mathcal{C}(\mathcal{E}_{\text{\textrm{mb}}})$ maximizes the number of target
bits per multiband transmission. By stretching the bands $\{\mathcal{E}_{i}\}$
into resource states $\{\sigma_{i}\}$, we find $\mathcal{C}(\mathcal{E}%
_{\text{\textrm{mb}}})\leq\sum_{i}E_{\mathrm{R}}(\sigma_{i})$. For
Choi-stretchable bands, this means $\mathcal{C}(\mathcal{E}_{\text{\textrm{mb}%
}})\leq\sum_{i}\Phi(\mathcal{E}_{i})$, giving the additive capacity
$\mathcal{C}(\mathcal{E}_{\text{\textrm{mb}}})=\sum_{i}\mathcal{C}%
(\mathcal{E}_{i})$ if they are distillable (see Methods).

\bigskip

\textbf{Ultimate limits of bosonic communications.} We now apply our method to
derive the ultimate rates for quantum and secure communication through bosonic
Gaussian channels. These channels are Choi-stretchable with an asymptotic
simulation involving $\rho_{\mathcal{E}}:=\lim_{\mu}\rho_{\mathcal{E}}^{\mu}$.
From Eqs.~(\ref{REE_weaker}) and~(\ref{th22}), we may write%
\begin{equation}
\mathcal{C}(\mathcal{E})\leq\Phi(\mathcal{E})\leq\underset{\mu\rightarrow
+\infty}{\lim\inf}S(\rho_{\mathcal{E}}^{\mu}||\tilde{\sigma}_{s}^{\mu})~,
\label{liminf}%
\end{equation}
\SP{for a suitable sequence of separable states $\tilde{\sigma}_{s}^{\mu}$.}

For Gaussian channels, the sequences in Eq.~(\ref{liminf}) involve Gaussian
states, for which we easily compute the relative entropy. In fact, for any two
Gaussian states, $\rho_{1}$ and $\rho_{2}$, we prove the general formula
$S(\rho_{1}||\rho_{2})=\Sigma(V_{1},V_{2})-S(\rho_{1})$, where $\Sigma$ is a
simple functional of their statistical moments (see Methods). After technical
derivations (see Supplementary Note~4), we then bound the two-way capacities
of all Gaussian channels, starting from the most important, the lossy channel.

\bigskip

\textbf{Fundamental rate-loss scaling in quantum optical communications.}
Optical communications through free-space links or telecom fibres are
inevitably lossy and the standard model to describe this scenario is the lossy
channel. This is a bosonic Gaussian channel characterized by a transmissivity
parameter $\eta$, which quantifies the fraction of input photons surviving the
channel. It can be represented as a beam splitter mixing the signals with
zero-temperature environment (background thermal noise is negligible at
optical and telecom frequencies).

For a lossy channel $\mathcal{E}_{\eta}$ with arbitrary transmissivity $\eta$
we apply our reduction method and compute the entanglement flux $\Phi
(\eta)\leq-\log_{2}(1-\eta)$. This coincides with the reverse coherent
information of this channel $I_{\mathrm{RC}}(\eta)$, first derived in
ref.~\cite{ReverseCAP}. Thus, we find that this channel is distillable and all
its two-way capacities are given by%
\begin{equation}
\mathcal{C}(\eta)=D_{2}(\eta)=Q_{2}(\eta)=K(\eta)=-\log_{2}(1-\eta).
\label{formCloss}%
\end{equation}
Interestingly, this capacity coincides with the maximum
discord~\cite{RMPdiscord} that can be distributed, since we may
write~\cite{DiscordQKD} $I_{\mathrm{RC}}(\eta)=D(B|A)$, where the latter is
the discord of the (asymptotic) Gaussian Choi matrix $\rho_{\mathcal{E}_{\eta
}}$~\cite{OptimalDIS}. We also prove the strict separation $Q_{2}(\eta
)>Q(\eta)$, where $Q$ is the unassisted quantum capacity~\cite{QC1,QC2}.

\begin{figure}[ptbh]
\begin{center}
\vspace{+0.1cm} \includegraphics[width=0.48\textwidth]{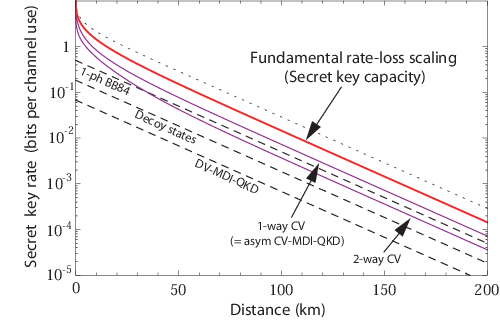}
\end{center}
\par
\vspace{-0.7cm}\caption{Ideal performances in QKD. We plot the secret key rate
(bits per channel use) versus Alice-Bob's distance (km) at\ the loss rate of
$0.2$~dB per km. The secret key capacity of the channel (red line) sets the
fundamental rate limit for point-to-point QKD in the presence of loss. Compare
this capacity with a previous non-achievable upperbound~\cite{TGW} (dotted
line). We then show the maximum rates that are potentially achievable by
current protocols, assuming infinitely long keys and\ ideal conditions, such
as unit detector efficiencies, zero dark count rates, zero intrinsic error,
unit error correction efficiency, zero excess noise (for CVs), and large
modulation (for CVs). In the figure, we see that ideal implementations of CV
protocols (purple lines) are not so far from the ultimate limit. In
particular, we consider: (i) One-way no-switching protocol~\cite{Wee},
coinciding with CV-MDI-QKD~\cite{CVMDIQKD,Carlo} in the most asymmetric
configuration (relay approaching Alice~\cite{Gae}). For high loss ($\eta
\simeq0$), the rate scales as $\eta/\ln4$, which is just $1/2$ of the
capacity. Same scaling for the one-way switching protocol of ref.~\cite{Fred};
(ii) Two-way protocol with coherent states and homodyne
detection~\cite{Twoway,Carlo2main} which scales as $\simeq\eta/(4\ln2)$ for
high loss (thermal noise is needed for two-way to beat one-way
QKD~\cite{Twoway}). For the DV protocols (dashed lines), we consider: BB84
with single-photon sources~\cite{BB84} with rate $\eta/2$; BB84 with weak
coherent pulses and decoy states~\cite{Scaranim} with rate $\eta/(2e)$; and
DV-MDI-QKD~\cite{MDI1,MDI2} with rate $\eta/(2e^{2})$. See Supplementary
Note~6\ for details on these ideal rates.}%
\label{ProtocolsPIC}%
\end{figure}

Expanding Eq.~(\ref{formCloss}) at high loss $\eta\simeq0$, we find
\begin{equation}
\mathcal{C}(\eta)\simeq\eta/\ln2\simeq1.44\eta~\text{(bits per channel use),}
\label{scalingFINAL}%
\end{equation}
or about $\eta$ nats per channel use. This completely characterizes the
fundamental rate-loss scaling which rules long-distance quantum optical
communications in the absence of quantum repeaters. It is important to remark
that our work also proves the achievability of this scaling. This is a major
advance with respect to existing literature, where previous studies with the
squashed entanglement~\cite{TGW} only identified a non-achievable upper
bound.
In Fig.~\ref{ProtocolsPIC}, we compare the scaling of Eq.~(\ref{scalingFINAL})
with the maximum rates achievable by current
QKD\ protocols.

The capacity in Eq.~(\ref{formCloss}) is also valid for two-way quantum
communication with lossy channels, assuming that $\eta$ is the maximum
transmissivity between the forward and feedback channels. It can also be
extended to a multiband lossy channel, for which we write $\mathcal{C}%
=-\sum_{i}\log_{2}(1-\eta_{i})$, where $\eta_{i}$ are the transmissivities of
the various bands or frequency components. For instance, for a multimode
telecom fibre with constant transmissivity $\eta$ and bandwidth $W$, we have%
\begin{equation}
\mathcal{C}=-W\log_{2}(1-\eta).
\end{equation}
Finally, note that free-space satellite communications may be modeled as a
fading lossy channel, i.e., an ensemble of lossy channels $\mathcal{E}%
_{\eta_{i}}$ with associated probabilities $p_{i}$~\cite{Satellite}. In
particular, slow fading can be associated with variations of satellite-Earth
radial distance~\cite{sat1,sat2}. For a fading lossy channel $\{\mathcal{E}%
_{\eta_{i}},p_{i}\}$, we may write
\begin{equation}
\mathcal{C}\leq-\sum_{i}p_{i}\log_{2}(1-\eta_{i})~. \label{bfading}%
\end{equation}

\bigskip

\textbf{Quantum communications with Gaussian noise.} The fundamental limit of
the lossy channel bounds the two-way capacities of all channels decomposable
as $\mathcal{E}=\mathcal{E}^{\prime\prime}\circ\mathcal{E}_{\eta}%
\circ\mathcal{E}^{\prime}$ where $\mathcal{E}_{\eta}$ is a lossy component
while $\mathcal{E}^{\prime}$ and $\mathcal{E}^{\prime\prime}$ are extra
channels. A channel $\mathcal{E}$\ of this type is stretchable with resource
state $\sigma=\rho_{\mathcal{E}_{\eta}}\neq\rho_{\mathcal{E}}$ and we may
write $\mathcal{C}(\mathcal{E})\leq-\log_{2}(1-\eta)$. For Gaussian channels,
such decompositions are known but we achieve tighter bounds if we directly
stretch them using their own Choi matrix.

Let us start from the thermal-loss channel, which can be modeled as a
beamsplitter with transmissivity $\eta$ in a thermal background with $\bar{n}$
mean photons. Its action on input quadratures $\mathbf{\hat{x}}=(\hat{q}%
,\hat{p})$ is given by $\mathbf{\hat{x}}\rightarrow\sqrt{\eta}\mathbf{\hat{x}%
}+\sqrt{1-\eta}\mathbf{\hat{x}}_{E}$ with $E$ being a thermal mode. This
channel is central for microwave
communications~\cite{Usenko10,thermal-PRL,thermal-PRA,thermal-2-PRA} but also
important for CV QKD at optical and telecom frequencies, where Gaussian
eavesdropping via entangling cloners results into a thermal-loss
channel~\cite{RMP}.

\begin{figure*}[ptbh]
\vspace{-0.2cm}
\par
\begin{center}
\includegraphics[width=0.97\textwidth] {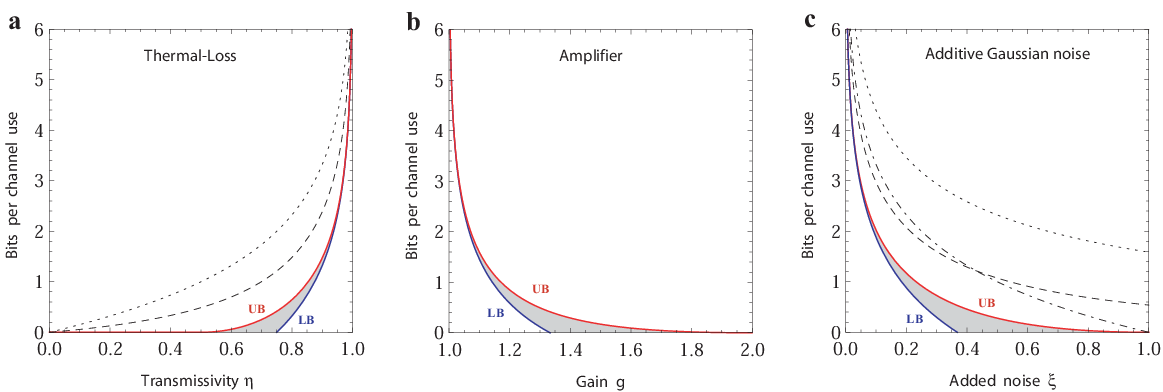}
\end{center}
\par
\vspace{-0.3cm}\caption{Two-way capacities for Gaussian channels in terms of
the relevant channel parameters. (\textbf{a})~Two-way capacity $\mathcal{C}%
(\eta,\bar{n})$ of the thermal-loss channel as a function of transmissivity
$\eta$ for $\bar{n}=1$ thermal photon. It is contained in the shadowed area
identified by the lower bound (LB) and upper bound (UB) of
Eq.~(\ref{CthermalLOSS}). Our upper bound is clearly tighter than those based
on the squashed entanglement, computed in ref.~\cite{TGW} (dotted) and
ref.~\cite{GEW} (dashed). Note that $\mathcal{C}(\eta,\bar{n})\simeq-\log
_{2}(1-\eta)-h(\bar{n})$ at high transmissivities. For $\bar{n}=0$ (lossy
channel) the shadowed region shrinks into a single line. (\textbf{b})~Two-way
capacity $\mathcal{C}(g,\bar{n})$ of the amplifier channel as a function of
the gain $g$ for $\bar{n}=1$ thermal photon. It is contained in the shadowed
specified by the bounds in Eq.~(\ref{CthermalAMP}). For small gains, we have
$\mathcal{C}(g,\bar{n})\simeq\log_{2}[g/(g-1)]-h(\bar{n})$. For $\bar{n}=0$
(quantum-limited amplifier) the shadowed region shrinks into a single line.
(\textbf{c})~Two-way capacity $\mathcal{C}(\xi)$ of the additive-noise
Gaussian channel with added noise $\xi$. It is contained in the shadowed
region specified by the bounds in Eq.~(\ref{CaddedCHANN}). For small noise, we
have $\mathcal{C}(\xi)\simeq-1/\ln2-\log_{2}\xi$. Our upper bound is much
tighter than those of ref.~\cite{TGW} (dotted), ref.~\cite{GEW} (dashed), and
ref.~\cite{HolevoWerner} (dot-dashed).}%
\label{FigTOTAL}%
\end{figure*}

For an arbitrary thermal-loss channel $\mathcal{E}_{\eta,\bar{n}}$\ we apply
our reduction method and compute the entanglement flux%
\begin{equation}
\Phi(\eta,\bar{n})\leq-\log_{2}\left[  (1-\eta)\eta^{\bar{n}}\right]
-h(\bar{n}), \label{LossUB}%
\end{equation}
for $\bar{n}<\eta/(1-\eta)$, while zero otherwise. Here we set
\begin{equation}
h(x):=(x+1)\log_{2}(x+1)-x\log_{2}x. \label{hEntropyMAIN}%
\end{equation}
Combining this result with the lower bound given by the reverse coherent
information~\cite{ReverseCAP}, we write the following inequalities for the
two-way capacity of this channel%
\begin{equation}
-\log_{2}(1-\eta)-h(\bar{n})\leq\mathcal{C}(\eta,\bar{n})\leq\Phi(\eta,\bar
{n}). \label{CthermalLOSS}%
\end{equation}
As shown in Fig.~\ref{FigTOTAL}a, the two bounds tend to coincide at
sufficiently high transmissivity. We clearly retrieve the previous result of
the lossy channel for $\bar{n}=0$.

Another important Gaussian channel is the quantum amplifier. This channel
$\mathcal{E}_{g,\bar{n}}$ is described by
\SP{$\hat{x}\to\sqrt{g}\,\hat{x}+\sqrt{g-1}\,Z\hat{x}_E,~Z:=\operatorname{diag}(1,-1)$,}
where $g>1$ is the gain
and $E$ is the thermal environment with $\bar{n}$ mean photons. We compute%
\begin{equation}
\Phi(g,\bar{n})\leq\log_{2}\left(  \dfrac{g^{\bar{n}+1}}{g-1}\right)
-h(\bar{n}),
\end{equation}
for $\bar{n}<(g-1)^{-1}$, while zero otherwise. Combining this result with the
coherent information~\cite{HolevoWerner}, we get
\begin{equation}
\log_{2}\left(  \frac{g}{g-1}\right)  -h(\bar{n})\leq\mathcal{C}(g,\bar
{n})\leq\Phi(g,\bar{n}), \label{CthermalAMP}%
\end{equation}
whose behavior is plotted in Fig.~\ref{FigTOTAL}b.

In the absence of thermal noise ($\bar{n}=0$), the previous channel describes
a quantum-limited amplifier $\mathcal{E}_{g}$, for which the bounds in
Eq.~(\ref{CthermalAMP}) coincide. This channel is therefore distillable and
its two-way capacities are%
\begin{equation}
\mathcal{C}(g)=D_{2}(g)=Q_{2}(g)=K(g)=-\log_{2}(1-g^{-1}). \label{Campli}%
\end{equation}
In particular, this proves that $Q_{2}(g)$ coincides with the unassisted
quantum capacity $Q(g)$~\cite{HolevoWerner,Wolf}. Note that a gain-$2$
amplifier can transmit at most $1$ qubit per use.

Finally, one of the simplest models of bosonic decoherence is the
additive-noise Gaussian channel~\cite{RMP}. This is the direct extension of
the classical model of a Gaussian channel to the quantum regime. It can be
seen as the action of a random Gaussian displacement over incoming states. In
terms of input-output transformations, it is described by $\mathbf{\hat{x}%
}\rightarrow\mathbf{\hat{x}}+(z,z)^{T}$ where $z$ is a classical Gaussian
variable with zero mean and variance $\xi\geq0$. For this channel
$\mathcal{E}_{\xi}$ we find the entanglement flux%
\begin{equation}
\Phi(\xi)\leq\frac{\xi-1}{\ln2}-\log_{2}\xi,
\end{equation}
for $\xi<1$, while zero otherwise. Including the lower bound given by the
coherent information~\cite{HolevoWerner}, we get%
\begin{equation}
-\frac{1}{\ln2}-\log_{2}\xi\leq\mathcal{C}(\xi)\leq\Phi(\xi)~.
\label{CaddedCHANN}%
\end{equation}
In Fig.~\ref{FigTOTAL}c, see its behavior and how the two bounds tend to
rapidly coincide for small added noise.

\bigskip

\textbf{Ultimate limits in qubit communications. }We now study the ultimate
rates for quantum communication, entanglement distribution and secret key
generation through qubit channels, with generalizations to any finite
dimension. For any DV channel $\mathcal{E}$ from dimension $d_{A}$ to
dimension $d_{B}$, we may write the dimensionality bound $\mathcal{C}%
(\mathcal{E})\leq\min\{\log_{2}d_{A},\log_{2}d_{B}\}$. This is because we may
always decompose the channel into $\mathcal{I}\circ\mathcal{E}$ (or
$\mathcal{E}\circ\mathcal{I}$), include $\mathcal{E}$ in Alice's (or Bob's)
LOs and stretch the identity map into a Bell state with dimension $d_{B}$ (or
$d_{A}$). For DV channels, we may also write the following simplified version
of our Theorem~\ref{addTHEO} (see Methods for proof).

\bigskip

\begin{proposition}
\label{simplePROP5}For a Choi-stretchable\ channel $\mathcal{E}$ in finite
dimension, we may write the chain%
\begin{equation}
K(\mathcal{E})=K(\rho_{\mathcal{E}})\leq E_{\mathrm{R}}^{\infty}%
(\rho_{\mathcal{E}})\leq E_{\mathrm{R}}(\rho_{\mathcal{E}})=E_{\mathrm{R}%
}(\mathcal{E}), \label{th33}%
\end{equation}
where $K(\rho_{\mathcal{E}})$ is the distillable key of $\rho_{\mathcal{E}}$.
\end{proposition}

\bigskip

\noindent In the following we provide our results for DV channels, with
technical details available in Supplementary Note~5.

\bigskip

\textbf{Pauli channels.~}A general error model for the transmission of qubits
is the Pauli channel
\begin{equation}
\mathcal{P}(\rho)=p_{0}\rho+p_{1}X\rho X+p_{2}Y\rho Y+p_{3}Z\rho Z,
\end{equation}
where $X$, $Y$, and $Z$ are Pauli operators~\cite{NielsenChuangm} and
$\mathbf{p}:=\{p_{k}\}$ is a probability distribution. It is easy to check
that this channel is Choi-stretchable and its Choi matrix is Bell-diagonal. We
compute its entanglement flux as%
\begin{equation}
\Phi(\mathcal{P})=1-H_{2}(p_{\max}),
\end{equation}
if $p_{\max}:=\max\{p_{k}\}\geq1/2$, while zero otherwise. Since the channel
is unital, we have that $I_{\mathrm{C}}(\mathcal{P})=I_{\mathrm{RC}%
}(\mathcal{P})=1-H(\mathbf{p})$, where $H$ is the Shannon entropy. Thus, the
two-way capacity of a Pauli channel satisfies%
\begin{equation}
1-H(\mathbf{p})\leq\mathcal{C}(\mathcal{P})\leq\Phi(\mathcal{P}).
\label{PauligenMAIN}%
\end{equation}
This can be easily generalized to arbitrary finite dimension (see
Supplementary Note~5).

\begin{figure*}[ptbh]
\vspace{-0.0cm}
\par
\begin{center}
\includegraphics[width=0.68\textwidth] {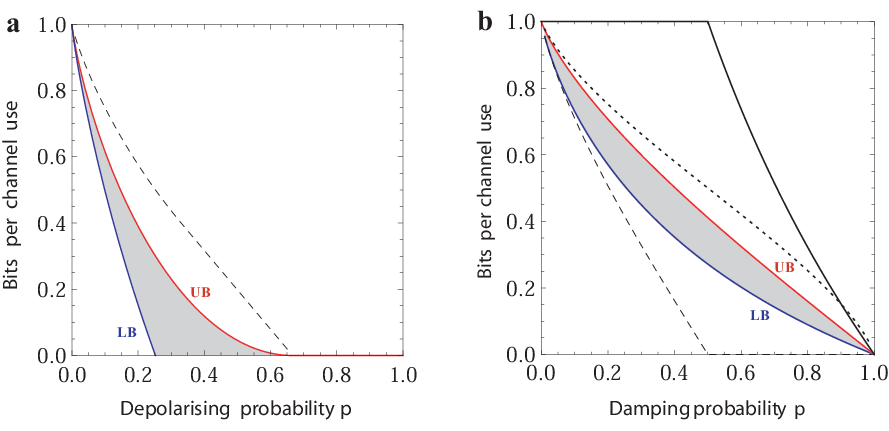} \vspace{0.3cm}
\end{center}
\par
\vspace{-0.5cm}\caption{Two-way capacities of basic qubit
channels.~(\textbf{a})~Two-way capacity of the depolarising channel
$\mathcal{P}_{\text{\textrm{depol}}}$ with arbitrary probability $p$. It is
contained in the shadowed region specified by the bounds in Eq.~(\ref{depmain}%
). We also depict the best known bound based on the squashed
entanglement~\cite{GEW} (dashed). (\textbf{b})~Two-way capacity of the
amplitude damping channel $\mathcal{E}_{\text{\textrm{damp}}}$ for arbitrary
damping probability $p$. It is contained in the shadowed area identified by
the lower bound (LB) of Eq.~(\ref{LBrevcoherAD}) and the upper bound (UB) of
Eq.~(\ref{UBmainSQUASH}). We also depict the bound of Eq.~(\ref{C2dampbb})
(upper solid line), which is good only at high dampings; and the bound
$C_{\mathrm{A}}(\mathcal{E}_{\text{\textrm{damp}}})/2$ of ref.~\cite{GEW}
(dotted line), which is computed from the entanglement-assisted classical
capacity $C_{\mathrm{A}}$. Finally, note the separation of the two-way
capacity $\mathcal{C}(\mathcal{E}_{\text{\textrm{damp}}})$\ from the
unassisted quantum capacity $Q(\mathcal{E}_{\text{\textrm{damp}}})$ (dashed
line).}%
\label{DVplots}%
\end{figure*}

Consider the depolarising channel, which is a Pauli channel shrinking the
Bloch sphere. With probability $p$, an input state becomes the maximally-mixed
state
\begin{equation}
\mathcal{P}_{\text{\textrm{depol}}}(\rho)=(1-p)\rho+pI/2. \label{depolQUBITs}%
\end{equation}
Setting $\kappa(p):=1-H_{2}\left(  3p/4\right)  $, we may then write%
\begin{equation}
\kappa(p)-\frac{3p}{4}\log_{2}3\leq\mathcal{C}(\mathcal{P}%
_{\text{\textrm{depol}}})\leq\kappa(p), \label{depmain}%
\end{equation}
for $p\leq2/3$, while $0$ otherwise (see Fig.~\ref{DVplots}a). The result can
be extended to any dimension $d\geq2$. A qudit depolarising channel is defined
as in Eq.~(\ref{depolQUBITs}) up to using the mixed state $I/d$. Setting
$f:=p(d^{2}-1)/d^{2}$ and $\kappa(d,p):=\log_{2}d-H_{2}(f)-f\log_{2}(d-1)$, we
find
\begin{equation}
\kappa(d,p)-f\log_{2}(d+1)\leq\mathcal{C}(\mathcal{P}_{\text{\textrm{depol}}%
})\leq\kappa(d,p), \label{depoldDIM}%
\end{equation}
for $p\leq d/(d+1)$, while zero otherwise.

Consider now the dephasing channel. This is a Pauli channel which deteriorates
quantum information without energy decay, as it occurs in spin-spin relaxation
or photonic scattering through waveguides. It is defined as
\begin{equation}
\mathcal{P}_{\text{\textrm{deph}}}(\rho)=(1-p)\rho+pZ\rho Z,
\end{equation}
where $p$ is the probability of a phase flip. We can easily check that the two
bounds of Eq.~(\ref{PauligenMAIN}) coincide, so that this channel is
distillable and its two-way capacities are
\begin{align}
\mathcal{C}(\mathcal{P}_{\text{\textrm{deph}}})  &  =D_{2}(\mathcal{P}%
_{\text{\textrm{deph}}})=Q_{2}(\mathcal{P}_{\text{\textrm{deph}}})\nonumber\\
&  =K(\mathcal{P}_{\text{\textrm{deph}}})=1-H_{2}(p).
\end{align}
Note that this also proves $Q_{2}(\mathcal{P}_{\text{\textrm{deph}}%
})=Q(\mathcal{P}_{\text{\textrm{deph}}})$, where the latter was derived in
ref.~\cite{degradable}.

For an arbitrary qudit with computational basis $\{|j\rangle\}$, the
generalized dephasing channel is defined as
\begin{equation}
\mathcal{P}_{\text{\textrm{deph}}}(\rho)=\sum_{i=0}^{d-1}P_{i}Z^{i}%
\rho(Z^{\dag})^{i},
\end{equation}
where $P_{i}$ is the probability of $i$ phase flips, with a single flip being
$Z\left\vert j\right\rangle =e^{ij2\pi/d}\left\vert j\right\rangle $. This
channel is distillable and its two-way capacities are functionals of
$\mathbf{P}=\{P_{i}\}$%
\begin{equation}
\mathcal{C}(\mathcal{P}_{\text{\textrm{deph}}})=\log_{2}d-H(\mathbf{P}).
\end{equation}

\bigskip

\textbf{Quantum erasure channel. }A simple decoherence model is the erasure
channel. This is described by%
\begin{equation}
\mathcal{E}_{\text{\textrm{erase}}}(\rho)=(1-p)\rho+p\left\vert e\right\rangle
\left\langle e\right\vert , \label{erasureDEF}%
\end{equation}
where $p$ is the probability of getting an orthogonal erasure state
$\left\vert e\right\rangle $. We already know that $Q_{2}(\mathcal{E}%
_{\text{\textrm{erase}}})=1-p$~\cite{ErasureChannelm}. Therefore we compute
the secret key capacity.

Following ref.~\cite{ErasureChannelm}, one shows that $D_{1}(\rho
_{\mathcal{E}_{\text{\textrm{erase}}}})\geq1-p$. In fact, suppose that Alice
sends halves of EPR states to Bob. A fraction $1-p$ will be perfectly
distributed. These good cases can be identified by Bob applying the
measurement $\{\left\vert e\right\rangle \left\langle e\right\vert
,I-\left\vert e\right\rangle \left\langle e\right\vert \}$ on each output
system, and communicating the results back to Alice in a single and final CC.
Therefore, they distill at least $1-p$ ebits per copy. It is then easy to
check that this channel is Choi-stretchable and we compute $\Phi
(\rho_{\mathcal{E}_{\text{\textrm{erase}}}})\leq1-p$. Thus, the erasure
channel is distillable and we may write
\begin{equation}
\mathcal{C}(\mathcal{E}_{\text{\textrm{erase}}})=K(\mathcal{E}%
_{\text{\textrm{erase}}})=1-p.
\end{equation}

In arbitrary dimension $d$, \ the generalized erasure channel is defined as in
Eq.~(\ref{erasureDEF}), where $\rho$ is now the state of a qudit and the
erasure state $\left\vert e\right\rangle $ lives in the extra $d+1$ dimension.
We can easily generalize the previous derivations to find that this channel is
distillable and
\begin{equation}
K(\mathcal{E}_{\text{\textrm{erase}}})=(1-p)\log_{2}d\text{.}
\label{erasureGEN}%
\end{equation}
Note that the latter result can also be obtained by computing the squashed
entanglement of the erasure channel, as shown by the independent derivation of
ref.~\cite{GEW}.

\bigskip

\textbf{Amplitude damping channel.~}Finally, an important model of decoherence
in spins or optical cavities is energy dissipation or amplitude
damping~\cite{Bose1,Bose2}. The action of this channel on a qubit is
\begin{equation}
\mathcal{E}_{\text{\textrm{damp}}}(\rho)=%
{\textstyle\sum\nolimits_{i=0,1}}
A_{i}\rho A_{i}^{\dagger},
\end{equation}
where $A_{0}:=\left\vert 0\right\rangle \left\langle 0\right\vert +\sqrt
{1-p}\left\vert 1\right\rangle \left\langle 1\right\vert $, $A_{1}:=\sqrt
{p}\left\vert 0\right\rangle \left\langle 1\right\vert $, and $p$ is the
damping probability. Note that $\mathcal{E}_{\text{\textrm{damp}}}$ is not
teleportation-covariant. However, it is decomposable as
\begin{equation}
\mathcal{E}_{\text{\textrm{damp}}}=\mathcal{E}_{\text{\textrm{CV}}%
\rightarrow\text{\textrm{DV}}}\circ\mathcal{E}_{\eta(p)}\circ\mathcal{E}%
_{\text{\textrm{DV}}\rightarrow\text{\textrm{CV}}},
\end{equation}
where $\mathcal{E}_{\text{\textrm{DV}}\rightarrow\text{\textrm{CV}}}$
teleports the original qubit into a single-rail bosonic
qubit~\cite{telereview}; then, $\mathcal{E}_{\eta(p)}$ is a lossy channel with
transmissivity $\eta(p):=1-p$; and $\mathcal{E}_{\text{\textrm{CV}}%
\rightarrow\text{\textrm{DV}}}$ teleports the single-rail qubit back to the
original qubit. Thus, $\mathcal{E}_{\text{\textrm{damp}}}$ is stretchable into
the asymptotic Choi matrix of the lossy channel $\mathcal{E}_{\eta(p)}$.\ This
shows that we need a dimension-independent theory even for stretching DV channels.

From Theorem~\ref{addTHEO} we get $\mathcal{C}(\mathcal{E}%
_{\text{\textrm{damp}}})\leq\Phi(\mathcal{E}_{\eta(p)})$, implying%
\begin{equation}
\mathcal{C}(\mathcal{E}_{\text{\textrm{damp}}})\leq\min\{1,-\log_{2}p\},
\label{C2dampbb}%
\end{equation}
while the reverse coherent information implies~\cite{RevCohINFO}%
\begin{equation}
\max_{u}\{H_{2}\left(  u\right)  -H_{2}\left(  up\right)  \}\leq
\mathcal{C}(\mathcal{E}_{\text{\textrm{damp}}}). \label{LBrevcoherAD}%
\end{equation}
The bound in Eq.~(\ref{C2dampbb}) is simple but only good for strong damping
($p>0.9$). A shown in Fig.~\ref{DVplots}b, we find a tighter bound using the
squashed entanglement, i.e.,
\begin{equation}
\SP{\mathcal{C}(\mathcal{E}_{\mathrm{damp}})
\leq \max_{0\leq\gamma\leq1}
\left\{
H_2\!\left[\gamma\left(1-\frac p2\right)\right]
-H_2\!\left(\frac{p\gamma}{2}\right)
\right\}.}\label{UBmainSQUASH}
\end{equation}

\section*{Discussion\label{Conclu}}

In this work we have established the ultimate rates for point-to-point quantum
communication, entanglement distribution and secret key generation at any
dimension, from qubits to bosonic systems. These limits provide the
fundamental benchmarks that only quantum repeaters may surpass. To achieve our
results we have designed a general reduction method for adaptive protocols,
based on teleportation stretching and the relative entropy of entanglement,
suitably extended to quantum channels. This method has allowed us to bound the
two-way capacities ($Q_{2}$, $D_{2}$ and $K$) with single-letter quantities,
establishing exact formulas for bosonic lossy channels, quantum-limited
amplifiers, dephasing and erasure channels, after about $20$ years since the
first studies~\cite{B2main,ErasureChannelm}.

In particular, we have characterized the fundamental rate-loss scaling which
affects any quantum optical communication, setting the ultimate achievable
rate for repeaterless QKD at $-\log_{2}(1-\eta)$ bits per channel use, i.e.,
about $1.44\eta$ bits per use at high loss. There are two remarkable aspects
to stress about this bound. First, it remains sufficiently tight even when we
consider input energy constraints (down to $\simeq1$ mean photon). 
\SP{Second, it can be reached by using 1-way CCs (see Supplementary Note~7 for a discussion of the classical communication cost).}

Our reduction method is very general and goes well beyond the
scope of this work. It has been already used to extend the results
to quantum repeaters. Ref.~\cite{networkPIRS} has showed how to
simplify the most general adaptive protocols of quantum and
private communication between two end-points of a repeater chain
and, more generally, of an arbitrary multi-hop quantum network,
where systems may be routed though single or multiple paths.
Depending on the type of routing, the end-to-end capacities are
determined by quantum versions of the widest path problem and the
max-flow min-cut theorem. More recently, teleportation stretching
has been also used to completely simplify adaptive protocols of
quantum parameter estimation and quantum channel
discrimination~\cite{LupoPIR}. See Supplementary Discussion for a
summary of our findings, other follow-up works and further
remarks.

\newpage

\section*{Methods\label{Tools}}

\textbf{Basics of bosonic systems and Gaussian states.} Consider $n$ bosonic
modes with quadrature operators $\hat{x}=(\hat{q}_{1},\dots,\hat{q}_{n}%
,\hat{p}_{1},\dots,\hat{p}_{n})^{T}$. The latter satisfy the canonical
commutation relations~\cite{arvind_real_1995}
\begin{equation}
\lbrack\hat{x},\hat{x}^{T}]=i\Omega,~~~\Omega:=%
\begin{pmatrix}
0 & 1\\
-1 & 0
\end{pmatrix}
\otimes I_{n}~,
\end{equation}
with $I_{n}$ being the $n\times n$\ identity matrix. An arbitrary multimode
Gaussian state $\rho(u,V)$, with mean value $u$\ and covariance matrix (CM)
$V$, can be written as~\cite{Banchim}%
\begin{equation}
\rho=\frac{\exp\left[  -\frac{1}{2}(\hat{x}-u)^{T}G(\hat{x}-u)\right]  }%
{\det\left(  V+i\Omega/2\right)  ^{1/2}}, \label{GformGAUSS}%
\end{equation}
where the Gibbs matrix $G$ is specified by
\begin{equation}
G=2i\Omega\,\coth^{-1}(2Vi\Omega). \label{gVFUN22}%
\end{equation}

Using symplectic transformations~\cite{RMP}, the CM $V$ can be decomposed into
the Williamson's form $\bigoplus\nolimits_{k=1}^{n}\nu_{k}I_{2}$ where the
generic symplectic eigenvalue $\nu_{k}$ satisfies the uncertainty principle
$\nu_{k}\geq1/2$. Similarly, we may write $\nu_{k}=\bar{n}_{k}+1/2$ where
$\bar{n}_{k}$ are thermal numbers, i.e., mean number of photons in each mode.
The von Neumann entropy of a Gaussian state can be easily computed as
\begin{equation}
S(\rho)=\sum_{k=1}^{n}h(\bar{n}_{k}),
\end{equation}
where $h(x)$ is given in Eq.~(\ref{hEntropyMAIN}).

A two-mode squeezed vacuum (TMSV) state $\Phi^{\mu}$ is a zero-mean pure
Gaussian state with CM%
\begin{equation}
V^{\mu}=\left(
\begin{array}
[c]{cc}%
\mu & c\\
c & \mu
\end{array}
\right)  \oplus\left(
\begin{array}
[c]{cc}%
\mu & -c\\
-c & \mu
\end{array}
\right)  ~,
\end{equation}
where $c:=\sqrt{\mu^{2}-1/4}$ and $\mu=\bar{n}+1/2$. Here $\bar{n}$ is the
mean photon number of the reduced thermal state associated with each mode $A$
and $B$. The Wigner function of a TMSV state $\Phi^{\mu}$ is the Gaussian
\begin{equation}
W[\Phi^{\mu}](x)=\pi^{-2}\exp\left[  -\frac{x^{T}(V^{\mu})^{-1}x}{2}\right]  ,
\label{WignerTMSV}%
\end{equation}
where $x:=(q_{A},q_{B},p_{A},p_{B})^{T}$. For large $\mu$, this function
assumes the delta-like expression~\cite{Samtele}%
\begin{equation}
W[\Phi^{\mu}](x)\rightarrow N~\delta(q_{A}-q_{B})~\delta(p_{A}+p_{B}),
\end{equation}
where $N$ is a normalization factor, function of the anti-squeezed quadratures
$q_{+}:=q_{A}+q_{B}$ and $p_{-}:=p_{A}-p_{B}$, such that $\int N(q_{+}%
,p_{-})~dq_{+}dp_{-}=1$. Thus, the infinite-energy limit of TMSV states
$\lim_{\mu}\Phi^{\mu}$ defines the asymptotic CV EPR\ state $\Phi$, realizing
the ideal EPR\ conditions $\hat{q}_{A}=\hat{q}_{B}$ for position and $\hat
{p}_{A}=-\hat{p}_{B}$ for momentum.

Finally, recall that single-mode Gaussian channels can be put in canonical
form~\cite{RMP}, so that their action on input quadratures $\hat{x}=(\hat
{q},\hat{p})^{T}$ is%
\begin{equation}
\hat{x}\rightarrow T\hat{x}+N\hat{x}_{E}+z~,
\end{equation}
where $T$ and $N$ are diagonal matrices, $E$ is an environmental mode with
$\bar{n}_{E}$ mean photons, and $z$ is a classical Gaussian variable, with
zero mean and CM\ $\xi I\geq0$.

\bigskip

\textbf{Relative entropy between Gaussian states.~}We now provide a simple
formula for the relative entropy between two arbitrary Gaussian states
$\rho_{1}(u_{1},V_{1})$ and $\rho_{2}(u_{2},V_{2})$ directly in terms of their
statistical moments. Because of this feature, our formula supersedes previous
expressions~\cite{Chenmain,Scheelm}. We have the following.

\begin{theorem}
\label{TheoRELATIVE} For two arbitrary multimode Gaussian states, $\rho
_{1}(u_{1},V_{1})$ and $\rho_{2}(u_{2},V_{2})$, the entropic functional
\begin{equation}
\Sigma:=-\mathrm{Tr}\left(  \rho_{1}\log_{2}\rho_{2}\right)
\label{Sigma_functional}%
\end{equation}
is given by%
\begin{equation}
\Sigma(V_{1},V_{2},\delta)=\frac{\ln\det\left(  V_{2}+\frac{i\Omega}%
{2}\right)  +\mathrm{Tr}(V_{1}G_{2})+\delta^{T}G_{2}\delta}{2\ln2},
\label{functional}%
\end{equation}
where $\delta:=u_{1}-u_{2}$ and $G:=g(V)$ as given in Eq.~(\ref{gVFUN22}). As
a consequence, the von Neumann entropy of a Gaussian state $\rho(u,V)$ is
equal to%
\begin{equation}
S(\rho):=-\mathrm{Tr}\left(  \rho\log_{2}\rho\right)  =\Sigma(V,V,0)~,
\label{entroTH}%
\end{equation}
and the relative entropy of two Gaussian states $\rho_{1}(u_{1},V_{1})$ and
$\rho_{2}(u_{2},V_{2})$ is given by%
\begin{align}
S(\rho_{1}||\rho_{2})  &  :=\mathrm{Tr}\left[  \rho_{1}(\log_{2}\rho_{1}%
-\log_{2}\rho_{2})\right] \nonumber\\
&  =-S(\rho_{1})-\mathrm{Tr}\left(  \rho_{1}\log_{2}\rho_{2}\right)
\nonumber\\
&  =-\Sigma(V_{1},V_{1},0)+\Sigma(V_{1},V_{2},\delta)~. \label{RelENTROPY}%
\end{align}

\end{theorem}

\textit{Proof}:~~The starting point is the use of the Gibbs-exponential form
for Gaussian states~\cite{Banchim} given in Eq.~(\ref{GformGAUSS}). Start with
zero-mean Gaussian states, which can be written as $\rho_{i}=Z_{i}^{-1}%
\exp[-\hat{x}^{T}G_{i}\hat{x}/2]$, where $G_{i}=g(V_{i})\ $is the Gibbs-matrix
and $Z_{i}=\det\left(  V_{i}+i\Omega/2\right)  ^{1/2}$ is the normalization
factor (with $i=1,2$). Then, replacing into the definition of $\Sigma$ given
in Eq.~(\ref{Sigma_functional}), we find
\begin{align}
(2\ln2)\Sigma &  =2\ln Z_{2}+\mathrm{Tr}\left(  \rho_{1}\hat{x}^{T}G_{2}%
\hat{x}\right) \nonumber\\
&  =\ln\det\left(  V_{2}+i\Omega/2\right) \nonumber\\
&  +\sum_{jk}\mathrm{Tr}\left(  \rho_{1}\hat{x}_{j}\hat{x}_{k}\right)
G_{2jk}~. \label{prLEO1}%
\end{align}
Using the commutator $\langle{[\hat{x}_{j},\hat{x}_{k}]}\rangle=i\Omega_{jk}$
and the anticommutator $\langle{\{\hat{x}_{j},\hat{x}_{k}\}}\rangle=2V_{jk}$,
we derive%
\begin{align}
\sum_{jk}\mathrm{Tr}\left(  \rho_{1}\hat{x}_{j}\hat{x}_{k}\right)  G_{2jk}  &
=\mathrm{Tr}\left[  \left(  V_{1}+\frac{i\Omega}{2}\right)  ^{T}G_{2}\right]
\label{GmeanV}\\
&  =\mathrm{Tr}\left(  V_{1}G_{2}\right)  ,\nonumber
\end{align}
where we also exploit the fact that $\mathrm{Tr}(\Omega G)=0$, because
$\Omega$ is antisymmetric and $G$ is symmetric (as $V$).

Let us now extend the formula to non-zero mean values (with difference
$\delta=u_{1}-u_{2}$). This means to perform the replacement $\hat
{x}\rightarrow\hat{x}-u_{2}$, so that%
\begin{align}
\mathrm{Tr}\left(  \rho_{1}\hat{x}_{j}\hat{x}_{k}\right)   &  \rightarrow
\mathrm{Tr}\left[  \rho_{1}(\hat{x}_{j}-u_{2j})(\hat{x}_{k}-u_{2k})\right]
\nonumber\\
&  =\mathrm{Tr}\left[  \rho_{1}(\hat{x}_{j}-u_{1j}+\delta_{j})(\hat{x}%
_{k}-u_{1k}+\delta_{k})\right] \nonumber\\
&  =\mathrm{Tr}\left[  \rho_{1}(\hat{x}_{j}-u_{1j})(\hat{x}_{k}-u_{1k}%
)\right]  +\delta_{j}\delta_{k}~.
\end{align}
By replacing this expression in Eq.~(\ref{GmeanV}), we get%
\begin{equation}
\sum_{jk}\mathrm{Tr}\left(  \rho_{1}\hat{x}_{j}\hat{x}_{k}\right)
G_{2jk}\rightarrow\mathrm{Tr}\left(  V_{1}G_{2}\right)  +\delta^{T}G_{2}%
\delta~. \label{prLEO2}%
\end{equation}
Thus, by combining Eqs.~(\ref{prLEO1}) and~(\ref{prLEO2}), we achieve
Eq.~(\ref{functional}). The other Eqs.~(\ref{entroTH}) and~(\ref{RelENTROPY})
are immediate consequences. This completes the proof of
Theorem~\ref{TheoRELATIVE}.

As discussed in ref.~\cite{Banchim}, the Gibbs-matrix $G$ becomes singular for
a pure state or, more generally, for a mixed state containing vacuum
contributions (i.e., with some of the symplectic eigenvalues equal to $1/2$).
In this case the Gibbs-exponential form must be used carefully by making a
suitable limit. Since $\Sigma$ is basis independent, we can perform the
calculations in the basis in which $V_{2}$, and therefore $G_{2}$, is
diagonal. In this basis%
\begin{equation}
\Sigma=\frac{1}{2}\sum_{k=1}^{n}\sum_{\pm}\alpha_{k}^{\pm}\log_{2}(v_{2k}%
\pm1/2)~, \label{e.singularsum}%
\end{equation}
where $\{v_{2k}\}$ is the symplectic spectrum of $V_{2}$, and
\SP{\begin{equation}
\alpha_k^\pm = 1\pm\left[
(V_1)_{k,k}+(V_1)_{k+n,k+n}
+\delta_k^2+\delta_{k+n}^2
\right],
\end{equation}
where $V_1$ and $\delta=u_1-u_2$ are expressed in the basis in which $V_2$ is diagonal.}
Now, if $v_{2k}=1/2$ for some $k$, then its contribution to the sum in
Eq.~\eqref{e.singularsum} is either zero or infinity.

\bigskip

\textbf{Basics of quantum teleportation.~}Ideal teleportation exploits an
ideal EPR state $\Phi_{AB}=\left\vert \Phi\right\rangle _{AB}\left\langle
\Phi\right\vert $ of systems $A$\ (for Alice) and $B$ (for Bob). In finite
dimension $d$, this is the maximally-entangled Bell state%
\begin{equation}
\left\vert \Phi\right\rangle _{AB}:=d^{-1/2}\sum_{i=0}^{d-1}\left\vert
i\right\rangle _{A}\left\vert i\right\rangle _{B}~. \label{BellstateMAIN}%
\end{equation}
In particular, it is the usual Bell state $(\left\vert 00\right\rangle
+\left\vert 11\right\rangle )/\sqrt{2}$ for a qubit. To teleport, we need to
apply a Bell detection $\mathcal{B}$ on the input system $a$ and the EPR
system $A$ (i.e., Alice's part of the EPR state). This detection corresponds
to projecting onto a basis of Bell states $\left\vert \Phi^{k}\right\rangle
_{aA}$ where the outcome $k$ takes $d^{2}$ values with probabilities
$p_{k}=d^{-2}$.

More precisely, the Bell detection is a positive-operator valued measure
(POVM) with operators
\begin{equation}
M_{k}=(U_{k}\otimes I)^{\dagger}\Phi_{aA}(U_{k}\otimes I)~,
\label{BellPOVMele}%
\end{equation}
where $\Phi_{aA}:=\left\vert \Phi\right\rangle _{aA}\left\langle
\Phi\right\vert $ is the Bell state as in Eq.~(\ref{BellstateMAIN}) and
$U_{k}$ is one of $d^{2}$ teleportation unitaries, corresponding to
generalized Pauli operators (described below). For any state $\rho$ of the
input system $a$, and outcome $k$ of the Bell detection, the other EPR system
$B$ (Bob's part)\ is projected onto $U_{k}\rho U_{k}^{\dagger}$. Once Alice
has communicated $k$ to Bob (feed-forward), he applies the correction unitary
$U_{k}^{-1}$ to retrieve the original state $\rho$ on its system $B$. Note
that this process also teleports all correlations that the input system $a$
may have with ancillary systems.

For CV systems ($d\rightarrow+\infty$), the ideal EPR\ source $\Phi_{AB}$\ can
be expressed as a TMSV state $\Phi^{\mu}$\ in the limit of infinite energy
$\mu\rightarrow+\infty$. The unitaries $U_{k}$ are phase-space displacements
$D(k)$ with complex amplitude $k$~\cite{telereview}. The CV\ Bell detection is
also energy-unbounded, corresponding to a projection onto the asymptotic EPR
state up to phase-space displacements $D(k)$. To deal with this, we need to
consider a finite-energy version of the measurement, defined as a
quasi-projection onto displaced versions of the TMSV state $\Phi^{\mu}$ with
finite parameter $\mu$. This defines a POVM $\mathcal{B}^{\mu}$ with
operators
\SP{\begin{equation}
M_{l,m}^{\mu}:=\pi^{-2}[D(-l)\otimes D(-m)]\Phi_{aA}^{\mu}[D(l)\otimes D(m)],~l,m \in \mathbb{C}.\label{BellFINITE}
\end{equation}
More precisely, $k=l-m^*$ is communicated, and the unused outcome is integrated out at fixed $k$, defining
$\widetilde{M}_k^\mu := \int d^2m\, M_{k+m^*,m}^\mu$. This normalized POVM yields a trace-preserving teleportation LOCC whose averaged channel converges strongly to the identity as \(\mu \to \infty\).}

Optically, this can be interpreted as applying a balanced beam-splitter
followed by two projections, one onto a position-squeezed state and the other
onto a momentum-squeezed state \SP{(both displaced and with finite squeezing)}. The ideal
CV\ Bell detection $\mathcal{B}$ is reproduced by taking the limit of
$\mu\rightarrow+\infty$ in Eq.~(\ref{BellFINITE}). Thus, CV\ teleportation
must always be interpreted \textit{a la} Braunstein and Kimble~\cite{Samtele},
so that we first consider finite resources $(\Phi^{\mu},\mathcal{B}^{\mu})$ to
compute the $\mu$-dependent output and then we take the limit of large $\mu$.

\bigskip

\textbf{Teleportation unitaries.~}Let us characterize the set of teleportation
unitaries $\mathbb{U}_{d}=\{U_{k}\}$ for a qudit of dimension $d$. First, let
us write $k$ as a multi-index $k=(a,b)$ with $a,b\in\mathbb{Z}_{d}%
:=\{0,\ldots,d-1\}$. The teleportation set is therefore composed of $d^{2}$
generalized Pauli operators $\mathbb{U}_{d}=\{U_{ab}\}$, where $U_{ab}%
:=X^{a}Z^{b}$. These are defined by introducing unitary (non-Hermitian)
operators%
\begin{equation}
X\left\vert j\right\rangle =\left\vert j\oplus1\right\rangle ~,~Z\left\vert
j\right\rangle =\omega^{j}\left\vert j\right\rangle ~, \label{Pauli_DEF}%
\end{equation}
where $\oplus$ is the modulo $d$ addition and
\begin{equation}
\omega:=\exp(i2\pi/d), \label{factorPAULIgen}%
\end{equation}
so that they satisfy the generalized commutation relation%
\begin{equation}
Z^{b}X^{a}=\omega^{ab}X^{a}Z^{b}. \label{genCOMM}%
\end{equation}

Note that any qudit unitary can be expanded in terms of these generalized
Pauli operators. We may construct the set of finite-dimensional displacement
operators $D(j,a,b):=\omega^{j}X^{a}Z^{b}$ with $j,a,b\in\mathbb{Z}_{d}$ which
form the finite-dimensional Weyl-Heisenberg group (or Pauli group). For
instance, for a qubit ($d=2$), we have $\mathbb{U}_{2}=\{I,X,XZ,Z\}$ and the
group $\pm1\times\left\{  I,X,XZ,Z\right\}  $. For a CV system ($d=+\infty$),
the teleportation set is composed of infinite displacement operators, i.e., we
have $\mathbb{U}_{\infty}=\{D(k)\}$, where $D(k)$ is a phase-space
displacement operator~\cite{RMP} with complex amplitude $k$. This set is the
infinite-dimensional Weyl-Heisenberg group.

It is important to note that, at any dimension (finite or infinite), the
teleportation unitaries satisfy
\begin{equation}
U_{k}U_{\ell}=e^{i\phi(k,l)}U_{f}, \label{groupUN0}%
\end{equation}
where $U_{f}$ is another teleportation unitary and $\phi(k,\ell)$ is a phase.
In fact, for finite $d$, let us write $k$ and $\ell$ as multi-indices, i.e.,
$k=(a,b)$ and $\ell=(s,t)$. From $U_{ab}=X^{a}Z^{b}=\sum_{n}\omega
^{nb}|{n\oplus a}\rangle\langle{n}|$, we see that $U_{ab}U_{st}=\omega
^{sb}U_{a{\oplus}s,b{\oplus}t}$. Then, for infinite $d$, we know that the
displacement operators satisfy $D(u)D(v)=\SP{e^{(uv^{\ast}-u^{\ast}v)/2}}D(u+v)$, for
any two complex amplitudes $u$ and $v$.

Now, let us represent a teleportation unitary as
\begin{equation}
\mathcal{U}_{g}(\rho):=U_{g}\rho U_{g}^{\dagger}. \label{groupUN1}%
\end{equation}
It is clear that we have $\mathcal{U}_{a,b}\circ\mathcal{U}_{s,t}%
=\mathcal{U}_{a\oplus s,b\oplus t}$ for DV systems, and $\mathcal{U}_{u}%
\circ\mathcal{U}_{v}=\mathcal{U}_{u+v}$ for CV systems. Therefore
$\mathcal{U}_{g}$ satisfies the group structure%
\begin{equation}
\mathcal{U}_{g}\circ\mathcal{U}_{h}=\mathcal{U}_{g\cdot h}~~(g,h\in G),
\label{groupUN2}%
\end{equation}
where $G$ is a product of two groups of addition modulo $d$ for DVs, while $G$
is the translation group for CVs. Thus, the (multi-)index of the teleportation
unitaries can be taken from the abelian group $G$.

\bigskip

\textbf{Teleportation-covariant channels.~}Let us a give a group
representation to the property of teleportation covariance\ specified by
Eq.~(\ref{tele-covariant}). Following Eq.~(\ref{groupUN1}), we may express an
arbitrary teleportation unitary as $\mathcal{U}_{g}(\rho):=U_{g}\rho
U_{g}^{\dagger}$ where $g\in G$. Calling $\mathcal{V}_{g}(\rho):=V_{g}\rho
V_{g}^{\dagger}$, we see that Eq.~(\ref{tele-covariant}) implies%
\begin{equation}
\mathcal{V}_{g}\circ\mathcal{V}_{h}\circ\mathcal{E}=\mathcal{E}\circ
\mathcal{U}_{g}\circ\mathcal{U}_{h}=\mathcal{E}\circ\mathcal{U}_{g\cdot
h}=\mathcal{V}_{g\cdot h}\circ\mathcal{E},
\end{equation}
so that $\mathcal{U}$ and $\mathcal{V}$ are generally-different unitary
representations of the same abelian group $G$. Thus, Eq.~(\ref{tele-covariant}%
) can also be written as%
\begin{equation}
\mathcal{V}_{h}^{\dagger}\circ\mathcal{E}\circ\mathcal{U}_{h}=\mathcal{E},
\label{stretchGGG}%
\end{equation}
for all $h\in G$, where $\mathcal{V}_{h}^{\dag}(\rho):=V_{h}^{\dagger}\rho
V_{h}=\mathcal{V}_{h^{-1}}(\rho)$.

The property of Eq.~(\ref{tele-covariant}) is certainly satisfied if the
channel is covariant with respect to the Weyl-Heisenberg group, describing the
teleportation unitaries in both finite- and infinite-dimensional Hilbert
spaces. This happens when the channel is dimension-preserving and we may set
$V_{k}=U_{k^{\prime}}$ for some $k^{\prime}$ in Eq.~(\ref{tele-covariant}).
Equivalently, this also means that $\mathcal{U}$ and $\mathcal{V}$ are exactly
the same unitary representation in Eq.~(\ref{stretchGGG}). We call
\textquotedblleft Weyl-covariant\textquotedblright\ these specific types of
tele-covariant channels.

In finite-dimension, a Weyl-covariant channel must necessarily be a Pauli
channel. In infinite-dimension, a Weyl-covariant channel \SP{is covariant under} 
displacements which is certainly a property of the bosonic Gaussian channels.
A simple channel which is tele-covariant but not Weyl-covariant is the erasure
channel. This is in fact dimension-altering (since it adds an orthogonal state
to the input Hilbert space) and the output correction unitaries to be used in
Eq.~(\ref{tele-covariant}) have the augmented form $V_{k}=U_{k}\oplus I$.
Hybrid channels, mapping DVs to CVs and viceversa, cannot be Weyl-covariant
but they may be tele-covariant. Finally, the amplitude damping channel is an
example of a channel which is not tele-covariant. See below for a schematic
classification.
\[%
\begin{tabular}
[c]{c|c|c}
& Tele-covariant & Weyl-covariant\\\hline
Pauli channels & Yes & Yes\\
Gaussian channels & Yes & Yes\\
Erasure channels & Yes & No\\
Hybrid channels & Yes/No & No\\
Amplitude damping & No & No
\end{tabular}
\ \
\]

Note that, for a quantum channel in finite dimension, we may easily re-write
Eq.~(\ref{tele-covariant}) in terms of an equivalent condition for the Choi
matrix. In fact, by evaluating the equality in Eq.~(\ref{tele-covariant})\ on
the EPR\ state $\Phi=\left\vert \Phi\right\rangle \left\langle \Phi\right\vert
$\ and using the property that $I\otimes U\left\vert \Phi\right\rangle
=U^{T}\otimes I\left\vert \Phi\right\rangle $, one finds
\begin{equation}
\rho_{\mathcal{E}}=(U_{k}^{\ast}\otimes V_{k})\rho_{\mathcal{E}}(U_{k}%
^{T}\otimes V_{k}^{\dagger}).
\end{equation}
Thus, a finite-dimensional $\mathcal{E}$ is tele-covariant if and only if, for
any teleportation unitary $U_{k}$, we may write%
\begin{equation}
\left[  \rho_{\mathcal{E}},U_{k}^{\ast}\otimes V_{k}\right]  =0~,
\label{ChoiLEO}%
\end{equation}
for another generally-different unitary $V_{k}$. There are finite-dimensional
channels satisfying conditions stronger than Eq.~(\ref{ChoiLEO}). For Pauli
channels, we may write $\left[  \rho_{\mathcal{E}},U_{k}^{\ast}\otimes
U_{k}\right]  =0$ for any $k$, i.e., the Choi matrix is invariant under
twirling operations restricted to the generators of the Pauli group
$\{U_{k}\}$. For depolarising channels, we may even write $\left[
\rho_{\mathcal{E}},U^{\ast}\otimes U\right]  =0$ for an arbitrary unitary $U$.
This means that the Choi matrix of a depolarising channel is an isotropic state.

\bigskip

\textbf{LOCC-averaging in teleportation stretching.} Consider an arbitrary
adaptive protocol described by some fundamental preparation of the local
registers $\rho_{\mathbf{a}}^{0}\otimes\rho_{\mathbf{b}}^{0}$ and a sequence
of adaptive LOCCs $\mathcal{L}:=\{\Lambda_{0},\ldots,\Lambda_{n}\}$. In
general, these LOs may involve measurements. Call $u_{i}$ the (vectorial)
outcome of Alice's and Bob's local measurements performed within the $i$th
adaptive LOCC, so that $\Lambda_{i}=\Lambda_{i}^{u_{i}}$. It is clear that
$\Lambda_{i}^{u_{i}}$ will be conditioned by measurements and outcomes of all
the previous LOCCs, so that a more precise notation will be $\Lambda
_{i|i-1,i-2\cdots}^{u_{i}}$ where the output\ $u_{i}$ is achieved with a
conditional probability $p(u_{i}|u_{i-1},u_{i-2}\ldots)$. After $n$
transmissions, we have a sequence of outcomes $\mathbf{u}=u_{0}\ldots u_{n}$
with joint probability%
\begin{equation}
p(\mathbf{u})=p(u_{0})p(u_{1}|u_{0})\ldots p(u_{n}|u_{n-1}\ldots),
\end{equation}
and a sequence of LOCCs
\begin{equation}
\mathcal{L}(\mathbf{u}):=\{\Lambda_{0}^{u_{0}},\Lambda_{1|0}^{u_{1}}%
,\ldots,\Lambda_{n|n-1\ldots}^{u_{n}}\}.
\end{equation}
The mean rate of the protocol is achieved by averaging the output state over
all possible outcomes $\mathbf{u}$, which is equivalent to considering the
output state generated by the trace-preserving LOCC-sequence $\mathcal{L}%
:=\sum_{\mathbf{u}}\mathcal{L}(\mathbf{u})$.

In fact, suppose that the (normalized) output state $\rho_{\mathbf{ab}}%
^{n}(\mathbf{u})$ generated by the conditional $\mathcal{L}(\mathbf{u})$ is
epsilon-close to a corresponding target state $\phi_{n}(\mathbf{u})$\ with
rate $R_{n}(\mathbf{u})$.\ This means that we have $D\left[  \rho
_{\mathbf{ab}}^{n}(\mathbf{u}),\phi_{n}(\mathbf{u})\right]  \leq\varepsilon$
in trace distance. The mean rate of the protocol $R_{n}=\left\langle
R_{n}(\mathbf{u})\right\rangle :=\sum_{\mathbf{u}}p(\mathbf{u})R_{n}%
(\mathbf{u})$ is associated with the average target state $\phi_{n}%
=\left\langle \phi_{n}(\mathbf{u})\right\rangle $. It is easy to show that
$\phi_{n}$ is approximated by the mean output state $\rho_{\mathbf{ab}}%
^{n}=\left\langle \rho_{\mathbf{ab}}^{n}(\mathbf{u})\right\rangle $ generated
by $\mathcal{L}$. In fact, by using the joint convexity of the trace
distance~\cite{NielsenChuangm}, we may write
\begin{equation}
D(\rho_{\mathbf{ab}}^{n},\phi_{n})\leq\sum_{\mathbf{u}}p(\mathbf{u})D\left[
\rho_{\mathbf{ab}}^{n}(\mathbf{u}),\phi_{n}(\mathbf{u})\right]  \leq
\varepsilon.
\end{equation}

Now we show that the LOCC-simulation of a channel $\mathcal{E}$ does not
change the average output state $\rho_{\mathbf{ab}}^{n}$ and this state can be
re-organized in a block form. The $i$th (normalized) conditional output
$\rho_{\mathbf{ab}}^{i}$ can be expressed in terms of the $i-1$th output
$\rho_{\mathbf{ab}}^{i-1}=\rho_{\mathbf{a}a_{i}\mathbf{b}}$ as follows%
\begin{equation}
\rho_{\mathbf{ab}}^{i}(u_{i}|u_{i-1}\ldots)=\frac{\Lambda_{i|i-1\cdots}%
^{u_{i}}\circ\mathcal{E}(\rho_{\mathbf{a}a_{i}\mathbf{b}})}{p(u_{i}%
|u_{i-1}\ldots)}, \label{repOO}%
\end{equation}
where $\mathcal{E}$ is meant as $\mathcal{I}_{\mathbf{a}}\otimes
\mathcal{E}_{a_{i}}\otimes\mathcal{I}_{\mathbf{b}}$ with $a_{i}$ being the
system transmitted. Thus, after $n$ transmissions, the conditional output
state is $\rho_{\mathbf{ab}}^{n}(\mathbf{u})=p(\mathbf{u})^{-1}\Lambda
_{\mathbf{u}}^{\mathcal{E}}(\rho_{\mathbf{a}}^{0}\otimes\rho_{\mathbf{b}}%
^{0})$, where%
\begin{equation}
\Lambda_{\mathbf{u}}^{\mathcal{E}}:=\Lambda_{n|n-1\ldots}^{u_{n}}%
\circ\mathcal{E}\circ\Lambda_{n-1|n-2\ldots}^{u_{n-1}}\circ\cdots\circ
\Lambda_{1|0}^{u_{1}}\circ\mathcal{E}\circ\Lambda_{0}^{u_{0}},
\end{equation}
and the average output state is given by%
\begin{equation}
\rho_{\mathbf{ab}}^{n}=\sum_{\mathbf{u}}p(\mathbf{u})\rho_{\mathbf{ab}}%
^{n}(\mathbf{u})=\bar{\Lambda}^{\mathcal{E}}(\rho_{\mathbf{a}}^{0}\otimes
\rho_{\mathbf{b}}^{0}),
\end{equation}
where $\bar{\Lambda}^{\mathcal{E}}:=\sum_{\mathbf{u}}\Lambda_{\mathbf{u}%
}^{\mathcal{E}}$.

For some LOCC $\mathcal{T}$ and resource state $\sigma$, let us write the
simulation
\begin{equation}
\mathcal{E}(\rho_{\mathbf{a}a_{i}\mathbf{b}})=\mathcal{T}(\rho_{\mathbf{a}%
a_{i}\mathbf{b}}\otimes\sigma)=\sum_{k}\mathcal{T}^{k}(\rho_{\mathbf{a}%
a_{i}\mathbf{b}}\otimes\sigma)~,
\end{equation}
where $\mathcal{T}^{k}(\rho):=(\mathbb{A}_{k}\otimes\mathbb{B}_{k}%
)\rho(\mathbb{A}_{k}\otimes\mathbb{B}_{k})^{\dagger}$ is Alice and Bob's
conditional LOCC with probability $p(k)$. For simplicity we omit other
technical labels that may describe independent local measurements or classical
channels, because they will also be averaged at the end of the procedure. \SP{Let
us introduce the vector $\mathbf{k}=k_{1}\ldots k_{n}$ where $k_{i}$
identifies a conditional LOCC $\mathcal{T}^{k_{i}}$ associated with the $i$th
transmission.}

By replacing the simulation in Eq.~(\ref{repOO}), we obtain%
\begin{equation}
\rho_{\mathbf{ab}}^{i}(u_{i}|u_{i-1}\ldots)=\frac{\Lambda_{i|i-1\cdots}%
^{u_{i}}\circ\mathcal{T}(\rho_{\mathbf{a}a_{i}\mathbf{b}}\otimes\sigma
)}{p(u_{i}|u_{i-1}\ldots)}.
\end{equation}
By iteration, the latter equation yields
\begin{equation}
\rho_{\mathbf{ab}}^{n}(\mathbf{u})=p(\mathbf{u})^{-1}\Lambda_{\mathbf{u}%
}^{\mathcal{T}}(\rho_{\mathbf{a}}^{0}\otimes\rho_{\mathbf{b}}^{0}\otimes
\sigma^{\otimes n}),
\end{equation}
where
\begin{align}
\Lambda_{\mathbf{u}}^{\mathcal{T}}  &  :=\Lambda_{n|n-1\cdots}^{u_{n}}%
\circ\mathcal{T}\circ\cdots\circ\mathcal{T}\circ\Lambda_{0}^{u_{0}}\nonumber\\
&  \SP{=\sum_{\mathbf{k}}\Lambda_{n|n-1\cdots}^{u_{n}}\circ
\mathcal{T}^{k_{n}}\circ\cdots\circ\mathcal{T}^{k_{1}}\circ\Lambda_{0}^{u_{0}
}.}
\end{align}
Therefore, the average output state of the original protocol may be
equivalently expressed in the form
\begin{equation}
\rho_{\mathbf{ab}}^{n}=\bar{\Lambda}^{\mathcal{T}}(\rho_{\mathbf{a}}%
^{0}\otimes\rho_{\mathbf{b}}^{0}\otimes\sigma^{\otimes n}),~~\bar{\Lambda
}^{\mathcal{T}}:=\sum_{\mathbf{u}}\Lambda_{\mathbf{u}}^{\mathcal{T}}~.
\end{equation}
Finally, we may include the preparation $\rho_{\mathbf{a}}^{0}\otimes
\rho_{\mathbf{b}}^{0}$ in the LOCC, so that we may write
\begin{equation}
\rho_{\mathbf{ab}}^{n}=\bar{\Lambda}(\sigma^{\otimes n})~.
\end{equation}

To extend this technical proof to CV systems, we perform the replacement
$\sum_{\mathbf{u}}\rightarrow\int d\mathbf{u}$ with the probabilities becoming
probability densities. Then, $\mathcal{T}$ and $\sigma$ may be both
asymptotic, i.e., defined as infinite-energy limits $\mathcal{T}:=\lim_{\mu
}\mathcal{T}^{\mu}$ and $\sigma:=\lim_{\mu}\sigma^{\mu}$ from corresponding
finite-versions $\mathcal{T}^{\mu}$ and $\sigma^{\mu}$. In this case, we
repeat the previous procedure for some $\mu$ and then we take the limit on the
output state $\rho_{\mathbf{ab}}^{n,\mu}$.

\bigskip

\textbf{Teleportation stretching with bosonic channels (more details on
Lemma~\ref{stretch_LEMMA}).} For a bosonic channel, the Choi matrix and the
ideal Bell detection are both energy-unbounded. Therefore, any Choi-based LOCC
simulation of these channels must necessarily be asymptotic. Here we discuss
in more detail how an asymptotic channel simulation $(\mathcal{T}%
,\sigma):=\lim_{\mu}(\mathcal{T}^{\mu},\sigma^{\mu})$ leads to an asymptotic
form of stretching as described in Lemma~\ref{stretch_LEMMA}. Any operation or
functional applied to $(\mathcal{T},\sigma)$ is implicitly meant to be applied
to the finite-energy simulation $(\mathcal{T}^{\mu},\sigma^{\mu})$, whose
output then undergoes the $\mu$-limit.

\begin{figure}[ptbh]
\vspace{-2.7cm}
\par
\begin{center} \vspace{2.1cm}
\includegraphics[width=0.40\textwidth]{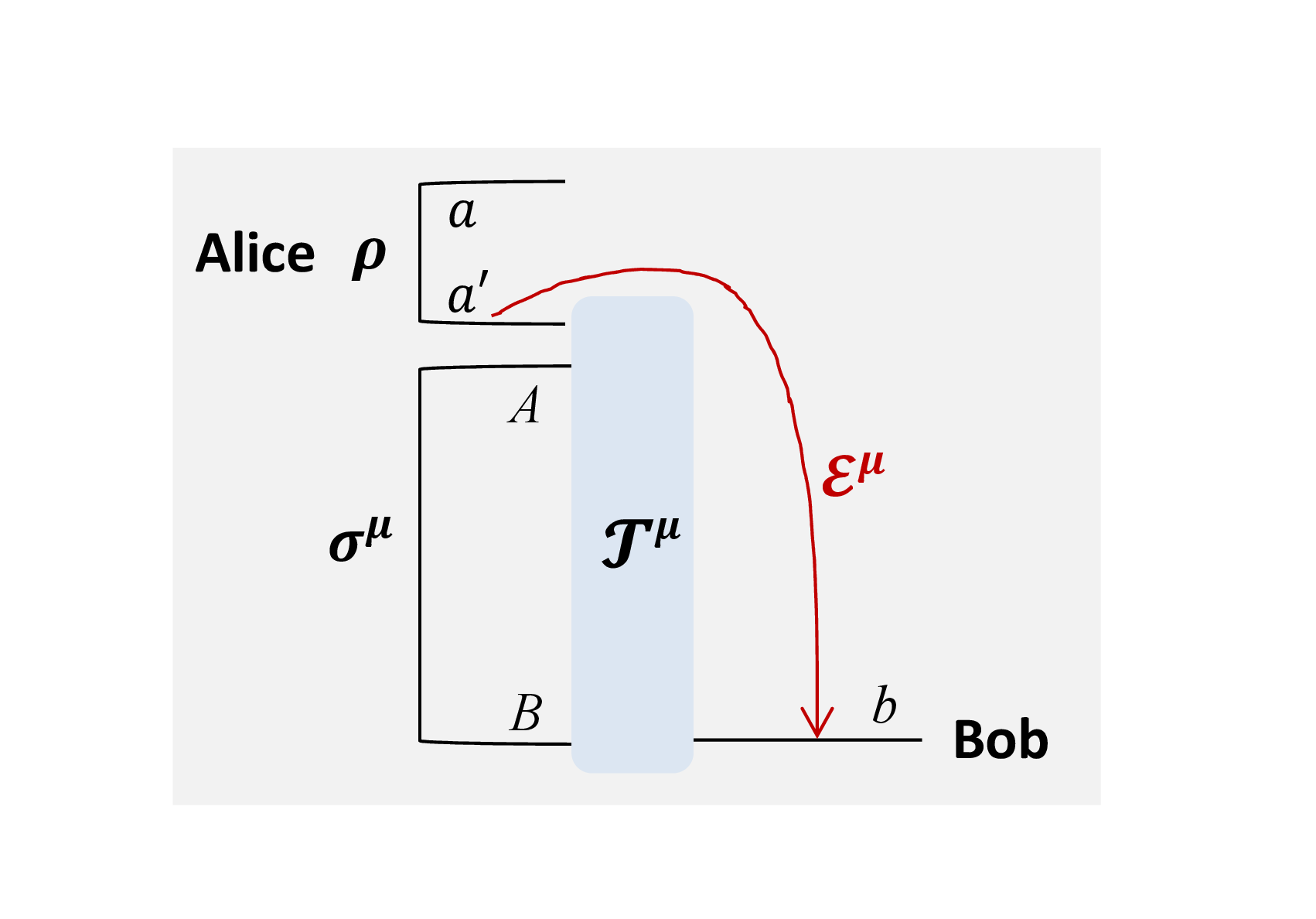}
\vspace{-0.4cm} \vspace{3.1cm} \vspace{-3.5cm}
\end{center}
\caption{Asymptotic LOCC\ simulation of bosonic channels. The finite-energy
LOCC\ simulation $(\mathcal{T}^{\mu},\sigma^{\mu})$ generates a teleportation
channel $\mathcal{E}^{\mu}$. Assume that $\mathcal{E}^{\mu}$ defines a target
bosonic channel $\mathcal{E}$ according to the pointwise limit in
Eq.~(\ref{pointwise}). Then, we say that the bosonic channel $\mathcal{E}$ has
asymptotic simulation $(\mathcal{T},\sigma):=\lim_{\mu}(\mathcal{T}^{\mu
},\sigma^{\mu})$. }%
\label{limitChannel}%
\end{figure}

Consider a bosonic channel $\mathcal{E}$ with asymptotic simulation
$(\mathcal{T},\sigma):=\lim_{\mu}(\mathcal{T}^{\mu},\sigma^{\mu})$. As
depicted in Fig.~\ref{limitChannel}, this means that there is a channel
$\mathcal{E}^{\mu}$ generated by $(\mathcal{T}^{\mu},\sigma^{\mu})$ such that
$\mathcal{E}:=\lim_{\mu}\mathcal{E}^{\mu}$ in the sense that
\begin{equation}
\Vert\mathcal{I}\otimes\mathcal{E}(\rho_{aa^{\prime}})-\mathcal{I}%
\otimes\mathcal{E}^{\mu}(\rho_{aa^{\prime}})\Vert\overset{\mu}{\rightarrow
}0~~\mathrm{for~any~}\rho_{aa^{\prime}}. \label{pointwise}%
\end{equation}
In other words, for any (energy-bounded) bipartite state $\rho_{aa^{\prime}}$,
whose $a^{\prime}$-part is propagated, the original channel output $\rho
_{ab}:=\mathcal{I}\otimes\mathcal{E}(\rho_{aa^{\prime}})$ and the simulated
channel output $\rho_{ab}^{\mu}:=\mathcal{I}\otimes\mathcal{E}^{\mu}%
(\rho_{aa^{\prime}})$ satisfy the limit
\begin{equation}
\Vert\rho_{ab}^{\mu}-\rho_{ab}\Vert\overset{\mu}{\rightarrow}0.
\label{limitMUMU}%
\end{equation}

By teleportation stretching, we may equivalently decompose the output state
$\rho_{ab}^{\mu}$ into the form%
\begin{equation}
\rho_{ab}^{\mu}=\bar{\Lambda}_{\mu}(\sigma^{\mu}), \label{lll1}%
\end{equation}
where $\bar{\Lambda}_{\mu}$ is a trace-preserving LOCC, which is includes both
$\mathcal{T}^{\mu}$ and the preparation of $\rho_{aa^{\prime}}$ (it is
trace-preserving because we implicitly assume that we average over all
possible measurements present in the simulation LOCC $\mathcal{T}^{\mu}$). By
taking the limit of $\mu\rightarrow+\infty$ in Eq.~(\ref{lll1}), the state
$\rho_{ab}^{\mu}$ becomes the channel output state $\rho_{ab}$ according to
Eq.~(\ref{limitMUMU}). Therefore, we have the limit
\begin{equation}
\Vert\rho_{ab}-\bar{\Lambda}_{\mu}(\sigma^{\mu})\Vert\overset{\mu}%
{\rightarrow}0,
\end{equation}
that we may compactly write as%
\begin{equation}
\rho_{ab}=\lim_{\mu}\bar{\Lambda}_{\mu}(\sigma^{\mu})~.
\end{equation}

Note that we may express Eq.~(\ref{pointwise}) in a different form. In fact,
consider the set of energy-constrained bipartite states $\mathcal{D}%
_{N}:=\{\rho_{aa^{\prime}}~|~\mathrm{Tr}(\hat{N}\rho_{aa^{\prime}})\leq N\}$,
where $\hat{N}$ is the total number operator. Then, for two bosonic channels,
$\mathcal{E}_{1}$ and $\mathcal{E}_{2}$, we may define the bounded diamond
norm%
\begin{equation}
\left\Vert \mathcal{E}_{1}-\mathcal{E}_{2}\right\Vert _{\diamond N}%
:=\sup_{\rho_{aa^{\prime}}\in\mathcal{D}_{N}}\Vert\mathcal{I}\otimes
\mathcal{E}_{1}(\rho_{aa^{\prime}})-\mathcal{I}\otimes\mathcal{E}_{2}%
(\rho_{aa^{\prime}})\Vert~. \label{defBBBB}%
\end{equation}
Using the latter definition and the fact that $\mathcal{D}_{N}$ is a compact
set, we have that the pointwise limit in Eq.~(\ref{pointwise}) implies the
following uniform limit
\begin{equation}
\left\Vert \mathcal{E}-\mathcal{E}^{\mu}\right\Vert _{\diamond N}\overset{\mu
}{\rightarrow}0\text{~~\textrm{for any }}N. \label{defBBNN}%
\end{equation}

The latter expression is useful to generalize the reasoning to the adaptive
protocol, with LOCCs applied before and after transmission. Consider the
output $\rho_{\mathbf{ab}}^{n}$ after $n$ adaptive uses of the channel
$\mathcal{E}$, and the simulated output $\rho_{\mathbf{ab}}^{n,\mu}$, which is
generated by replacing $\mathcal{E}$ with the imperfect channel $\mathcal{E}%
^{\mu}$. Explicitly, we may write
\begin{equation}
\rho_{\mathbf{ab}}^{n}=\Lambda_{n}\circ\mathcal{E}\circ\Lambda_{n-1}%
\cdots\circ\Lambda_{1}\circ\mathcal{E}(\rho_{\mathbf{ab}}^{0}),
\end{equation}
with its approximate version
\begin{equation}
\rho_{\mathbf{ab}}^{n,\mu}=\Lambda_{n}\circ\mathcal{E}^{\mu}\circ\Lambda
_{n-1}\cdots\circ\Lambda_{1}\circ\mathcal{E}^{\mu}(\rho_{\mathbf{ab}}^{0}),
\end{equation}
where it is understood that $\mathcal{E}$ and $\mathcal{E}^{\mu}$ are applied
to system $a_{i}$ in the $i$-th transmission, i.e., $\mathcal{E}%
=\mathcal{I}_{\mathbf{a}}\otimes\mathcal{E}_{a_{i}}\otimes\mathcal{I}%
_{\mathbf{b}}$.

Assume that the mean photon number of the total register states $\rho
_{\mathbf{ab}}^{n}$\ and $\rho_{\mathbf{ab}}^{n,\mu}$\ is bounded by some
large but yet finite value $N(n)$. For instance, we may consider a sequence
$N(n)=N(0)+nt$, where $N(0)$ is the initial photon contribution and $t$ is the
channel contribution, which may be negative for energy-decreasing channels
(like the thermal-loss channel) or positive for energy-increasing channels
(like the quantum amplifier). We then prove%
\begin{equation}
\left\Vert \rho_{\mathbf{ab}}^{n}-\rho_{\mathbf{ab}}^{n,\mu}\right\Vert
\leq\sum_{i=0}^{n-1}\left\Vert \mathcal{E}-\mathcal{E}^{\mu}\right\Vert
_{\diamond N(i)}~. \label{tossll}%
\end{equation}
In fact, for $n=2$, we may write
\begin{align}
&  \Vert\rho_{\mathbf{ab}}^{2}-\rho_{\mathbf{ab}}^{2,\mu}\Vert\nonumber\\
&  =\Vert\Lambda_{2}\circ\mathcal{E}\circ\Lambda_{1}\circ\mathcal{E}%
(\rho_{\mathbf{ab}}^{0})-\Lambda_{2}\circ\mathcal{E}^{\mu}\circ\Lambda
_{1}\circ\mathcal{E}^{\mu}(\rho_{\mathbf{ab}}^{0})\Vert\nonumber\\
&  \overset{(1)}{\leq}\Vert\mathcal{E}\circ\Lambda_{1}\circ\mathcal{E}%
(\rho_{\mathbf{ab}}^{0})-\mathcal{E}^{\mu}\circ\Lambda_{1}\circ\mathcal{E}%
^{\mu}(\rho_{\mathbf{ab}}^{0})\Vert\nonumber\\
&  \overset{(2)}{\leq}\Vert\mathcal{E}\circ\Lambda_{1}\circ\mathcal{E}%
(\rho_{\mathbf{ab}}^{0})-\mathcal{E}\circ\Lambda_{1}\circ\mathcal{E}^{\mu
}(\rho_{\mathbf{ab}}^{0})\Vert\nonumber\\
&  +\Vert\mathcal{E}\circ\Lambda_{1}\circ\mathcal{E}^{\mu}(\rho_{\mathbf{ab}%
}^{0})-\mathcal{E}^{\mu}\circ\Lambda_{1}\circ\mathcal{E}^{\mu}(\rho
_{\mathbf{ab}}^{0})\Vert\nonumber\\
&  \overset{(3)}{\leq}\Vert\mathcal{E}(\rho_{\mathbf{ab}}^{0})-\mathcal{E}%
^{\mu}(\rho_{\mathbf{ab}}^{0})\Vert\nonumber\\
&  +\Vert\mathcal{E}[\Lambda_{1}\circ\mathcal{E}^{\mu}(\rho_{\mathbf{ab}}%
^{0})]-\mathcal{E}^{\mu}[\Lambda_{1}\circ\mathcal{E}^{\mu}(\rho_{\mathbf{ab}%
}^{0})]\Vert\nonumber\\
&  \overset{(4)}{\leq}\Vert\mathcal{E}-\mathcal{E}^{\mu}\Vert_{\Diamond
N(0)}+\Vert\mathcal{E}-\mathcal{E}^{\mu}\Vert_{\Diamond N(1)}~, \label{casen2}%
\end{align}
where: (1)~we use monotonicity under $\Lambda_{2}$; (2)~we use the triangle
inequality; (3)~we use monotonicity with respect to $\mathcal{E}\circ
\Lambda_{1}$; and (4)~we use the definition of Eq.~(\ref{defBBBB}) assuming
$a^{\prime}=a_{i}$ and the energy bound $N(n)$. Generalization to arbitrary
$n$ is just a technicality.

By using Eq.~(\ref{defBBNN}) we may write that, for any bound $N(n)$ and
$\varepsilon\geq0$, there is a sufficiently large $\mu$ such that $\left\Vert
\mathcal{E}-\mathcal{E}^{\mu}\right\Vert _{\diamond N(n)}\leq\varepsilon$, so
that Eq.~(\ref{tossll}) becomes%
\begin{equation}
\left\Vert \rho_{\mathbf{ab}}^{n}-\rho_{\mathbf{ab}}^{n,\mu}\right\Vert \leq
n\varepsilon~. \label{fromg}%
\end{equation}
By applying teleportation stretching we derive $\rho_{\mathbf{ab}}^{n,\mu
}=\bar{\Lambda}_{\mu}(\sigma^{\mu\otimes n})$, where $\bar{\Lambda}_{\mu}$
includes the original LOCCs $\Lambda_{i}$ and the teleportation LOCCs
$\mathcal{T}^{\mu}$. Thus, Eq.~(\ref{fromg}) implies%
\begin{equation}
\left\Vert \rho_{\mathbf{ab}}^{n}-\bar{\Lambda}_{\mu}(\sigma^{\mu\otimes
n})\right\Vert \leq n\varepsilon,
\end{equation}
or, equivalently, $\left\Vert \rho_{\mathbf{ab}}^{n}-\bar{\Lambda}_{\mu
}(\sigma^{\mu\otimes n})\right\Vert \overset{\mu}{\rightarrow}0$.

Therefore, given an adaptive protocol with arbitrary register energy, and
performed $n$ times through a bosonic channel $\mathcal{E}$ with asymptotic
simulation, we may write its output state as the (trace-norm) limit%
\begin{equation}
\rho_{\mathbf{ab}}^{n}=\lim_{\mu}\bar{\Lambda}_{\mu}(\sigma^{\mu\otimes n}).
\label{expr222}%
\end{equation}
This means that we may formally write the asymptotic stretching $\bar{\Lambda
}(\sigma^{\otimes n}):=\lim_{\mu}\bar{\Lambda}_{\mu}(\sigma^{\mu\otimes n}%
)$\ for an asymptotic channel simulation $(\mathcal{T},\sigma):=\lim_{\mu
}(\mathcal{T}^{\mu},\sigma^{\mu})$.

\bigskip

\textbf{One-shot REE\ bound (Theorem~\ref{addTHEO}).} The main steps for
proving Eq.~(\ref{th11}) are already given in the main text. Here we provide
more details of the formalism for the specific case of bosonic channels,
involving asymptotic simulations $(\mathcal{T},\sigma):=\lim_{\mu}%
(\mathcal{T}^{\mu},\sigma^{\mu})$. Given the asymptotic stretching of the
output state $\rho_{\mathbf{ab}}^{n}$ as in Eq.~(\ref{expr222}), the
simplification of the REE bound $E_{\mathrm{R}}(\rho_{\mathbf{ab}}^{n})$
explicitly goes as follows
\SP{\begin{align}
E_{\mathrm R}(\rho_{\mathbf{ab}}^{n})
&=\inf_{\sigma_s}S(\rho_{\mathbf{ab}}^{n}\Vert\sigma_s) \nonumber\\
&\overset{(1)}{\leq}\inf_{\sigma_s^\mu}\liminf_{\mu\to+\infty}
S\!\left[\bar\Lambda_\mu(\sigma^{\mu\otimes n})
\,\middle\Vert\,\bar\Lambda_\mu(\sigma_s^\mu)\right] \nonumber\\
&\overset{(2)}{\leq}\inf_{\sigma_s^\mu}\liminf_{\mu\to+\infty}
S\!\left(\sigma^{\mu\otimes n}\middle\Vert\sigma_s^\mu\right) \nonumber \\
&\overset{(3)}{=}E_{\mathrm R}(\sigma^{\otimes n}),
\end{align}
where $\sigma_s^\mu$ ranges over all sequences of separable
states. Here: (1) we use joint
lower semicontinuity of relative entropy~\cite{Arakinew} and data
processing~\cite{Lindbladnew}; (2) we use monotonicity of relative entropy under trace-preserving
LOCCs; and (3) we use the asymptotic REE definition in Eq.~(4). For completeness, we clarify~(1). For fixed $\sigma_s^\mu$, let us write
$r_\mu=\bar\Lambda_\mu(\sigma^{\mu\otimes n})$ and
$t_\mu=\bar\Lambda_\mu(\sigma_s^\mu)$, so $r_\mu\to r:=\rho_{\mathbf{ab}}^n$
in trace norm. Along a subsequence realizing finite
$L=\liminf_\mu S(r_\mu\Vert t_\mu)$, extract a weak operator limit
$t\geq0$, $\operatorname{Tr} t\leq1$. For increasing finite-rank local projections
$P_m=P_m^A\otimes P_m^B\uparrow I$, the channels
$X\mapsto P_mXP_m\oplus\operatorname{Tr}[(I-P_m)X]$ have trace-norm convergent
outputs for each fixed $m$. Data processing and joint lower
semicontinuity yield
$p_m S(P_mrP_m/p_m\Vert P_mtP_m/q_m)\leq L$,
where $p_m=\operatorname{Tr}(P_mr)\to1$ and finiteness ensures
$q_m=\operatorname{Tr}(P_mt)>0$ for large $m$.
By finite-dimensional closedness, the normalized compressions of $t$
are separable; they converge in trace norm to $s=t/\operatorname{Tr} t$.
A final application of lower semicontinuity gives
$E_{\mathrm R}(r)\leq S(r\Vert s)\leq L$.
Taking the infimum over $\sigma_s^\mu$ provides (1).}

Thus, from Theorem~\ref{TheoMAIN}, we may write the following upper bound for
the two-way capacity of a bosonic channel%
\begin{equation}
\mathcal{C}(\mathcal{E})\leq E_{\mathrm{R}}^{\bigstar}(\mathcal{E})\leq
\lim_{n}n^{-1}E_{\mathrm{R}}(\sigma^{\otimes n})=E_{\mathrm{R}}^{\infty
}(\sigma). \label{ENTfluxBOSONIC}%
\end{equation}
The supremum over all adaptive protocols which defines $E_{\mathrm{R}%
}^{\bigstar}(\mathcal{E})$ disappears in the right hand side of
Eq.~(\ref{ENTfluxBOSONIC}). The resulting bound applies to both
energy-constrained protocols and the limit of energy-unconstrained protocols.
The proof of the further condition $E_{\mathrm{R}}^{\infty}(\sigma)\leq
E_{\mathrm{R}}(\sigma)$ in Eq.~(\ref{th11}) comes from the subadditivity of
the REE over tensor product states. This subadditivity also holds for a tensor
product of asymptotic states; it is proven by restricting the minimization on
tensor-product sequences $\sigma_{s}^{\mu\otimes n}$ in the corresponding
definition of the REE.

Let us now prove Eq.~(\ref{th22}). The two inequalities in Eq.~(\ref{th22})
are simply obtained by using $\sigma=\rho_{\mathcal{E}}$ for a
Choi-stretchable channel (where the Choi matrix is intended to be asymptotic
for a bosonic channel). Then we show the equality $E_{\mathrm{R}}%
(\rho_{\mathcal{E}})=E_{\mathrm{R}}(\mathcal{E})$. By restricting the
optimization in $E_{\mathrm{R}}(\mathcal{E})$ to an input EPR state $\Phi$, we
get the direct part $E_{\mathrm{R}}(\mathcal{E})\geq E_{\mathrm{R}}%
(\rho_{\mathcal{E}})$ as already noticed in Eq.~(\ref{channelREEmain}). For
CVs, this means to choose an asymptotic EPR state $\Phi:=\lim_{\mu}\Phi^{\mu}%
$, so that%
\begin{equation}
\mathcal{I}\otimes\mathcal{E}(\Phi):=\lim_{\mu}\mathcal{I}\otimes
\mathcal{E}(\Phi^{\mu})=\lim_{\mu}\rho_{\mathcal{E}}^{\mu}:=\rho_{\mathcal{E}%
},
\end{equation}
and therefore
\begin{equation}
E_{\mathrm{R}}(\mathcal{E})\geq E_{\mathrm{R}}(\rho_{\mathcal{E}}%
):=\inf_{\sigma_{s}^{\mu}}\underset{\mu\rightarrow+\infty}{\lim\inf}~S\left(
\rho_{\mathcal{E}}^{\mu}~||~\sigma_{s}^{\mu}\right)  .
\end{equation}

For the converse part, consider first DVs. By applying teleportation
stretching to a single use of the channel $\mathcal{E}$, we may write
$\mathcal{I}\otimes\mathcal{E}(\rho)=\bar{\Lambda}(\rho_{\mathcal{E}})$ for a
trace-preserving LOCC $\bar{\Lambda}$. Then, the monotonicity of the REE leads
to%
\begin{equation}
E_{\mathrm{R}}(\mathcal{E})=\sup_{\rho}E_{\mathrm{R}}[\mathcal{I}%
\otimes\mathcal{E}(\rho)]=\sup_{\rho}E_{\mathrm{R}}[\bar{\Lambda}%
(\rho_{\mathcal{E}})]\leq E_{\mathrm{R}}(\rho_{\mathcal{E}}).
\end{equation}
For CVs, we have an asymptotic stretching $\mathcal{I}\otimes\mathcal{E}
(\rho)=\lim_{\mu}\sigma^{\mu}$ where $\sigma^{\mu}:=\bar{\Lambda}_{\mu}%
(\rho_{\mathcal{E}}^{\mu})$. Therefore, we may write
\begin{align}
E_{\mathrm{R}}[\mathcal{I}\otimes\mathcal{E}(\rho)]  &  \SP{\le}\inf_{\sigma_{s}%
^{\mu}}\underset{\mu\rightarrow+\infty}{\lim\inf}S(\sigma^{\mu}||\sigma
_{s}^{\mu})\nonumber\\
&  \leq\inf_{\sigma_{s}^{\mu}}\underset{\mu\rightarrow+\infty}{\lim\inf}%
S[\bar{\Lambda}_{\mu}(\rho_{\mathcal{E}}^{\mu})||\bar{\Lambda}_{\mu}%
(\sigma_{s}^{\mu})]\nonumber\\
&  \leq\inf_{\sigma_{s}^{\mu}}\underset{\mu\rightarrow+\infty}{\lim\inf}%
S(\rho_{\mathcal{E}}^{\mu}||\sigma_{s}^{\mu})=E_{\mathrm{R}}(\rho
_{\mathcal{E}}).
\end{align}
Since this is true for any $\rho$, it also applies to the supremum and,
therefore, to the channel's REE $E_{\mathrm{R}}(\mathcal{E})$.

\bigskip

\textbf{One-shot REE bound for DV channels (Proposition~\ref{simplePROP5}).
}At finite dimension, we may first use teleportation stretching to derive
$K(\mathcal{E})\leq K(\rho_{\mathcal{E}})$ and then apply any upper bound to
the distillable key $K(\rho_{\mathcal{E}})$, among which the REE\ bound has
the best performance. Consider a key-generation protocol described by a
sequence $\mathcal{L}$ of adaptive LOCCs (implicitly assumed to be averaged).
If the protocol is implemented over a Choi-stretchable channel $\mathcal{E}$
in finite dimension $d$, its stretching allows us to write the output as
$\rho_{\mathbf{ab}}^{n}=\bar{\Lambda}\left(  \rho_{\mathcal{E}}^{\otimes
n}\right)  $ for a trace-preserving LOCC $\bar{\Lambda}$. Since any
LOCC-sequence $\mathcal{L}$ is transformed into $\bar{\Lambda}$, any
key-generation protocol through $\mathcal{E}$ becomes a key distillation
protocol over copies of the Choi matrix $\rho_{\mathcal{E}}$. For large $n$,
this means $K(\mathcal{E})\leq K(\rho_{\mathcal{E}})$.

To derive the opposite inequality, consider Alice sending EPR states through
the channel, so that the shared output will be $\rho_{\mathcal{E}}^{\otimes
n}$. There exists an optimal LOCC on these states which reaches the
distillable key $K(\rho_{\mathcal{E}})$ for large $n$. This is a specific
key-generation protocol over $\mathcal{E}$, so that we may write
$K(\rho_{\mathcal{E}})\leq K(\mathcal{E})$. Thus, for a $d$-dimensional
Choi-stretchable channel, we find%
\begin{equation}
K(\mathcal{E})=K(\rho_{\mathcal{E}})\leq E_{\mathrm{R}}^{\infty}%
(\rho_{\mathcal{E}}), \label{sssrrr}%
\end{equation}
where we also exploit the fact that the distillable key of a DV state is
bounded by its regularized REE~\cite{KD1}. It is also clear that
$E_{\mathrm{R}}^{\infty}(\rho_{\mathcal{E}})\leq E_{\mathrm{R}}(\rho
_{\mathcal{E}})=E_{\mathrm{R}}(\mathcal{E})$, where the latter equality is
demonstrated in the proof of Theorem~\ref{addTHEO}.

Note that $K(\mathcal{E})=K(\rho_{\mathcal{E}})$ cannot be directly written
for a bosonic channel, because its Choi matrix $\rho_{\mathcal{E}}$ is
energy-unbounded, so that its distillable key $K(\rho_{\mathcal{E}})$ is not
well-defined. By contrast, we know how to extend $E_{\mathrm{R}}^{\infty}%
(\rho_{\mathcal{E}})$ to bosonic channels and show $K(\mathcal{E})\leq
E_{\mathrm{R}}^{\infty}(\rho_{\mathcal{E}})$ at any dimension:\ This is the
more general procedure of Theorem~\ref{addTHEO} which first exploits the
general REE bound $K(\mathcal{E})\leq E_{\mathrm{R}}^{\bigstar}(\mathcal{E})$
and then simplifies $E_{\mathrm{R}}^{\bigstar}(\mathcal{E})\leq E_{\mathrm{R}%
}^{\infty}(\rho_{\mathcal{E}})$ by means of teleportation stretching at any dimension.

\bigskip

\textbf{Two-way quantum communication.~}Our method can be extended to more
complex forms of quantum communication. In fact, our weak converse theorem can
be applied to any scenario where two parties produce an output state by means
of an adaptive protocol. All the details of the protocol are contained in the
LOCCs $\mathcal{L}$ which are collapsed into $\bar{\Lambda}$ by teleportation
stretching and then discarded using the REE.

\begin{figure}[ptbh]
\vspace{-2.8cm}
\par
\begin{center}
\vspace{+0.9cm} \includegraphics[width=0.45\textwidth]{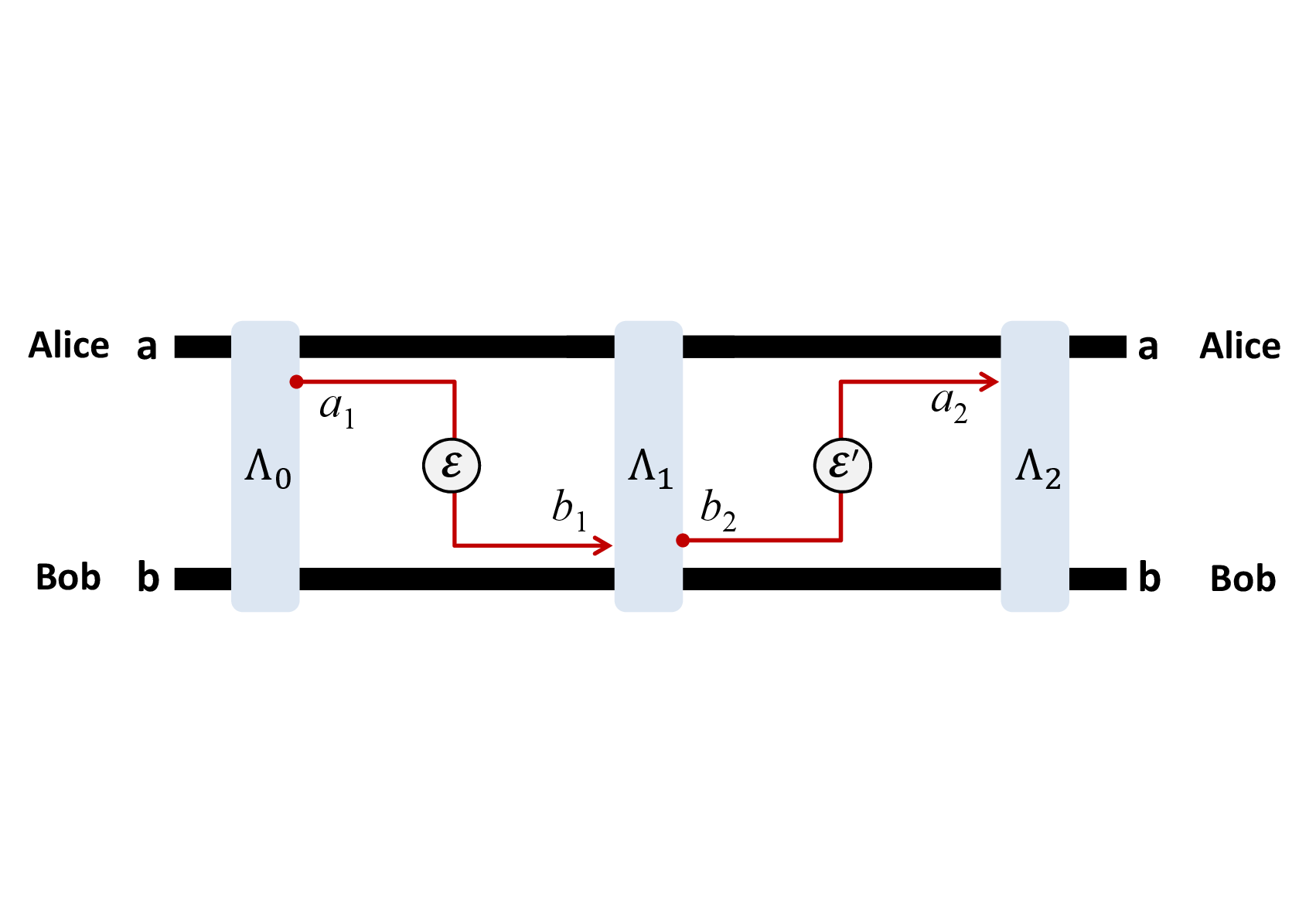}
\vspace{+1.3cm} \vspace{-3.5cm}
\end{center}
\caption{Adaptive protocol for two-way quantum or private communication. The
protocol employs a forward channel $\mathcal{E}$ and backward channel
$\mathcal{E}^{\prime}$. Transmissions are alternated with adaptive LOCCs
$\mathcal{L}=\{\Lambda_{0},\Lambda_{1},\Lambda_{2},\ldots\}$.}%
\label{feedbackCH}%
\end{figure}

Consider the scenario where Alice and Bob send systems to each other by
choosing between two possible channels, $\mathcal{E}$ (forward)\ or
$\mathcal{E}^{\prime}$ (backward), and performing adaptive LOCC after each
single transmission (see also Fig.~\ref{feedbackCH}). The capacity
$\mathcal{C}(\mathcal{E},\mathcal{E}^{\prime})$ is defined as the maximum
number of target bits distributed per individual transmission, by using one of
the two channels $\mathcal{E}$ and $\mathcal{E}^{\prime}$, and assuming LOs
assisted by unlimited two-way CCs.

In general, the feedback transmission may occur a fraction $p$ of the rounds,
with associated capacity
\begin{equation}
\mathcal{C}(p,\mathcal{E},\mathcal{E}^{\prime})\geq(1-p)\mathcal{C}%
(\mathcal{E})+p\mathcal{C}(\mathcal{E}^{\prime}). \label{bb1}%
\end{equation}
The lower bound is a convex combination of the individual capacities of the
two channels, which is achievable by using independent LOCC-sequences for the
two channels.

Assume that $(\mathcal{E},\mathcal{E}^{\prime})$ are stretchable into the pair
of resource states $(\sigma,\sigma^{\prime})$. Then, we can stretch the
protocol and decompose the output state as%
\begin{equation}
\rho_{\mathbf{ab}}^{n}=\bar{\Lambda}\left[  \sigma^{\otimes n(1-p)}%
\otimes\sigma^{\prime\otimes np}\right]  , \label{two-way}%
\end{equation}
where the tensor exponents $n(1-p)$ and $np$ are integers for suitably large
$n$ (it is implicitly understood that we consider suitable limits in the
bosonic case). Using the monotonicity of the REE under trace-preserving LOCCs
and its subadditivity over tensor products, we write%
\begin{align}
E_{\mathrm{R}}(\rho_{\mathbf{ab}}^{n})  &  \leq E_{\mathrm{R}}[\sigma^{\otimes
n(1-p)}\otimes\sigma^{\prime\otimes np}]\nonumber\\
&  \leq n(1-p)E_{\mathrm{R}}(\sigma)+npE_{\mathrm{R}}(\sigma^{\prime}).
\end{align}

As previously said, our weak converse theorem can be applied to any adaptive
protocol where two parties finally share a bipartite state $\rho_{\mathbf{ab}%
}^{n}$. Thus, we may write%
\begin{align}
\mathcal{C}(p,\mathcal{E},\mathcal{E}^{\prime})  &  \leq\sup_{\mathcal{L}%
}\underset{n\rightarrow+\infty}{\lim}\frac{E_{\mathrm{R}}(\rho_{\mathbf{ab}%
}^{n})}{n}\nonumber\\
&  \leq(1-p)E_{\mathrm{R}}(\sigma)+pE_{\mathrm{R}}(\sigma^{\prime})~.
\label{bb2}%
\end{align}
From Eqs.~(\ref{bb1}) and~(\ref{bb2}), we find that $\mathcal{C}%
(\mathcal{E},\mathcal{E}^{\prime})=\max_{p}\mathcal{C}(p,\mathcal{E}%
,\mathcal{E}^{\prime})$ must satisfy%
\begin{equation}
\max\{\mathcal{C}(\mathcal{E}),\mathcal{C}(\mathcal{E}^{\prime})\}\leq
\mathcal{C}(\mathcal{E},\mathcal{E}^{\prime})\leq\max\{E_{\mathrm{R}}%
(\sigma),E_{\mathrm{R}}(\sigma^{\prime})\}.
\end{equation}
For Choi-stretchable channels, this means%
\begin{equation}
\max\{\mathcal{C}(\mathcal{E}),\mathcal{C}(\mathcal{E}^{\prime})\}\leq
\mathcal{C}(\mathcal{E},\mathcal{E}^{\prime})\leq\max\{\Phi(\mathcal{E}%
),\Phi(\mathcal{E}^{\prime})\}.
\end{equation}
In particular, if the two channels are distillable, i.e., $\mathcal{C}%
(\mathcal{E})=\Phi(\mathcal{E})$ and $\mathcal{C}(\mathcal{E}^{\prime}%
)=\Phi(\mathcal{E}^{\prime})$, then we may write%
\begin{equation}
\mathcal{C}(\mathcal{E},\mathcal{E}^{\prime})=\max\{\mathcal{C}(\mathcal{E}%
),\mathcal{C}(\mathcal{E}^{\prime})\},
\end{equation}
and the optimal strategy (value of $p$) corresponds to using the channel with
maximum capacity.

Note that we may also consider a two-way quantum communication protocol where
the forward and backward transmissions occur simultaneously, and
correspondingly define a capacity that quantifies the maximum number of target
bits which are distributed in each double communication, forward \textit{and}
backward (instead of each single transmission, forward \textit{or} backward).
However, this case can be considered as a double-band quantum channel.

\bigskip

\textbf{Multiband quantum channel.~}Consider the communication scenario where
Alice and Bob can exploit a multiband quantum channel, i.e., a quantum channel
whose single use involves the simultaneous transmission of $m$ distinct
systems. In practice, this channel $\mathcal{E}_{\text{\textrm{mb}}}$ is
represented by a set of $m$ independent channels or bands $\{\mathcal{E}%
_{1},\ldots,\mathcal{E}_{m}\}$, i.e., it can be written as%
\begin{equation}
\mathcal{E}_{\text{\textrm{mb}}}=%
{\textstyle\bigotimes\nolimits_{i=1}^{m}}
\mathcal{E}_{i}~. \label{multibandGEN}%
\end{equation}
For instance, the bands may be bosonic Gaussian channels associated with
difference frequencies.

In this case, the adaptive protocol is modified in such a way that\ each
(multiband) transmission involves Alice simultaneously sending $m$ quantum
systems to Bob. These $m$ input systems may be in a generally-entangled state,
which may also involve correlations with the remaining systems in Alice's
register. Before and after each multiband transmission, the parties perform
adaptive LOCCs on their local registers $\mathbf{a}$ and $\mathbf{b}$. The
multiband protocol is therefore characterized by a LOCC\ sequence
$\mathcal{L}=\{\Lambda_{0},\ldots,\Lambda_{n}\}$ after $n$ transmissions.

The definition of the generic two-way capacity is immediately extended to a
multiband channel. This capacity quantifies the maximum number of target bits
that are distributed (in parallel) for each multiband transmission by means of
adaptive protocols. It must satisfy
\begin{equation}
\mathcal{C}(\mathcal{E}_{\text{\textrm{mb}}})\geq\sum_{i=1}^{m}\mathcal{C}%
(\mathcal{E}_{i})~, \label{broadb1}%
\end{equation}
where the lower bound is the sum of the two-way capacities of the single bands
$\mathcal{E}_{i}$. This lower bound is obtained by using adaptive LOCCs which
are independent between different $\mathcal{E}_{i}$, and considering an output
state of the form $\otimes_{i}\rho_{\mathbf{ab}}^{n,i}$ where $\rho
_{\mathbf{ab}}^{_{n,i}}$ is the output associated with $\mathcal{E}_{i}$.

Now consider an adaptive protocol performed over a multiband channel, whose
$m$ bands $\{\mathcal{E}_{i}\}$\ are stretchable into $m$ resources states
$\{\sigma_{i}\}$. By teleportation stretching, we find that Alice and Bob's
output state can be decomposed in the form%
\begin{equation}
\rho_{\mathbf{ab}}^{n}=\bar{\Lambda}\left(  \otimes_{i=1}^{m}~\sigma
_{i}^{\otimes n}\right)  .
\end{equation}
(It is understood that the formulation is asymptotic for bosonic channels.)
This previous decomposition leads to
\begin{equation}
E_{\mathrm{R}}(\rho_{\mathbf{ab}}^{n})\leq\sum_{i=1}^{m}E_{\mathrm{R}}%
(\sigma_{i}^{\otimes n}).
\end{equation}
Using our weak converse theorem, we can then write%
\begin{align}
\mathcal{C}(\mathcal{E}_{\text{\textrm{mb}}})  &  \leq\sup_{\mathcal{L}%
}\underset{n\rightarrow+\infty}{\lim}\frac{E_{\mathrm{R}}(\rho_{\mathbf{ab}%
}^{n})}{n}\nonumber\\
&  \leq\sum_{i=1}^{m}E_{\mathrm{R}}^{\infty}(\sigma_{i})\leq\sum_{i=1}%
^{m}E_{\mathrm{R}}(\sigma_{i}). \label{broadb2}%
\end{align}

Combining Eqs.~(\ref{broadb1}) and~(\ref{broadb2}) we may then write%
\begin{equation}
\sum_{i=1}^{m}\mathcal{C}(\mathcal{E}_{i})\leq\mathcal{C}(\mathcal{E}%
_{\text{\textrm{mb}}})\leq\sum_{i=1}^{m}E_{\mathrm{R}}(\sigma_{i}).
\end{equation}
For Choi-stretchable bands, this means
\begin{equation}
\sum_{i=1}^{m}\mathcal{C}(\mathcal{E}_{i})\leq\mathcal{C}(\mathcal{E}%
_{\text{\textrm{mb}}})\leq\sum_{i=1}^{m}\Phi(\mathcal{E}_{i}).
\end{equation}
Finally, if the bands are distillable, i.e., $\mathcal{C}(\mathcal{E}%
_{i})=\Phi(\mathcal{E}_{i})$, then we find the additive result%
\begin{equation}
\mathcal{C}(\mathcal{E}_{\text{\textrm{mb}}})=\sum_{i=1}^{m}\mathcal{C}%
(\mathcal{E}_{i})~.
\end{equation}


\bigskip

\textbf{Acknowledgments}.~We acknowledge support from the EPSRC\ via the `UK
Quantum Communications Hub' (EP/M013472/1) and `qDATA' (EP/L011298/1), and
from the ERC (Starting Grant 308253 PACOMANEDIA). In alphabetic order we would
like to thank G. Adesso, S. Bose, S. L. Braunstein, M. Christandl, T. P. W.
Cope, R. Demkowicz-Dobrzanski, K. Goodenough, F. Grosshans, S. Guha,
M.\ Horodecki, L. Jiang, J. Kolodynski, S. Lloyd, C. Lupo, L. Maccone, S.
Mancini, A. M\"{u}ller-Hermes, B. Schumacher, G. Spedalieri, G. Vallone, M.
Westmoreland, M. M. Wilde and A. Winter.

\newpage

\newcounter{S} \setcounter{section}{0} \setcounter{subsection}{0}
\setcounter{figure}{0}\setcounter{equation}{0}
\renewcommand{\thefigure}{\arabic{figure}} \renewcommand{\figurename}{Supplementary Figure}

\renewcommand{\theequation}{S\arabic{equation}}
\renewcommand{\bibnumfmt}[1]{[S#1]} {} \renewcommand{\citenumfont}[1]{S#1}
\renewcommand{\thesection}{Supplementary Note \arabic{section}} \renewcommand\refname{Supplementary References}

\onecolumngrid
\bigskip

\bigskip

\begin{center}
{\huge Supplementary Notes}
\end{center}

\section{Preliminary technical tools\label{app1}}

\subsection*{Truncation of infinite-dimensional Hilbert spaces}

In the following it will be useful to use truncation tools which enables us to
connect continuous-variable (CV)\ and discrete-variable (DV) states. Consider
$m$ bosonic modes with Hilbert space ${\mathcal{H}}^{\otimes m}$ and space of
density operators $\mathcal{D}({\mathcal{H}}^{\otimes m})$. Then, consider the
energy operator $\hat{H}=\sum_{i=1}^{m}\hat{N}_{i}$ (with $\hat{N}_{i}$ being
the number operator of mode $i$)\ and the following compact set of
energy-constrained\ states~\cite{Hol03a}
\begin{equation}
\mathcal{D}_{E}({\mathcal{H}}^{\otimes m}):=\{\rho\in\mathcal{D}({\mathcal{H}%
}^{\otimes m})~|~\mathrm{Tr}(\rho\hat{H})\leq E\}.
\end{equation}
It is easy to show that every such state is essentially supported on a
finite-dimensional Hilbert space.

\bigskip

\begin{lemma}
\label{lemmaDARIA}Consider an energy-constrained $m$-mode bosonic state
$\rho\in\mathcal{D}_{E}({\mathcal{H}}^{\otimes m})$. There exists a
finite-dimensional projector $P_{d}$ which projects this state onto a
$d$-dimensional support of the $m$-mode Hilbert space with probability
\begin{equation}
\mathrm{Tr}(\rho P_{d})\geq1-\gamma,~~\gamma:=\frac{E}{\sqrt[m]{d}-1}.
\label{newLL}%
\end{equation}
Correspondingly, the trace distance between the original state $\rho$ and the
$d$-dimensional truncated state%
\begin{equation}
\delta:=\frac{P_{d}\rho P_{d}}{\mathrm{Tr}(\rho P_{d})} \label{deltaABtr}%
\end{equation}
satisfies the inequality%
\begin{equation}
D(\rho,\delta):=\frac{1}{2}\left\Vert \rho-\delta\right\Vert \leq\sqrt{\gamma
}. \label{lemmaEQ}
\end{equation}

\end{lemma}

\bigskip

\textit{Proof}:~~Let us arrange the degenerate eigenvalues of $\hat{H}$ in
increasing order as $h_{0}\leq h_{1}\leq\ldots\leq h_{n}\leq\ldots$. Each
eigenvalue is computed as $\sum_{i=1}^{m}N_{i}$ where $N_{i}$ is the photon
number of mode $i$. The corresponding eigenstates are of the type $|\tilde
{h}_{n}\rangle=\left\vert N_{1}\right\rangle \otimes\ldots\otimes\left\vert
N_{m}\right\rangle $. For instance%
\begin{align}
|\tilde{h}_{0}\rangle &  =\left\vert 0\right\rangle \otimes\left\vert
0\right\rangle \otimes\ldots\otimes\left\vert 0\right\rangle ,~~~(h_{0}%
=0),\nonumber\\
|\tilde{h}_{1}\rangle &  =\left\vert 1\right\rangle \otimes\left\vert
0\right\rangle \otimes\ldots\otimes\left\vert 0\right\rangle ,~~~(h_{1}%
=1),\nonumber\\
|\tilde{h}_{2}\rangle &  =\left\vert 0\right\rangle \otimes\left\vert
1\right\rangle \otimes\ldots\otimes\left\vert 0\right\rangle ,~~~(h_{2}%
=1),\nonumber\\
&  \vdots~~~~~~~~~~~~~~~~~~~~~~~~~~~~~~~~~~~~~~~\vdots
\end{align}
Note that $n+1$ counts the dimension of the truncated Hilbert space and we
have $h_{n}\leq n$, because of the degeneracy of the eigenvalues. Since
$h_{n}$ is the total number of photons in all $m$ modes, we have that each
mode can have at most dimension $(h_{n}+1)$, so that we may write the upper
bound $n\leq n+1\leq(h_{n}+1)^{m}$ or, equivalently,
\begin{equation}
h_{n}\geq\sqrt[m]{n}-1 \label{hnccoo}%
\end{equation}
Then, we proceed as in Refs.~\cite{Hol03a,Darianoa}. Denote by $P_{n}%
:=|\tilde{h}_{n}\rangle\langle\tilde{h}_{n}|$ the eigenprojector associated
with $|\tilde{h}_{n}\rangle$. For dimension $d$, we consider the truncation
projector
\begin{equation}
P_{d}:=\sum_{n=0}^{d-1}P_{n}~. \label{truncPROJ}%
\end{equation}
Therefore, for all $|\psi\rangle\in{\mathcal{H}}$, we may write
\begin{equation}
\langle\psi|h_{d}(I-P_{d})|\psi\rangle=\langle\psi|\left[  h_{d}\sum
_{n=d}^{\infty}P_{n}\right]  |\psi\rangle\leq\langle\psi|\left[  \sum
_{n=d}^{\infty}h_{n}P_{n}\right]  |\psi\rangle\leq\langle\psi|\hat{H}%
|\psi\rangle.
\end{equation}
This implies that, for all $\rho\in\mathcal{D}_{E}({\mathcal{H}})$, we have
\begin{equation}
\mathrm{Tr}\left[  \rho(I-P_{d})\right]  \leq\frac{1}{h_{d}}\mathrm{Tr}%
(\rho\hat{H})\leq\frac{E}{h_{d}}~. \label{eq:energy05}%
\end{equation}
According to Eq.~(\ref{hnccoo}), we may write $h_{d}\geq\sqrt[m]{d}-1$, so
that
\begin{equation}
\frac{E}{h_{d}}\leq\gamma:=\frac{E}{\sqrt[m]{d}-1},
\end{equation}
which proves Eq.~(\ref{newLL}). The proof of Eq.~(\ref{lemmaEQ}) is a simple
modification of the one given by ref.~\cite{Darianoa}.~$\square$

\bigskip

Note that we may derive a similar result in terms of a truncation channel,
i.e., by means of a completely positive trace-preserving (CPTP) map.

\bigskip

\begin{lemma}
\label{truncaLEMMA}Consider an energy-constrained $m$-mode bosonic state
$\rho\in\mathcal{D}_{E}({\mathcal{H}}^{\otimes m})$. There exists a truncation
channel $\mathbb{T}_{d}$ which maps the state $\rho$ into a truncated state
$\tilde{\rho}$ defined over a $d$-dimensional support of the $m$-mode Hilbert
space, such that
\begin{equation}
D(\rho,\tilde{\rho})\leq\sqrt{\gamma}+\gamma, \label{Dclose}%
\end{equation}
where $\gamma$ is defined in Eq.~(\ref{newLL}).
\end{lemma}

\bigskip

\textit{Proof:}~~For any multimode energy-constrained bosonic state $\rho
\in\mathcal{D}_{E}({\mathcal{H}}^{\otimes m})$, we may define the following
(non-local) truncation map%
\begin{equation}
\tilde{\rho}:=\mathbb{T}_{d}(\rho)=\sum_{i=0,1}\mathcal{E}_{i}\left(  \Pi
_{i}\rho\Pi_{i}^{\dagger}\right)  , \label{truncmapEXP}%
\end{equation}
where $\Pi_{0}:=P_{d}$ and $\Pi_{1}:=I-P_{d}$, while for any projected state
$\sigma$ we have either the identity channel $\mathcal{E}_{0}(\sigma)=\sigma$
or the collapsing map $\mathcal{E}_{1}(\sigma)=\rho_{0}$, where $\rho_{0}$ is
an arbitrary fixed state within the $d$-dimensional support. Setting
$p:=\mathrm{Tr}(\rho P_{d})$, we may write%
\begin{equation}
\tilde{\rho}=p\delta+(1-p)\rho_{0},
\end{equation}
where $\delta$ is defined in Eq.~(\ref{deltaABtr}). Note that $S_{0}%
:=\left\Vert \rho-\rho_{0}\right\Vert \leq2$. Then, by exploiting the
convexity of the trace norm, we may write%
\begin{equation}
D(\rho,\tilde{\rho})=\frac{1}{2}\left\Vert \rho-\tilde{\rho}\right\Vert
\leq\frac{p}{2}\left\Vert \rho-\delta\right\Vert +\frac{1-p}{2}S_{0}\leq
p\sqrt{\gamma}+1-p\leq\sqrt{\gamma}+\gamma,
\end{equation}
where we have also used $p\leq1$ and Lemma~\ref{lemmaDARIA}.~$\square$

\subsection*{Local CV-DV mappings}

It is easy to modify the previous truncation tools to make them bipartite and
local, i.e., based on LOs assisted by (generally two-way) CCs. Suppose that
Alice and Bob share a CV\ bipartite state $\rho_{\mathbf{ab}}$, where Alice's
local system $\mathbf{a}$ contains $m_{\mathbf{a}}$\ modes and Bob's local
system $\mathbf{b}$ contains $m_{\mathbf{b}}$\ modes. Then, we may analyze how
this state is transformed by a truncation channel which is based on LOCC. In
fact, we may state the following.

\bigskip

\begin{lemma}
\label{LOCCtruncationChannel}Consider an energy-constrained bosonic state
$\rho_{\mathbf{ab}}\in\mathcal{D}_{E}({\mathcal{H}}^{\otimes m_{\mathbf{a}}%
}\otimes{\mathcal{H}}^{\otimes m_{\mathbf{b}}})$ where Alice (Bob) has
$m_{\mathbf{a}}$ ($m_{\mathbf{b}}$)\ modes. \SP{Set $m:=m_a+m_b$.} There is an LOCC truncation
channel $\mathbb{T}_{d}^{\otimes}$ (local with respect to the bipartition
$m_{\mathbf{a}}+m_{\mathbf{b}}$)\ which maps $\rho_{\mathbf{ab}}$ into a
truncated state $\tilde{\rho}_{\mathbf{ab}}$ defined over a $d\times
d$-dimensional support and such that
\begin{equation}
D(\rho_{\mathbf{ab}},\tilde{\rho}_{\mathbf{ab}})\leq\sqrt{\gamma}%
+\gamma,~~\gamma:=\frac{E}{\SP{d^{1/m}}-1}.
\end{equation}
The implementation of such truncation channel needs two bits of CC between
Alice and Bob.
\end{lemma}

\bigskip

\textit{Proof:}~~Assuming the bipartition of modes $m=m_{A}+m_{B}$, let us
write the energy operator as $\hat{H}=\hat{H}_{A}+\hat{H}_{B}$, where
\begin{equation}
\hat{H}_{A}=\sum_{i=1}^{m_{A}}\hat{N}_{i},~~\hat{H}_{B}=\sum_{i=1+m_{A}%
}^{m_{A}+m_{B}}\hat{N}_{i},
\end{equation}
with $\hat{N}_{i}$ being the number operator of the $i$-th mode, with
eigenstates $\left\vert N_{i}\right\rangle $. Let us arrange the eigenvalues
$h_{n}$ of $\hat{H}$ in increasing order $h_{0}\leq h_{1}\leq\ldots$ The
corresponding eigenstates are of the type $|\tilde{h}_{n}\rangle=\left\vert
N_{1}\right\rangle \otimes\ldots\otimes\left\vert N_{m}\right\rangle $. Call
$h_{k}^{A}$ ($h_{l}^{B}$)\ the eigenvalues of $\hat{H}_{A}$ ($\hat{H}_{B}$).
Correspondingly, we have eigenstates of the type
\begin{align}
|\tilde{h}_{k}^{A}\rangle &  =\left\vert N_{1}\right\rangle \otimes
\ldots\otimes\left\vert N_{m_{A}}\right\rangle ,\\
|\tilde{h}_{l}^{B}\rangle &  =\left\vert N_{1+m_{A}}\right\rangle
\otimes\ldots\otimes\left\vert N_{m_{A+}m_{B}}\right\rangle .
\end{align}

It is clear that, given an arbitrary $|\tilde{h}_{n}\rangle$, we may always
decompose it as $|\tilde{h}_{n}\rangle=|\tilde{h}_{k}^{A}\rangle\otimes
|\tilde{h}_{l}^{B}\rangle$ for some pair of labels $k$ and $l$. For this
reason, any set of $d$ states $\{|\tilde{h}_{n}\rangle\}$ for the $m$ modes
can certainly be represented by a tensor product of $d\times d$ states
suitably chosen within the local sets $\{|\tilde{h}_{k}^{A}\rangle\}$ and
$\{|\tilde{h}_{l}^{B}\rangle\}$. As a consequence, the support of a
$d$-dimensional projector as in Eq.~(\ref{truncPROJ}) is always contained in
the support of a local $d\times d$ projector $P_{d}^{\otimes}=P_{d}^{A}\otimes
P_{d}^{B}$, where%
\begin{equation}
P_{d}^{A}:=\sum_{k=0}^{d-1}|\tilde{h}_{k}^{A}\rangle\langle\tilde{h}_{k}%
^{A}|,~~P_{d}^{B}:=\sum_{l=0}^{d-1}|\tilde{h}_{l}^{B}\rangle\langle\tilde
{h}_{l}^{B}|,
\end{equation}
for some suitable choice of $\{|\tilde{h}_{k}^{A}\rangle\}$ and $\{|\tilde
{h}_{l}^{B}\rangle\}$.

This implies that there always exists a local projector $P_{d}^{\otimes}$ for
which we may write
\begin{equation}
\mathrm{Tr}(\rho_{\mathbf{ab}}P_{d}^{\otimes})\geq\mathrm{Tr}(\rho
_{\mathbf{ab}}P_{d})\geq1-\gamma,~~\gamma=\frac{E}{\SP{d^{1/m}}-1},\label{localPRequation}
\end{equation}
where we have also used Lemma~\ref{lemmaDARIA}. Set $p:=\mathrm{Tr}%
(\rho_{\mathbf{ab}}P_{d})$ and $p^{\prime}:=\mathrm{Tr}(\rho_{\mathbf{ab}%
}P_{d}^{\otimes})$, so that we have truncated states
\begin{equation}
\delta_{\mathbf{ab}}=p^{-1}P_{d}\rho_{\mathbf{ab}}P_{d},~~\delta_{\mathbf{ab}%
}^{\prime}=p^{\prime-1}P_{d}^{\otimes}\rho_{\mathbf{ab}}P_{d}^{\otimes}.
\end{equation}
\SP{It is easy to check that
\begin{equation}
\|\rho_{ab}-\delta'_{ab}\|\leq 2\sqrt{\gamma},
\end{equation}
where we use Eq.~\eqref{localPRequation} and the projection bound underlying Eq.~\eqref{lemmaEQ}.}

In order\ to construct the LOCC\ truncation channel, let us consider the local
POVM $\Pi_{ij}:=\Pi_{i}^{\mathbf{a}}\otimes\Pi_{j}^{\mathbf{b}}$ where
\begin{equation}
\Pi_{0}^{\mathbf{a}(\mathbf{b})}=P_{d}^{\mathbf{a}(\mathbf{b})},~~\Pi
_{1}^{\mathbf{a}(\mathbf{b})}=I^{\mathbf{a}(\mathbf{b})}-P_{d}^{\mathbf{a}%
(\mathbf{b})}.
\end{equation}
The parties apply these projections and then they communicate their outcomes
to each other, employing one bit of classical information for each one-way CC.
If both parties project onto the local $d$-dimensional support then they apply
an identity channel; if one of them projects outside this local support, they
both apply a damping channel which maps any input into a fixed state within
the support (which can always be chosen as the vacuum state).

More precisely, we define the LOCC truncation channel%
\begin{equation}
\mathbb{T}_{d}^{\otimes}(\rho_{\mathbf{ab}})=\sum_{i,j=0,1}\mathcal{E}%
_{ij}\left(  \Pi_{ij}\rho_{\mathbf{ab}}\Pi_{ij}^{\dagger}\right)  ,
\end{equation}
where $\Pi_{ij}$ is the local POVM defined above and%
\begin{equation}
\mathcal{E}_{ij}=\left\{
\begin{array}
[c]{c}%
\mathcal{I}_{\mathbf{a}}\otimes\mathcal{I}_{\mathbf{b}}\text{~~~for }i=j=0\\
\\
\mathcal{E}_{\mathbf{a}}^{\ast}\otimes\mathcal{E}_{\mathbf{b}}^{\ast
}\text{~~~otherwise,~~~~}%
\end{array}
\right.
\end{equation}
where channel $\mathcal{E}_{\mathbf{a}(\mathbf{b})}^{\ast}$ provides an
$m_{\mathbf{a}}$- ($m_{\mathbf{b}}$-) mode vacuum state\ for any input.

It is clear that the result is a truncated state $\tilde{\rho}_{\mathbf{ab}%
}:=\mathbb{T}_{d}^{\otimes}(\rho_{\mathbf{ab}})$ where each set of modes
$\mathbf{a}$ and $\mathbf{b}$ is supported in a $d$-dimensional Hilbert space.
In particular, we have%
\begin{equation}
\tilde{\rho}_{\mathbf{ab}}=p^{\prime}\delta_{\mathbf{ab}}^{\prime
}+(1-p^{\prime})\left\vert 0\right\rangle _{\mathbf{ab}}\left\langle
0\right\vert ~.
\end{equation}
Using the convexity of the trace norm, we get%
\begin{equation}
D(\rho_{\mathbf{ab}},\tilde{\rho}_{\mathbf{ab}})=\frac{1}{2}\left\Vert
\rho_{\mathbf{ab}}-\tilde{\rho}_{\mathbf{ab}}\right\Vert \leq\frac{p^{\prime}%
}{2}\left\Vert \rho_{\mathbf{ab}}-\delta_{\mathbf{ab}}^{\prime}\right\Vert
+\frac{1-p^{\prime}}{2}\left\Vert \rho_{\mathbf{ab}}-\left\vert 0\right\rangle
_{\mathbf{ab}}\left\langle 0\right\vert \right\Vert \leq p^{\prime}%
\sqrt{\gamma}+1-p^{\prime}\leq\sqrt{\gamma}+\gamma,
\end{equation}
which concludes the proof.$~\square$

\bigskip

Finally, note that LOCC channels from DVs to CVs can be constructed by using
hybrid quantum teleportation~\cite{telereviewa}. For instance, a polarisation
qubit $\alpha\left\vert \uparrow\right\rangle _{a}+\beta\left\vert
\downarrow\right\rangle _{a}$ can be teleported onto a single-rail qubit,
which is the bosonic subspace spanned by the vacuum $\left\vert 0\right\rangle
_{b}$ and the single-photon state $\left\vert 1\right\rangle _{b}$. It is
sufficient to build a hyper-entangled Bell state $\left\vert \uparrow
\right\rangle _{a^{\prime}}\left\vert 1\right\rangle _{b}+\left\vert
\downarrow\right\rangle _{a^{\prime}}\left\vert 0\right\rangle _{b}$ and apply
a discrete variable Bell detection on qubits $a$ and $a^{\prime}$. This
teleports $a$\ onto the bosonic mode $b$, up to Pauli operators (suitably
re-written in terms of the ladder operators) that can be undone from the
output state. Such procedure can be readily extended to teleport qudits into
bosonic modes in a LOCC fashion.

\section{Lower bound at any dimension\label{app2}}

\subsection*{Coherent and reverse coherent information of a quantum channel}

Consider a quantum channel $\mathcal{E}$ applied to some input state $\rho
_{A}$ of system $A$. Let us introduce the purification $|\psi\rangle_{RA}$ of
$\rho_{A}$ by means of an auxiliary system $R$. We can therefore consider the
output $\rho_{RB}=\mathcal{I}\otimes\mathcal{E}(|\psi\rangle_{RA}\langle
\psi|)$. By definition, the coherent information is~\cite{QC1a,QC2a}%
\begin{equation}
I_{\mathrm{C}}(\mathcal{E},\rho_{A})=I(A\rangle B)_{\rho_{RB}}=S(\rho
_{B})-S(\rho_{RB})~,
\end{equation}
where $\rho_{B}:=\mathrm{Tr}_{R}(\rho_{RB})$ and $S(\rho):=-\mathrm{Tr}%
(\rho\log_{2}\rho)$ is the von Neumann entropy. Similarly, the reverse
coherent information is given by~\cite{RevCohINFOa,ReverseCAPa}%
\begin{equation}
I_{\mathrm{RC}}(\mathcal{E},\rho_{A})=I(A\langle B)_{\rho_{RB}}=S(\rho
_{R})-S(\rho_{RB})~,
\end{equation}
where $\rho_{R}:=\mathrm{Tr}_{B}(\rho_{RB})$.

When the input state $\rho_{A}$ is a maximally-mixed state, its purification
is a maximally-entangled state $\Phi_{RA}$, so that $\rho_{RB}$ is the Choi
matrix of the channel, i.e., $\rho_{\mathcal{E}}$. We then define the coherent
information of the channel as
\begin{equation}
I_{\mathrm{C}}(\mathcal{E})=I(A\rangle B)_{\rho_{\mathcal{E}}}~.
\end{equation}
Similarly, its reverse coherent information is
\begin{equation}
I_{\mathrm{RC}}(\mathcal{E})=I(A\langle B)_{\rho_{\mathcal{E}}}~.
\end{equation}

Note that for unital channels, i.e., channels preserving the identity
$\mathcal{E}(I)=I$, we have $I_{\mathrm{C}}(\mathcal{E})=I_{\mathrm{RC}%
}(\mathcal{E})$. This is just a consequence of the fact that, the reduced
states $\rho_{A}$ and $\rho_{R}$ of a maximally entangled state $\Phi_{RA}$ is
a maximally-mixed state $I/d$, where $d$ is the dimension of the Hilbert space
(including the limit for $d\rightarrow+\infty$). If the channel is unital,
also the reduced state $\rho_{B}=\mathcal{E}(\rho_{A})$ is maximally-mixed. As
a result, $S(\rho_{B})=S(\rho_{A})=S(\rho_{R})$ and we may write
$I_{\mathrm{C}}(\mathcal{E})=I_{\mathrm{RC}}(\mathcal{E}):=I_{\mathrm{(R)C}%
}(\mathcal{E})$.

In the specific case of discrete-variable systems ($d<+\infty$), we have
$S(\rho_{R})=\log_{2}d$ and therefore%
\begin{equation}
I_{\mathrm{(R)C}}(\mathcal{E})=\log_{2}d-S(\rho_{\mathcal{E}})~.
\label{unitalGEN}%
\end{equation}
In particular, for unital qubit channels ($d=2$), one has%
\begin{equation}
I_{\mathrm{(R)C}}(\mathcal{E})=1-S(\rho_{\mathcal{E}})~. \label{unitalRC}%
\end{equation}
The latter two formulas will be exploited to compute the coherent information
of discrete-variable channels.

The coherent information is an achievable rate for \textit{forward} one-way
entanglement distillation. Similarly, the reverse coherent information is an
achievable rate for \textit{backward} one-way entanglement distillation (i.e.,
assisted by a single and final CC from Bob to Alice). In fact, thanks to the
hashing inequality~\cite{DWrates2a}, we may write
\begin{equation}
\max\{I_{\mathrm{C}}(\mathcal{E}),I_{\mathrm{RC}}(\mathcal{E})\}=\max
\{I(A\rangle B)_{\rho_{\mathcal{E}}},I(A\langle B)_{\rho_{\mathcal{E}}}\}\leq
D_{1}(\rho_{\mathcal{E}}). \label{hashAPP}%
\end{equation}

\subsection*{Hashing inequality in infinite dimension}

The hashing inequality is known to be valid for finite-dimensional quantum
systems. It is easy to extend this inequality to energy-constrained bosonic
states by exploiting the continuity of the (reverse) coherent information in
the limit of infinite dimension. Consider the state $\rho_{AB}$ of two bosonic
modes, each mode having $\leq\bar{n}$ mean photons. Then, we may apply a
projector $P_{d}$ generating a $d$-dimensional truncated state $\delta_{AB}$
such that (see Lemma~\ref{lemmaDARIA})%
\begin{equation}
D(\rho_{AB},\delta_{AB})\leq\sqrt{\gamma},~~\gamma=\frac{2\bar{n}}{\sqrt{d}%
-1}.
\end{equation}

According to ref.~\cite[Lemma~17]{WinterCONa}, the trace-distance condition
$D(\rho,\delta)\leq\sqrt{\gamma}<1/6$\ implies that the coherent information
$I(A\rangle B)=-S(A|B)$ satisfies%
\begin{equation}
\left\vert I(A\rangle B)_{\rho}-I(A\rangle B)_{\delta}\right\vert \leq
16\sqrt{\gamma}\log_{2}\left[  \frac{2e(\bar{n}+1)}{1-\sqrt{\gamma}}\right]
+32H_{2}(3\sqrt{\gamma})~, \label{condEnt}%
\end{equation}
where $H_{2}$ the binary Shannon entropy%
\begin{equation}
H_{2}(p):=-p\log_{2}p-(1-p)\log_{2}(1-p). \label{binSHANNON}%
\end{equation}
For any $\bar{n}$, the limit $d\rightarrow+\infty$ implies that $\gamma
\rightarrow0$ and therefore%
\begin{equation}
\left\vert I(A\rangle B)_{\rho}-I(A\rangle B)_{\delta}\right\vert
\rightarrow0~.
\end{equation}
An equivalent result holds for the reverse coherent information $I(A\langle
B)=-S(B|A)$.

Thus for any $\bar{n}$, the coherent and reverse coherent information are
continuous in the limit of infinite dimension. This means that the hashing
inequality~\cite{DWrates2a} is extended to bosonic systems with constrained
energy. In other words, $I(A\rangle B)_{\rho}$ ($I(A\langle B)_{\rho}$)
represents an achievable rate for the distillable entanglement of the
energy-bounded bosonic state $\rho$ via forward (backward) CCs.

\subsection*{Extension to energy-unbounded Choi matrices of bosonic Gaussian
channels}

For bosonic systems, the ideal EPR\ state $\Phi$ is defined as the limit of
two-mode squeezed vacuum (TMSV) states $\Phi^{\mu}$, where $\mu=\bar{n}+1/2$
is sent to infinity (here $\bar{n}$ is the mean photon number in each
mode)~\cite{RMPa}. Thus, the Choi matrix of a Gaussian channel is defined as
the asymptotic operator $\rho_{\mathcal{E}}:=\lim_{\mu}\rho_{\mathcal{E}}%
^{\mu}$ where $\rho_{\mathcal{E}}^{\mu}:=\mathcal{I}\otimes\mathcal{E}%
(\Phi^{\mu})$. Correspondingly, the computation of the (reverse) coherent
information of the channel is performed as a limit, i.e., we have%
\begin{align}
I_{\mathrm{C}}(\mathcal{E})  &  =I(A\rangle B)_{\rho_{\mathcal{E}}}:=\lim
_{\mu}I(A\rangle B)_{\rho_{\mathcal{E}}^{\mu}}~,\label{Icv1}\\
I_{\mathrm{RC}}(\mathcal{E})  &  =I(A\langle B)_{\rho_{\mathcal{E}}}%
:=\lim_{\mu}I(A\langle B)_{\rho_{\mathcal{E}}^{\mu}}~. \label{Icv2}%
\end{align}

As we will see afterwards in the technical derivations of~\ref{app4}, for
bosonic Gaussian channels the functionals $I(A\rangle B)_{\rho_{\mathcal{E}%
}^{\mu}}$ and $I(A\langle B)_{\rho_{\mathcal{E}}^{\mu}}$ are \SP{continuous, asymptotically monotonic and bounded in $\mu$ (with the exception of limiting cases, such as the identity channel).}
Therefore, the previous limits are finite and
we can continuously extend the hashing inequality of Eq.~(\ref{hashAPP}) to
the asymptotic Choi matrix $\rho_{\mathcal{E}}$ of a Gaussian channel, for
which we may set $D_{1}(\rho_{\mathcal{E}}):=\lim_{\mu}D_{1}(\rho
_{\mathcal{E}}^{\mu})$.

\section{Upper bound at any dimension\label{app3}}

We provide alternate proofs of the weak converse theorem (Theorem~1 in the
main paper). The first proof relies on an exponential growth for the total
dimension of the private state~\cite{KD1a,KD2a,RenesSmith} (which is justified
by well-known arguments~\cite{Matthias1a,Matthias2a}). The second proof relies
on an exponential growth for the total energy. Finally, the third proof does
not have any of the previous assumptions; in particular, it only depends on
the \textquotedblleft key part\textquotedblright\ of the private state. The
first and third proofs are first given for DV channels and then extended to
CV\ channels by means of truncation arguments (see~\ref{app1} for full
details). The second proof simultaneously applies to both DV and CV channels,
by means of embedding arguments. Besides truncation and embedding, the other
main ingredients are basic properties of the trace norm and the relative
entropy of entanglement (REE)~\cite{RMPrelenta}, the \textquotedblleft
asymptotic continuity\textquotedblright\ of the REE~\cite{Donald,Synaka}, and
the REE upper bound for the distillable key of a quantum
state~\cite{KD1a,KD2a}.

\subsection*{First proof of the weak converse theorem}

Let us start by assuming that the output state $\rho_{\mathbf{ab}}^{n}$ in
Alice and Bob's registers has total finite dimension $d_{\mathbf{ab}}$. Given
$\rho_{\mathbf{ab}}^{n}$ and $\phi_{n}$ such that $\left\Vert \rho
_{\mathbf{ab}}^{n}-\phi_{n}\right\Vert \leq\varepsilon\leq1/3$, we may write
the Fannes-type inequality~\cite{Donald}
\begin{equation}
E_{\mathrm{R}}(\phi_{n})\leq E_{\mathrm{R}}(\rho_{\mathbf{ab}}^{n}%
)+2\varepsilon\log_{2}d_{\mathbf{ab}}+f(\varepsilon)~,
\end{equation}
where $f(\varepsilon):=4\varepsilon-2\varepsilon\log_{2}\varepsilon$. This
result is also known as asymptotic continuity\ of the REE. An alternate
version states that $\left\Vert \rho_{\mathbf{ab}}^{n}-\phi_{n}\right\Vert
\leq\varepsilon\leq1/2$ implies~\cite{Synaka}%
\begin{equation}
E_{\mathrm{R}}(\phi_{n})\leq E_{\mathrm{R}}(\rho_{\mathbf{ab}}^{n}%
)+4\varepsilon\log_{2}d_{\mathbf{ab}}+2H_{2}(\varepsilon)~, \label{SynakNEW}%
\end{equation}
where $H_{2}$ is the binary Shannon entropy. Note that the total dimension
$d_{\mathbf{ab}}$ of the output state may always be considered to be greater
than or equal to the dimension $d_{\mathrm{P}}$ of the private state. The
latter involves two key systems (with total dimension $d_{\mathrm{K}}^{2}$)
and a shield system (with total dimension $d_{\mathrm{S}}$), so that
$d_{\mathrm{P}}=d_{\mathrm{K}}^{2}d_{\mathrm{S}}$. The logarithm of the
dimension $d_{\mathrm{K}}$ determines the key rate, while the extra
dimension$\ d_{\mathrm{S}}$ is needed to shield the key and can be assumed to
grow exponentially in $n$ (see the next subsection \textquotedblleft Private
states and size of the shield system\textquotedblright\ for full details on
this\ secondary technical issue).

According to ref.~\cite{KD1a}, we may write
\begin{equation}
E_{\mathrm{R}}(\phi_{n})\geq K(\phi_{n})\SP{\ge}\log_{2}d_{\mathrm{K}}:=nR_{n}%
^{\varepsilon},
\end{equation}
where $K(\phi_{n})$ is the distillable key of $\phi_{n}$. Therefore, from
Eq.~(\ref{SynakNEW}), we find%
\begin{equation}
R_{n}^{\varepsilon}\leq\frac{E_{\mathrm{R}}(\rho_{\mathbf{ab}}^{n}%
)+4\varepsilon\log_{2}d_{\mathbf{ab}}+2H_{2}(\varepsilon)}{n}~. \label{Rkey22}%
\end{equation}
For some sufficiently high $\alpha\geq2$, let us set
\begin{equation}
\log_{2}d_{\mathbf{ab}}\leq\alpha nR_{n}^{\varepsilon}~. \label{trunc}%
\end{equation}
Then the previous inequality becomes%
\begin{equation}
R_{n}^{\varepsilon}\leq\frac{E_{\mathrm{R}}(\rho_{\mathbf{ab}}^{n}%
)+2H_{2}(\varepsilon)}{n(1-4\varepsilon\alpha)}~. \label{keyEQ}%
\end{equation}
Asymptotically in $n$, we therefore get%
\begin{equation}
\lim_{n}R_{n}^{\varepsilon}\leq\frac{1}{1-4\varepsilon\alpha}\lim
n^{-1}E_{\mathrm{R}}(\rho_{\mathbf{ab}}^{n})~.
\end{equation}
For $\varepsilon\rightarrow0$, we derive
\begin{equation}
\lim_{n}R_{n}\leq\lim n^{-1}E_{\mathrm{R}}(\rho_{\mathbf{ab}}^{n})~,
\end{equation}
whose optimization over adaptive protocols leads to the following weak
converse bound for the key generation capacity
\begin{equation}
K(\mathcal{E}):=\sup_{\mathcal{L}}\lim_{n}R_{n}\leq E_{\mathrm{R}}^{\bigstar
}(\mathcal{E}):=\sup_{\mathcal{L}}\lim n^{-1}E_{\mathrm{R}}(\rho_{\mathbf{ab}%
}^{n})~. \label{towritt}%
\end{equation}

When $\rho_{\mathbf{ab}}^{n}$ is a CV\ bosonic state, we may consider an LOCC
truncation channel $T^{\otimes}$\ which maps the state into a DV state
$\tilde{\rho}_{\mathbf{ab}}^{n}=T^{\otimes}(\rho_{\mathbf{ab}}^{n})$ supported
in a subspace with cut-off $\alpha$, so that the effective dimension is
$2^{\alpha nR_{n}^{\varepsilon}}$ as in Eq.~(\ref{trunc}). This CV-to-DV
mapping is large enough to leave the private state $\phi_{n}$ invariant, i.e.,
$\phi_{n}=T^{\otimes}(\phi_{n})$. Because $\left\Vert \tilde{\rho
}_{\mathbf{ab}}^{n}-\phi_{n}\right\Vert \leq\left\Vert \rho_{\mathbf{ab}}%
^{n}-\phi_{n}\right\Vert \leq\varepsilon$, we can then repeat the previous
derivation and write Eq.~(\ref{towritt}) for $\tilde{\rho}_{\mathbf{ab}}^{n}$.
Then, we introduce the upper-bound $E_{\mathrm{R}}(\tilde{\rho}_{\mathbf{ab}%
}^{n})\leq E_{\mathrm{R}}(\rho_{\mathbf{ab}}^{n})$, which derives from the
monotonicity of the REE under trace-preserving LOCCs (such as $T^{\otimes}$).
For clarity, this derivation can be broken down into the following steps%
\begin{equation}
E_{\mathrm{R}}(\tilde{\rho}_{\mathbf{ab}}^{n})\overset{(1)}{=}S(\tilde{\rho
}_{\mathbf{ab}}^{n}||\tilde{\sigma}_{s}^{\text{\textrm{opt}}})\overset
{(2)}{\leq}S(\tilde{\rho}_{\mathbf{ab}}^{n}||\sigma_{s}^{\prime})\overset
{(3)}{\leq}S(\rho_{\mathbf{ab}}^{n}||\sigma_{s}^{\text{\textrm{opt}}%
})=E_{\mathrm{R}}(\rho_{\mathbf{ab}}^{n})~,
\end{equation}
where (1)~we use the optimal separable state $\tilde{\sigma}_{s}%
^{\text{\textrm{opt}}}$ which is the closest to $\tilde{\rho}_{\mathbf{ab}%
}^{n}$ in terms of relative entropy; (2)~we introduce the non-optimal
separable state $\sigma_{s}^{\prime}=T^{\otimes}(\sigma_{s}%
^{\text{\textrm{opt}}})$, where $\sigma_{s}^{\text{\textrm{opt}}}$ is the
separable state closest to $\rho_{\mathbf{ab}}^{n}$ (because $T^{\otimes}$ is
a LOCC, it preserves the separability of input states); and (3)~we exploit the
fact that the relative entropy cannot increase under trace-preserving LOCCs,
which holds in arbitrary dimension~\cite{RMPrelenta,HolevoBOOKa}. Thus, we may
write Eq.~(\ref{towritt}) where $E_{\mathrm{R}}(\rho_{\mathbf{ab}}^{n})$ is
directly computed on the bosonic state $\rho_{\mathbf{ab}}^{n}$.

\subsection*{Private states and size of the shield system}

Let us discuss here the secondary technical detail related with the size of
the shield system which appears in the definition of a private state. Consider
a finite dimensional system of dimension $d_{\mathrm{K}}$ and basis $\left\{
\left\vert i\right\rangle \right\}  _{i=0}^{d_{\mathrm{K}}-1}$. A private
state between Alice and Bob can be written in the form~\cite{KD1a,KD2a}
\begin{equation}
\phi_{ABA^{\prime}B^{\prime}}=U(\Phi_{AB}\otimes\chi_{A^{\prime}B^{\prime}%
})U^{\dagger}, \label{privateSTATEform}%
\end{equation}
where $AB$ is the total key system\ in the maximally entangled state%
\begin{equation}
\Phi_{AB}=\left\vert \Phi\right\rangle _{AB}\left\langle \Phi\right\vert
,~~\left\vert \Phi\right\rangle _{AB}:=d_{\mathrm{K}}^{-1/2}\sum
_{i=0}^{d_{\mathrm{K}}-1}\left\vert i\right\rangle _{A}\left\vert
i\right\rangle _{B}, \label{EPRfinite}%
\end{equation}
while $A^{\prime}B^{\prime}$ is the shield system\ in a state $\chi
_{A^{\prime}B^{\prime}}$ protecting the key from eavesdropping. In
Eq.~(\ref{privateSTATEform}), the unitary $U$ is a controlled-unitary known as
\textquotedblleft twisting unitary\textquotedblright\ which takes the
form~\cite{KD2a}%
\begin{equation}
U=\sum_{i,j=0}^{d_{\mathrm{K}}-1}\left\vert i\right\rangle _{A}\left\langle
i\right\vert \otimes\left\vert j\right\rangle _{B}\left\langle j\right\vert
\otimes U_{A^{\prime}B^{\prime}}^{ij}, \label{twistingUNI}%
\end{equation}
with $U_{A^{\prime}B^{\prime}}^{ij}$ being arbitrary unitary operators.

One can prove that a dilation of a private state into an environment $E$
(owned by Eve) must take the form~\cite{KD2a}%
\begin{equation}
\phi_{ABA^{\prime}B^{\prime}E}=(U\otimes I_{E})(\Phi_{AB}\otimes
\chi_{A^{\prime}B^{\prime}E})(U\otimes I_{E})^{\dagger},
\label{privateSTATEextended}%
\end{equation}
with$\ \chi_{A^{\prime}B^{\prime}}=\mathrm{Tr}_{E}(\chi_{A^{\prime}B^{\prime
}E})$. Note that one can also equivalently write%
\begin{equation}
\phi_{ABA^{\prime}B^{\prime}E}=d_{\mathrm{K}}^{-1}\sum_{i,j=0}^{d_{\mathrm{K}%
}-1}\left\vert ii\right\rangle _{AB}\left\langle jj\right\vert \otimes
U_{A^{\prime}B^{\prime}}^{ii}\chi_{A^{\prime}B^{\prime}E}(U_{A^{\prime
}B^{\prime}}^{jj})^{\dag}.
\end{equation}
By making local measurements on the key system $AB$ and tracing out the shield
system $A^{\prime}B^{\prime}$, Alice and Bob retrieve the ideal
classical-classical-quantum (ccq) state~\cite{KD2a}
\begin{equation}
\tau_{ABE}=d_{\mathrm{K}}^{-1}\sum_{i=0}^{d_{\mathrm{K}}-1}\left\vert
i\right\rangle _{A}\left\langle i\right\vert \otimes\left\vert i\right\rangle
_{B}\left\langle i\right\vert \otimes\tau_{E}, \label{targetidealQKD}%
\end{equation}
with $\tau_{E}$ arbitrary. More precisely, one shows~\cite{KD2a} that
$\tau_{E}^{i}=\tau_{E}$ for any $i$ in Eq.~(\ref{targetidealQKD}). The shared
randomness in the final classical $AB$ system provides $\log_{2}d_{\mathrm{K}%
}$ secret-key bits. Thus, the dimension $d_{\mathrm{K}}$ of each key system
defines the number of secret-key bits (i.e., the rate of the protocol), while
the dimension $d_{\mathrm{S}}$\ of the shield system can in principle be
arbitrary. The total dimension of the private state is $d_{\mathrm{P}%
}=d_{\mathrm{K}}^{2}d_{\mathrm{S}}$.

In a key distillation protocol, where Alice and Bob start from $n$ shared
copies $\rho_{AB}^{\otimes n}$ and apply LOCCs to approximate a private state,
the size of the shield $d_{\mathrm{S}}$ grows with the number of classical
bits exchanged in their CCs. In fact, Eve may store all these bits in her
local register and a private state can be approximated by the parties only if
the dimension of Eve's register is smaller than the dimension of the shield
system. This is implied by Eq.~(\ref{targetidealQKD}) as explained in
ref.~\cite[Section~III]{KD2a}.

\bigskip

Now we may ask: \textit{Is the shield size }$d_{\mathrm{S}}$\textit{
super-exponential in }$n$\textit{? }The answer is \textit{no} for DV systems.

\bigskip

This was originally proven in ref.~\cite{Matthias1a} and also discussed in
ref.~\cite{Matthias2a}. This result also holds for key distribution through
memoryless channels at any dimension (finite or infinite). Let us remark that,
despite the proof may appear involved, it is actually a trivial modification
of the one in ref.~\cite[Appendix~A (arXiv v3 version)]{Matthias1a}. It is
based on the fact that one can\ always design an approximate protocol where
key distribution through $n$ uses of a finite- or infinite-dimensional channel
is broken down into $m$\ identical and independent $n_{0}$-long sub-protocols.
These sub-protocols provide $m$ copies, which are truncated, measured and
whose shields may be discarded. The effective increase of the shield size will
then come from one-way key distillation of these output copies, which has an
exponential contribution in $m<n$. In the following, we report this adaptation
(with all details) only for the sake of completeness.

\bigskip

\begin{lemma}
[Shield size. Trivially adapted from ref.~\cite{Matthias1a}]%
\label{shieldliminf}Consider $n$ uses of an adaptive key generation protocol
through a quantum channel at any dimension (finite or infinite). Without loss
of generality, we can assume that the effective dimension of the shield system
$d_{\mathrm{S}}$ grows in such a way that $\lim\inf_{n}(d_{\mathrm{S}}/c^{n})$
is a constant for some $c\geq1$.
\end{lemma}

\bigskip

\textit{Proof}:~~Let us represent Alice's and Bob's local registers as
$\mathbf{a}=AA^{\prime}$ and $\mathbf{b}=BB^{\prime}$, where $A$ and $B$ are
the local key systems, while $A^{\prime}$ and $B^{\prime}$ are the local
shield systems. Denote Eve's register by $E$. Even if all these systems are
infinite-dimensional for bosonic channels, the key systems $A$ and $B$ of the
private state $\phi_{AA^{\prime}BB^{\prime}E}^{n_{0}}$ have finite-dimensional
support, as we can see from Eq.~(\ref{privateSTATEextended}). Then consider an
arbitrary adaptive key generation protocol $\mathcal{P}_{\text{\textrm{key}}%
}^{n}$, with key rate $R$ and communication cost that is not necessarily
linear in the number $n$ of channel uses (this cost may even be singular,
i.e., involving an infinite number of classical bits per channel use). For any
$\varepsilon>0$, there is a sufficiently large integer $n_{0}$ such that its
output $\rho_{AA^{\prime}BB^{\prime}E}^{n_{0}}$ satisfies
\begin{equation}
\left\Vert \rho_{AA^{\prime}BB^{\prime}E}^{n_{0}}-\phi_{AA^{\prime}BB^{\prime
}E}^{n_{0}}\right\Vert \leq\varepsilon, \label{closMAP}%
\end{equation}
where $\phi_{AA^{\prime}BB^{\prime}E}^{n_{0}}$\ is a (dilated) private state
with $l_{n_{0}}$ secret bits such that%
\begin{equation}
R_{n_{0}}:=\frac{l_{n_{0}}}{n_{0}}\geq R-\varepsilon~. \label{raten0protocol}%
\end{equation}

Assume that the restricted $n_{0}$-adaptive protocol is repeated $m$ times, so
that the total number of channel uses can be written as $n=n_{0}m$.
Correspondingly, Alice and Bob's output state will be equal to the tensor
product $(\rho_{AA^{\prime}BB^{\prime}}^{n_{0}})^{\otimes m}$ with
$\rho_{AA^{\prime}BB^{\prime}}^{n_{0}}=\mathrm{Tr}_{E}(\rho_{AA^{\prime
}BB^{\prime}E}^{n_{0}})$. Now, assume that the parties measure their key
systems in the computational basis $\left\vert i\right\rangle _{A}%
\otimes\left\vert j\right\rangle _{B}$ while truncating any outcome outside
the finite-dimensional $2^{l_{n_{0}}}\times2^{l_{n_{0}}}$ support
$\mathcal{S}_{\text{\textrm{key}}}$\ of the private state's key system. They
then discard their shield systems. This means that they apply the LOCC channel%
\begin{equation}
\mathbb{L}^{\otimes}(\rho_{AA^{\prime}BB^{\prime}}^{n_{0}})=\mathrm{Tr}%
_{A^{\prime}B^{\prime}}\left[  \sum_{i,j}\mathcal{E}_{ij}\left(  \Pi_{ij}%
\rho_{AA^{\prime}BB^{\prime}}^{n_{0}}\Pi_{ij}^{\dagger}\right)  \right]  ,
\end{equation}
where $\Pi_{ij}:=\Pi_{i}^{A}\otimes\Pi_{j}^{B}$ projects onto the local
computational bases, while the conditional channel $\mathcal{E}_{ij}$ is%
\begin{equation}
\mathcal{E}_{ij}=\left\{
\begin{array}
[c]{c}%
\mathcal{I}_{AB}~~~\text{for~}i,j\in\lbrack0,2^{l_{n_{0}}}-1]\\
\\
\mathcal{E}_{e}^{A}\otimes\mathcal{E}_{e}^{B}\text{~~~otherwise,~~~~~~~}%
\end{array}
\right.
\end{equation}
where $\mathcal{E}_{e}$ is a map replacing any input with an erasure state
$\left\vert e\right\rangle $ orthogonal to the support $\mathcal{S}%
_{\text{\textrm{key}}}$. (Note that the contribution of the extra dimension of
$\left\vert e\right\rangle $ to the output state is completely negligible
since $2^{l_{n_{0}}}$ is very large).

The action of $\mathbb{L}^{\otimes}$ on the (dilated) output state
$\rho_{AA^{\prime}BB^{\prime}E}^{n_{0}}$ is such that we achieve a truncated
ccq state $\tilde{\rho}_{ABE}^{n_{0}}:=(\mathbb{L}^{\otimes}\otimes
\mathcal{I}_{E})(\rho_{AA^{\prime}BB^{\prime}E}^{n_{0}})$, where the key
systems $A$ and $B$ are classical and finite-dimensional. Similarly, the
action on the (dilated) private state provides the ideal ccq target state
$\tau_{ABE}^{n_{0}}:=(\mathbb{L}^{\otimes}\otimes I_{E})(\phi_{AA^{\prime
}BB^{\prime}E}^{n_{0}})$ which corresponds to Eq.~(\ref{targetidealQKD}) with
$\log_{2}d_{\mathrm{K}}=l_{n_{0}}$. Also note that this classicalization and
truncation step just needs two bits of CC to be implemented: These bits are
needed to identify those instances where the measurement of the other party
falls outside the support (this classical overhead is clearly linear in the
number $m$ of blocks).

All the procedure is a trivial modification of the one in
ref.~\cite[Appendix~A (arXiv v3 version)]{Matthias1a}. Here we implement it in
a coherent way and we include CV systems, for which $\mathbb{L}^{\otimes}$
measures the key systems within the finite-dimensional support of the private
state, while collapsing any contribution from the remaining part of the
infinite-dimensional Hilbert space. It is easy to check that $\mathbb{L}%
^{\otimes}$ can also be implemented in two subsequent steps: First a
truncation channel into a $(2^{l_{n_{0}}}+1)\times(2^{l_{n_{0}}}+1)$\ subspace
and then a measurement channel in the computational bases.

Using the monotonicity of the trace distance under channels (at any
dimension), from Eq.~(\ref{closMAP}) we may write
\begin{equation}
\left\Vert \tilde{\rho}_{ABE}^{n_{0}}-\tau_{ABE}^{n_{0}}\right\Vert
\leq\varepsilon. \label{epsclosef}%
\end{equation}
Let us consider the reduced state $\tilde{\rho}_{AB}^{n_{0}}=\mathrm{Tr}%
_{E}(\tilde{\rho}_{ABE}^{n_{0}})$. Given many copies $(\tilde{\rho}%
_{AB}^{n_{0}})^{\otimes m}$, Alice and Bob may apply one-way key distillation
at an achievable rate (secret bits per block) given by the Devetak-Winter (DW)
rate~\cite{DWrates2a}%
\begin{equation}
R_{\tilde{\rho}}^{\text{\textrm{DW}}}:=I(A:B)-I(A:E)=S(A|E)-S(A|B),
\label{DWprova2}%
\end{equation}
where $I(:)$ is the quantum mutual information (equal to the classical mutual
information on classical systems $A$ and $B$), and $S(|)$ is the conditional
von Neumann entropy, with all quantities being computed on the extended output
$\tilde{\rho}:=\tilde{\rho}_{ABE}^{n_{0}}$. Note that the DW rates are
achievable rates at any dimension (finite or infinite).

Let us set $\tau:=\tau_{ABE}^{n_{0}}$. We then compute the difference%
\begin{equation}
R_{\tau}^{\text{\textrm{DW}}}-R_{\tilde{\rho}}^{\text{\textrm{DW}}}%
\leq\left\vert S(A|E)_{\tau}-S(A|E)_{\tilde{\rho}}\right\vert +\left\vert
S(A|B)_{\tilde{\rho}}-S(A|B)_{\tau}\right\vert \leq8\varepsilon\log_{2}%
\dim\mathcal{H}_{A}+4H_{2}(\varepsilon), \label{fannesALI}%
\end{equation}
where $H_{2}$ is the binary Shannon entropy. In Eq.~(\ref{fannesALI}), we have
used the Alicki-Fannes' inequality for the conditional quantum
entropy~\cite{AlickFANN} which is valid for any $||\tilde{\rho}-\tau
||\leq\varepsilon<1$ as in Eq.~(\ref{epsclosef}).

Note that Eq.~(\ref{fannesALI}) only contains the dimension of Alice's Hilbert
space $\mathcal{H}_{A}$ (truncated in each sub-protocol), while the Hilbert
spaces of Bob and Eve do not have any restriction of their dimensionality.
Because $R_{\tau}^{\text{\textrm{DW}}}=l_{n_{0}}$ and $\dim\mathcal{H}%
_{A}=2^{l_{n_{0}}}$, we may then write%
\begin{equation}
R_{\tilde{\rho}}^{\text{\textrm{DW}}}\geq(1-8\varepsilon)l_{n_{0}}%
-4H_{2}(\varepsilon),
\end{equation}
exactly as in ref.~\cite[Appendix~A (arXiv v3 version)]{Matthias1a}.
Therefore, by dividing the latter equation by $n_{0}$, one gets the average
rate (per channel use)%
\begin{equation}
\tilde{R}:=\frac{1}{n_{0}}R_{\tilde{\rho}}^{\text{\textrm{DW}}}\geq
(1-8\varepsilon)(R-\varepsilon)-\frac{4H_{2}(\varepsilon)}{n_{0}}.
\label{approachingR}%
\end{equation}

It is now important to note that Alice and Bob can achieve the average DW rate
$\tilde{R}$ using an amount of one-way CC which is linear in the\ block number
$m<n$. In fact, the communication cost (bits per block) associated with the
one-way key distillation of Alice and Bob's copies $(\tilde{\rho}_{AB}^{n_{0}%
})^{\otimes m}$ is equal to the conditional (Shannon) entropy $S(A|B)$ between
the two classical finite-dimensional systems $A$ and $B$~\cite{DWrates2a}.
This overhead is bounded by $\log_{2}\dim\mathcal{H}_{A,B}=l_{n_{0}}$
classical bits per block, so that it scales at most linearly as $ml_{n_{0}}$.
Therefore, by decreasing $\varepsilon$, we get a sequence of protocols whose
classical communication scales linearly in $m$ while their rates approach $R$
according to Eq.~(\ref{approachingR}). Correspondingly, the size of the shield
grows at most exponentially in $m$.

Let us take a closer look at the dynamics of the shield. Within each block,
the shield size may increase super-exponentially (even to infinite) but then
this size collapses to zero at the end of each block (after $n_{0}$ uses) once
the parties have generated their finite-dimensional cc-state $\tilde{\rho
}_{AB}^{n_{0}}$. For this reason, there is no surviving contribution to shield
coming from the $m$ sub-protocols. The only contribution to the shield size
is\ that (exponential) coming from the protocol of one-way key distillation on
the output finite-dimensional copies. Thus, at values of $n=n_{0}m$ for
integer $m$, the dimension $d_{\mathrm{S}}$ scales as an exponential function,
i.e., it is bounded by $c^{n}$ for some $c\geq1$. If we look at the shield
dynamics for every $n$, then we may always replace the limit by an inferior
limit, i.e., we may always say that $\lim\inf_{n}(d_{\mathrm{S}}/c^{n})$ is a
constant (with the infimum reached by the sequence of points $n=n_{0}%
m$).~$\square$

\subsection*{Second proof of the weak converse theorem}

This second proof simultaneously applies to DV\ and CV systems, and relies on
the physical assumption that the energy of the output state grows at most
exponentially in the number of channel uses. Consider bosonic modes, since any
DV system can be unitarily embedded into a CV system (operation which does not
change the trace distance). In general, we assume $m_{\mathbf{a}}$ modes at
Alice's side and $m_{\mathbf{b}}$ modes at Bob's side (recall that the
parties' local registers may be composed of a countable set of quantum
systems). Assume that the output state $\rho_{\mathbf{ab}}^{n}$ and the target
state $\phi_{\mathbf{ab}}^{n}$ have mean photon numbers bounded by $E_{n}$,
where we may set $E_{n}\leq2^{cn}$ for some constant $c$.
\SP{Set $m:=m_a+m_b<\infty$, with $m$ and $c$ fixed
throughout the limits $n\to\infty$ and $\varepsilon\to0$.}

Let us apply a LOCC truncation channel $\mathbb{T}_{d}^{\otimes}$, local with
respect to Alice and Bob's bipartition of modes $m_{\mathbf{a}}+m_{\mathbf{b}%
}$, which truncates Alice's and Bob's local Hilbert spaces to finite dimension
\SP{$d=E_{n}^{2m}$} (other choices are possible). This means that the truncated
states $\rho_{\mathbf{ab}}^{n,d}:=\mathbb{T}_{d}^{\otimes}(\rho_{\mathbf{ab}%
}^{n})$ and $\phi_{\mathbf{ab}}^{n,d}:=\mathbb{T}_{d}^{\otimes}(\phi
_{\mathbf{ab}}^{n})$ satisfies (see Lemma~\ref{LOCCtruncationChannel})%
\begin{equation}
||\rho_{\mathbf{ab}}^{n,d}-\rho_{\mathbf{ab}}^{n}||,||\phi_{\mathbf{ab}}%
^{n,d}-\phi_{\mathbf{ab}}^{n}||\leq2(\sqrt{\gamma}+\gamma),~~\gamma
=\frac{E_{n}}{E_{n}^{2}-1}.
\end{equation}
Because $\left\Vert \rho_{\mathbf{ab}}^{n}-\phi_{\mathbf{ab}}^{n}\right\Vert
\leq\varepsilon$, we can apply the triangle inequality and find%
\begin{equation}
||\rho_{\mathbf{ab}}^{n,d}-\phi_{\mathbf{ab}}^{n,d}||\leq\varepsilon
+\varepsilon^{\prime},~~\varepsilon^{\prime}:=4(\sqrt{\gamma}+\gamma
)=O(E_{n}^{-1/2}).
\end{equation}
Now the asymptotic continuity of the REE~\cite{Synaka} leads to%
\begin{equation}
E_{\mathrm{R}}(\phi_{\mathbf{ab}}^{n,d})\leq E_{\mathrm{R}}(\rho_{\mathbf{ab}%
}^{n,d})+\SP{16m}(\varepsilon+\varepsilon^{\prime})\log_{2}E_{n}+2H_{2}%
(\varepsilon+\varepsilon^{\prime}), \label{previousINtwo}%
\end{equation}
where we use the fact that the total dimension of the truncated states is
$d_{\mathbf{ab}}=d^{2}=\SP{E_{n}^{4m}}$.

In Eq.~(\ref{previousINtwo}) we may replace $E_{\mathrm{R}}(\rho_{\mathbf{ab}%
}^{n,d})\leq E_{\mathrm{R}}(\rho_{\mathbf{ab}}^{n})$ due to the fact that the
REE is monotonic under $\mathbb{T}_{d}^{\otimes}$ and invariant under
embedding local unitaries. We may also replace $\log_{2}E_{n}\leq cn$ and
$E_{\mathrm{R}}(\phi_{\mathbf{ab}}^{n,d})\geq K(\phi_{\mathbf{ab}}%
^{n,d})\SP{\ge}nR_{n}^{\varepsilon}(E_{n})$, where the energy-constrained key rate
must satisfy $\lim_{n}R_{n}^{\varepsilon}(E_{n})=\lim_{n}R_{n}^{\varepsilon}$,
with $R_{n}^{\varepsilon}$ being the (finite) key rate associated with
$\phi_{\mathbf{ab}}^{n}$. Therefore, we may write%
\begin{equation}
nR_{n}^{\varepsilon}(E_{n})\leq E_{\mathrm{R}}(\rho_{\mathbf{ab}}%
^{n})+\SP{16m}(\varepsilon+\varepsilon^{\prime})cn+2H_{2}(\varepsilon+\varepsilon
^{\prime}).
\end{equation}
\SP{Dividing} by $n$ and taking the limit for $n\rightarrow+\infty$, we get%
\begin{equation}
\lim_{n}R_{n}^{\varepsilon}\leq\lim_{n}n^{-1}E_{\mathrm{R}}(\rho_{\mathbf{ab}%
}^{n})+\SP{16m}\varepsilon c.
\end{equation}
Finally, by taking the limit of $\varepsilon\rightarrow0$, we find%
\begin{equation}
\lim_{n}R_{n}\leq\lim_{n}n^{-1}E_{\mathrm{R}}(\rho_{\mathbf{ab}}^{n})~,
\end{equation}
\SP{which establishes the REE weak-converse bound for adaptive protocols
with a fixed finite number of output modes and the stated exponential
bound on the mean photon numbers of the output and target states.
The third proof below establishes the unrestricted bound for
arbitrary adaptive protocols.}

\subsection*{Third proof of the weak converse theorem}

Let us now give a final proof which is completely independent from the
dimensionality of the shield system in the private state. We start from the DV
case and then we prove the CV case by resorting to truncation arguments. After
$n$ uses of a DV quantum channel $\mathcal{E}$, an adaptive key-generation
protocol has an output $\rho_{\mathbf{ab}}^{n}=\rho_{\mathbf{ab}}%
(\mathcal{E}^{\otimes n})$ such that
\begin{equation}
\left\Vert \rho_{\mathbf{ab}}^{n}-\phi_{\mathbf{ab}}^{n}\right\Vert
\leq\varepsilon, \label{epsclos}%
\end{equation}
where $\phi_{\mathbf{ab}}^{n}$ is a private state. Let us write the local
registers as $\mathbf{a}=AA^{\prime}$ and $\mathbf{b}=BB^{\prime}$, with $AB$
being the key part (with dimension $d_{\mathrm{K}}\times d_{\mathrm{K}}$) and
$A^{\prime}B^{\prime}$ being the shield. By definition of private state, we
have
\begin{equation}
\phi_{\mathbf{ab}}^{n}=\phi_{ABA^{\prime}B^{\prime}}^{n}=U(\Phi_{AB}%
^{n}\otimes\chi_{A^{\prime}B^{\prime}})U^{\dagger}, \label{pstatePRR}%
\end{equation}
where $U$ is a twisting unitary,$\ \chi_{A^{\prime}B^{\prime}}\ $is a state of
the shield, and $\Phi_{AB}^{n}$ is a Bell state with $\log d_{\mathrm{K}%
}=nR_{n}^{\varepsilon}$ secret bits.

Let us \textquotedblleft untwist\textquotedblright\ the output state
$\rho_{\mathbf{ab}}^{n}=\rho_{ABA^{\prime}B^{\prime}}^{n}$ and then take the
partial trace over the shield system $A^{\prime}B^{\prime}$. This means to
consider%
\begin{equation}
\rho_{AB}^{n}=\mathrm{Tr}_{A^{\prime}B^{\prime}}\left(  U^{\dagger}%
\rho_{ABA^{\prime}B^{\prime}}^{n}U\right)  :=\mathcal{W}(\rho_{ABA^{\prime
}B^{\prime}}^{n}). \label{unTW}%
\end{equation}
Trace norm is \SP{non-increasing} under partial trace and invariant under
unitaries, so that Eq.~(\ref{epsclos}) implies%
\begin{equation}
\left\Vert \rho_{AB}^{n}-\Phi_{AB}^{n}\right\Vert \leq\varepsilon.
\label{synackAPP}%
\end{equation}

Following ref.~\cite{KD2a}, let us consider the set $T$ of bipartite states
$\sigma_{AB}$ which are defined by $\sigma_{AB}=\mathcal{W}(\sigma
_{ABA^{\prime}B^{\prime}})$ where $\sigma_{ABA^{\prime}B^{\prime}}$ is an
arbitrary separable state (with respect to Alice and Bob's bipartition
$AA^{\prime}$ and $BB^{\prime}$). One may define the relative entropy distance
from this set as%
\begin{equation}
E_{\mathrm{R}}^{T}(\rho)=\inf_{\sigma\in T}S(\rho||\sigma)~.
\end{equation}
Because the set $T$ is \SP{convex} and contains the maximally mixed
state~\cite{KD2a}, this distance is asymptotically continuous, i.e., the
condition $\left\Vert \rho_{1}-\rho_{2}\right\Vert \leq\varepsilon<1/2$
implies~\cite{Synaka}%
\begin{equation}
\left\vert E_{\mathrm{R}}^{T}(\rho_{1})-E_{\mathrm{R}}^{T}(\rho_{2}%
)\right\vert \leq4\varepsilon\log_{2}d+2H_{2}(\varepsilon)~,
\end{equation}
where $d$\ is the total dimension of the Hilbert space and $H_{2}$ is the
binary Shannon entropy.

By applying this property to Eq.~(\ref{synackAPP}) with $d_{AB}=d_{\mathrm{K}%
}^{2}$, we then get%
\begin{equation}
E_{\mathrm{R}}^{T}(\Phi_{AB}^{n})\leq E_{\mathrm{R}}^{T}(\rho_{AB}%
^{n})+8\varepsilon\log_{2}d_{\mathrm{K}}+2H_{2}(\varepsilon). \label{bnm0}%
\end{equation}
Now we exploit two observations. The first is that
\begin{equation}
E_{\mathrm{R}}^{T}(\Phi_{AB}^{n})\geq\log d_{\mathrm{K}}=nR_{n}^{\varepsilon
}~, \label{bnm1}%
\end{equation}
as shown in ref.~\cite[Lemma~7]{KD2a}. Then, we also have
\begin{equation}
E_{\mathrm{R}}^{T}(\rho_{AB}^{n})\leq E_{\mathrm{R}}(\rho_{ABA^{\prime
}B^{\prime}}^{n}):=E_{\mathrm{R}}(\rho_{\mathbf{ab}}^{n})~. \label{bnm2}%
\end{equation}
In fact, this is proven by the following chain of (in)equalities%
\begin{equation}
E_{\mathrm{R}}^{T}(\rho_{AB}^{n})\overset{(1)}{\leq}S[\rho_{AB}^{n}%
||\mathcal{W}(\sigma_{ABA^{\prime}B^{\prime}})]\overset{(2)}{=}S[\mathcal{W}%
(\rho_{ABA^{\prime}B^{\prime}}^{n})||\mathcal{W}(\sigma_{ABA^{\prime}%
B^{\prime}})]\overset{(3)}{\leq}S(\rho_{ABA^{\prime}B^{\prime}}^{n}%
|\sigma_{ABA^{\prime}B^{\prime}})\overset{(4)}{=}E_{\mathrm{R}}(\rho
_{ABA^{\prime}B^{\prime}}^{n}), \label{bnm3}%
\end{equation}
where (1)~we use some arbitrary state $\sigma_{AB}\in T$, (2)~we use
Eq.~(\ref{unTW}), (3)~we use the fact that the relative entropy is monotonic
under partial trace and invariant under unitaries, and finally (4)~we may
always choose the separable state $\sigma_{ABA^{\prime}B^{\prime}}$ to be the
one which is the closest to $\rho_{ABA^{\prime}B^{\prime}}^{n}$ in relative
entropy (so that it defines its REE).

Using Eqs.~(\ref{bnm1}) and~(\ref{bnm2}) into Eq.~(\ref{bnm0}), we find%
\begin{equation}
nR_{n}^{\varepsilon}\leq E_{\mathrm{R}}(\rho_{\mathbf{ab}}^{n})+8\varepsilon
\log_{2}d_{\mathrm{K}}+2H_{2}(\varepsilon). \label{leadEQ}%
\end{equation}
Because $\log_{2}d_{\mathrm{K}}=nR_{n}^{\varepsilon}$, this leads to
\begin{equation}
R_{n}^{\varepsilon}\leq\frac{E_{\mathrm{R}}(\rho_{\mathbf{ab}}^{n}%
)+2H_{2}(\varepsilon)}{n(1-8\varepsilon)}~,
\end{equation}
so that, for large $n$, we may write%
\begin{equation}
\lim_{n}R_{n}^{\varepsilon}\leq\frac{1}{1-8\varepsilon}\lim_{n}n^{-1}%
E_{\mathrm{R}}(\rho_{\mathbf{ab}}^{n})~.
\end{equation}
By taking the limit of $\varepsilon\rightarrow0$, we then find
\begin{equation}
\lim_{n}R_{n}\leq\lim_{n}n^{-1}E_{\mathrm{R}}(\rho_{\mathbf{ab}}^{n}).
\end{equation}
Finally, by optimizing over all adaptive protocols $\mathcal{L}$, we establish
the weak converse bound for the two-way key-generation capacity of the channel%
\begin{equation}
K(\mathcal{E}):=\sup_{\mathcal{L}}\lim_{n}R_{n}\leq\sup_{\mathcal{L}}\lim
_{n}n^{-1}E_{\mathrm{R}}(\rho_{\mathbf{ab}}^{n})~.
\end{equation}

We now consider the CV case, i.e., a bosonic channel $\mathcal{E}$. In this
case, the private state $\phi_{\mathbf{ab}}^{n}=\phi_{ABA^{\prime}B^{\prime}%
}^{n}$ of Eq.~(\ref{pstatePRR}) is still built on a finite-dimensional
$d_{\mathrm{K}}\times d_{\mathrm{K}}$ Bell state $\Phi_{AB}^{n}$ containing
$\log d_{\mathrm{K}}=nR_{n}^{\varepsilon}$ secret bits. This Bell state may
equivalently be thought to be embedded into a CV system where it is supported
within a $d_{\mathrm{K}}\times d_{\mathrm{K}}$ subspace of the
infinite-dimensional Hilbert space. The shield state $\chi_{A^{\prime
}B^{\prime}}$ can be an arbitrary CV state and the twisting $U$ is an
arbitrary control-unitary as in Eq.~(\ref{twistingUNI}) but where the target
unitaries $U_{A^{\prime}B^{\prime}}^{ij}$ are defined on a CV state.

Let us apply an LOCC truncation channel $\mathbb{T}_{d_{\mathrm{K}}}^{\otimes
}$ to the key systems $A$ and $B$, so that their Hilbert spaces are truncated
to finite dimension $d_{\mathrm{K}}\times d_{\mathrm{K}}$. Clearly, we have
the invariance $\phi_{ABA^{\prime}B^{\prime}}^{n}=\mathbb{T}_{d_{\mathrm{K}}%
}^{\otimes}\otimes\mathcal{I}_{A^{\prime}B^{\prime}}(\phi_{ABA^{\prime
}B^{\prime}}^{n})$, while we set $\rho_{ABA^{\prime}B^{\prime}}%
^{n,d_{\mathrm{K}}}:=\mathbb{T}_{d_{\mathrm{K}}}^{\otimes}\otimes
\mathcal{I}_{A^{\prime}B^{\prime}}(\rho_{ABA^{\prime}B^{\prime}}^{n})$, where
$\mathcal{I}_{A^{\prime}B^{\prime}}$ is an identity channel acting\ on the
shield systems. By using the monotonicity of the trace norm under channels, we
may write%
\begin{equation}
||\rho_{ABA^{\prime}B^{\prime}}^{n,d_{\mathrm{K}}}-\phi_{ABA^{\prime}%
B^{\prime}}^{n}||\leq\varepsilon~.
\end{equation}
As before, let us define a channel $\mathcal{W}$ which untwists and
partial-traces the states as in Eq.~(\ref{unTW}). This channel provides the
$d_{\mathrm{K}}\times d_{\mathrm{K}}$ states $\tilde{\rho}_{AB}^{n}%
=\mathcal{W}(\rho_{ABA^{\prime}B^{\prime}}^{n,d_{\mathrm{K}}})$ and $\Phi
_{AB}^{n}=\mathcal{W}(\phi_{ABA^{\prime}B^{\prime}}^{n})$, for which we may
write (using monotonicity)%
\begin{equation}
||\tilde{\rho}_{AB}^{n}-\Phi_{AB}^{n}||\leq\varepsilon~.
\end{equation}

Consider now the set $T$ of states defined by $\sigma_{AB}=\mathcal{W}%
(\sigma_{ABA^{\prime}B^{\prime}})$, where $\sigma_{ABA^{\prime}B^{\prime}}$ is
an arbitrary separable state (with respect to the bipartition $AA^{\prime}$
and $BB^{\prime}$) where the key-part $AB$ has dimension $d_{\mathrm{K}}\times
d_{\mathrm{K}}$ while the shield-part $A^{\prime}B^{\prime}$ is
infinite-dimensional. The set $T$ is \SP{convex} and contains the
maximally mixed state. Thus, the relative entropy distance $E_{\mathrm{R}}%
^{T}(\rho)=\inf_{\sigma\in T}S(\rho||\sigma)$ is asymptotically continuous.
This means that we may write%
\begin{equation}
E_{\mathrm{R}}^{T}(\Phi_{AB}^{n})\leq E_{\mathrm{R}}^{T}(\tilde{\rho}_{AB}%
^{n})+8\varepsilon\log_{2}d_{\mathrm{K}}+2H_{2}(\varepsilon). \label{bnm4}%
\end{equation}
Now we derive%
\begin{equation}
E_{\mathrm{R}}^{T}(\tilde{\rho}_{AB}^{n})\overset{(1)}{\leq}E_{\mathrm{R}%
}(\rho_{ABA^{\prime}B^{\prime}}^{n,d_{\mathrm{K}}})\overset{(2)}{\leq
}E_{\mathrm{R}}(\rho_{ABA^{\prime}B^{\prime}}^{n}):=E_{\mathrm{R}}%
(\rho_{\mathbf{ab}}^{n}), \label{bnm5}%
\end{equation}
where (1) follows the derivation given in Eq.~(\ref{bnm3}), and (2) comes from
the monotonicity of the REE under $\mathbb{T}_{d_{\mathrm{K}}}^{\otimes
}\otimes\mathcal{I}_{A^{\prime}B^{\prime}}$. By replacing Eqs.~(\ref{bnm1})
and~(\ref{bnm5}) into Eq~(\ref{bnm4}), we find Eq.~(\ref{leadEQ}) where
$\rho_{\mathbf{ab}}^{n}$ is a now a CV\ state. The remainder of the proof is
the same as before.

\section{Technical derivations for bosonic Gaussian channels\label{app4}}

\subsection*{Basic tools for continuous variables}

Let us consider $n$ bosonic modes with quadrature operators $\hat{x}=(\hat
{q}_{1},\dots,\hat{q}_{n},\hat{p}_{1},\dots,\hat{p}_{n})^{T}$ and canonical
commutation relations~\cite{arvind_real_1995a}
\begin{equation}
\lbrack\hat{x},\hat{x}^{T}]=i\Omega,~~~\Omega:=%
\begin{pmatrix}
0 & 1\\
-1 & 0
\end{pmatrix}
\otimes I~,
\end{equation}
with $I$ being the $n\times n$\ identity matrix. An arbitrary multimode
Gaussian state $\rho(u,V)$, with mean value $u$\ and covariance matrix (CM)
$V$, can be written as~\cite{Banchi}%
\begin{equation}
\rho=\frac{\exp\left[  -\frac{1}{2}(\hat{x}-u)^{T}G(\hat{x}-u)\right]  }%
{\det\left(  V+i\Omega/2\right)  ^{1/2}}, \label{Gexpression}%
\end{equation}
where the Gibbs matrix $G$ is specified by
\begin{equation}
G=2i\Omega\,\coth^{-1}(2Vi\Omega). \label{gVFUN2}%
\end{equation}

The CM of a Gaussian state can be decomposed by using Williamson's
theorem~\cite{RMPa}. This provides the symplectic spectrum $\{\nu_{1}%
,\ldots,\nu_{n}\}$ which must satisfy the uncertainty principle $\nu_{k}%
\geq1/2$. Similarly, we may write $\nu_{k}=\bar{n}_{k}+1/2$ where $\bar{n}%
_{k}$ are thermal numbers, i.e., mean number of photons in each mode. The von
Neumann entropy of a Gaussian state can be easily computed as
\begin{equation}
S(\rho)=\sum_{k}s(\nu_{k})=\sum_{k}h(\bar{n}_{k}), \label{vonGauss}%
\end{equation}
where%
\begin{equation}
\left\{
\begin{array}
[c]{c}%
s(\nu):=\left(  \nu+\tfrac{1}{2}\right)  \log_{2}\left(  \nu+\tfrac{1}%
{2}\right)  -\left(  \nu-\tfrac{1}{2}\right)  \log_{2}\left(  \nu-\tfrac{1}%
{2}\right)  ,\\
\\
h(\bar{n}):=\left(  \bar{n}+1\right)  \log_{2}\left(  \bar{n}+1\right)
-\bar{n}\log_{2}\bar{n}.~~~~~~~~~~~~~~~~~~
\end{array}
\right.  \label{hFUNCTaa}%
\end{equation}

The most typical Gaussian state of two modes $A$ and $B$ is a two-mode
squeezed thermal state. This has zero-mean and CM\ of the form%
\begin{equation}
V=\left(
\begin{array}
[c]{cc}%
a & c\\
c & b
\end{array}
\right)  \oplus\left(
\begin{array}
[c]{cc}%
a & -c\\
-c & b
\end{array}
\right)  ~, \label{Vnform}%
\end{equation}
with arbitrary $a,b\geq1/2$ and $c$ satisfying the condition%
\begin{equation}
c\leq c_{\text{\textrm{max}}}:=\min\left\{  \sqrt{\left(  a-\tfrac{1}%
{2}\right)  \left(  b+\tfrac{1}{2}\right)  },\sqrt{\left(  a+\tfrac{1}%
{2}\right)  \left(  b-\tfrac{1}{2}\right)  }\right\}  .
\end{equation}
These bona-fide conditions can be checked using the tools in
Refs.~\cite{bonafidea,bonafide2a} adapted to our different notation. For a CM
as in Eq.~(\ref{Vnform}), separability corresponds to
\begin{equation}
c\leq c_{\text{\textrm{sep}}}:=\sqrt{\left(  a-\tfrac{1}{2}\right)  \left(
b-\tfrac{1}{2}\right)  }~. \label{sepmaxCON}%
\end{equation}
Thus, at any fixed $a$ and $b$, the maximally-correlated but still separable
Gaussian state is given by imposing the boundary condition
$c=c_{\text{\textrm{sep}}}$. It is easy to check that this state contains the
maximum correlations among the separable states, e.g., as quantified by its
(unrestricted, generally non-Gaussian) quantum discord~\cite{OptimalDISa}.

For $c_{\text{\textrm{sep}}}<c\leq c_{\text{\textrm{max}}}$ in
Eq.~(\ref{Vnform}), the Gaussian state is entangled. A specific case is the
TMSV state $\Phi^{\mu}$ with CM of the form%
\begin{equation}
V^{\mu}=\left(
\begin{array}
[c]{cc}%
\mu & c\\
c & \mu
\end{array}
\right)  \oplus\left(
\begin{array}
[c]{cc}%
\mu & -c\\
-c & \mu
\end{array}
\right)  ,~~c:=\sqrt{\mu^{2}-1/4},~~\mu\geq1/2. \label{EPR_CMa}%
\end{equation}
As already discussed, for $\mu\rightarrow\infty$, this state describes the
asymptotic CV EPR\ state $\Phi$, realizing the ideal EPR conditions $\hat
{q}_{A}=\hat{q}_{B}$ and $\hat{p}_{A}=-\hat{p}_{B}$.

A Gaussian channel is a CPTP\ map which transforms Gaussian states into
Gaussian states. Single-mode Gaussian channels can be greatly simplified by
means of input-output unitaries. In fact, these can always be put in canonical
form~\cite{RMPa} whose general action on input quadratures $\hat{x}=(\hat
{q},\hat{p})^{T}$ is given by%
\begin{equation}
\hat{x}\rightarrow T\hat{x}+N\hat{x}_{E}+z~, \label{genFORM}%
\end{equation}
where $T$ and $N$ are diagonal matrices, $E$ is an environmental mode with
$\bar{n}$ mean photons, and $z$ is a classical Gaussian variable, with zero
mean and CM\ $\xi I$\ where $\xi\geq0$. All Gaussian channels are
teleportation-covariant and, therefore, Choi-stretchable (with an asymptotic
Choi matrix). Teleportation-covariance is given by the fact that any
displacement of the input\ $\hat{x}\rightarrow\hat{x}+d_{k}$ is mapped into a
displacement $Td_{k}$ on the output.

Depending on the specific canonical form we have different expressions in
Eq.~(\ref{genFORM}). We have:

\begin{itemize}
\item The thermal-loss channel $\mathcal{E}_{\text{\textrm{loss}}}(\eta
,\bar{n})$ with transmissivity $0\leq\eta\leq1$ and $\bar{n}$ thermal photons.
This\ is described by\
\begin{equation}
\hat{x}\rightarrow\sqrt{\eta}\hat{x}+\sqrt{1-\eta}\hat{x}_{E}~.
\end{equation}
For $\bar{n}=0$, the channel $\mathcal{E}_{\text{\textrm{loss}}}%
(\eta):=\mathcal{E}_{\text{\textrm{loss}}}(\eta,0)$\ is called pure-loss
channel or just \textquotedblleft lossy channel\textquotedblright.

\item The amplifier channel $\mathcal{E}_{\text{\textrm{amp}}}(\eta,\bar{n})$
with gain $\eta>1$ and $\bar{n}$ thermal photons (in the main text we use the
letter $g$ for the gain). This\ corresponds to the transformation
\begin{equation}
\SP{\hat{x}\to\sqrt{\eta}\,\hat{x}+\sqrt{\eta-1}\,Z\hat{x}_E,~~Z:=\operatorname{diag}(1,-1).}
\end{equation}
For $\bar{n}=0$, the channel $\mathcal{E}_{\text{\textrm{amp}}}(\eta
):=\mathcal{E}_{\text{\textrm{amp}}}(\eta,0)$ is called \textquotedblleft
quantum-limited amplifier\textquotedblright.

\item The additive-noise Gaussian channel $\mathcal{E}_{\text{\textrm{add}}%
}(\xi)$, which simply corresponds to%
\begin{equation}
\hat{x}\rightarrow\hat{x}+z.
\end{equation}

\item Finally, there are other secondary forms. One is the conjugate of the
amplifier, which is described by $\hat{x}\rightarrow\sqrt{-\eta}Z\hat{x}%
+\sqrt{1-\eta}\hat{x}_{E}$, where $\eta<0$ and $Z=\mathrm{diag}(1,-1)$ is the
reflection matrix. Then, other pathological forms~\cite{RMPa}: The $A_{2}%
$-form, which is a `half' depolarising channel and corresponds to $\hat
{x}\rightarrow(\hat{q},0)^{T}+\hat{x}_{E}$; and the $B_{1}$-form, which is
described by $\hat{x}\rightarrow\hat{x}+(0,\hat{p}_{v})^{T}$ where $v$ is the vacuum.
\end{itemize}

\subsection*{Coherent and reverse coherent information of a Gaussian channel}

Here we discuss the computation of the (reverse) coherent information for the
most important single-mode Gaussian channels, i.e., the thermal-loss channel,
the amplifier channel and the additive-noise Gaussian channel. Compactly,
their action on input quadratures is given by
\SP{\begin{equation}
\hat{x}\to\sqrt{\eta}\,\hat{x}
+\sqrt{|1-\eta|}\,Z_\eta\hat{x}_E+z,
\end{equation}
where $Z_\eta=I$ for $0\le\eta\le1$ and $Z_\eta=Z$ for $\eta>1$. Here $\eta \ge 0$ is the transmission (or gain),}
$E$ is the environmental mode
in a thermal state with $\bar{n}$ mean photons, and $z$ is a classical
Gaussian variable with CM\ $\xi I\geq0$. The Choi matrix $\rho_{\mathcal{E}}$
of this Gaussian channel $\mathcal{E}=\mathcal{E}(\eta,\bar{n},\xi)$ is
defined as an asymptotic limit. At the input we consider a sequence of TMSV
states $\Phi^{\mu}$ with CM as in Eq.~(\ref{EPR_CMa}). Then, at the output, we
get a sequence of finite-energy Gaussian states
\begin{equation}
\rho_{\mathcal{E}}^{\mu}:=\mathcal{I}\otimes\mathcal{E}(\Phi^{\mu}),
\label{seqnorm}%
\end{equation}
whose limit defines $\rho_{\mathcal{E}}:=\lim_{\mu}\rho_{\mathcal{E}}^{\mu}$.
The quasi-Choi matrices $\rho_{\mathcal{E}}^{\mu}$ are zero-mean Gaussian
states with CM%
\begin{equation}
V^{\mu}(\eta,\bar{n},\xi)=\left(
\begin{array}
[c]{cc}%
\mu & \gamma\\
\gamma & \beta
\end{array}
\right)  \oplus\left(
\begin{array}
[c]{cc}%
\mu & -\gamma\\
-\gamma & \beta
\end{array}
\right)  ,~~\beta:=\eta\mu+\left\vert 1-\eta\right\vert \left(  \bar{n}%
+\frac{1}{2}\right)  +\xi,~~\gamma:=\sqrt{\eta(\mu^{2}-1/4)}.
\label{CMeigtocom}%
\end{equation}

Let us consider the symplectic eigenvalues of the output CM in
Eq.~(\ref{CMeigtocom}), which are given by~\cite{RMPa}
\begin{equation}
\nu_{\pm}=\sqrt{\frac{\Delta\pm\sqrt{\Delta^{2}-4\det V^{\mu}}}{2}}%
,~~\Delta:=\mu^{2}+\beta^{2}-2\gamma^{2}.
\end{equation}
Using the formula of the von Neumann entropy for Gaussian states and the
definitions of the coherent information $I_{\mathrm{C}}$ and reverse coherent
information $I_{\mathrm{RC}}$, we may write
\begin{equation}
I_{\mathrm{C}}(\mathcal{E},\Phi^{\mu})=I(A\rangle B)_{\rho_{\mathcal{E}}^{\mu
}}=s(\beta)-s(\nu_{-})-s(\nu_{+}),~~I_{\mathrm{RC}}(\mathcal{E},\Phi^{\mu
})=I(A\langle B)_{\rho_{\mathcal{E}}^{\mu}}=s(\mu)-s(\nu_{-})-s(\nu_{+}),
\label{revcoh11}%
\end{equation}
where function $s(\cdot)$ is given in Eq.~(\ref{hFUNCTaa}).

It is easy to see that these quantities are continuous and \SP{asymptotically monotonic in $\mu$ (nondecreasing wherever they are positive),}
for any fixed values of $\eta$, $\bar{n}$ and $\xi$. For instance, for the
lossy channel ($0\leq\eta\leq1$, $\bar{n}=\xi=0$), we simply have%
\begin{equation}
I(A\rangle B)_{\rho_{\mathcal{E}}^{\mu}}=s\left[  \frac{1-\eta}{2}+\eta
\mu\right]  -s\left[  \frac{\eta}{2}+(1-\eta)\mu\right]  ,~~I(A\langle
B)_{\rho_{\mathcal{E}}^{\mu}}=s(\mu)-s\left[  \frac{\eta}{2}+(1-\eta
)\mu\right]  . \label{revcohFINITE}%
\end{equation}
Thus, the limit for $\mu\rightarrow+\infty$ in the expressions of
Eq.~(\ref{revcoh11}) is regular and finite. The asymptotic values represent
the coherent and reverse coherent information of the considered Gaussian
channels, i.e., we have
\begin{equation}
I_{\mathrm{C}}(\mathcal{E})=I(A\rangle B)_{\rho_{\mathcal{E}}}:=\lim_{\mu
}I(A\rangle B)_{\rho_{\mathcal{E}}^{\mu}},~~I_{\mathrm{RC}}(\mathcal{E}%
)=I(A\langle B)_{\rho_{\mathcal{E}}}:=\lim_{\mu}I(A\langle B)_{\rho
_{\mathcal{E}}^{\mu}},
\end{equation}
as already defined in Eqs.~(\ref{Icv1}) and~(\ref{Icv2}). Correspondingly, the
hashing inequality can be safely extended to the limit, i.e., from
\begin{equation}
\max\{I(A\rangle B)_{\rho_{\mathcal{E}}^{\mu}},I(A\langle B)_{\rho
_{\mathcal{E}}^{\mu}}\}\leq D_{1}(\rho_{\mathcal{E}}^{\mu}),
\end{equation}
we may write
\begin{equation}
\max\{I_{\mathrm{C}}(\mathcal{E}),I_{\mathrm{RC}}(\mathcal{E})\}\leq
D_{1}(\rho_{\mathcal{E}}):=\lim_{\mu}D_{1}(\rho_{\mathcal{E}}^{\mu}).
\end{equation}

For the thermal-loss channel, the best lower bound is the reverse coherent
information, given by~\cite{ReverseCAPa}%
\begin{equation}
I_{\mathrm{RC}}(\eta,\bar{n})=-\log_{2}\left(  1-\eta\right)  -h(\bar{n}),
\label{RCthermal}%
\end{equation}
where $h(\cdot)$ is the entropic function defined in Eq.~(\ref{hFUNCTaa}). In
particular, for a lossy channel ($\bar{n}=0$), one has
\begin{equation}
I_{\mathrm{RC}}(\eta)=-\log_{2}\left(  1-\eta\right)  . \label{RCpureloss}%
\end{equation}
For the amplifier channel, the best lower bound is given by the coherent
information, which is equal to~\cite{HolevoWernera,ReverseCAPa}%
\begin{equation}
I_{\mathrm{C}}(\eta,\bar{n})=\log_{2}\left(  \frac{\eta}{\eta-1}\right)
-h(\bar{n}), \label{cohamplifierch}%
\end{equation}
and becomes%
\begin{equation}
I_{\mathrm{C}}(\eta)=\log_{2}\left(  \frac{\eta}{\eta-1}\right)  ,
\label{cohqlimited}%
\end{equation}
for the quantum-limited amplifier ($\bar{n}=0$).
The coherent information and reverse coherent information of the
additive-noise Gaussian channel coincide. We have~\cite{HolevoWernera}%
\begin{equation}
I_{\mathrm{C}}(\xi)=I_{\mathrm{RC}}(\xi)=-\log_{2}\xi-\frac{1}{\ln2}.
\label{cadditiveCH}%
\end{equation}

Due to the hashing inequality, the quantities $I_{\mathrm{C}}(\mathcal{E})$
and $I_{\mathrm{RC}}(\mathcal{E})$ are achievable rates for one-way
entanglement distillation. Therefore, they also represent achievable rates for
key generation, just because an ebit is a particular type of secret bit. In
particular, ref.~\cite{ReverseCAPa} proved that $I_{\mathrm{RC}}(\mathcal{E})$
is an achievable lower bound for quantum key distribution (QKD) through a
Gaussian channel without the need of preliminary entanglement distillation. In
fact, $I_{\mathrm{RC}}(\mathcal{E})$ can be computed as the asymptotic key
rate of a coherent protocol where:

\bigskip

(i) Alice prepares TMSV states $\Phi_{AA^{\prime}}^{\mu}$ sending $A^{\prime}$
to Bob;

\bigskip

(ii) Bob heterodynes each output mode $B$ and sends final CCs back to Alice;

\bigskip

(iii) Alice measures all her modes $A$ by means of an optimal coherent
detection that reaches the Holevo bound.

\bigskip

\noindent The achievable rate of this coherent protocol is given by a
Devetak-Winter rate $R_{\text{\textrm{DW}}}$~\cite{DWrates2a}. Because Eve
holds the entire purification of Alice and Bob's Gaussian output state
$\rho_{\mathcal{E}}^{\mu}$ and Bob's detections are rank-1 measurements,\ this
rate is equal to the reverse coherent information~\cite{ReverseCAPa}
$R_{\text{\textrm{DW}}}=I(A\langle B)_{\rho_{\mathcal{E}}^{\mu}}$ computed on
Alice and Bob's output. Then, by taking the limit of $\mu\rightarrow+\infty$,
one obtains $K(\mathcal{E})\geq I_{\mathrm{RC}}(\mathcal{E})$.

\subsection*{How to compute the entanglement flux of a Gaussian channel}

Here we discuss how to compute the entanglement flux of a single-mode Gaussian
channel (in canonical form). We provide the general recipe and then we go into
details of the specific channels in the next subsections. The entanglement
flux of a Gaussian channel $\mathcal{E}$ satisfies%
\begin{equation}
\Phi(\mathcal{E})\leq\underset{\mu\rightarrow+\infty}{\lim\inf}S(\rho
_{\mathcal{E}}^{\mu}||\tilde{\sigma}_{s}^{\mu})~, \label{liminfaa}%
\end{equation}
where $\rho_{\mathcal{E}}^{\mu}$ is a sequence of quasi-Choi matrices as
defined in Eq.~(\ref{seqnorm}) with CMs as in Eq.~(\ref{CMeigtocom}), while
$\tilde{\sigma}_{s}^{\mu}$ is a suitable sequence of separable Gaussian states.

For any $\mu$, we choose a separable Gaussian state $\tilde{\sigma}_{s}^{\mu}$
with CM $\tilde{V}^{\mu}(\eta,\bar{n},\xi)$ as in Eq.~(\ref{CMeigtocom}) but
with the replacement
\begin{equation}
\gamma\rightarrow\sqrt{(\mu-1/2)(\beta-1/2)},
\end{equation}
for the off-diagonal term. At fixed marginals $\mu$ and $\beta$, this is the
most-correlated separable Gaussian state that we can build according to
Eqs.~(\ref{Vnform}) and~(\ref{sepmaxCON}); it has maximum (non-Gaussian)
discord~\cite{OptimalDISa} and minimizes the relative entropy $S(\rho
_{\mathcal{E}}^{\mu}||\tilde{\sigma}_{s}^{\mu})$ as long as $\rho
_{\mathcal{E}}^{\mu}$ is an entangled state. In the specific case where the
channel $\mathcal{E}$ is entanglement-breaking, then $\rho_{\mathcal{E}}^{\mu
}$ becomes separable and we can trivially pick $\tilde{\sigma}_{s}^{\mu}%
=\rho_{\mathcal{E}}^{\mu}$, which gives $S(\rho_{\mathcal{E}}^{\mu}%
||\tilde{\sigma}_{s}^{\mu})=0$.

In general, we are left with the analytical calculation of the relative
entropy $S(\rho_{\mathcal{E}}^{\mu}||\tilde{\sigma}_{s}^{\mu})$ between two
Gaussian states. This can be done in terms of their statistical moments
according to our formula for the REE between two arbitrary multimode Gaussian
states, which is given in the \textquotedblleft Methods\textquotedblright%
\ section of our paper. For $S(\rho_{\mathcal{E}}^{\mu}||\tilde{\sigma}%
_{s}^{\mu})$ we find regular expressions with a well-defined limit, so that we
can put $\lim\inf_{\mu}=\lim_{\mu}$ in Eq.~(\ref{liminfaa}). We provide full
algebraic details below for the various Gaussian channels.

\subsection*{Entanglement flux of a thermal-loss channel}

Consider a thermal-loss channel $\mathcal{E}_{\text{\textrm{loss}}}(\eta
,\bar{n})$ with transmissivity $0\leq\eta\leq1$ and thermal number $\bar{n}$,
so that thermal noise has variance $\omega=\bar{n}+1/2$. For $\bar{n}\geq
\eta(1-\eta)^{-1}$, this channel is entanglement-breaking and we have
$\Phi(\eta,\bar{n})=0$. For $\bar{n}<\eta(1-\eta)^{-1}$ we compute the
relative entropy $S^{\mu}:=S\left(  \rho_{\mathcal{E}}^{\mu}||\tilde{\sigma
}_{s}^{\mu}\right)  $ from the CMs $V^{\mu}(\eta\leq1,\bar{n},0)$ and
$\tilde{V}^{\mu}(\eta\leq1,\bar{n},0)$ of the zero-mean Gaussian states
$\rho_{\mathcal{E}}^{\mu}$ and $\tilde{\sigma}_{s}^{\mu}$. Using our formula
for the relative entropy between Gaussian states, we get%
\begin{equation}
S^{\mu}=-S_{1}+\frac{\Delta}{2\ln2}+\frac{1}{2}\log_{2}\left\{  \frac{2\mu
-1}{4}[2\omega-1+2\eta(\mu-\omega)]\right\}  ,
\end{equation}
where $S_{1}$ is the contribution of the von Neumann entropy, while the other
two terms come from the entropic functional $\Sigma(V^{\mu},\tilde{V}^{\mu
},0)$ (see Methods for its definition). Term $\Delta$ is analytical but too
cumbersome to be reported here.

By expanding for large $\mu$, we may write
\begin{equation}
\Delta\rightarrow2\left[  1-2\omega\coth^{-1}\left(  \frac{1+\eta}{\eta
-1}\right)  \right]  +O(\mu^{-1}),~~S_{1}\rightarrow h(\bar{n})+\log
_{2}\left[  e(1-\eta)\mu\right]  +O(\mu^{-1}),
\end{equation}
and%
\begin{equation}
\frac{1}{2}\log_{2}\left\{  \frac{2\mu-1}{4}[2\omega-1+2\eta(\mu
-\omega)]\right\}  \rightarrow\log_{2}\mu\sqrt{\eta}+O(\mu^{-1})~.
\end{equation}
Taking the limit $S^{\infty}=\lim\inf_{\mu}S^{\mu}=\lim_{\mu}S^{\mu}$, we get%
\begin{equation}
S^{\infty}=-\log_{2}\left[  (1-\eta)\eta^{\bar{n}}\right]  -h(\bar{n})~.
\label{S1app}%
\end{equation}
As a result, by replacing in Eq.~(\ref{liminfaa}), we find that the
entanglement flux of a thermal-loss channel $\mathcal{E}_{\text{\textrm{loss}%
}}(\eta,\bar{n})$ satisfies%
\begin{equation}
\Phi(\eta,\bar{n})\leq\Phi_{\text{\textrm{loss}}}(\eta,\bar{n}):=\left\{
\begin{array}
[c]{l}%
-\log_{2}\left[  (1-\eta)\eta^{\bar{n}}\right]  -h(\bar{n})~\text{~for~}%
\bar{n}<\frac{\eta}{1-\eta},\\
\\
0\text{~~~\ otherwise.}%
\end{array}
\right.  \label{thermalUBa}%
\end{equation}

The thermal bound in Eq.~(\ref{thermalUBa}) is clearly tighter than previous
bounds based on the squashed entanglement, such as the \textquotedblleft
Takeoka-Guha-Wilde\textquotedblright\ (TGW) thermal bound~\cite{TGWa}%
\begin{equation}
K_{\text{\textrm{TGW}}}=\log_{2}\left[  \frac{(1-\eta)\bar{n}+1+\eta}%
{(1-\eta)\bar{n}+1-\eta}\right]  ~, \label{TGWthermal}%
\end{equation}
and its improved version~\cite{GEWa}. However, $\Phi_{\text{\textrm{loss}}}$
does not generally coincide with the achievable lower-bound~\cite{ReverseCAPa}
given by the reverse coherent information of the channel [see
Eq.~(\ref{RCthermal})]. Thus, the generic two-way capacity of the thermal-loss
channel satisfies the sandwich relation
\begin{equation}
-\log_{2}\left(  1-\eta\right)  -h(\bar{n})\leq\mathcal{C}%
_{\text{\textrm{loss}}}(\eta,\bar{n})\leq\Phi_{\text{\textrm{loss}}}(\eta
,\bar{n}). \label{sandAAA}%
\end{equation}
It is easy to check that, for a lossy channel ($\bar{n}=0$), the bounds
Eq.~(\ref{sandAAA}) coincide, therefore establishing%
\begin{equation}
\mathcal{C}_{\text{\textrm{loss}}}(\eta)=-\log_{2}\left(  1-\eta\right)  ~.
\label{lossssy}%
\end{equation}

\subsection*{Relation with quantum discord}

The result of Eq.~(\ref{lossssy}) sets the fundamental limit for secret-key
generation, entanglement distribution and quantum communication in bosonic
lossy channels. For high loss it provides the fundamental rate-loss scaling of
$1.44\eta$ bits per channel use. This also coincides with the maximum discord
that can be distributed to the parties in a single use of the channel. In
fact, we may write the reverse coherent information of a (bosonic) channel
$\mathcal{E}$ as~\cite{DiscordQKDa} $I(A\langle B)_{\rho_{\mathcal{E}}%
}=D(B|A)-E_{\mathrm{f}}(B,E)$, where $D(B|A)$ is the quantum
discord~\cite{RMPdiscorda} of Alice and Bob's (asymptotic)\ Choi matrix
$\rho_{\mathcal{E}}$, while $E_{\mathrm{f}}(B,E)$ is the entanglement of
formation between Bob and Eve. Because a lossy channel $\mathcal{E}%
_{\text{\textrm{loss}}}:=\mathcal{E}_{\text{\textrm{loss}}}(\eta,0)$ is
dilated into a beamsplitter with a vacuum environment, we have $E_{\mathrm{f}%
}(B,E)=0$. Thus, for a lossy channel, we simultaneously have $I(A\langle
B)_{\rho_{\mathcal{E}_{\text{\textrm{loss}}}}}=D(B|A)$ and $\mathcal{C}%
_{\text{\textrm{loss}}}(\eta)=I(A\langle B)_{\rho_{\mathcal{E}%
_{\text{\textrm{loss}}}}}$. These relations lead to
\begin{equation}
\mathcal{C}_{\text{\textrm{loss}}}(\eta)=D(B|A)~,
\end{equation}
where $D(B|A)$ is the quantum discord of the (asymptotic) Gaussian Choi matrix
$\rho_{\mathcal{E}_{\text{\textrm{loss}}}}$~\cite{OptimalDISa}. In particular,
this discord can be computed as Gaussian discord~\cite{GerryDa,ParisDa}.

\subsection*{Full calculation details for the lossy channel}

For the sake of completeness, we provide the specific details of the
computation of the relative entropy $S^{\mu}$ for the specific case of a lossy
channel. After some algebra, we achieve
\begin{equation}
S^{\mu}=\log_{2}\left[  \left(  \mu-\frac{1}{2}\right)  \sqrt{\eta}\right]
-s\left[  (1-\eta)\mu+\frac{\eta}{2}\right]  +\frac{\Delta}{2\ln2}~,
\label{limitHEREa}%
\end{equation}
where%
\begin{equation}
\Delta:=\frac{c-(2\mu-1)(1-\eta)a}{b}\coth^{-1}\left[  \frac{(1-\eta
)(1-2\mu)-a}{2}\right]  -\frac{c+(2\mu-1)(1-\eta)a}{b}\coth^{-1}\left[
\frac{(1-\eta)(1-2\mu)+a}{2}\right]  ,
\end{equation}
and
\begin{align}
a  &  :=\sqrt{1-(6-\eta)\eta+4\mu\lbrack1+(4-\eta)\eta+(1-\eta)^{2}\mu]},\\
b  &  :=\sqrt{8\mu+(2\mu-1)[4\eta+(2\mu-1)(1-\eta)^{2}]},\\
c  &  :=2\eta(2\mu-1)\left(  2\sqrt{4\mu^{2}-1}-1-2\mu\right)  -\eta^{2}%
(2\mu-1)^{2}-(1+2\mu)^{2}.
\end{align}

We now insert the expression of $\Delta$ in Eq.~(\ref{limitHEREa}) and we take
the limit for $\mu\rightarrow+\infty$. This limit is defined (i.e., $\lim
\inf_{\mu}=\lim_{\mu}$) and we get%
\begin{equation}
S^{\infty}=\lim_{\mu\rightarrow+\infty}S^{\mu}=-\log_{2}(1-\eta)~.
\label{limRESULTa}%
\end{equation}
We can show this limit step-by-step. First note that, for large $\nu$, we
have
\begin{equation}
s(\nu)\rightarrow\log_{2}e\nu+O(\nu^{-1})~.
\end{equation}
Thus, in the limit of $\mu\rightarrow+\infty$, the first two terms in the RHS
of Eq.~(\ref{limitHEREa}) become%
\begin{align}
\log_{2}\left[  \left(  \mu-\frac{1}{2}\right)  \sqrt{\eta}\right]   &
\rightarrow\log_{2}\left(  \mu\sqrt{\eta}\right)  +O(\mu^{-1}),\label{lim1a}\\
-s\left[  (1-\eta)\mu+\frac{\eta}{2}\right]   &  \rightarrow-\log_{2}%
[e(1-\eta)\mu]+O(\mu^{-1}). \label{lim2a}%
\end{align}
Then, it is easy to show that, for $\mu\rightarrow+\infty$, we have
\begin{align}
\Delta &  \rightarrow\left[  -4(1-\eta)\mu+O(\mu^{0})\right]  \coth
^{-1}\left[  -2(1-\eta)\mu+O(\mu^{0})\right]  -\left[  -2+O(\mu^{-1})\right]
\coth^{-1}\left[  \frac{1+\eta}{1-\eta}+O(\mu^{-1})\right] \nonumber\\
&  \rightarrow2-\ln\eta+O(\mu^{-1})~. \label{lim3a}%
\end{align}
In conclusion, by using Eqs.~(\ref{lim1a}), (\ref{lim2a}) and~(\ref{lim3a})
into Eq.~(\ref{limitHEREa}), we obtain the final result in
Eq.~(\ref{limRESULTa}).

\subsection*{Entanglement flux of a quantum amplifier}

Consider an amplifier channel $\mathcal{E}_{\text{\textrm{amp}}}(\eta,\bar
{n})$ with gain $\eta>1$ and thermal number $\bar{n}$, so that thermal noise
has variance $\omega=\bar{n}+1/2$. For $\bar{n}\geq(\eta-1)^{-1}$ this channel
is entanglement breaking and therefore $\Phi(\eta,\bar{n})=0$. For $\bar
{n}<(\eta-1)^{-1}$ we compute the relative entropy $S^{\mu}:=S\left(
\rho_{\mathcal{E}}^{\mu}||\tilde{\sigma}_{s}^{\mu}\right)  $ from the CMs
$V^{\mu}(\eta>1,\bar{n},0)$ and $\tilde{V}^{\mu}(\eta>1,\bar{n},0)$ of the
zero-mean Gaussian states $\rho_{\mathcal{E}}^{\mu}$ and $\tilde{\sigma}%
_{s}^{\mu}$. Up to terms $O(\mu^{-1})$, we get%
\begin{equation}
S(\rho_{\mathcal{E}}^{\mu})\rightarrow h(\bar{n})+\log_{2}e(\eta
-1)\mu,~~-\mathrm{Tr}\left(  \rho_{\mathcal{E}}^{\mu}\log_{2}\tilde{\sigma
}_{s}^{\mu}\right)  \rightarrow\frac{\ln(\eta\mu^{2})+2+4\omega\coth
^{-1}\left(  \frac{\eta+1}{\eta-1}\right)  }{2\ln2}.
\end{equation}
For large $\mu$ we therefore obtain%
\begin{equation}
S^{\infty}=\log_{2}\left(  \dfrac{\eta^{\bar{n}+1}}{\eta-1}\right)  -h(\bar
{n}).
\end{equation}

Thus we find
\begin{equation}
\Phi(\eta,\bar{n})\leq\Phi_{\text{\textrm{amp}}}(\eta,\bar{n}):=\left\{
\begin{array}
[c]{l}%
\log_{2}\left(  \dfrac{\eta^{\bar{n}+1}}{\eta-1}\right)  -h(\bar
{n})~\text{~for~}\bar{n}<(\eta-1)^{-1},\\
\\
0\text{~~~\ otherwise.}%
\end{array}
\right.  \label{ampliBOUND}%
\end{equation}

In general, $\Phi_{\text{\textrm{amp}}}(\eta,\bar{n})$ does not coincide with
the best known lower bound which is given by the coherent information of the
channel in Eq.~(\ref{cohamplifierch}). Thus, the two-way capacity of a quantum
amplifier channel satisfies
\begin{equation}
\log_{2}\left(  \dfrac{\eta}{\eta-1}\right)  -h(\bar{n})\leq\mathcal{C}%
_{\text{\textrm{amp}}}(\eta,\bar{n})\leq\Phi_{\text{\textrm{amp}}}(\eta
,\bar{n}).
\end{equation}
It is easy to check that, for the quantum-limited amplifier ($\bar{n}=0$), the
previous upper and lower bounds coincide, thus determining its two-way
capacity%
\begin{equation}
\mathcal{C}_{\text{\textrm{amp}}}(\eta)=\log_{2}\left(  \dfrac{\eta}{\eta
-1}\right)  . \label{ampliMAIN}%
\end{equation}
Thus, $\mathcal{C}_{\text{\textrm{amp}}}(\eta)$ turns out to coincide with the
unassisted quantum capacity $Q_{\text{\textrm{amp}}}(\eta)$%
~\cite{HolevoWernera,Wolfa}. The result of Eq.~(\ref{ampliMAIN}) sets the
fundamental limit for key generation, entanglement distribution and quantum
communication with amplifiers. A trivial consequence is that infinite
amplification is useless for communication since $\mathcal{C}%
_{\text{\textrm{amp}}}(\infty)\rightarrow0$. For an amplifier with typical
gain $2$, the maximum achievable rate for quantum communication is just $1$
qubit per use.

\subsection*{Entanglement flux of an additive-noise Gaussian channel}

Consider an additive-noise Gaussian channel $\mathcal{E}_{\text{\textrm{add}}%
}(\xi)$ with noise variance $\xi\geq0$. For $\xi\geq1$ this channel is
entanglement breaking and therefore we have $\Phi(\xi)=0$. For $\xi<1$ we
compute the relative entropy $S^{\mu}:=S\left(  \rho_{\mathcal{E}}^{\mu
}||\tilde{\sigma}_{s}^{\mu}\right)  $ from the CMs $V^{\mu}(1,0,\xi)$ and
$\tilde{V}^{\mu}(1,0,\xi)$ of the zero-mean Gaussian states $\rho
_{\mathcal{E}}^{\mu}$ and $\tilde{\sigma}_{s}^{\mu}$. Discarding terms
$O(\mu^{-1})$, we get%
\begin{equation}
S(\rho_{\mathcal{E}}^{\mu})\rightarrow\log_{2}(e^{2}\xi\mu),~~-\mathrm{Tr}%
\left(  \rho_{\mathcal{E}}^{\mu}\log_{2}\tilde{\sigma}_{s}^{\mu}\right)
\rightarrow\frac{\ln\left[  \frac{(2\mu-1)(2\xi+2\mu-1)}{4}\right]  +2(1+\xi
)}{2\ln2}.
\end{equation}
which leads to%
\begin{equation}
S^{\infty}=\lim\inf_{\mu}S^{\mu}=\lim_{\mu}S^{\mu}=\frac{\xi-1}{\ln2}-\log
_{2}\xi~\text{.} \label{AdditiveUB}%
\end{equation}
Thus we find
\begin{equation}
\Phi(\xi)\leq\Phi_{\text{\textrm{add}}}(\xi):=\left\{
\begin{array}
[c]{l}%
\frac{\xi-1}{\ln2}-\log_{2}\xi~~\text{~for~}\xi<1,\\
\\
0\text{~~~\ otherwise.}%
\end{array}
\right.
\end{equation}
The best lower bound is its coherent information $I_{\mathrm{C}}(\xi)$ of
Eq.~(\ref{cadditiveCH}), so that the two-way capacity satisfies%
\begin{equation}
-1/\ln2-\log_{2}\xi\leq\mathcal{C}_{\text{\textrm{add}}}(\xi)\leq
\Phi_{\text{\textrm{add}}}(\xi)~.
\end{equation}

It is interesting to note how quantum communication rapidly degrades when we
compose quantum channels. For instance, a quantum-limited amplifier with gain
$2$ can transmit $Q_{2}=1$ qubit per use from Alice to Bob. This is the same
amount which can be transmitted from Bob to Charlie, through a lossy channel
with transmissivity $1/2$. By using Bob as a quantum repeater, Alice can
therefore transmit at least $1$ qubit per use to Charlie. If we remove Bob and
we compose the two channels, we obtain an additive-noise Gaussian channel with
variance $\xi=1/2$, which is limited to $Q_{2}\lesssim0.278$ qubits per use.

\subsection*{Secondary canonical forms}

For the conjugate of the amplifier it is easy to check that this channel is
always entanglement-breaking, so that it has zero flux and, therefore, zero
two-way capacity $\mathcal{C}=0$. The $A_{2}$-form~\cite{RMPa}, which is a
`half' depolarising channel, is also an entanglement-breaking channel, so that
$\Phi=\mathcal{C}=0$. Finally, for the \textquotedblleft
pathological\textquotedblright\ $B_{1}$-form~\cite{RMPa}, we find the trivial
bound $\Phi=+\infty$.

\section{Technical derivations for discrete-variable channels\label{app5}}

Given a discrete-variable channel $\mathcal{E}$ in dimension $d$, we can
easily derive its Choi matrix $\rho_{\mathcal{E}}=I\otimes\mathcal{E}(\Phi)$
from the maximally-entangled state
\begin{equation}
\Phi=\frac{1}{\sqrt{d}}\sum_{i=0}^{d-1}|ii\rangle,
\end{equation}
where $\{\left\vert 0\right\rangle ,\ldots,\left\vert i\right\rangle
,\ldots,\left\vert d-1\right\rangle \}$ is the computational basis of the
qudit. We write the spectral decomposition
\begin{equation}
\rho_{\mathcal{E}}=\sum_{k}p_{k}|\varphi_{k}\rangle\langle\varphi_{k}|,
\end{equation}
where $\mathbf{p}=\{p_{k}\}$ are the eigenvalues of the Choi matrix. The von
Neumann entropy is simply equal to the Shannon entropy of the previous
eigenvalues, i.e.,
\begin{equation}
S(\rho_{\mathcal{E}})=H(\mathbf{p}):=-\sum_{k}p_{k}\log_{2}p_{k}.
\end{equation}

From the Choi matrix we may compute the coherent and reverse coherent
information of the channel.\ In particular, for unital channels, these
quantities coincide and are given by the simple formula in
Eq.~(\ref{unitalGEN}), i.e.,
\begin{equation}
I_{\mathrm{C}}(\mathcal{E})=I_{\mathrm{RC}}(\mathcal{E})=\log_{2}%
d-S(\rho_{\mathcal{E}})=\log_{2}d-H(\mathbf{p}). \label{tocomputeRCI}%
\end{equation}
To compute the entanglement flux of the channel (upper bound), recall that we
have%
\begin{equation}
\Phi(\mathcal{E}):=E_{\mathrm{R}}(\rho_{\mathcal{E}})\leq S(\rho_{\mathcal{E}%
}||\tilde{\sigma}_{s})~,
\end{equation}
for some suitable separable state $\tilde{\sigma}_{s}$. Let us write its
spectral decomposition%
\begin{equation}
\tilde{\sigma}_{s}=\sum_{k}s_{k}|\lambda_{k}\rangle\langle\lambda_{k}|,
\end{equation}
where $|\lambda_{k}\rangle$ ($s_{k}$) are the orthogonal eigenstates
(eigenvalues) of $\tilde{\sigma}_{s}$. We may then write%
\begin{equation}
S(\rho_{\mathcal{E}}||\tilde{\sigma}_{s})=-S(\rho_{\mathcal{E}})-\mathrm{Tr}%
\left(  \rho_{\mathcal{E}}\log_{2}\tilde{\sigma}_{s}\right)  =-H(\mathbf{p}%
)-\sum_{k}\langle\lambda_{k}|\rho_{\mathcal{E}}|\lambda_{k}\rangle\log
_{2}s_{k}~. \label{SrelCALCOLO}%
\end{equation}

The separable state $\tilde{\sigma}_{s}$ may be constructed by applying the
channel $I\otimes\mathcal{E}$ to the input separable state%
\begin{equation}
\sigma_{s}=\frac{1}{d}\sum_{i=0}^{d-1}|ii\rangle\left\langle ii\right\vert ,
\end{equation}
so that we have the output%
\begin{equation}
\tilde{\sigma}_{s}=\frac{1}{d}\sum_{i=0}^{d-1}|i\rangle\langle i|\otimes
\mathcal{E}(|i\rangle\langle i|). \label{sepCANDIDATO}%
\end{equation}
This specific choice will be optimal in some cases and suboptimal in others.

\subsection*{Erasure channel in arbitrary finite dimension}

Consider a qudit in arbitrary dimension $d$ with computational basis
$\{\left\vert i\right\rangle \}$\ (results can be easily specified to the case
of a qubit $d=2$). The erasure channel replaces an incoming qudit state $\rho$
with an orthogonal erasure state $|e\rangle$ with some probability $p$. In
other words, we have the action
\begin{equation}
\mathcal{E}_{\text{\textrm{erase}}}(\rho)=(1-p)\rho+p|e\rangle\langle e|~.
\end{equation}
The simplicity of this channel relies in the fact that the input states either
are perfectly transmitted or they are lost (while in other quantum channels,
the input states are all transmitted into generally-different outputs). This
feature allows one to apply simple reasonings such as those in
ref.~\cite{ErasureChannel} which determined the $Q_{2}$ of this channel (more
precisely, the $Q_{2}$ of the qubit erasure channel, but the extension to
arbitrary $d$ is trivial).

It is easy to see that this channel is teleportation-covariant (and therefore
Choi-stretchable). In fact, any input unitary $U$ applied to the state $\rho$
is mapped into an output augmented unitary $U\oplus I$, i.e., we may write
\begin{equation}
\mathcal{E}_{\text{\textrm{erase}}}(U\rho U^{\dagger})=(U\oplus I)\mathcal{E}%
_{\text{\textrm{erase}}}(\rho)(U\oplus I)^{\dagger}.
\end{equation}
Let us write the Kraus decomposition of this channel%
\begin{equation}
\mathcal{E}_{\text{\textrm{erase}}}(\rho)=A\rho A^{\dagger}+\sum_{i=0}%
^{d-1}A_{i}\rho A_{i}^{\dagger},
\end{equation}
where $A:=\sqrt{1-p}I$ (with $I$ being the $d\times d$ identity) and
$A_{i}:=\sqrt{p}|{e}\rangle{\langle{i}|}$. We then compute its Choi matrix
\begin{equation}
\rho_{\mathcal{E}_{\text{\textrm{erase}}}}=(1-p){\Phi}+\frac{p}{d}%
(I\otimes|e\rangle\langle e|). \label{choiDec}%
\end{equation}

Note that $\mathrm{Tr}[{\Phi(}I\otimes|e\rangle\langle e|)]=0$, so that
Eq.~(\ref{choiDec}) is the spectral decomposition of $\rho_{\mathcal{E}}$ over
two orthogonal subspaces, where ${\Phi}$ has eigenvalue $1-p$, and
$I\otimes|e\rangle\langle e|$ is degenerate with $d$ eigenvalues equal to
$p/d$. Therefore, it is easy to compute the von Neumann entropy, which is%
\begin{equation}
S\left(  \rho_{\mathcal{E}_{\text{\textrm{erase}}}}\right)  =-(1-p)\log
_{2}(1-p)-p\log_{2}\left(  \frac{p}{d}\right)  .
\end{equation}
To compute the entanglement flux of the channel, we consider the separable
state $\tilde{\sigma}_{s}$ in Eq.~(\ref{sepCANDIDATO}), which here becomes
\begin{equation}
\tilde{\sigma}_{s}=\frac{1}{d}\sum_{i=0}^{d-1}\left[  (1-p)|ii\rangle\langle
ii|+p|i,e\rangle\langle i,e|\right]  .
\end{equation}
We have now all the elements to be used in Eq.~(\ref{SrelCALCOLO}), which
provides%
\begin{equation}
\Phi(\mathcal{E}_{\text{\textrm{erase}}})\leq S(\rho_{\mathcal{E}%
_{\text{\textrm{erase}}}}||\tilde{\sigma}_{s})=(1-p)\log_{2}d. \label{combERS}%
\end{equation}

For the lower bound, one can easily check that the coherent and reverse
coherent information of this channel are not sufficient to reach the upper
bound, since we get%
\begin{equation}
I_{\mathrm{C}}(\mathcal{E}_{\text{\textrm{erase}}})=(1-2p)\log_{2}%
d,~~I_{\mathrm{RC}}(\mathcal{E}_{\text{\textrm{erase}}})=(1-p)\log_{2}%
d-H_{2}(p),
\end{equation}
where the extra term $H_{2}(p)$ is the binary Shannon entropy. Note that these
quantities are achievable rates for one-way entanglement distribution but not
necessarily the optimal rates. Indeed it is easy to find a strategy based on
one-way backward CCs which reaches $(1-p)\log_{2}d$. This follows the same
reasoning of ref.~\cite{ErasureChannel}.

Alice can send halves of EPR\ states to Bob in large $n$ uses of the channel.
A fraction $1-p$ will be perfectly distributed. The identification of these
good cases can be done by Bob performing a dichotomic POVM $\{|e\rangle\langle
e|,I-|e\rangle\langle e|\}$ on each received system and communicating to Alice
which instances were perfectly transmitted. At that point Alice and Bob
possess $n(1-p)$ EPR states with $\log_{2}d$ ebits each. On average this gives
a rate of $(1-p)\log_{2}d$ ebits per channel use. Thus, one may write
\begin{equation}
D_{1}(\rho_{\mathcal{E}_{\text{\textrm{erase}}}})\geq(1-p)\log_{2}d~,
\end{equation}
whose combination with Eq.~(\ref{combERS}) provides%
\begin{equation}
\mathcal{C}(\mathcal{E}_{\text{\textrm{erase}}})=D_{2}(\mathcal{E}%
_{\text{\textrm{erase}}})=Q_{2}(\mathcal{E}_{\text{\textrm{erase}}%
})=K(\mathcal{E}_{\text{\textrm{erase}}})=\Phi(\mathcal{E}%
_{\text{\textrm{erase}}})=(1-p)\log_{2}d.
\end{equation}
Since the two-way quantum capacity of the erasure channel is already
known~\cite{ErasureChannel}, our novel result regards the determination of its
secret key capacity
\begin{equation}
K(\mathcal{E}_{\text{\textrm{erase}}})=(1-p)\log_{2}d.
\end{equation}
It is clear that, for qubits, we have $K(\mathcal{E}_{\text{\textrm{erase}}%
})=1-p$.

\subsection*{Qubit Pauli channels}

Consider a Pauli channel $\mathcal{P}$ acting on a qubit state $\rho$. The
Kraus representation of this channel is%
\begin{equation}
\mathcal{P}(\rho)=\sum_{k=0}^{3}p_{k}P_{k}\rho P_{k}^{\dagger}=p_{0}\rho
+p_{1}X\rho X+p_{2}Y\rho Y+p_{3}Z\rho Z,
\end{equation}
where $\mathbf{p}:=\{p_{k}\}$ is a probability distribution and $P_{k}%
\in\{I,X,Y,Z\}$ are Pauli operators, with $I$ the identity and%
\begin{equation}
X:=\left(
\begin{array}
[c]{cc}%
0 & 1\\
1 & 0
\end{array}
\right)  ,~Y:=\left(
\begin{array}
[c]{cc}%
0 & -i\\
i & 0
\end{array}
\right)  ,~Z:=\left(
\begin{array}
[c]{cc}%
1 & 0\\
0 & -1
\end{array}
\right)  .
\end{equation}

It is easy to check that a Pauli channel is teleportation-covariant and,
therefore, Choi-stretchable. Teleportation covariance simply comes from the
fact that the Pauli operators (qubit teleportation unitaries) either commute
or anticommute with the other Pauli operators (Kraus operators of the
channel). For a Pauli channel we can also write the stronger condition
\begin{equation}
\left[  \rho_{\mathcal{P}},P_{k}^{\ast}\otimes P_{k}\right]  =0~~\text{for any
}k\text{,}%
\end{equation}
i.e., its Choi matrix is invariant under twirling operations restricted to the
generators of the Pauli group. In fact, the Choi matrix of a Pauli channel is
Bell-diagonal, i.e., it has spectral decomposition%
\begin{equation}
\rho_{\mathcal{P}}=\sum_{k=0}^{3}p_{k}\Phi_{k},
\end{equation}
where the eigenvalues $p_{k}$\ are the channel probabilities, and the
eigenvectors $\Phi_{k}$\ are the four Bell states%
\begin{equation}
\left\{  \frac{|00\rangle\pm|11\rangle}{\sqrt{2}},~\frac{|10\rangle
\pm|01\rangle}{\sqrt{2}}\right\}  .
\end{equation}

It is clear that $S(\rho_{\mathcal{P}})=H(\mathbf{p})$. Then, using the
separable state $\tilde{\sigma}_{s}$ as in Eq.~(\ref{sepCANDIDATO}), we derive
the following upper bound for the entanglement flux of this channel%
\begin{equation}
\Phi(\mathcal{P})\leq1-H(\mathbf{p})+H_{2}(p_{1}+p_{2}). \label{PauliGEN}%
\end{equation}
Since a Pauli channel is unital, its (reverse) coherent information is just
given by $I_{\mathrm{(R)C}}(\mathcal{P})=1-H(\mathbf{p})$. Therefore, the
two-way capacity of a Pauli channel with arbitrary distribution $\mathbf{p}%
:=\{p_{k}\}$ must satisfy%
\begin{equation}
1-H(\mathbf{p})\leq\mathcal{C}(\mathcal{P})\leq1-H(\mathbf{p})+H_{2}%
(p_{1}+p_{2}). \label{PauliFIRST}%
\end{equation}

Latter result can be made stronger by exploiting the fact that $\rho
_{\mathcal{P}}$ is Bell-diagonal. For any such a state we can compute the REE
by using the formula of ref.~\cite{VedFORM}. In fact, let us set
$p_{\mathrm{max}}:=\max\{p_{k}\}$, then we may write
\begin{equation}
E_{\mathrm{R}}(\rho_{\mathcal{P}})=%
\begin{cases}
1-H_{2}(p_{\mathrm{max}}) & \text{if }p_{\mathrm{max}}\geq\frac{1}{2}\\
0 & \text{otherwise.}%
\end{cases}
\end{equation}
Thus, we have the tighter upper bound%
\begin{equation}
1-H(\mathbf{p})\leq\mathcal{C}(\mathcal{P})\leq\Phi(\mathcal{P})=%
\begin{cases}
1-H_{2}(p_{\mathrm{max}}) & \text{if }p_{\mathrm{max}}\geq\frac{1}{2}\\
0 & \text{otherwise.}%
\end{cases}
. \label{strettoPAULI}%
\end{equation}
In the following subsections, we specialize this result to depolarising and
dephasing channels.

\subsection*{Qubit depolarising channel}

This is a Pauli channel with probability distribution
\begin{equation}
\mathbf{p}=\left\{  1-\frac{3p}{4},\frac{p}{4},\frac{p}{4},\frac{p}%
{4}\right\}  ,
\end{equation}
so that we have%
\begin{equation}
\mathcal{P}_{\text{\textrm{depol}}}(\rho)=\left(  1-\frac{3p}{4}\right)
\rho+\frac{p}{4}(X\rho X+Y\rho Y+Z\rho Z)=(1-p)\rho+p\frac{I}{2}~.
\end{equation}
Let us set
\begin{equation}
\kappa(p):=1-H_{2}\left(  \frac{3p}{4}\right)  .
\end{equation}
Then, from Eq.~(\ref{strettoPAULI}), we derive the following bounds for the
two-way capacity of the depolarising channel%
\begin{equation}
\kappa(p)-\frac{3p}{4}\log_{2}3\leq\mathcal{C}(\mathcal{P}%
_{\text{\textrm{depol}}})\leq\kappa(p),
\end{equation}
for $p\leq2/3$, while $\mathcal{C}(\mathcal{P}_{\text{\textrm{depol}}})=0$ otherwise.

\subsection*{Qubit dephasing channel}

This is a Pauli channel with probability distribution $\mathbf{p}%
=\{1-p,0,0,p\}$, so that we have
\begin{equation}
\mathcal{P}_{\text{\textrm{deph}}}(\rho)=(1-p)\rho+pZ\rho Z.
\end{equation}
It is easy to see that $H(\mathbf{p})=H_{2}(p_{\mathrm{max}})=H_{2}(p)$, so
that Eq.~(\ref{strettoPAULI}) leads to
\begin{equation}
\mathcal{C}(\mathcal{P}_{\text{\textrm{deph}}})=D_{2}(\mathcal{P}%
_{\text{\textrm{deph}}})=Q_{2}(\mathcal{P}_{\text{\textrm{deph}}%
})=K(\mathcal{P}_{\text{\textrm{deph}}})=\Phi(\mathcal{P}_{\text{\textrm{deph}%
}})=1-H_{2}(p),
\end{equation}
which also coincides with the unassisted quantum capacity of this channel
$Q(\mathcal{P}_{\text{\textrm{deph}}})$~\cite{degradablea}.

\subsection*{Pauli channels in arbitrary finite dimension}

Let us now consider Pauli channels $\mathcal{P}_{d}$ in arbitrary dimension
$d\geq2$. These qudit channels are also called \textquotedblleft Weyl
channels\textquotedblright\ and they have Kraus representation
\begin{equation}
\mathcal{P}_{d}(\rho)=\sum_{a,b=0}^{d-1}p_{ab}(X^{a}Z^{b})\rho(X^{a}%
Z^{b})^{\dagger},
\end{equation}
where $p_{ab}$ is a probability distribution for $a,b\in\mathbb{Z}%
_{d}:=\{0,1,\ldots,d-1\}$. Here $X$ and $Z$ are generalized Pauli operators
whose action on the computational basis $\{\left\vert j\right\rangle \}$\ is
given by%
\begin{equation}
X\left\vert j\right\rangle =\left\vert j\oplus1\right\rangle ~,~Z\left\vert
j\right\rangle =\omega^{j}\left\vert j\right\rangle ~, \label{PauligenaPP}%
\end{equation}
where $\oplus$ is the modulo $d$ addition and
\begin{equation}
\omega:=\exp(i2\pi/d).
\end{equation}
These operators satisfy the generalized commutation relation%
\begin{equation}
Z^{b}X^{a}=\omega^{ab}X^{a}Z^{b}.
\end{equation}
Not only for $d=2$ (qubits) but also at any $d\geq2$ a Pauli channel is teleportation-covariant.

The channel's Choi matrix $\rho_{\mathcal{P}_{d}}$ is Bell-diagonal with
eigenvalues $\{p_{ab}\}$, so that we may write its von Neumann entropy in
terms of the Shannon entropy as follows%
\begin{equation}
S(\rho_{\mathcal{P}_{d}})=H(\{p_{ab}\}).
\end{equation}
Note that the Choi matrix can also be written as%
\begin{equation}
\rho_{\mathcal{P}_{d}}=\frac{1}{d}\sum_{a,b,j,k}^{d-1}p_{ab}(I\otimes
X^{a}Z^{b})|jj\rangle\langle kk|(I\otimes X^{a}Z^{b})^{\dagger}=\frac{1}%
{d}\sum_{a,b,j,k}^{d-1}p_{ab}~\omega^{b(j-k)}|j,j\oplus a\rangle\langle
k,k\oplus a|.
\end{equation}
Then, let us consider a separable state $\tilde{\sigma}_{s}$ which is
constructed as in Eq.~(\ref{sepCANDIDATO}). This state can be re-written as
\begin{equation}
\tilde{\sigma}_{s}=\frac{1}{d}\sum_{a,b,i=0}^{d-1}p_{ab}|i,i\oplus
a\rangle\langle i,i\oplus a|.
\end{equation}

By applying Eq.~(\ref{SrelCALCOLO}), we find
\begin{equation}
\Phi(\mathcal{P}_{d})\leq\log_{2}d-H(\{p_{ab}\})+H(\{p_{a}\}),
\end{equation}
where $p_{a}:=\sum_{b=0}^{d-1}p_{ab}$. Since the $d$-dimensional Pauli channel
is unital, we may also write $I_{\mathrm{(R)C}}(\mathcal{P}_{d})=\log
_{2}d-H(\{p_{ab}\})$, so that we derive the following bounds for its two-way
capacity
\begin{equation}
\log_{2}d-H(\{p_{ab}\})\leq\mathcal{C}(\mathcal{P}_{d})\leq\log_{2}%
d-H(\{p_{ab}\})+H(\{p_{a}\}),
\end{equation}
which generalizes Eq.~(\ref{PauliFIRST}) to arbitrary dimension $d$. In the
following two subsections, we consider the specific cases of the depolarising
and dephasing channels in arbitrary finite dimension $d$.

\subsection*{Depolarising channel in arbitrary finite dimension}

Consider a depolarising channel acting on a qudit with dimension $d\geq2$.
This channel can be written as%
\begin{equation}
\mathcal{P}_{d\text{\textrm{-depol}}}(\rho)=(1-p)\rho+p\frac{I}{d}=A\rho
A^{\dagger}+\sum_{i,j=0}^{d-1}A_{ij}\rho A_{ij}^{\dagger},
\end{equation}
where $A=\sqrt{1-p}I$ and $A_{ij}=\sqrt{p/d}|{i}\rangle\langle{j}|$. Its Choi
matrix is the isotropic state%
\begin{equation}
\rho_{\mathcal{P}_{d\text{\textrm{-depol}}}}=(1-p)|{\Phi}\rangle\langle{\Phi
}|+\frac{p}{d^{2}}I\otimes I,
\end{equation}
satisfying the twirling condition%
\begin{equation}
\left[  \rho_{\mathcal{P}_{d\text{\textrm{-depol}}}},U^{\ast}\otimes U\right]
=0,
\end{equation}
for any qudit unitary $U$.

The REE of an isotropic state can be evaluated exactly by using the formula of
ref.~\cite{Synak}. Thus we can exactly compute the entanglement flux of the
$d$-dimensional depolarising channel. Let us set%
\begin{equation}
f:=\frac{d^{2}-1}{d^{2}}p,~~\kappa(d,p):=\log_{2}d-H_{2}\left(  f\right)
-f\log_{2}(d-1).
\end{equation}
Then, we may write the following expression%
\begin{equation}
\Phi(\mathcal{P}_{d\text{\textrm{-depol}}})=E_{\mathrm{R}}\left(
\rho_{\mathcal{P}_{d\text{\textrm{-depol}}}}\right)  =%
\begin{cases}
\kappa(d,p) & \text{if }p\leq\frac{d}{d+1},\\
0 & \text{otherwise.}%
\end{cases}
\end{equation}

Because the depolarising channel is unital, we may use Eq.~(\ref{tocomputeRCI}%
) to compute its (reverse) coherent information. We specifically find%
\begin{equation}
I_{\mathrm{(R)C}}(\mathcal{P}_{d\text{\textrm{-depol}}})=\log_{2}%
d-H_{2}\left(  f\right)  -f\log_{2}(d^{2}-1)=\kappa(d,p)-f\log_{2}(d+1).
\end{equation}
Thus, the two-way capacity of this channel must satisfy the bounds%
\begin{equation}
\kappa(d,p)-f\log_{2}(d+1)\leq\mathcal{C}(\mathcal{P}_{d\text{\textrm{-depol}%
}})\leq\kappa(d,p),
\end{equation}
for $p\leq d/(d+1)$, while zero otherwise.

\subsection*{Dephasing channel in arbitrary finite dimension}

Consider a generalized dephasing channel affecting a qudit in arbitrary
dimension $d\geq2$. This channel has Kraus
representation~\cite{depha1,depha2}
\begin{equation}
\mathcal{P}_{d\text{\textrm{-deph}}}(\rho)=\sum_{i=0}^{d-1}P_{i}Z^{i}%
\rho(Z^{\dag})^{i},~~,
\end{equation}
where $Z$ is the generalized Pauli (phase-flip) operator defined in
Eq.~(\ref{PauligenaPP}), and $P_{i}$ is the probability of $i$ phase flips.

The channel's Choi matrix is
\begin{equation}
\rho_{\mathcal{P}_{d\text{\textrm{-deph}}}}=\sum_{mjl}\frac{P_{m}}{d}%
\exp\left[  \frac{2i\pi}{d}(j-l)m\right]  |jj\rangle\langle ll|.
\end{equation}
By diagonalizing, we find $d$ non-zero eigenvalues $\mathbf{P}:=\{P_{0}%
,\ldots,P_{d-1}\}$, so that the Von Neumann entropy is given by
\begin{equation}
S(\rho_{\mathcal{P}_{d\text{\textrm{-deph}}}})=H(\mathbf{P}). \label{VonNe}%
\end{equation}
The separable state $\tilde{\sigma}_{s}$ in Eq.~(\ref{sepCANDIDATO}) turns out
to be diagonal in the computational basis and takes the form%
\begin{equation}
\tilde{\sigma}_{s}=\sum_{i=0}^{d-1}\frac{1}{d}|ii\rangle\langle ii|~.
\end{equation}
Thus, using Eq.~(\ref{SrelCALCOLO}), we find
\begin{equation}
\Phi(\mathcal{P}_{d\text{\textrm{-deph}}})\leq S(\rho_{\mathcal{P}%
_{d\text{\textrm{-deph}}}}||\tilde{\sigma}_{s})=\log_{2}d-H(\mathbf{P}).
\end{equation}

Since this channel is unital, from Eq.~(\ref{tocomputeRCI}) we have that its
(reverse) coherent information is $I_{\mathrm{(R)C}}(\mathcal{P}%
_{d\text{\textrm{-deph}}})=\log_{2}d-H(\mathbf{P})$, so that lower and upper
bounds coincide. This means that this channel is distillable and its two-way
capacity is equal to
\begin{equation}
C(\mathcal{P}_{d\text{\textrm{-deph}}})=D_{2}(\mathcal{P}%
_{d\text{\textrm{-deph}}})=Q_{2}(\mathcal{P}_{d\text{\textrm{-deph}}%
})=K(\mathcal{P}_{d\text{\textrm{-deph}}})=\Phi(\mathcal{P}%
_{d\text{\textrm{-deph}}})=\log_{2}d-H(\mathbf{P}).
\end{equation}

\subsection*{Amplitude damping channel}

The amplitude damping channel describes the process of energy dissipation
through spontaneous emission in a two-level system. Its application to an
input qubit state is defined by the Kraus representation%
\begin{equation}
\mathcal{E}_{\text{\textrm{damp}}}(\rho)=%
{\textstyle\sum\nolimits_{i=0,1}}
A_{i}\rho A_{i}^{\dagger}, \label{damp1}%
\end{equation}
where%
\begin{equation}
A_{0}:=\left\vert 0\right\rangle \left\langle 0\right\vert +\sqrt
{1-p}\left\vert 1\right\rangle \left\langle 1\right\vert ,~~A_{1}:=\sqrt
{p}\left\vert 0\right\rangle \left\langle 1\right\vert , \label{damp3}%
\end{equation}
and $p$ is the probability of damping. This channel is not
teleportation-covariant. In fact, because we have%
\begin{equation}
\left\vert 0\right\rangle \left\langle 0\right\vert \rightarrow\left\vert
0\right\rangle \left\langle 0\right\vert ,~~\left\vert 1\right\rangle
\left\langle 1\right\vert \rightarrow p\left\vert 0\right\rangle \left\langle
0\right\vert +(1-p)\left\vert 1\right\rangle \left\langle 1\right\vert ,
\end{equation}
there is no unitary $U$ able to realize $U\mathcal{E}_{\text{\textrm{damp}}%
}(\left\vert 0\right\rangle \left\langle 0\right\vert )U^{\dagger}%
=\mathcal{E}_{\text{\textrm{damp}}}(X\left\vert 0\right\rangle \left\langle
0\right\vert X)$ for Pauli operator $X$.

The amplitude damping channel can be decomposed as
\begin{equation}
\mathcal{E}_{\text{\textrm{damp}}}=\mathcal{E}_{\text{\textrm{CV}}%
\rightarrow\text{\textrm{DV}}}\circ\mathcal{E}_{\eta(p)}\circ\mathcal{E}%
_{\text{\textrm{DV}}\rightarrow\text{\textrm{CV}}},
\end{equation}
where $\mathcal{E}_{\text{\textrm{DV}}\rightarrow\text{\textrm{CV}}}$ is an
identity mapping from the original qubit (e.g. a spin) to a single-rail qubit,
which is the subspace of a bosonic mode spanned by the vacuum and the single
photon states; then, $\mathcal{E}_{\eta(p)}$ is a lossy channel with
transmissivity $\eta(p):=1-p$; finally, $\mathcal{E}_{\text{\textrm{CV}%
}\rightarrow\text{\textrm{DV}}}$ is an identity mapping from the single-rail
qubit to the original qubit. Note that the two mappings can be performed via
perfect hybrid teleportation and the middle lossy channel preserves the
$2$-dimensional effective Hilbert space of the system. \

Thanks to this decomposition, we can include $\mathcal{E}_{\text{\textrm{DV}%
}\rightarrow\text{\textrm{CV}}}$ in Alice's LOs and $\mathcal{E}%
_{\text{\textrm{CV}}\rightarrow\text{\textrm{DV}}}$ into Bob's LOs. The middle
lossy channel $\mathcal{E}_{\eta(p)}$ can therefore be stretched into its
asymptotic Choi matrix $\rho_{\mathcal{E}_{\eta(p)}}$. Overall, this means
that the amplitude damping channel can be stretched into the asymptotic
resource state $\sigma=\rho_{\mathcal{E}_{\eta(p)}}$ by means of an asymptotic
simulation. By applying teleportation stretching, we therefore reduce the
output of an adaptive protocol to the form%
\begin{equation}
\rho_{\mathbf{ab}}^{n}:=\rho_{\mathbf{ab}}(\mathcal{E}_{\text{\textrm{damp}}%
}^{\otimes n})=\bar{\Lambda}\left(  \rho_{\mathcal{E}_{\eta(p)}}^{\otimes
n}\right)  ,
\end{equation}
where both $\bar{\Lambda}$ and $\rho_{\mathcal{E}_{\eta(p)}}$ are intended as
asymptotic limits. Thus, our reduction method provides the upper bound%
\begin{equation}
\mathcal{C}(\mathcal{E}_{\text{\textrm{damp}}})\leq\Phi\left[  \mathcal{E}%
_{\eta(p)}\right]  =-\log_{2}p.
\end{equation}
We can combine the latter result with the fact that we cannot exceed the
logarithm of the dimension of the input Hilbert space (see this simple
\textquotedblleft dimensionality bound\textquotedblright\ in the main text, in
the discussion just before Proposition~5). This leads to
\begin{equation}
\mathcal{C}(\mathcal{E}_{\text{\textrm{damp}}})\leq\min\{1,-\log_{2}p\}.
\label{REEubdamp}%
\end{equation}

The best lower bound is given by optimizing the reverse coherent information
over the input states $\rho_{u}=\mathrm{diag}(1-u,u)$ for $0\leq u\leq1$. In
fact, we have~\cite{RevCohINFOa}%
\begin{equation}
I_{\mathrm{RC}}(p):=\max_{u}I_{\mathrm{RC}}(\mathcal{E}_{\text{\textrm{damp}}%
},\rho_{u})=\max_{u}\{H_{2}\left(  u\right)  -H_{2}\left(  up\right)  \}.
\label{LBampDAMP}%
\end{equation}
This is an achievable lower bound for entanglement distribution assisted by a
final round of backward CCs. Note that this is strictly higher than the
$Q_{1}=Q$ of the channel, which is given by~\cite{RevCohINFOa}%
\begin{equation}
Q_{1}(\mathcal{E}_{\text{\textrm{damp}}})=\max_{u}\{H_{2}[u(1-p)]-H_{2}\left(
up\right)  \}. \label{Q1damp}%
\end{equation}
Thus, in total, we may write
\begin{equation}
I_{\mathrm{RC}}(p)\leq\mathcal{C}(\mathcal{E}_{\text{\textrm{damp}}})\leq
\min\{1,-\log_{2}p\}, \label{C2damp}%
\end{equation}
which is shown in Supplementary Fig.~\ref{dampAPP}a. See the next section for
the derivation of a tighter upper bound which is based on the squashed
entanglement.\begin{figure}[ptbh]
\begin{center}
\vspace{-0.0cm} \includegraphics[width=0.70\textwidth]{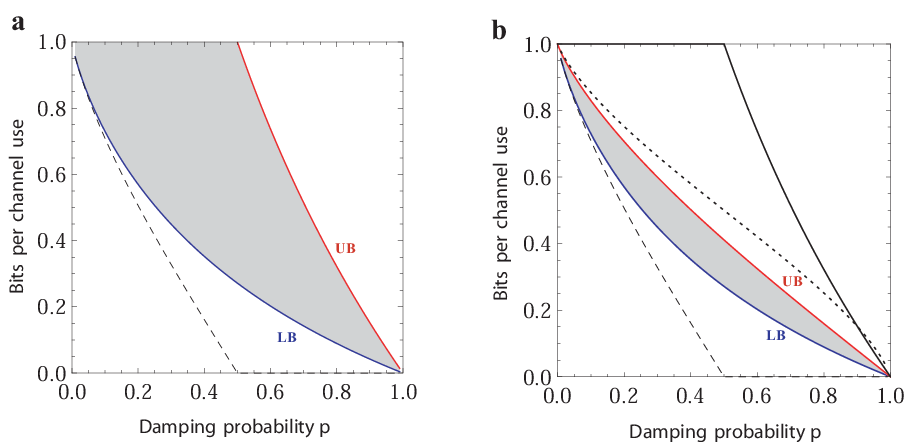}
\vspace{-0.5cm}
\end{center}
\caption{Two-way capacity of the amplitude damping channel $\mathcal{E}%
_{\text{\textrm{damp}}}$ with probability $p$. (\textbf{a})~The two-way
capacity $\mathcal{C}(\mathcal{E}_{\text{\textrm{damp}}})$ is contained in the
shadowed area identified by the lower bound (LB) and the upper bound (UB) of
Eq.~(\ref{C2damp}). Note the separation from the unassisted quantum capacity
$Q$ of the channel (dashed). (\textbf{b})~More precisely, $\mathcal{C}%
(\mathcal{E}_{\text{\textrm{damp}}})$ is contained in the shadowed area
identified by the lower bound (LB) and the upper bound (UB) of
Eq.~(\ref{C2wayampDAMP}). We also plot the unassisted quantum capacity $Q$ of
the channel (dashed), the REE upper bound of Eq.~(\ref{REEubdamp}) (solid),
and the bound of ref.~\cite{GEWa}\ (dotted).}%
\label{dampAPP}%
\end{figure}

\subsection*{Amplitude damping channel: Upper bound based on the squashed
entanglement}

An alternative upper bound for the two-way capacity of a quantum channel is
its squashed entanglement, i.e., we may write~\cite{SquashChannela}
\begin{equation}
\mathcal{C}(\mathcal{E})\leq E_{\text{\textrm{sq}}}(\mathcal{E}).
\end{equation}
The squashed entanglement of an arbitrary channel $\mathcal{E}$, from system
$A$ to system $B$, is defined as~\cite{SquashChannela}%
\begin{equation}
E_{\text{\textrm{sq}}}(\mathcal{E}):=\frac{1}{2}\max_{\rho_{A}}\inf
_{V_{C\rightarrow EF}}[S(B|E)_{\omega}+S(B|F)_{\omega}], \label{SQ}%
\end{equation}
where $\rho_{A}$ is an arbitrary input state, and $\omega$ is the global
output state%
\begin{equation}
\omega_{BEF}:=V_{C\rightarrow EF}[U_{A\rightarrow BC}^{\mathcal{E}}(\rho
_{A})],
\end{equation}
with $U_{A\rightarrow BC}^{\mathcal{E}}$ being an isometric extension of
$\mathcal{E}$ and $V_{C\rightarrow EF}$ being an arbitrary \textquotedblleft
squashing isometry\textquotedblright.

In Eq.~(\ref{SQ}), the terms in the brackets are conditional von Neumann
entropies computed over $\omega_{BEF}$, i.e.,
\begin{equation}
S(B|E)_{\omega}=S(BE)_{\omega}-S(E)_{\omega},~~S(B|F)_{\omega}=S(BF)_{\omega
}-S(F)_{\omega}. \label{entr2}%
\end{equation}
Then note that the most general input state reads%
\begin{equation}
\rho_{A}=\left(
\begin{array}
[c]{cc}%
1-\gamma & c^{\ast}\\
c & \gamma
\end{array}
\right)  ,
\end{equation}
where $\gamma\in\lbrack0,1]$ is the population of the excited state
$|1\rangle$, while the off-diagonal term $|c|\leq\sqrt{(1-\gamma)\gamma}$
accounts for coherence. Thus, the maximization in Eq.~(\ref{SQ}) is mapped
into a maximization over parameters $\gamma$ and $c$.

Let us compute the squashed entanglement of the amplitude damping channel
$\mathcal{E}_{\text{damp}}$. Recall that its action is described by
Eq.~(\ref{damp1}) with Kraus operators as in Eq.~(\ref{damp3}). In the
computational basis $\{\left\vert 00\right\rangle ,\left\vert 01\right\rangle
,\left\vert 10\right\rangle ,\left\vert 11\right\rangle \}$, the unitary
dilation of $\mathcal{E}_{\text{damp}}$ is therefore given by the following
matrix
\begin{equation}
U_{p}=\left(
\begin{array}
[c]{cccc}%
1 & 0 & 0 & 0\\
0 & \sqrt{1-p} & \sqrt{p} & \\
0 & -\sqrt{p} & \sqrt{1-p} & 0\\
0 & 0 & 0 & 1
\end{array}
\right)  ,
\end{equation}
so that we may write
\begin{equation}
\mathcal{E}_{\text{\textrm{damp}}}(\rho_{A})=\text{\textrm{Tr}}_{C}[U_{p}%
(\rho_{A}\otimes|0\rangle\langle0|_{C})U_{p}^{\dagger}], \label{damp4}%
\end{equation}
where $C$ is an environmental qubit prepared in the fundamental state
$|0\rangle$. It is clear that Eq.~(\ref{damp4}) expresses the isometric
extension of the channel, i.e., it corresponds to $\mathcal{E}%
_{\text{\textrm{damp}}}(\rho_{A})=\mathrm{Tr}_{C}[U_{A\rightarrow
BC}^{_{\text{\textrm{damp}}}}(\rho_{A})]$.

As a squashing channel we consider another amplitude damping channel but with
damping probability equal to $1/2$, so that its unitary dilation is
$V=U_{1/2}$. In other words, we consider the squashing isometry
$V_{C\rightarrow EF}=\left[  U_{C\rightarrow EF}^{_{\text{\textrm{damp}}}%
}\right]  _{p=1/2}$ (so that we are more precisely deriving an upper bound of
the squashed entanglement of the channel). Let us derive the global output
state $\omega_{BEF}$\ step-by-step.

The state of systems $B$ and $C$ at the output of the dilation $U_{p}$ is
given by
\begin{equation}
\rho_{BC}:=U_{p}(\rho_{A}\otimes|0\rangle\langle0|_{C})U_{p}^{\dagger}=\left(
\begin{array}
[c]{cccc}%
1-\gamma & \sqrt{p}c^{\ast} & \sqrt{1-p}c^{\ast} & 0\\
c\sqrt{p} & p\gamma & \sqrt{1-p}\sqrt{p}\gamma & 0\\
c\sqrt{1-p} & \sqrt{1-p}\sqrt{p}\gamma & (1-p)\gamma & 0\\
0 & 0 & 0 & 0
\end{array}
\right)  .
\end{equation}
Now the system $C$ is sent through the squashing amplitude damping channel
with probability $1/2$. At the output of the dilation $U_{1/2}$ we have the
final output state%
\begin{equation}
\omega_{BEF}=(I_{B}\otimes U_{1/2})\rho_{BC}\otimes|0\rangle\langle
0|_{F}(I_{B}\otimes U_{1/2})^{\dagger}=\left(
\begin{array}
[c]{cccccccc}%
1-\gamma & \frac{\sqrt{p}c^{\ast}}{\sqrt{2}} & \frac{\sqrt{p}c^{\ast}}%
{\sqrt{2}} & 0 & \sqrt{1-p}c^{\ast} & 0 & 0 & 0\\
\frac{c\sqrt{p}}{\sqrt{2}} & \frac{p\gamma}{2} & \frac{p\gamma}{2} & 0 &
\frac{\sqrt{(1-p)p}\gamma}{\sqrt{2}} & 0 & 0 & 0\\
\frac{c\sqrt{p}}{\sqrt{2}} & \frac{p\gamma}{2} & \frac{p\gamma}{2} & 0 &
\frac{\sqrt{(1-p)p}\gamma}{\sqrt{2}} & 0 & 0 & 0\\
0 & 0 & 0 & 0 & 0 & 0 & 0 & 0\\
c\sqrt{1-p} & \frac{\sqrt{(1-p)p}\gamma}{\sqrt{2}} & \frac{\sqrt{(1-p)p}%
\gamma}{\sqrt{2}} & 0 & \gamma-p\gamma & 0 & 0 & 0\\
0 & 0 & 0 & 0 & 0 & 0 & 0 & 0\\
0 & 0 & 0 & 0 & 0 & 0 & 0 & 0\\
0 & 0 & 0 & 0 & 0 & 0 & 0 & 0
\end{array}
\right)  .
\end{equation}

We now proceed with the calculation of the entropies in Eq.~(\ref{entr2}),
which are obtained from the eigenvalues of the reduced states $\rho_{BE}$,
$\rho_{BF}$, $\rho_{E}$ and $\rho_{F}$. We obtain
\begin{equation}
\rho_{E}=\rho_{F}=\left(
\begin{array}
[c]{cc}%
1-\frac{p\gamma}{2} & \frac{\sqrt{p}c^{\ast}}{\sqrt{2}}\\
\frac{c\sqrt{p}}{\sqrt{2}} & \frac{p\gamma}{2}%
\end{array}
\right)  ,
\end{equation}
with eigenvalues%
\begin{equation}
\lambda_{1,2}=\frac{1}{2}\left(  1\pm\sqrt{2|c|^{2}p+(p\gamma-1)^{2}}\right)
. \label{lambda12}%
\end{equation}
The eigenvalues of $\rho_{BE}$ and $\rho_{BF}$ are too complicated to be
reported here but it is easy to check that, exactly as for $\lambda_{1,2}$ in
previous Eq.~(\ref{lambda12}), their dependence on $c$ is just through the
modulus $|c|$, so that we can choose $c$ to be real without losing generality.

Because $c$ is real, we also have that the entropic functional $\digamma
(\rho)=S(B|E)_{\omega}+S(B|F)_{\omega}$ computed over the input state $\rho$
is exactly the same as that computed over the state $Z\rho Z$, with $Z$ being
the phase-flip Pauli operator. Using the latter observation, together with the
concavity of the conditional quantum entropy, one simply has
\begin{equation}
\digamma(\rho)=\frac{\digamma(\rho)+\digamma(Z\rho Z)}{2}\leq\digamma\left(
\frac{\rho+Z\rho Z}{2}\right)  =\digamma(\bar{\rho}),
\end{equation}
where $\bar{\rho}$ is diagonal. This means that we may reduce the maximization
to diagonal input states ($c=0$).

As a result, we may just consider
\begin{equation}
\rho_{E}=\rho_{F}=\left(
\begin{array}
[c]{cc}%
1-\frac{p\gamma}{2} & 0\\
0 & \frac{p\gamma}{2}%
\end{array}
\right)  ,
\end{equation}
with eigenvalues
\begin{equation}
\lambda_{1}=\frac{p\gamma}{2},~\lambda_{2}=1-\frac{p\gamma}{2},
\end{equation}
and%
\begin{equation}
\rho_{BE}=\rho_{BF}=\left(
\begin{array}
[c]{cccc}%
\frac{1}{2}(p-2)\gamma+1 & 0 & 0 & 0\\
0 & \frac{p\gamma}{2} & \frac{\sqrt{(1-p)p}\gamma}{\sqrt{2}} & 0\\
0 & \frac{\sqrt{(1-p)p}\gamma}{\sqrt{2}} & \gamma-p\gamma & 0\\
0 & 0 & 0 & 0
\end{array}
\right)  ,
\end{equation}
with eigenvalues%
\begin{equation}
\nu_{1}=\frac{\gamma}{2}(2-p),~\nu_{2}=1-\nu_{1},~\nu_{3}=\nu_{4}=0.
\end{equation}

From the previous eigenvalues, we compute the conditional quantum entropies in
Eq.~(\ref{entr2}). Thus, we find that the squashed entanglement of the
amplitude damping channel must satisfy the bound
\begin{equation}
E_{\text{\textrm{sq}}}(\mathcal{E}_{\text{\textrm{damp}}})\leq\max_{\gamma
}\left\{  H_{2}(\nu_{1})-H_{2}(\lambda_{1})\right\}  , \label{SQ1}%
\end{equation}
where $H_{2}$ is the binary Shannon entropy of Eq.~(\ref{binSHANNON}). 
\SP{The function
$H_2(\nu_1)-H_2(\lambda_1)$ is concave in $\gamma$ but is not generally
symmetric. Substituting Eqs.~(S228) and (S230) into Eq.~(S231) gives
\begin{equation}
E_{\mathrm{sq}}(\mathcal{E}_{\mathrm{damp}})
\leq \max_{0\leq\gamma\leq1}
\left\{
H_2\!\left[\gamma\left(1-\frac p2\right)\right]
-H_2\!\left(\frac{p\gamma}{2}\right)
\right\}.\label{UBampDAMP}
\end{equation}}

If we choose a squashing amplitude damping channel with generic probability of
damping $\eta$ and we repeat the calculation from the beginning we obtain the
following bound for the squashed entanglement
\begin{equation}
E_{\text{\textrm{sq}}}(\mathcal{E}_{\text{\textrm{damp}}})\leq\frac{1}{2}%
\max_{\gamma}\min_{\eta}\left\{  H_{2}(\gamma-p\gamma\eta)+\right.  \left.
H_{2}\left[  \gamma(1-p+p\eta)\right]  -H_{2}\left[  p\gamma(1-\eta)\right]
-H_{2}(p\gamma\eta)\right\}  . \label{settingg}%
\end{equation}
The minimum of the function inside the curly bracket is for $\eta=1/2$, so our
choice of a balanced amplitude damping channel as a squashing channel is now
justified. Note that the sub-optimal choice $\eta=0$ corresponds to use the
identity as squashing channel; correspondingly, the right hand side of
Eq.~(\ref{settingg}) becomes half of the entanglement-assisted classical
capacity $C_{\mathrm{A}}$ of the amplitude damping channel, i.e.,
\begin{equation}
E_{\text{\textrm{sq}}}(\mathcal{E}_{\text{\textrm{damp}}})\leq\frac{1}%
{2}C_{\mathrm{A}}(\mathcal{E}_{\text{\textrm{damp}}})=\frac{1}{2}\max_{\gamma
}\left\{  H_{2}(\gamma)+H_{2}\left[  \gamma(1-p)\right]  -H_{2}(p\gamma
)\right\}  .
\end{equation}

In conclusion, combining the lower bound of Eq.~(\ref{LBampDAMP}) and the
upper bound of Eq.~(\ref{UBampDAMP}), we find that the two-way capacity of the
amplitude damping channel is within the sandwich%
\begin{equation}
\max_{u}\{H_{2}\left(  u\right)  -H_{2}\left(  up\right)  \}\leq
\mathcal{C}(\mathcal{E}_{\text{\textrm{damp}}})
\le 
\SP{\max_{0\leq\gamma\leq1}
\left\{
H_2\!\left[\gamma\left(1-\frac p2\right)\right]
-H_2\!\left(\frac{p\gamma}{2}\right)
\right\}.}
\label{C2wayampDAMP}%
\end{equation}
This is shown in Supplementary Fig.~\ref{dampAPP}b, which also contains a
comparison with the previous upper bound based on the REE. Note that, for high
damping ($p\simeq1$), the upper bound in Eq.~(\ref{C2wayampDAMP}) provides the
scaling of \SP{$\lesssim 0.804(1-p)$}
bits per channel use, while
Eq.~(\ref{REEubdamp}) provides the scaling of $\lesssim1.44(1-p)$ bits per
channel use.

\section{Maximum rates achievable by current QKD protocols\label{app6}}

We consider the state of the art in high-rate QKD, by analyzing the maximum
rates which are achievable by current practical protocols in CVs and DVs. We
assume the optimal asymptotic case of infinitely long keys, so that
finite-size effects are negligible. We also assume ideal parameters. For CVs
this means: Unit detector efficiency, zero excess noise, large modulation and
unit reconciliation efficiency. For DVs this means: Unit detector
efficiencies, zero dark count rates, zero intrinsic error, unit error
correction efficiency, and no other internal loss in the devices. Note that
all the following results are already present in the literature or are easily
derivable from those in the literature. They are given to the reader for the
sake of completeness.

\subsection*{Continuous-variable protocols}

$\bullet~$No-switching protocol~\cite{Weea}. This is the practical
CV\ protocol with the highest secret key rate. It is based on coherent states
and heterodyne detection. In reverse reconciliation (RR), its maximum secret
key rate over a lossy channel with transmissivity $\eta$ is equal to
\begin{equation}
R_{\text{\textrm{no-switch}}}=\log_{2}\left[  \frac{\eta}{e(1-\eta)}\right]
+s\left(  \frac{2-\eta}{2\eta}\right)  , \label{CV1way}%
\end{equation}
where $s(\cdot)$ is the entropic function given in Eq.~(\ref{hFUNCTaa}). For
high loss ($\eta\simeq0$), it scales as $\simeq\eta/2\ln2$, which is $1/2$ of
the secret key capacity.

$\bullet~$Switching protocol~\cite{GG02a,Freda}. This was the first practical
CV\ protocol. It is based on coherent states and homodyne detection (with
switching between the two quadratures). In RR, it reaches the rate
\begin{equation}
R_{\text{\textrm{switch}}}=\frac{1}{2}\log_{2}\left(  \frac{1}{1-\eta}\right)
,
\end{equation}
which is $1/2$ of the secret key capacity. For high loss, it clearly scales as
the previous protocol.

$\bullet~$CV measurement-device-independent (MDI)
protocol~\cite{CVMDIQKDa,Correspondencea}. This is based on coherent states
sent to an untrusted relay implementing a CV\ Bell detection. Alice-relay
channel has transmissivity $\eta_{A}$\ and Bob-relay channel has
transmissivity $\eta_{B}$, so that the total Alice-Bob channel transmissivity
is $\eta=\eta_{A}\eta_{B}$. In the symmetric configuration with the relay
perfectly in the middle ($\eta_{A}=\eta_{B}$)~\cite{CVMDIQKDa,Carloa}, it has
maximum rate%
\begin{equation}
R_{\text{\textrm{CVMDI-sym}}}=\log_{2}\left[  \frac{\eta}{e^{2}(1-\sqrt{\eta
})}\right]  +s\left(  \frac{1}{\sqrt{\eta}}-\frac{1}{2}\right)  .
\end{equation}
In the asymmetric configuration ($\eta_{A}\neq\eta_{B}$), it has maximum rate%
\begin{equation}
R_{\text{\textrm{CVMDI-asym}}}=s\left(  \frac{1}{\eta_{B}}-\frac{1}{2}\right)
-s\left(  \frac{2-\eta_{A}-\eta_{B}}{2|\eta_{A}-\eta_{B}|}\right)  +\log
_{2}\left(  \frac{\eta_{A}\eta_{B}}{e|\eta_{A}-\eta_{B}|}\right)  .
\end{equation}
In particular, in the most asymmetric configuration, where the relay coincides
with Alice ($\eta_{A}=1$)~\cite{CVMDIQKDa,Gaea}, we recover the one-way rate
of Eq.~(\ref{CV1way}).

$\bullet~$CV\ two-way protocols~\cite{Twowaya}. In the first main variant, Bob
sends coherent states to Alice, who randomly displaces their amplitudes before
sending them back to Bob for heterodyne detection. In RR (Bob as encoder),
this protocol has maximum rate%
\begin{equation}
R_{\text{\textrm{2way-het}}}=\frac{1}{2}\left\{  s\left[  \frac{2-\eta
+\eta^{2}}{2\eta(1+\eta)}\right]  +\log_{2}\left[  \frac{\eta(1+\eta
)}{e(1-\eta)}\right]  \right\}  .
\end{equation}
In the second main variant, the protocol runs as before except that Bob's
measurement is homodyne detection (with switching between the quadratures). In
RR, it has maximum rate~\cite{Carlo2ww}%
\begin{equation}
R_{\text{\textrm{2way-hom}}}=\frac{1}{4}\log_{2}\left(  \frac{1+\eta^{2}%
}{1-\eta}\right)  .
\end{equation}
It is easy to check that both the variants scale as $\simeq\eta/4\ln2$ for
high loss. Despite the fact that two-way protocols have lower key rates than
one-way protocols in a lossy channel, they are more robust when excess noise
is present. In this case, one considers the \textquotedblleft security
threshold\textquotedblright\ of the protocol which is defined as the maximum
tolerable excess noise above which the rate becomes negative. Two-way
protocols have higher security thresholds than one-way
protocols~\cite{Twowaya,Carlo2ww}.

\subsection*{Discrete-variable protocols}

Here we consider various DV protocols. As said before, we assume the optimal
asymptotic case of infinitely long keys and also ideal parameters, which here
means: Unit detector efficiencies, zero dark count rates, zero intrinsic
error, unit error correction efficiency, and no other internal loss in the
devices. Under these assumptions, we consider the ideal BB84 protocol with
single photon sources~\cite{BB84a}, the BB84 with weak coherent pulses and
decoy states~\cite{Decoya,Scarania}, and DV-MDI-QKD~\cite{MDI1a,MDI2a}.

Let us consider the BB84 protocol~\cite{BB84a} assuming that Alice's source
generates perfect single-photon pulses. The general formula of the key rate
can be found in ref.~\cite{Scarania}. It reduces to the following expression%
\begin{equation}
R=\bar{R}\left\{  \left[  1-H_{2}\left(  Q\right)  \right]  -\delta
(Q)\right\}  , \label{KAPPA}%
\end{equation}
where $H_{2}$ is the binary Shannon entropy. In Eq.~(\ref{KAPPA}),
$\delta(Q)=f~H_{2}(Q)$ is a function accounting for the leak of information
from imperfect error correction, $f\geq1$ is the efficiency of the classical
error correction codes, $Q$ is the total error rate (QBER), and $\bar{R}$ is
the total detection rate after quantum communication (the raw key). Under
ideal conditions of zero dark-count rates, unit efficiency detectors, perfect
visibility, and perfect classical error correction $(f=1)$, one has $Q=0$ and
obtains the following maximum rate $R_{\text{\textrm{BB84-1ph}}}=\eta/2$,
setting the maximum rate for the current DV protocols.

A realistic photon source is a device emitting attenuated coherent pulses. In
this case, the performance of the protocol depends on an additional parameter
which is the intensity of the source. In the BB84 protocol, with weak coherent
pulses and decoy states~\cite{Decoya}, Alice randomly changes the intensity
$\mu$ of the pulses, and reveals publicly their values during the final
classical communication. In this way Eve cannot adapt her attacks during the
quantum communication. The $\mu$-dependent key rate of the protocol is given
by~\cite{Scarania}%
\begin{equation}
R^{\mu}=\bar{R}\left\{  Y_{0}^{\mu}+Y_{1}^{\mu}\left[  1-H_{2}\left(
\frac{Q^{\mu}}{Y_{1}^{\mu}}\right)  \right]  -\delta(Q^{\mu})\right\}  ,
\end{equation}
where $Q^{\mu}$ is the $\mu$-dependent\ QBER, and $Y_{n}^{\mu}=R_{n}^{\mu
}/\bar{R}$ is the ratio between the $\mu$-dependent\ detection rate
$R_{n}^{\mu}$, associated to Alice sending $n$ photons, and the total
detection rate $\bar{R}$. Assuming ideal conditions, one finds $R^{\mu
}=e^{-\mu}\eta\mu/2$. The optimal key rate is obtained by maximizing over the
intensities, i.e., $R=\max_{\mu}R^{\mu}$. It is easy to check that the optimum
is given by $\mu=1$ and the maximum key rate becomes
$R_{\text{\textrm{BB84-decoy}}}=\eta/(2e)$.

Finally consider DV-MDI-QKD. The general expression of the key rate is given
by the following expression~\cite{MDI2a}%
\begin{equation}
R=P_{Z}^{11}Y_{Z}^{11}\left[  1-H_{2}(e_{Z}^{11})\right]  -G_{Z}~\delta
(Q_{Z}), \label{DV-MDI}%
\end{equation}
where $P_{Z}^{11}=\mu_{A}\mu_{B}\exp[-(\mu_{A}+\mu_{B})]$ is the joint
probability that both emitters (with intensities $\mu_{A}$ and $\mu_{B}$)
generate a single-photon pulse. The quantity $Y_{Z}^{11}$ gives the gain in
the $Z$-basis (one assumes $Y_{X}^{11}=Y_{Z}^{11}$ for the $X$-basis), and
$e_{Z}^{11}$ is the error rate in the $Z$-basis. Finally, the quantity $G_{Z}$
describes the gain and $Q_{Z}$ the QBER, both in the $Z$-basis. Under ideal
conditions, the $\mu$-dependent key rate becomes%
\begin{equation}
R^{\mu_{A}\mu_{B}}=\frac{1}{2}e^{-(\mu_{A}+\mu_{B})}\eta_{A}\eta_{B}\mu_{A}%
\mu_{B},
\end{equation}
where $\eta_{A}$ and $\eta_{B}$ are the transmissivities of Alice's and Bob's
channels. It is easy to check that the maximum is taken for $\mu_{A}=\mu
_{B}=1$, providing $R_{\text{\textrm{DV-MDI}}}=\eta/(2e^{2})$.


\section{Other aspects: Energy constraints and cost of the CC}

\subsection*{Input energy constraints}

It is important to remark that the two-way capacities that we computed for
Gaussian channels are bounded quantities, which do not diverge even if the
maximum is achieved in the limit of infinite input energy (excluding the case
of a pathological canonical form). In fact, one may consider an alphabet of
input states whose mean number of photons is capped at some finite value
$\bar{N}$. This assumption automatically defines a hard-constrained two-way
capacity $\mathcal{C}(\mathcal{E},\bar{N})$. For a bosonic Gaussian channel,
$\mathcal{C}(\mathcal{E},\bar{N})$ is increasing in $\bar{N}$\ but also
upper-bounded by the entanglement flux of the channel $\Phi(\mathcal{E})$. (In
fact, note that all the procedure of teleportation stretching still applies if
we enforce an input energy constraint for the adaptive protocols. For instance
the constraint can be realized by a pinching map which is then absorbed in
Alice's LOs). As a result, the asymptotic limit of the unconstrained capacity
$\mathcal{C}(\mathcal{E}):=\lim_{\bar{N}}\mathcal{C}(\mathcal{E},\bar{N})$ is
finite. This is clearly true for $Q_{2}(\mathcal{E})$, $D_{2}(\mathcal{E})$
and $K(\mathcal{E})$, but the situation would be different for the two-way
classical capacity of the channel.

Another possibility is imposing a \textquotedblleft soft
constraint\textquotedblright\ on the input energy. This means to fix the
average number of photons at the input to some finite value $\bar{m}$. In this
case, it is interesting to see that our \textquotedblleft
unconstrained\textquotedblright\ upper bounds remain sufficiently tight even
in the presence of such an energy constraint. The best way to show this is
considering our main result for the lossy channel with arbitrary
transmissivity $\eta$, for which we have proven that
\begin{equation}
Q_{2}(\eta)=D_{2}(\eta)=K(\eta)=\Phi(\eta)=-\log_{2}(1-\eta).
\end{equation}
Even if we constrain the input to $\bar{m}$\ mean photons, it is easy to show that:

\begin{description}
\item[(1)] The unconstrained bound $\Phi(\eta)$ is still very tight, since it
is rapidly approached from below by the reverse coherent information computed
at finite energy;

\item[(2)] The unconstrained bound $\Phi(\eta)$ remains tighter than other
constrained bounds based on the squashed entanglement, even when $\bar{m}$ is
of the order of a few photons.
\end{description}

\noindent Let us start with point (1). From Eq.~(\ref{revcohFINITE}), we see
that the reverse coherent information associated with a lossy channel and a
TMSV state is%
\begin{equation}
I_{\mathrm{RC}}(\bar{m},\eta)=h(\bar{m})-h\left[  (1-\eta)\bar{m}\right]  ,
\label{revcoh22}%
\end{equation}
which is obtained by setting $\mu=\bar{m}+1/2$ in Eq.~(\ref{revcohFINITE}) and
using the $h$-function of Eq.~(\ref{hFUNCTaa}). In Supplementary
Fig.~\ref{loeppic}, we see that $I_{\mathrm{RC}}(\bar{m},\eta)$ rapidly
approaches the unconstrained upper bound $\Phi(\eta)$ already for $\bar
{m}\simeq1-5$ photons.\begin{figure}[ptbh]
\begin{center}
\vspace{-0.1cm} \includegraphics[width=0.90\textwidth]{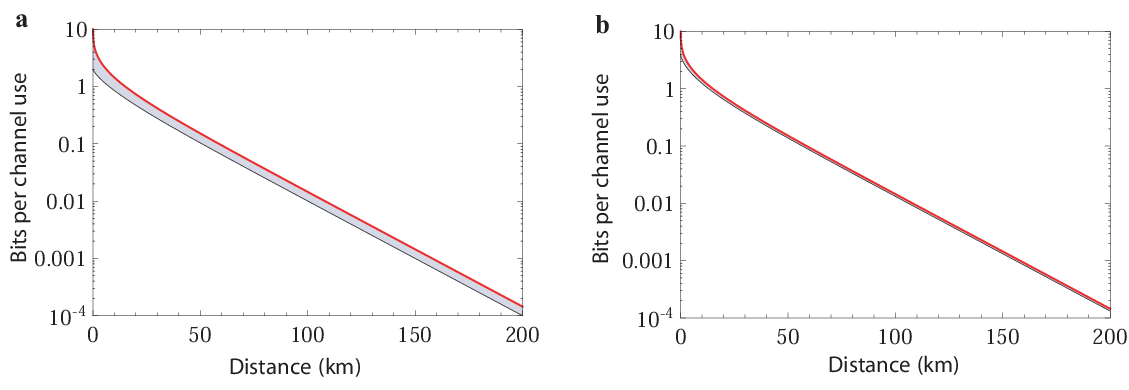}
\vspace{-0.5cm}
\end{center}
\caption{Study of the tightness of our unconstrained upper bound for the lossy
channel under the assumption of energy-constrained inputs. (\textbf{a})~We
plot the unconstrained upper bound $\Phi(\eta)=-\log_{2}(1-\eta)$ (upper red
line) and the constrained lower bound $I_{\mathrm{RC}}(\bar{m},\eta)$ (lower
black line) given by the reverse coherent information in Eq.~(\ref{revcoh22})
assuming $\bar{m}=1$ mean photons at the input. Both are plotted in terms of
the distance (km) assuming the standard loss rate of $0.2$dB/km. The
constrained two-way capacity of the lossy channel is in the middle dark area.
(\textbf{b}) Same as in (\textbf{a}) but now with $\bar{m}=5$ mean photons. We
see how the unconstrained upper bound is rapidly reached already with a few
photons.}%
\label{loeppic}%
\end{figure}

Let us now discuss point (2). We compare the unconstrained upper bound
$\Phi(\eta)$ with the unconstrained TGW upper bound for the lossy
channel~\cite{TGWa}%
\begin{equation}
K_{\text{\textrm{TGW}}}(\eta)=\log_{2}\left(  \frac{1+\eta}{1-\eta}\right)  ,
\label{KTGWW}%
\end{equation}
and its energy-constrained version
\begin{equation}
K_{\text{\textrm{TGW}}}(\eta,\bar{m})=h\left[  \frac{(1+\eta)\bar{m}}%
{2}\right]  -h\left[  \frac{(1-\eta)\bar{m}}{2}\right]  . \label{TGWfinite}%
\end{equation}
(Note that the latter was just a partial result~\cite{TGWa} used to derive the
bound in Eq.~(\ref{KTGWW}) for $\bar{m}\rightarrow+\infty$).

In Supplementary Fig.~\ref{tgwcomp} we clearly see that $\Phi(\eta)$ not only
is tighter than $K_{\text{\textrm{TGW}}}(\eta)$ but also outperforms the
constrained version $K_{\text{\textrm{TGW}}}(\eta,\bar{m})$ for all input
energies down to one mean photon. This is certainly true in the regime of
intermediate-long distances ($>25$~km), where DV-QKD protocols have ideal
performances at one mean photon per channel use. At short distances
($<25$~km), energy constraints do not really have so much practical value
since we can efficiently use highly-modulated CV-QKD whose number of photons
is high enough to approach the asymptotic infinite-energy behavior. In
general, note that CV-QKD protocols with highly-modulated Gaussian states can
be used at any distance. Their performance is not limited by the input energy,
but critically depends on the efficiency of the output detection scheme and
the quality of the data-processing (reconciliation efficiency).
\begin{figure}[ptbh]
\vspace{-0.0cm}
\par
\begin{center}
\includegraphics[width=0.45\textwidth]{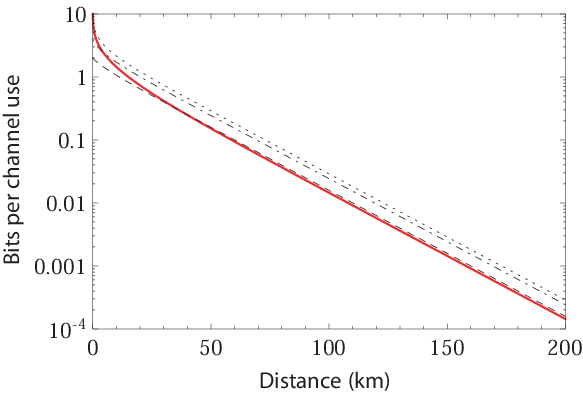} \vspace{+3.2cm}
\vspace{-0.2cm} \vspace{-3.3cm}
\end{center}
\caption{Comparison with previous bounds based on the squashed entanglement.
We compare the unconstrained upper bound $\Phi(\eta)=-\log_{2}(1-\eta)$ (solid
red line) with the unconstrained TGW bound (dotted), and the constrained TGW
bound for $\bar{m}=5$ mean photons (dashed-dotted) and $\bar{m}=1$ mean
photons (dashed line). Bounds are plotted in terms of distance (km) assuming
the standard loss rate of $0.2$dB/km. Note that $\Phi(\eta)$ remains the
tighter upper bound even if we constrain the input energy down to one mean
photon. This is true everywhere, except for short distances (where the energy
constraint is not so interesting since we can efficiently use highly-modulated
CV-QKD).}%
\label{tgwcomp}%
\end{figure}

\subsection*{Cost of classical communication}

It is important to discuss the cost associated with the CCs. In fact, in order
to achieve its performance, an optimal protocol will need a certain number of
classical bits per channel use. Furthermore, the physical transmission of
these bits is ultimately restricted by the speed of light. It is therefore
essential to consider these aspects in order to translate a capacity, which is
expressed in terms of target-bits\ (e.g. secret bits) per channel use, into a
practical throughput, which is expressed in terms of target-bits per second.
Consider the case of a bosonic lossy channel which is the most important for
quantum optical communications.

By definition, an adaptive protocol is assisted by unlimited and two-way CCs.
This is a very general formulation but it has an issue for practical
applications: An adaptive protocol, which may be optimal in terms of
target-bits per channel use, may have zero throughput in terms of target-bits
per second, just due the fact that its implementation may require infinite
rounds of feed-forward and feedback CCs in each channel use. The existence of
such protocol is not excluded by the TGW bounds~\cite{TGWa} of
Eqs.~(\ref{KTGWW}) and~(\ref{TGWfinite}), which are non-tight and do not have
control on the CCs. By contrast, this problem is completed solved by our bound.

In fact, for any distillable channel $\mathcal{E}$ (e.g., bosonic lossy
channel, quantum-limited amplifier, dephasing or erasure channel), the generic
two-way capacity $\mathcal{C}(\mathcal{E})$ is equal to $D_{1}(\rho
_{\mathcal{E}})$, which is the entanglement distillable from the Choi matrix
of the channel by means of one-way CCs (forward, from Alice to Bob, or
backward, from Bob to Alice). This means that an optimal protocol achieving
the capacity is non-adaptive and it does not involve infinite rounds of CCs,
but just a single round of forward or backward CCs.

For the specific case of a bosonic lossy channel, with transmissivity $\eta$,
we find that an optimal key-generation protocol, achieving the repeaterless
bound $K(\eta)=-\log_{2}(1-\eta)$, can be implemented by using backward CCs.
In fact, as already discussed in~\ref{app4}, an optimal key-generation
protocol is the following: Alice prepares TMSV states $\Phi_{AA^{\prime}}%
^{\mu}$ sending $A^{\prime}$ to Bob; Bob heterodynes each output mode, with
outcome $Y$, and sends final CCs back to Alice; Alice measures all her modes
$A$ by means of an optimal coherent detection. Taking the limit for large
$\mu$, the key rate of the parties achieves the bound $K(\eta)$.

\SP{For this reverse-reconciliation protocol, we consider the following
formal continuous-outcome quantity:
\begin{equation}
\gamma_{\mathrm{CC}}:=S(Y|A)
=S(Y)-[S(A)-S(A|Y)].
\end{equation}
Here $S(Y)=H(Y)$ is the differential entropy of Bob's heterodyne
outcomes, while $S(A)$ and
$S(A|Y)$ are the von Neumann entropies of Alice's reduced state
$\rho_A$ and conditional state $\rho_{A|Y}$. These quantities are
computable for finite $\mu$. By taking the limit for large $\mu$,
we obtain the following loose upper bound:
\begin{equation}
\gamma_{\mathrm{CC}}(\eta)
\leq \log_2(2\pi e)\approx 4.09,\label{CCnnn}
\end{equation}
which provides a formal continuous-outcome cost estimate
that excludes resolution-dependent digitization.}


\section{Advances in\label{app7} channel simulation}

The idea of channel simulation was originally introduced by
Bennett-DiVincenzo-Smolin-Wootters (BDSW)~\cite{B2} as a simple modification
of the original teleportation protocol. Instead of performing standard
teleportation by using a Bell state, one may consider an arbitrary mixed state
as a resource. As a result, the effect of teleportation is not an identity map
(transfer operator) but a noisy channel from the input to the output. BDSW
introduced this teleportation-simulation argument to simulate DV channels that
preserve the finite dimension $d$ of the input Hilbert space $\mathcal{H}^{d}%
$, also known as the \textquotedblleft tight\textquotedblright%
\ case~\cite{WernerTELE}. Let us discuss the BDSW simulation in more detail.

Consider a mixed state $\sigma$ of two qudits, $A$ and $B$, both having
dimension $d$, i.e., their joint Hilbert space is $\mathcal{H}_{A}^{d}%
\otimes\mathcal{H}_{B}^{d}$. The \textquotedblleft teleportation
channel\textquotedblright\ associated with the density operator $\sigma
\in\mathcal{D}(\mathcal{H}_{A}^{d}\otimes\mathcal{H}_{B}^{d})$ is the
dimension-preserving quantum channel $\mathcal{T}_{\sigma}:\mathcal{D}%
(\mathcal{H}^{d})\rightarrow\mathcal{D}(\mathcal{H}^{d})$, which is given by
teleporting an input $d$-dimensional qudit by using the resource state
$\sigma$. The procedure goes as follows. Alice measures qudit $A$ and input
qudit $a$ in a Bell detection, whose outcome $k\in\{0,\ldots,d^{2}-1\}$ is
associated with a qudit Pauli unitary $U_{k}$. This detection projects Bob's
qudit $B$ onto a $k$-dependent state. Once the outcome $k$\ is communicated to
Bob, he applies the Pauli correction $U_{k}^{-1}$ to qudit $B\ $thus
retrieving the final state on the output qudit $b$. The average over all
outcomes $k$ defines the teleportation channel $\mathcal{T}_{\sigma}$ from the
states of $a$ to those of $b$.

BDSW~\cite[Section~V]{B2} also recognized that a Pauli channel $\mathcal{E}$
(there called \textquotedblleft generalized depolarizing
channel\textquotedblright) can be simulated by teleporting over its Choi
matrix $\rho_{\mathcal{E}}$, so that $\mathcal{E}=\mathcal{T}_{\rho
_{\mathcal{E}}}$. This particular case was later re-considered in
ref.~\cite{HoroTEL} as a property of mutual reproducibility between mixed
states and quantum channels. In a few words, we may store a channel
$\mathcal{E}$ into its Choi matrix $\rho_{\mathcal{E}}$ (by sending half of an
EPR state), and then recover the channel back by performing teleportation over
$\rho_{\mathcal{E}}$. At this point, a natural question to ask is the following:

\bigskip

\noindent\textit{Can we generate other DV channels (beyond Pauli) using the
teleportation-simulation of BDSW~\cite[Section~V]{B2}? }

\bigskip

\noindent The answer is \textit{no}. In fact, ref.~\cite{SougatoBowen} showed
that the standard teleportation protocol (based on Bell detection and Pauli
corrections) performed over an arbitrary $d\times d$ state $\sigma$ can only
simulate\ a quantum channel of the form%
\begin{equation}
\mathcal{T}_{\sigma}(\rho)=\sum_{ab}\mathrm{Tr}(\sigma M_{ab})~U_{(-a)b}%
^{\dagger}~\rho~U_{(-a)b}~, \label{PauliasTELE}%
\end{equation}
where $M_{ab}:=(U_{ab}\otimes I)^{\dagger}\left\vert \Phi\right\rangle
\left\langle \Phi\right\vert (U_{ab}\otimes I)$ are the POVM elements of the
Bell detection (with $\left\vert \Phi\right\rangle $ being a $d$-dimensional
Bell state), and $U_{ab}$ are Pauli operators. This is clearly a
$d$-dimensional Pauli channel. The possibility to generate other DV channels
relies on a stronger modification of the original teleportation protocol,
where we allow for more general quantum operations~\cite{WernerTELE,Albeverio}
and also for the possibility of varying the dimension of the Hilbert space.
Recently, ref.~\cite{Leung} considered a generalization of the
teleportation-simulation argument for DV channels, using tools from
ref.~\cite{WernerTELE} and moving the first steps into the study of
teleportation covariance. Similarly, ref.~\cite{Niset} moved the first steps
in the simulation of single-mode Gaussian channels by employing Gaussian
resources and the standard CV teleportation protocol~\cite{SamKimble}.

In our paper we provide the most general and rigorous formulation. In fact, we
remove all the assumptions regarding the dimension of the quantum systems
which may also vary through the channel. Thus we may tele-simulate, DV
channels, CV channels and even hybrid channels, i.e., mappings between DVs and
CVs. More generally, our simulation is not limited to teleportation-LOCCs
(i.e., Bell detection and unitary corrections), but considers completely
general LOCCs which may also be asymptotic, i.e., defined as suitable
sequences. Furthermore, the simulating LOCCs may also include portions of the
channels (i.e., we may decompose a channel $\mathcal{E}$ as $\mathcal{E}%
_{\text{2}} \circ\mathcal{\tilde{E}} \circ\mathcal{E}_{\text{1}}$ and include
$\mathcal{E}_{\text{1}}$ and $\mathcal{E}_{\text{2}}$ in the LOCCs). For all
these reasons, we may simulate \textit{any} quantum channel at any dimension.
As discussed in the main text, the best case is when the simulation can be
done directly on the channel's Choi matrix. To identify this case we introduce
the criterion of teleportation-covariance at any dimension, finite or infinite.

Note that ours is the most general simulation to be used in quantum/private
communication, which is a setting where two remote parties can only apply
LOCCs. In this regard, it is different and more precise than the channel
simulation realized by using a deterministic version~\cite{qSIMa} of a
programmable quantum gate array (PQGA)~\cite{Processors,Port-based}. This is
also known as \textquotedblleft quantum simulation\textquotedblright%
~\cite{RafalJanek} and considers the simulation of \textquotedblleft
programmable channels\textquotedblright\ by means of \textit{joint}
operations. A programmable channel is defined as a (finite-dimensional)
channel $\mathcal{E}$ that can be simulated as
\begin{equation}
\mathcal{E}(\rho)=\Omega(\rho\otimes\sigma_{\mathcal{E}}),
\label{programmablech}%
\end{equation}
for a universal joint quantum operation $\Omega$ and some programme state
$\sigma_{\mathcal{E}}$. This clearly fails to catch the LOCC structure which
is essential for protocols of quantum/private communication. Furthermore, this
type of simulation has not been developed into an asymptotic version (via
CV\ teleportation), which is clearly needed for the representation of bosonic
channels. Finally, the universal character of the operation $\Omega$ restricts
the class of channels that can be simulated (universality implies that we
cannot include portions of the channel in the operation, missing a procedure
that allows one to simulate all channels). Our LOCC-simulation of channels
solves all these issues.

\bigskip

To conclude, we give the timeline of the previous main contributions before
our formulation of channel simulation:

\begin{description}
\item[1996] BDSW introduces the teleportation-simulation of Pauli
channels~\cite[Section~V]{B2}

\item[1997] Nielsen and Chuang introduce the PQGA~\cite{Processors}

\item[1998] Braunstein and Kimble design a realistic protocol for CV
teleportation~\cite{SamKimble}

\item[1999] Horodeckis consider the BDSW simulation for channel
reproducibility~\cite{HoroTEL}

\item[2001] Bowen and Bose show that the BDSW simulation can only simulate
Pauli channels~\cite{SougatoBowen}

\item[2001] Werner discusses generalized teleportation
protocols~\cite{WernerTELE}

\item[2008] Ji et al. use a deterministic PQGA to simulate certain
DV\ channels in the context of quantum metrology~\cite{qSIMa}

\item[2009] Niset et al. simulate Gaussian channels in the context of one-way
Gaussian entanglement distillation~\cite{Niset}

\item[2015] Leung and Matthews first discuss teleportation covariance in
connection with the simulation of DV channels~\cite{Leung}

\item[2015-6] The present paper rigorously generalizes the idea of
teleportation-simulation to CV systems (bosonic channels). More generally, it
introduces the LOCC-simulation of any channel at any dimension (including
asymptotic simulations), and identifies the criterion of
teleportation-covariance at any dimension (finite or infinite).
\end{description}

\section{Advances in protocol reduction}

Teleportation stretching is a general method to reduce adaptive protocols into
corresponding block protocols achieving exactly the same original task.
Furthermore, it may be applied to any channel at any dimension, finite or
infinite, thanks to our development of the tool of channel simulation
(see~\ref{app7}). In terms of reduction of protocols, a precursory but very
restricted argument was given in BDSW~\cite[Section~V]{B2}. Here we discuss
this preliminary argument and we point out the main and non-trivial advances
brought by our general formulation.

BDSW showed how to transform a quantum communication (QC) protocol, through a
finite-dimensional Pauli channel $\mathcal{E}$, into an entanglement
distillation (ED) protocol, implemented over mixed states $\sigma$. The
connection was established by interpreting $\mathcal{E}$ as the teleportation
channel generated by $\sigma$ (which can be taken to be a Choi-matrix for a
Pauli channel). This allowed them to prove
\begin{equation}
Q_{1}(\mathcal{E})\leq D_{1}(\sigma), \label{ub1BENNETT}%
\end{equation}
for protocols based on 1-way CCs. They also realized that the argument could
be applied to transform QC\ protocols based on 2-way CCs, so that they
implicitly extended the previous result to the following inequality
\begin{equation}
Q_{2}(\mathcal{E})\leq D_{2}(\sigma). \label{ubBENNETT}%
\end{equation}
An explicit proof for Eq.~(\ref{ubBENNETT}) is reported in Supplementary
Fig.~\ref{Bennett1}.

\begin{figure}[ptbh]
\vspace{-2.6cm}
\par
\begin{center}
\vspace{-0.3cm} \vspace{0.3cm} \vspace{1.5cm} \vspace{-1.8cm}
\includegraphics[width=0.90\textwidth]{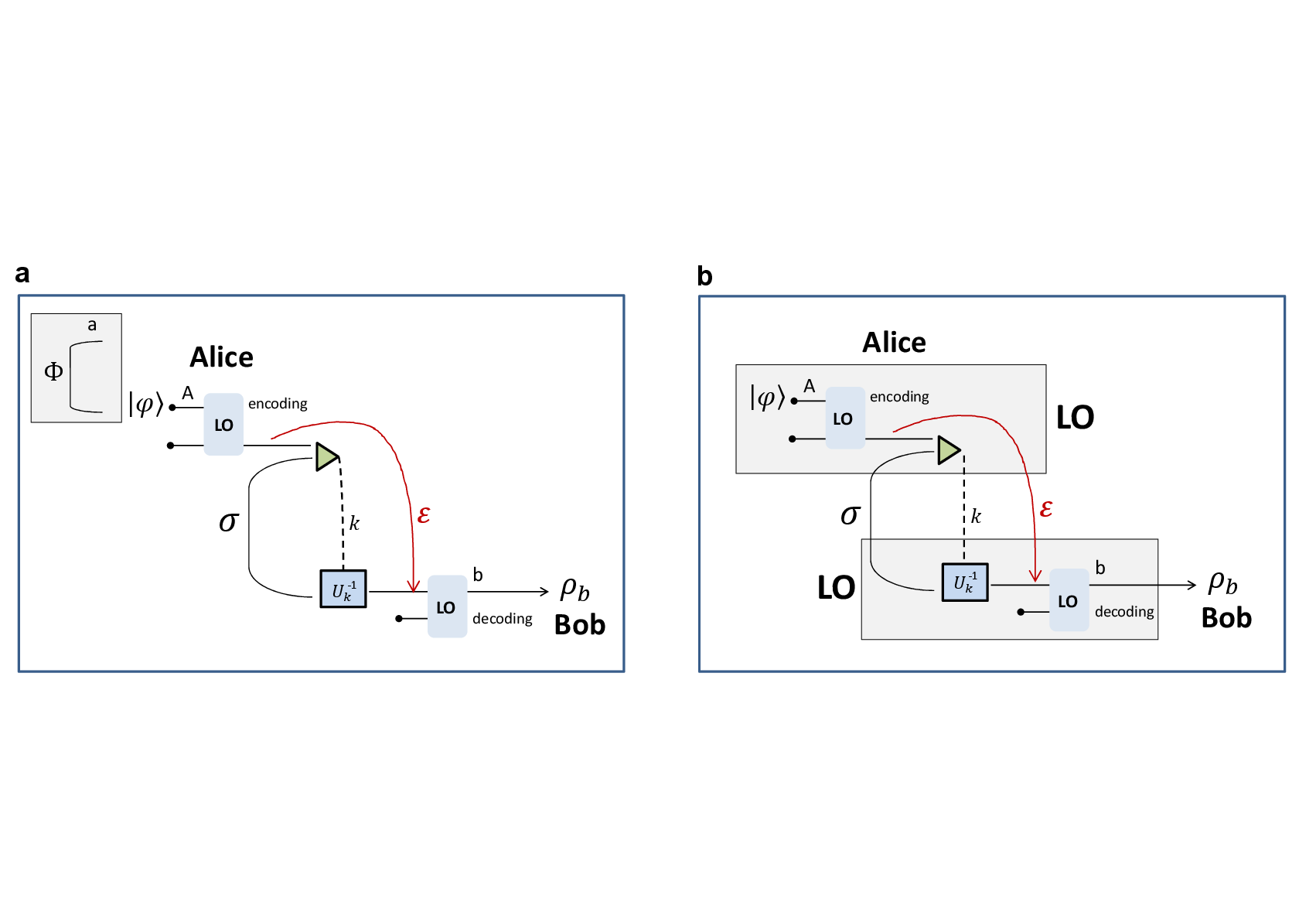} \vspace{-1.3cm}
\vspace{-0.6cm} \vspace{-1.4cm}
\end{center}
\caption{Different approaches to reduce quantum communication.~(\textbf{a}%
)~Precursory BDSW reduction argument~\cite[Section~V]{B2}, explicitly
considered for 2-way CCs. This may be described in 3 steps. (1)\textbf{~}%
Suppose that Alice and Bob implement a QC protocol for transmitting qubits
from system $A$ to system $b$ \ by means of channel $\mathcal{E}$ (red curvy
line). In the upper\ LO, Alice applies a suitable quantum error correcting
code (QECC) $\Lambda_{\text{\textrm{enc}}}^{m\rightarrow n}$ to encode an
$m$-qubit logical state $|\varphi^{(m)}\rangle$ into an $n$-qubit codeword
which is sent through $\mathcal{E}^{\otimes n}$. In the lower LO, Bob applies
a decoding operation $\Lambda_{\text{\textrm{dec}}}^{n\rightarrow m}$, so that
$\Lambda_{\text{\textrm{dec}}}^{n\rightarrow m}\circ\mathcal{E}^{\otimes
n}\circ\Lambda_{\text{\textrm{enc}}}^{m\rightarrow n}$ tends to the identity,
and the $n$-use output state $\rho_{b}^{n}$ approximates $|\varphi
^{(m)}\rangle\langle\varphi^{(m)}|$. In the general case, we assume that the
previous LOs are assisted by unlimited two-way CCs between Alice and Bob. By
optimizing over all QECCs and in the limit of infinite channel uses, one
defines the two-way quantum capacity $Q_{2}(\mathcal{E})$. (2) Notice that
Alice can use the QECC to send part of $m$\ ebits (see the Bell\ state $\Phi$
in the grey box), so that Alice and Bob share an output state $\rho_{ab}^{n}$
which approximates $\Phi^{\otimes m}$. Assuming an asymptotic and optimal
QECC, each ebit is reliably shared at the quantum capacity rate $Q_{2}%
(\mathcal{E})$. (3)~Finally, assume that the channel $\mathcal{E}$ can be
described by teleportation over the resource state $\sigma$. Any entanglement
distribution strategy through channel $\mathcal{E}$ can therefore be seen as a
specific protocol of entanglement distillation applied to the copies of
$\sigma$. This observation leads to $Q_{2}(\mathcal{E})\leq D_{2}(\sigma)$.
(\textbf{b})~Different re-organization of the quantum operations in
teleportation stretching. When we apply teleportation stretching to a QC
protocol, we directly reduce the output state as follows $\rho_{b}%
(\mathcal{E}^{\otimes n})=\bar{\Lambda}(\sigma^{\otimes n})$, for a
trace-preserving LOCC $\bar{\Lambda}$ which is not connected with ED, but
collapses the preparation $|\varphi^{(m)}\rangle\langle\varphi^{(m)}|$, the
encoding/decoding maps, and all the teleportation operations. This is not
asymptotic but done for any $n$.}%
\label{Bennett1}%
\end{figure}


Let us now compare teleportation stretching with the precursory BDSW argument.
We identify a number of non-trivial differences and advances.

\begin{enumerate}
\item \textbf{Finite-size decomposition and connection with REE}. The BDSW
reduction argument was formulated in an asymptotic fashion, i.e., for large
$n$, which is sufficient to prove Eqs.~(\ref{ub1BENNETT}) and (\ref{ubBENNETT}%
). Teleportation stretching regards any $n$, and gives the finite-size
decomposition of the output $\bar{\Lambda}(\sigma^{\otimes n})$ for a
trace-preserving LOCC $\bar{\Lambda}$ collapsing all the adaptive LOCCs. The
finite-size decomposition $\bar{\Lambda}(\sigma^{\otimes n})$ could have not
been exploited by BDSW, due to missing tools for the simplification of
$\bar{\Lambda}$. This simplification is today achieved by combining
teleportation stretching with the REE, which is the key insight giving
applicability to the technique.

\item \textbf{Task preserving}. The BDSW reduction argument was specifically
formulated to transform a QC protocol into an ED protocol, therefore changing
the task of the original protocol. In teleportation stretching, we maintain
the task. In the example of Supplementary Fig.~\ref{Bennett1}b, we show the
different re-organization of the quantum operations of the QC protocol.
Teleportation stretching would directly reduce the output of the QC protocol
as follows $\rho_{b}(\mathcal{E}^{\otimes n})=\bar{\Lambda}(\sigma^{\otimes
n})$, for a trace-preserving LOCC $\bar{\Lambda}$ which is not connected with
ED but collapses the preparation $|\varphi^{(m)}\rangle\langle\varphi^{(m)}|$,
the encoding/decoding maps, and the teleportation operations.

\item \textbf{Any task}. Maintaining the task and output of the original
protocol is crucial, because the reduction can now be applied to any kind of
adaptive protocol, not just quantum communication, but any other task,
including key generation (considered in this paper) and parameter
estimation/channel discrimination (considered in ref.~\cite{LupoPIRa}). This
aspect is also important in order to extend the procedure to more complex
scenarios, from two-way quantum communication to the presence of quantum
repeaters in arbitrary network topologies~\cite{networkPIRSa}.

\item \textbf{Any channel and dimension}. The BDSW reduction argument was
given for the restricted class of Pauli channel. Teleportation stretching is
formulated for any channel at any dimension (finite or infinite). This is
non-trivial because it involves the use of asymptotic simulations for
fundamental channels such as the bosonic Gaussian channels and the amplitude
damping channel. In general, we may write an output decomposition of the type
$\lim_{\mu}\bar{\Lambda}_{\mu}(\sigma^{\mu\otimes n})$ for sequences of
trace-preserving LOCCs $\bar{\Lambda}_{\mu}$ and resource states $\sigma^{\mu
}$.
\end{enumerate}

\bigskip

In the literature, we can also find another type of adaptive-to-block
reduction, which is based on the use of a deterministic PQGA. It is known that
a PQGA\ can simulate an arbitrary unitary or channel in a probabilistic
way~\cite{Processors}. However, as discussed in~\ref{app7}, one may also
define a class of programmable channels for which the PQGA works
deterministically: These are (finite-dimensional) channels $\mathcal{E}$ that
can be simulated as in Eq.~(\ref{programmablech}) for a universal
generally-joint quantum operation $\Omega$ and a programme state
$\sigma_{\mathcal{E}}$. It is easy to check that, in a protocol, this
\textquotedblleft quantum simulation\textquotedblright~\cite{RafalJanek} leads
to an output decomposition of the type $Q(\sigma_{\mathcal{E}}^{\otimes n})$,
where $Q$ is a joint quantum operation for Alice and Bob. Clearly this is not
suitable for quantum/private communication, where the parties are restricted
to LOCCs and, therefore, both the channel simulation and the adaptive-to-block
reduction must maintain the LOCC structure of the original protocol.
Furthermore, it lacks an asymptotic formulation which is needed for bosonic
channels and also the flexibility to include portions of the channels in the
simulating operations (these are elements introduced by our approach). It is
worth to mention that the quantum simulation plays a role for the
simplification of adaptive protocols in quantum metrology and channel
discrimination, where the parties are close (they are indeed the same entity)
and may therefore apply joint unitaries and joint measurements. See
refs.~\cite{Lorenzoa,LupoPIRa}.

\section{Advances in bounding two-way capacities}

By simulating Pauli channels, BDSW showed how to reduce a quantum
communication protocol into an entanglement distillation protocol. By
combining this argument with an opposite implication, they were able to show
that, for a Pauli channel $\mathcal{E}$, one may write $Q_{1}(\mathcal{E}%
)=D_{1}(\rho_{\mathcal{E}})$, which was implicitly extended to
\begin{equation}
Q_{2}(\mathcal{E})=D_{2}(\rho_{\mathcal{E}}). \label{BDSW96}%
\end{equation}
The latter result is not exploitable for computing the two-way quantum
capacity $Q_{2}$ unless one identifies simple (and tight) upper bounds for
$D_{2}$. Such elements were missing in 1996 but today we can exploit powerful tools.

Using today's knowledge, the simplest approach is to combine Eq.~(\ref{BDSW96}%
) with the fact that $D_{2}(\rho_{\mathcal{E}})\leq K(\rho_{\mathcal{E}})$
(since an ebit is a particular type of secret-bit) and the REE\ upper bound on
the distillable key of quantum states~\cite{KD1a}, so that $K(\rho
_{\mathcal{E}})\leq E_{\mathrm{R}}^{\infty}(\rho_{\mathcal{E}})$. All this
leads us to write
\begin{equation}
Q_{2}(\mathcal{E})\leq E_{\mathrm{R}}^{\infty}(\rho_{\mathcal{E}})\leq
E_{\mathrm{R}}(\rho_{\mathcal{E}})~. \label{tri1}%
\end{equation}
Our work shows the bound of Eq.~(\ref{tri1}) for any finite-dimensional
Choi-stretchable channel. In particular, we show that the single-letter REE
bound of~Eq.~(\ref{tri1}) is tight for dephasing and erasure channels.

The next non-trivial generalization is moving from quantum to \textit{private}
communication. In this regard, the notions of private
capacities~\cite{Devetaka} and private states~\cite{KD1a,KD2a} were available
well after 1996. Note that we may consider the secret-key capacity $K$, which
is the number of secret bits which are distributed between the parties (via
adaptive protocols), and the two-way private capacity $P_{2}$, which is the
maximum rate at which classical messages can be securely encoded and
transmitted~\cite{Devetaka}. Because of the unlimited two-way CCs and the
one-time pad, we have $P_{2}=K$. For a finite-dimensional Choi-stretchable
channel $\mathcal{E}$, it is easy to write the equivalence
\begin{equation}
P_{2}(\mathcal{E})=K(\mathcal{E})=K(\rho_{\mathcal{E}})~. \label{keyEXT}%
\end{equation}
The simplest way to show this is to apply teleportation stretching to reduce
adaptive key-generation protocols, which leads to $K(\mathcal{E}%
)=K(\rho_{\mathcal{E}})$ as in Proposition~6 of our main text. An alternate
way is to show $P_{2}(\mathcal{E})=K(\rho_{\mathcal{E}})$\ by means of a
suitable extension of the BDSW reduction argument. In fact, for a
finite-dimensional Choi-stretchable channel, we may transform a protocol of
private communication~\cite{Devetaka} through $\mathcal{E}$ into a protocol of
key-distillation~\cite{KD1a,KD2a} over the Choi matrix $\rho_{\mathcal{E}}$,
so that $P_{2}(\mathcal{E})\leq K(\rho_{\mathcal{E}})$. The latter bound is
achievable by a protocol where Alice transmits part of Bell states, so that
the parties distill a key from the output Choi matrices, which is then used to
send the message via the one-time pad. Note that these extensions from quantum
to private communication, and from entanglement to key distillation were not
available in 1996, which is why Eq.~(\ref{keyEXT}) can only be written today.
At the same time, it is surprising that Eq.~(\ref{keyEXT}) was never written
before our work, with many of the tools being available since 2005.

Now it is very important to observe that both Eqs.~(\ref{tri1})
and~(\ref{keyEXT}) cannot be used to investigate the most important setting
for quantum/private communication, which is the bosonic one. Furthermore, they
miss to provide single-letter bounds for other DV channels which involve
asymptotic simulations (e.g., amplitude damping). For these important reasons,
it is necessary to develop a general theory which is dimension-independent and
applicable to channels of any dimension, finite or infinite. This is the main
content of our Theorem~5 in the main text. This states that, for any channel
$\mathcal{E}$ stretchable into a resource state $\sigma$\ (even
asymptotically), we may write%
\begin{equation}
\mathcal{C}(\mathcal{E})\leq E_{\mathrm{R}}^{\infty}(\sigma)\leq
E_{\mathrm{R}}(\sigma), \label{genh1}%
\end{equation}
where $\mathcal{C}(\mathcal{E})$ is any among the two-way capacities
$Q_{2}(\mathcal{E})=D_{2}(\mathcal{E})\leq P_{2}(\mathcal{E})=K(\mathcal{E})$.
In particular, for a Choi stretchable channel ($\sigma=\rho_{\mathcal{E}}$),
we have%
\begin{equation}
\mathcal{C}(\mathcal{E})\leq E_{\mathrm{R}}^{\infty}(\rho_{\mathcal{E}})\leq
E_{\mathrm{R}}(\rho_{\mathcal{E}}). \label{genh2}%
\end{equation}
Recall that the proof of Eq.~(\ref{genh1}) relies on the following steps:

\begin{itemize}
\item First the derivation of the REE\ bound $\mathcal{C}(\mathcal{E})\leq
E_{\mathrm{R}}^{\bigstar}(\mathcal{E})$ for any channel $\mathcal{E}$ at any
dimension (weak converse theorem)

\item Second, the adaptive-to-block reduction by teleportation stretching\ at
any dimension, which decomposes the output of an arbitrary adaptive protocol
into $\bar{\Lambda}(\sigma^{\otimes n})$ or a suitable asymptotic form.
\end{itemize}

\noindent Because the REE is a functional which is monotonic under
trace-preserving LOCCs and subadditive over tensor products, we may then
derive Eq.~(\ref{genh1}). It is clear that this procedure can be adapted to
simplify any functional which is monotonic under LOCCs, which includes the
Rains bound~\cite{Rainsa,Rains2a} and entanglement monotones.


\bigskip

\bigskip

\section*{SUPPLEMENTARY DISCUSSION}

\subsection*{Schematic summary of our key findings}

\noindent\textbf{(1)}~We have designed an
\textbf{adaptive-to-block reduction} method which reduces any
adaptive protocol for quantum communication, entanglement
distribution and key generation to the computation of a
single-letter quantity. This is possible by combining the
following two main ingredients:

\begin{description}
\item[(1.1)] \textbf{Channel's REE}. We have extended the notion of relative
entropy of entanglement (REE)\ from states to channels. In
particular, we have shown that the two-way capacity
$\mathcal{C}(\mathcal{E})$ of any channel
$\mathcal{E}$ is upperbounded by a suitably-defined REE bound $E_{\mathrm{R}%
}^{\bigstar}(\mathcal{E})$.

\item[(1.2)] \textbf{LOCC simulation and teleportation stretching}. We have
introduced the most general form of simulation of a quantum
channel within a quantum/private communication scenario. This is
based on arbitrary LOCCs (even asymptotic) and can be used to
stretch an arbitrary channel $\mathcal{E}$ into a resource state
$\sigma$. By exploiting this simulation, we have shown how to
reduce an adaptive protocol (achieving an arbitrary task) into a
block form, so that its output can be decomposed as
$\bar{\Lambda}(\sigma^{\otimes n})$ for a trace-preserving LOCC
$\bar{\Lambda}$. This is valid at any dimension (finite or
infinite) and can be extended to more complex communication
scenarios.
\end{description}

\noindent Thus, the insight of our entire reduction method is the
combination of (1.1) and (1.2). `REE+teleportation stretching'
allows us to exploit the properties of the REE (monotonicity,
subadditivity) and simplify
$E_{\mathrm{R}}^{\bigstar}(\mathcal{E})$ into a single-letter
quantity so that we may write $\mathcal{C}(\mathcal{E})\leq
E_{\mathrm{R}}(\sigma)$ for any $\sigma$-stretchable channel. This
is valid at any dimension.

\bigskip

\noindent\textbf{(2)~Teleportation covariance}. At any dimension
(finite or infinite), we have identified a simple criterion
(teleportation covariance) which allows us to find those channels
which are stretchable into their Choi matrices (Choi-stretchable
channels). For these channels, we may write
$\mathcal{C}(\mathcal{E})\leq E_{\mathrm{R}}(\rho_{\mathcal{E}})$,
with the latter being the entanglement flux of the channel.

\bigskip

\noindent\textbf{(3)~Tight bounds and two-way capacities}. We have
shown that the entanglement flux is the tightest upper bound for
the two-way capacities of many quantum channels at any dimension,
including Pauli, erasure and bosonic Gaussian channels. In
particular, we have established the two-way capacities ($Q_{2}$,
$D_{2}$ and $K$) of the bosonic lossy channel, the quantum-limited
amplifier, and the dephasing channel in arbitrary finite
dimension, plus the secret key capacity $K$ of the erasure channel
in arbitrary finite dimension. All these capacities have extremely
simple formulas. For our calculations we have derived a simple
formula for the relative entropy between two arbitrary Gaussian
states.

\bigskip

\noindent\textbf{(4)~Fundamental rate-loss tradeoff}. We have
finally characterized the rate-loss tradeoff affecting quantum
optical communications, so that the rate of repeaterless QKD\ is
restricted to $1.44\eta$ bits per channel use at long distances.
This rate is achievable with one-way CCs and provides the maximum
throughput of a point-to-point QKD protocol.


\subsection*{Recent developments in quantum and private communications}

\subsubsection*{Repeater-assisted capacities and multi-hop
networks}

As also mentioned in the discussion of the main text, an important
generalization of the results has been achieved in ref.~\cite{networkPIRSa}%
\ with the study and determination of repeater-assisted capacities
in the presence of unlimited two-way CCs. Ref.~\cite{networkPIRSa}
establishes the ultimate rates for transmitting quantum
information, distributing entanglement and secret keys in
repeater-assisted quantum communications, under the most
fundamental decoherence models for both discrete and continuous
variable systems, including lossy channels, quantum-limited
amplifiers, dephasing and erasure channels. These capacities are
derived considering the most general adaptive protocols of quantum
and private communication between the two end-points of a repeater
chain and, more generally, of an arbitrarily-complex quantum
network or internet, where systems may be routed though single or
multiple paths. Methodology combines tools from quantum
information and classical network theory. Converse results are
derived by introducing a tensor-product representation for a
quantum communication network, where quantum channels are replaced
by their Choi matrices. Exploiting this representation and
suitable entanglement cuts of the network, one can upperbound the
end-to-end capacities by means of the relative entropy of
entanglement. Achievability of the bounds is proven by combining
point-to-point quantum communications with classical network
algorithms, so that optimal routing strategies are found by
determining the widest path and the maximum flow in the network.
In this way, ref.~\cite{networkPIRSa} extends both the widest path
problem and the max-flow min-cut theorem from classical to quantum
communications.

\subsubsection*{Single-hop networks (broadcast, multiple-access and
interference channels)}

Ref.~\cite{Multipointa}\ investigates the maximum rates for
transmitting quantum information, distributing entanglement and
secret keys in a single-hop multipoint network, with the
assistance of unlimited two-way classical communication among all
the parties. Ref.~\cite{Multipointa} first considers a sender
directly communicating with an arbitrary number of receivers, so
called quantum broadcast channel. In this case, it provides a
simple analysis in the bosonic setting considering quantum
broadcasting through a sequence of beamsplitters. This specific
case has been also investigated in ref.~\cite{Takeo} where the use
of our method (REE+teleportation stretching) has led to the
determination of the capacity region of the lossy broadcast
channel. Then, ref.~\cite{Multipointa} also considers the
multipoint setting where an arbitrary number of senders directly
communicate with a single receiver, so called quantum
multiple-access channel. Finally, ref.~\cite{Multipointa} studies
the general case of a quantum interference channel where an
arbitrary number of senders directly communicate with an arbitrary
number of receivers. Upper bounds are formulated for quantum
systems of arbitrary dimension, so that they can be applied to
many different physical scenarios involving multipoint quantum and
private communication.

\subsubsection*{Improving the lower bound for the thermal-loss
channel}

It remains an open problem to determine the two-way capacities of
several
channels, most notably that of the thermal-loss channel $\mathcal{E}%
_{\text{\textrm{loss}}}(\eta,\bar{n})$. Here we have shown lower-
and upper-bounds in Eq.~(\ref{sandAAA}). Recently,
ref.~\cite{CarloSPIE}\ has studied the specific case of the
secret-key capacity $K(\eta,\bar{n})$ of this channel
investigating a region where the lower-bound given by the reverse
coherent information can be beaten. This is possible by resorting
to a Gaussian\ QKD protocol based on trusted-noise detection.
However, the improved lower bound is still far from closing the
gap.

\subsubsection*{Strong converse rates for private communication}

Let us start by recalling that the first version of our paper
appeared on the arXiv in October 2015~\cite{firstP}. It originally
contained the main result for the bosonic lossy channels. The
other results for DV\ and CV channels were given in a second
paper, uploaded on the arXiv in December 2015~\cite{secondP}.
These two papers were later merged into a single contribution,
which is the present manuscript. In late February 2016, about 4
months after our first arXiv submission, other authors (Wilde,
Tomamichel and Berta) uploaded an interesting follow-up
paper~\cite{WildeFollowup} exploiting our methodology and showing
that our weak converse bounds for Choi-stretchable channels are
also strong converse bounds. In private communication, a weak
converse bound means that \textit{perfect} secret keys cannot be
established at rates above the bound. A strong converse bound is a
refinement according to which even \textit{imperfect} secret keys
($\varepsilon$-secure with $\varepsilon>0$) cannot be shared at
rates above the bound in the limit of large uses. Let us discuss this follow-up paper in more detail; we refer to its arXiv version~2 (updated in September 2016).

\bigskip

(1)~\textit{Methodology employed}.~These authors do not adopt the
terminology previously established by our work (teleportation
stretching, stretchable channels etc.) but one can check that they
exploit our methodology to achieve some of their most important
results. In fact, following our work, these authors:

\begin{itemize}
\item Consider a notion of channel's REE to bound key generation

\item Exploit teleportation stretching to simplify adaptive protocols for key
generation/private communication.
\end{itemize}

\noindent In a few words, they adopt our general method of
adaptive-to-block reduction (REE+teleportation stretching). This
is what allows them to derive computable single-letter bounds for
two-way assisted private communication.

To be more precise, these authors first define \textquotedblleft
classical pre- and post-processing (CPPP)
protocols\textquotedblright. These are \textit{non}-adaptive
protocols where the remote parties are limited to a single round
of initial LOCCs and another single round of final LOCCs. In this
context, they derive strong converse rates for CPPP-assisted
private communication over arbitrary channels (see Theorem~13 of ref.~\cite[version 2]{WildeFollowup}%
). To demonstrate results which are valid for adaptive protocols
with unlimited two-way CCs (but over suitable channels), they need
to employ REE+teleportation stretching; this allows them to prove
their Theorems~12 and~19 (these are strong converse versions of
our Theorem~5 stated in the main text).\ Thanks to these theorems,
they can show that our upper-bounds for dephasing, erasure and
Gaussian channels are strong converse rates (see Propositions~22,
23 and Theorem~24 in ref.~\cite[version 2]{WildeFollowup}). In
particular, the explicit computations for Gaussian channels make
use of the formula for the relative entropy between Gaussian
states.

\bigskip

(2)~\textit{Protocol reduction}.~As discussed in Supplementary
Note~9, our adaptive-to-block reduction (valid at any dimension)
must not be naively confused with the precursory but restricted
arguments of BDSW~\cite[Section~V]{B2}. The finite-$n$
decomposition $\Lambda_{A^{n} B^{n} \rightarrow K_{A} K_{B}}
(\omega_{AB}^{\otimes n})$ written in ref.~\cite[version
2]{WildeFollowup} (and its implicit extension to asymptotic
simulations) represents the reduction of an adaptive
\textit{key-generation} protocol which has been demonstrated in
our manuscript for the first time (not in BDSW which treats
quantum communication protocols and their asymptotic
$n\rightarrow+\infty$ performance). It is also clear that our
approach is different from ref.~\cite{Niset}, which does not
consider any reduction of adaptive protocols and not even key
generation (in fact, ref.~\cite{Niset} specifically deals with
one-way entanglement distillation, furthermore restricted to
Gaussian operations). For this reason, BDSW or ref.~\cite{Niset}
cannot play any direct role in the proof of Theorem~24 of
ref.~\cite[version 2]{WildeFollowup} specifically dealing with key
generation over CV\ channels.

\bigskip

(3)~\textit{Notation, teleportation-simulability and covariant
channels}.~Some definitions and results of ref.~\cite[version
2]{WildeFollowup} may be greatly simplified once they are
connected with our previous terminology. For instance,
ref.~\cite[version 2]{WildeFollowup} defines \textquotedblleft
teleportation-simulable\textquotedblright\ channels based on BDSW.
As explained in Supplementary Note~8, the original teleportation
simulation designed by BDSW is restricted to Pauli
channels~\cite{SougatoBowen}. Even admitting later
generalizations~\cite{WernerTELE}, the definition of
\textquotedblleft teleportation-simulable\textquotedblright\
channels given in~\cite[version 2]{WildeFollowup} does not
include: (i)~CV channels and~(ii) the possibility of more general
LOCCs beyond teleportation (including asymptotic simulations).
Thus, this definition is a very special case of what we call
$\sigma$-stretchable channels. In particular, what
ref.~\cite[version 2]{WildeFollowup} calls a \textquotedblleft
teleportation-simulable channel with associated
[Choi-matrix] state $\omega_{AB}=\mathcal{N}_{A^{\prime}\rightarrow B}%
(\Phi_{AA^{\prime}})$\textquotedblright\ corresponds to a
Choi-stretchable channel (in finite dimension). Ref.~\cite[version
2]{WildeFollowup} also considers DV channels which are
\textquotedblleft covariant\textquotedblright\ (see Definition~1
of ref.~\cite[version 2]{WildeFollowup}). One can check that
covariant channels must be (DV) teleportation-covariant channels
(contrary to the former, the latter definition catches the
important physical connection with teleportation).

Then, Proposition~2 of ref.~\cite[version 2]{WildeFollowup} states
that covariant channels are teleportation simulable with an
associated Choi-matrix state. This can be proven almost
immediately by using our Proposition~2 of the main text, where we
state that $(*)$ teleportation-covariant channels are
Choi-stretchable. In fact, one may just write the following:

\begin{align*}
\text{Covariant channel}  &  \Rightarrow\text{(DV)
teleportation-covariant
channel}\\
&\overset{(\ast)}{\Rightarrow} \text{(DV) Choi-stretchable channel simulable by teleportation}%
\\
&  \Rightarrow\text{Teleportation simulable with an associated
Choi-matrix
state.}%
\end{align*}

\noindent Note that our Proposition~2 is more general, being valid
for both DV and CV channels (and it was already present in earlier
arXiv versions, but with a slightly different terminology).

\bigskip

\subsection*{Further remarks}

\subsubsection*{Shield system}
In earlier arXiv versions of our manuscript, we proved our weak
converse theorem by exploiting an (at most) exponential growth of
the dimensionality of the shield system in the private state. This
corresponds to the first proof in~\ref{app3}. This assumption on
the shield size is correct and fully justified by the argument of
refs.~\cite{Matthias1a,Matthias2a} which may be applied to both DV
and CV channels, as presented in Lemma~\ref{shieldliminf}
of~\ref{app3} for the sake of completeness. Despite the
correctness of this approach, in later arXiv versions we have also
provided two additional proofs, alternative but essentially
equivalent to the first one (with exactly the same conclusions).
Our second proof relies on an exponential increase of the mean
number of photons in the private state, while our third proof is
independent from the shield system. See~\ref{app3} for full
details. It is clear that these proofs are all \textit{complete
proofs} which do not need any confirmation or validation by
follow-up works (including ref.~\cite{WildeFollowup}).

\subsubsection*{Simulation and stretching of bosonic channels}

In March 2016, several months after our manuscript was available
on the arXiv, an author uploaded a paper~\cite{Namiki} discussing
some mathematical aspects associated with our treatment of
teleportation stretching with bosonic channels. Let us briefly
give some background before clarifying that these mathematical
aspects were already taken into account and addressed in our arXiv
version 2 of December 2015~\cite{firstS}.

Teleportation stretching of bosonic channels involves the use of
an asymptotic CV EPR state $\Phi$, defined as the limit of TMSV
states $\Phi^{\mu}$. As a consequence, we have to consider the
following steps: (i) We first perform an imperfect stretching of
the protocol based on a finite-energy TMSV state $\Phi^{\mu}$;
(ii) we compute the relevant functionals on the finite-energy
decomposition of the output; and (iii) we take the infinite-energy
limit $\mu\rightarrow+\infty$ on the final result. This is
actually a standard procedure in any calculus with a delta
function, which is implicitly meant to be a limit of test
functions. This is also why the Vaidman teleportation
protocol~\cite{Vaidman} (based on an asymptotic delta-like CV EPR
state) has to be implicitly replaced by the Braunstein-Kimble
protocol~\cite{SamKimble}, where the resource state is a TMSV
state $\Phi^{\mu}$ and the infinite-energy limit is computed at
the end on the fidelity.

Such a basic argument was already present in our earlier arXiv
versions. Already in December 2015~\cite{firstS} we stated that,
for bosonic channels, one needs to relax the condition of infinite
energy and replace the asymptotic CV EPR state $\Phi$ by a
sequence of TMSV states $\Phi^{\mu}$, defining a
sequence of Choi-approximating states $\rho_{\mathcal{E}}^{\mu}:=\mathcal{I}%
\otimes\mathcal{E}(\Phi^{\mu})$. The latter states are then used
to compute the relative entropy of entanglement before taking the
limit for large $\mu$; see Eq.~(9) and corresponding text of
ref.~\cite{firstS}. Therefore, our treatment of bosonic channels
was already rigorous and correct well before ref.~\cite{Namiki}.
However, we have also realized that these non-trivial steps were
too implicit. For this reason, we have decided to fully expand the
specific treatment of bosonic channels in more recent arXiv
versions of our manuscript. Furthermore, in order to be completely
rigorous, we have also accounted for the fact that the CV Bell
detection also needs to be approximated by a suitable limit of
finite-energy measurements.


\newpage


\begin{thebibliography}{999}                                                                                              %


\bibitem {NielsenChuangm}Nielsen, M. A. \& Chuang, I. L. Quantum computation
and quantum information (Cambridge University Press, Cambridge, 2000).

\bibitem {RMP}Weedbrook, C. \textit{et al.} Gaussian quantum information.
\textit{Rev. Mod. Phys. }\textbf{84}, 621--669 (2012).

\bibitem {HolevoBOOK}Holevo, A. Quantum Systems, Channels, Information: A
Mathematical Introduction (De Gruyter, Berlin-Boston, 2012).

\bibitem {BB84}Bennett, C. H. \& Brassard, G. Quantum cryptography: Public key
distribution and coin tossing.\ \textit{Proc. IEEE International Conf. on
Computers, Systems, and Signal Processing}, \textit{Bangalore}, pp. 175--179 (1984).

\bibitem {GisinQKD}Gisin, N., Ribordy, G., Tittel, W. \& Zbinden, H. Quantum
cryptography. \textit{Rev. Mod. Phys.} \textbf{74}, 145-196 (2002).

\bibitem {Scaranim}Scarani, V. \textit{et al. }The security of practical
quantum key distribution. \textit{Rev. Mod. Phys.} \textbf{81}, 1301--1350 (2009).

\bibitem {Kimble2008}Kimble, H. J. The quantum internet. \textit{Nature}
\textbf{453}, 1023--1030 (2008).

\bibitem {HybridINTERNET}Pirandola, S. \& Braunstein, S. L. Unite to build a
quantum internet. \textit{Nature} \textbf{532}, 169--171 (2016).

\bibitem {telereview}Pirandola, S. \textit{et al.} Advances in quantum
teleportation. \textit{Nature Photon.} \textbf{9}, 641--652 (2015).

\bibitem {Ulrikreview}Andersen, U. L., Neergaard-Nielsen, J. S., van Loock, P.
\& Furusawa, A. Hybrid discrete- and continuous-variable quantum information.
\textit{Nature Phys.} \textbf{11}, 713--719 (2015).

\bibitem {Briegel98}Briegel, H.-J., D\"{u}r, W., Cirac, J. I. \& Zoller, P.
Quantum repeaters: The role of imperfect local operations in quantum
communication. \textit{Phys. Rev. Lett.} \textbf{81}, 5932-5935 (1998).

\bibitem {Fred}Grosshans, F. \textit{et al.\ }Quantum key distribution using
gaussian-modulated coherent states. \textit{Nature} \textbf{421}, 238-241 (2003).

\bibitem {CVMDIQKD}Pirandola, S. \textit{et al.} High-rate
measurement-device-independent quantum cryptography. \textit{Nature Photon.}
\textbf{9}, 397-402 (2015).

\bibitem {VedFORMm}Vedral, V., Plenio, M. B., Rippin, M. A. \& Knight, P. L.
Quantifying Entanglement. \textit{Phys. Rev. Lett.} \textbf{78}, 2275-2279 (1997).

\bibitem {Pleniom}Vedral, V. \& Plenio, M. B. Entanglement measures and
purification procedures. \textit{Phys. Rev. A} \textbf{57}, 1619--1633 (1998).

\bibitem {RMPrelent}Vedral, V. The role of relative entropy in quantum
information theory. \textit{Rev. Mod. Phys. }\textbf{74}, 197--234 (2002).

\bibitem {tele93}Bennett, C. H., Brassard, G., Crepeau, C., Jozsa, R., Peres,
A. \& Wootters, W. K. Teleporting an unknown quantum state via dual classical
and Einstein-Podolsky-Rosen channels. \textit{Phys. Rev. Lett.} \textbf{70},
1895--1899 (1993).

\bibitem {Samtele}Braunstein, S. L. \& Kimble, H. J. Teleportation of
Continuous Quantum Variables. \textit{Phys. Rev. Lett.} \textbf{80}, 869--872 (1998).

\bibitem {Samtele2}Braunstein, S. L., D'Ariano, G. M., Milburn, G. J. \&
Sacchi, M. F. Universal teleportation with a twist. \textit{Phys. Rev. Lett.}
\textbf{84}, 3486--3489 (2000).

\bibitem {ErasureChannelm}Bennett, C. H., DiVincenzo, D. P., \& Smolin, J. A.
Capacities of Quantum Erasure Channels. \textit{Phys. Rev. Lett.} \textbf{78},
3217-3220 (1997).

\bibitem {ReverseCAP}Pirandola, S., Garc\'{\i}a-Patr\'{o}n, R., Braunstein, S.
L. \& Lloyd, S. Direct and reverse secret-key capacities of a quantum channel.
\textit{Phys. Rev. Lett.} \textbf{102}, 050503 (2009).

\bibitem {RevCohINFO}Garc\'{\i}a-Patr\'{o}n, R., Pirandola, S., Lloyd, S. \&
Shapiro, J. H. Reverse coherent information. \textit{Phys. Rev. Lett.}
\textbf{102}, 210501 (2009).

\bibitem {TGW}Takeoka, M., Guha, S. \& Wilde, M.M. Fundamental rate-loss
tradeoff for optical quantum key distribution. \textit{Nat. Commun.}
\textbf{5}, 5235 (2014).

\bibitem {Matthias2}Christandl, M. The Structure of Bipartite Quantum States:
Insights from Group Theory and Cryptography. \textit{PhD thesis, University of
Cambridge, 2006.}

\bibitem {DWrates2}Devetak, I. \& Winter, A. Distillation of secret key and
entanglement from quantum states. \textit{Proc. R. Soc. A} \textbf{461},
207--235 (2005).

\bibitem {Devetak}Devetak, I. The private classical capacity and quantum
capacity of a quantum channel. \textit{IEEE Trans. Info. Theory }\textbf{51},
44--55 (2005).

\bibitem {Lutken}Curty, M., Lewenstein, M. \& L\"{u}tkenhaus, N. Entanglement
as a Precondition for Secure Quantum Key Distribution. \textit{Phys. Rev.
Lett. }\textbf{92}, 217903 (2004).

\bibitem {KD1}Horodecki, K., Horodecki, M., Horodecki, P. \& Oppenheim, J.
Secure key from bound entanglement. \textit{Phys. Rev. Lett.} \textbf{94},
160502 (2005).

\bibitem {QC1}Schumacher, B. \& Nielsen, M. A. Quantum data processing and
error correction. \textit{Phys. Rev. A} \textbf{54}, 2629--2635 (1996).

\bibitem {QC2}Lloyd, S. Capacity of the noisy quantum channel. \textit{Phys.
Rev. A} \textbf{55}, 1613--1622 (1997).

\bibitem {MatthiasMAIN}Christiandl, M., Schuch, N. \& Winter, A. Entanglement
of the antisymmetric state. \textit{Comm. Math. Phys.} \textbf{311}, 397--422 (2012).

\bibitem {B2main}Bennett, C. H., DiVincenzo, D. P., Smolin, J. A. \& Wootters,
W. K. Mixed-state entanglement and quantum error correction. \textit{Phys.
Rev. A} \textbf{54}, 3824--3851 (1996).

\bibitem {HoroTELm}Horodecki, M., Horodecki, P. \& Horodecki, R. General
teleportation channel, singlet fraction, and quasidistillation. \textit{Phys.
Rev. A} \textbf{99}, 1888--1898 (1999).

\bibitem {SougatoBowenm}Bowen, G. \& Bose, S. Teleportation as a Depolarizing
Quantum Channel, Relative Entropy and Classical Capacity. \textit{Phys. Rev.
Lett.} \textbf{87}, 267901 (2001).

\bibitem {Albeveriom}Albeverio, S., Fei, S.-M. \& Yang, W.-L. Optimal
teleportation based on Bell measurements. \textit{Phys. Rev. A} \textbf{66},
012301 (2002).

\bibitem{Alex} M\"{u}ller-Hermes, A. Transposition in Quantum Information
Theory. \textit{Master Thesis, Technische Universit\"{a}t
M\"{u}nchen} (2012).

\bibitem {Leungm}Leung, D. \& Matthews, W. On the power of ppt-preserving and
nonsignalling codes. \textit{IEEE Transactions on Information Theory}
\textbf{61}, 4486--4499 (2015).

\bibitem {WernerTELEmain}Werner, R. F. All teleportation and dense coding
schemes. \textit{J. Phys. A} \textbf{34}, 7081--7094 (2001).

\bibitem {NisetMAIN}J. Niset, J., Fiurasek, J. \& Cerf, N. J. No-go theorem
for Gaussian quantum error correction. \textit{Phys. Rev. Lett.} \textbf{102},
120501 (2009).

\bibitem {qSIM}Ji, Z., Wang, G., Duan, R., Feng, Y. \& Ying, M. Parameter
estimation of quantum channels.\textit{ IEEE Trans. Inform. Theory}
\textbf{54}, 5172--5185 (2008).

\bibitem {Processorsm}Nielsen, M. A. \& Chuang, I. L. Programmable Quantum
Gate Arrays. \textit{Phys. Rev. Lett.} \textbf{79}, 321--324 (1997).

\bibitem {RMPdiscord}Modi, K. \textit{et al.} The classical-quantum boundary
for correlations: Discord and related measures.\ \textit{Rev. Mod. Phys.}
\textbf{84}, 1655--1707 (2012).

\bibitem {DiscordQKD}Pirandola, S. Quantum discord as a resource for quantum
cryptography. \textit{Sci. Rep.} \textbf{4}, 6956 (2014).

\bibitem {OptimalDIS}Pirandola, S., Spedalieri, G., Braunstein, S.L., Cerf, N.
J. \& Lloyd, S. Optimality of Gaussian discord. \textit{Phys. Rev. Lett.}
\textbf{113}, 140405 (2014).

\bibitem {Wee}Weedbrook, C. \textit{et al.} Quantum cryptography without
switching. \textit{Phys. Rev. Lett.} \textbf{93}, 170504 (2004).

\bibitem {Carlo}Ottaviani, C., Spedalieri, G., Braunstein, S.L. \& Pirandola,
S. Continuous-variable quantum cryptography with an untrusted relay: Detailed
security analysis of the symmetric configuration. \textit{Phys. Rev. A}
\textbf{91}, 022320 (2015).

\bibitem {Gae}Spedalieri, G. \textit{et al.} Quantum cryptography with an
ideal local relay. \textit{Proceedings of the SPIE Security + Defence 2015
conference on Quantum Information Science and Technology}, Toulouse, France
(21-24 September 2015). Paper 9648-47. See also Preprint at
https://arxiv.org/abs/1509.01113 (2015).

\bibitem {Twoway}Pirandola, S., Mancini, S., Lloyd, S. \& Braunstein, S. L.
Continuous-variable quantum cryptography using two-way quantum communication.
\textit{Nature Phys.} \textbf{4}, 726--730 (2008).

\bibitem {Carlo2main}Ottaviani, C. \& Pirandola S. General immunity and
superadditivity of two-way Gaussian quantum cryptography. \textit{Sci. Rep}.
\textbf{6}, 22225 (2016).

\bibitem {MDI1}Braunstein, S. L. \& Pirandola, S. Side-channel-free quantum
key distribution. \textit{Phys. Rev. Lett.} \textbf{108}, 130502 (2012).

\bibitem {MDI2}Lo, H.-K., Curty, M. \& Qi, B. Measurement-Device-Independent
Quantum Key Distribution. \textit{Phys. Rev. Lett.} \textbf{108}, 130503 (2012).

\bibitem {Satellite}Hosseinidehaj, N. \& Malaney, R. Gaussian entanglement
distribution via satellite. \textit{Phys. Rev. A} \textbf{91}, 022304 (2015).

\bibitem {sat1}Vallone, G. \textit{et al.} Experimental Satellite Quantum
Communications. \textit{Phys. Rev. Lett.} \textbf{115}, 040502 (2015).

\bibitem {sat2}Dequal, D. \textit{et al. }Experimental single-photon exchange
along a space link of 7000 km. \textit{Phys. Rev. A} \textbf{93}, 010301(R) (2016).

\bibitem {Usenko10}Usenko, V. C. \& Filip, R. \ Feasibility of
continuous-variable quantum key distribution with noisy coherent states.
\textit{Phys. Rev. A} \textbf{81}, 022318 (2010).

\bibitem {thermal-PRL}Weedbrook, C. \textit{et al.} Quantum cryptography
approaching the classical limit. \textit{Phys. Rev. Lett.} \textbf{105},
110501 (2010).

\bibitem {thermal-PRA}Weedbrook, C. \textit{et al.} Continuous-variable
quantum key distribution using thermal states. \textit{Phys. Rev. A}
\textbf{86}, 022318 (2012).

\bibitem {thermal-2-PRA}Weedbrook, C., Ottaviani, C. \& Pirandola, S. Two-way
quantum cryptography at different wavelengths. \textit{Phys. Rev. A}
\textbf{89}, 012309 (2014).

\bibitem {GEW}Goodenough, K., Elkouss, D. \& Wehner, S. Assessing the
performance of quantum repeaters for all phase-insensitive Gaussian bosonic
channels.\ Preprint at https://arxiv.org/abs/1511.08710v1 (2015).

\bibitem {HolevoWerner}Holevo, A. S. \& Werner, R. F. Evaluating capacities of
bosonic Gaussian channels. \textit{Phys. Rev. A} \textbf{63}, 032312 (2001).

\bibitem {Wolf}Wolf, M. M., P\'{e}rez-Garc\'{\i}a, D. \& Giedke, G. Quantum
Capacities of Bosonic Channels. \textit{Phys. Rev. Lett.} \textbf{98}, 130501 (2007).

\bibitem {degradable}Devetak, I. \& Shor, P.W. The Capacity of a Quantum
Channel for Simultaneous Transmission of Classical and Quantum
Information,\ \textit{Commun. Math. Phys.} \textbf{256}, 287--303 (2005).

\bibitem {Bose1}Bose, S. Quantum Communication through an Unmodulated Spin
Chain. \textit{Phys. Rev. Lett.} \textbf{20}, 207901 (2003).

\bibitem {Bose2}Bose, S., Bayat, A., Sodano, P., Banchi, L. \& Verrucchi, P.
Spin Chains as Data Buses, Logic Buses and Entanglers in \textquotedblleft
Quantum State Transfer and Network Engineering\textquotedblright\ (Springer
Berlin Heidelberg, 2014), pp 1-37.

\bibitem {networkPIRS}Pirandola, S. Capacities of repeater-assisted quantum
communications. Preprint at https://arxiv.org/abs/1601.00966 (2016).

\bibitem {LupoPIR}Pirandola, S. \& Lupo, C. Ultimate precision of adaptive
noise estimation. Preprint at https://arxiv.org/abs/1609.02160 (2016).

\bibitem {arvind_real_1995}Arvind, B. Dutta, N. Mukunda, \& R. Simon. The real
symplectic groups in quantum mechanics and optics. \textit{Pramana}
\textbf{45}, 471--497 (1995).

\bibitem {Banchim}Banchi, L., Braunstein, S.L. \& Pirandola, S. Quantum
fidelity for arbitrary Gaussian states. \textit{Phys. Rev. Lett.}
\textbf{115}, 260501 (2015).

\bibitem {Chenmain}Chen, X.-y. Gaussian relative entropy of
entanglement.\textit{ Phys. Rev. A} \textbf{71}, 062320 (2005).

\bibitem {Scheelm}Scheel, S., \& Welsch D.-G. Entanglement generation and
degradation by passive optical devices. \textit{Phys. Rev. A} \textbf{64},
063811 (2001).
\SP{\bibitem {Arakinew} Araki, H. Relative Entropy for States of von Neumann Algebras II. \textit{Publ. RIMS Kyoto Univ.} \textbf{13}, 173--192 (1977).
\bibitem{Lindbladnew} Lindblad, G. Completely positive maps and entropy inequalities. \textit{Commun. Math. Phys.} \textbf{40}, 147--151 (1975).}


\end{thebibliography}

\begin{thebibliography}{99999}                                                                                            %
\section*{Supplementary References}


\bibitem {Hol03a}Holevo, A. S. Entanglement-assisted capacity of constrained
channels. \textit{Probab. Theory Appli.} \textbf{48}, 243--255
(2004).

\bibitem {Darianoa}D'Ariano, G. M., Kretschmann, D., Schlingemann, D. \&
Werner, R. F. Reexamination of quantum bit commitment: The
possible and the impossible. \textit{Phys. Rev. A} \textbf{76},
032328 (2007).

\bibitem {telereviewa}Pirandola, S. \textit{et al}. Advances in quantum
teleportation. \textit{Nature Photon}. \textbf{9}, 641-652 (2015).

\bibitem {QC1a}Schumacher, B. \& Nielsen, M. A. Quantum data processing and
error correction. \textit{Phys. Rev. A} \textbf{54}, 2629--2635
(1996).

\bibitem {QC2a}Lloyd, S. Capacity of the noisy quantum channel. \textit{Phys.
Rev. A} \textbf{55}, 1613--1622 (1997).

\bibitem {RevCohINFOa}Garc\'{\i}a-Patr\'{o}n, R., Pirandola, S., Lloyd, S. \&
Shapiro, J. H. Reverse coherent information. \textit{Phys. Rev.
Lett.} \textbf{102}, 210501 (2009).

\bibitem {ReverseCAPa}Pirandola, S., R. Garc\'{\i}a-Patr\'{o}n, Braunstein, S.
L. \& Lloyd, L. Direct and reverse secret-key capacities of a
quantum channel. \textit{Phys. Rev. Lett.} \textbf{102}, 050503
(2009).

\bibitem {DWrates2a}Devetak, I. \& Winter, A. Distillation of secret key and
entanglement from quantum states. \textit{Proc. R. Soc. A}
\textbf{461}, 207--235 (2005).

\bibitem {WinterCONa}Winter, A. Tight uniform continuity bounds for quantum
entropies: conditional entropy, relative entropy distance and
energy constraints. \textit{Commun. Math. Phys.} \textbf{347},
291--313 (2016).

\bibitem {RMPa}Weedbrook, C. \textit{et al.} Gaussian quantum information.
\textit{Rev. Mod. Phys.} \textbf{84}, 621--669 (2012).

\bibitem {KD1a}Horodecki, K., Horodecki, M., Horodecki, P. \& Oppenheim, J.
Secure key from bound entanglement. \textit{Phys. Rev. Lett.}
\textbf{94}, 160502 (2005).

\bibitem {KD2a}Horodecki, K., Horodecki, M., Horodecki, P. \& Oppenheim, J.
General Paradigm for Distilling Classical Key From Quantum States.
\textit{IEEE Trans. Inf. Theory} \textbf{55}, 1898--1929 (2009).

\bibitem {RenesSmith}Renes, J. \& Smith, G. Noisy Processing and Distillation
of Private Quantum States. \textit{Phys. Rev. Lett.} \textbf{98},
020502 (2007).

\bibitem {Matthias1a}Christiandl, M., Ekert, A., Horodecki, M., Horodecki, P.,
Oppenheim, J. \& Renner, R. Unifying classical and quantum key
distillation. \textit{Lecture Notes in Computer Science
}\textbf{4392}, 456--478 (2007). See also preprint at
https://arxiv.org/abs/quant-ph/0608199v3 (2006).

\bibitem {Matthias2a}Christiandl, M., Schuch, N. \& Winter, A. Entanglement of
the antisymmetric state. \textit{Comm. Math. Phys.} \textbf{311},
397--422 (2012).

\bibitem {RMPrelenta}Vedral, V. The role of relative entropy in quantum
information theory. \textit{Rev. Mod. Phys. }\textbf{74}, 197--234
(2002).

\bibitem {Donald}Donald, M. J.\ \& Horodecki, M. Continuity of Relative
Entropy of Entanglement. \textit{Phys. Lett. A }\textbf{264},
257--260 (1999).

\bibitem {Synaka}Synak-Radtke, B. \& Horodecki, M. On asymptotic continuity of
functions of quantum states.\ \textit{J. Phys. A: Math. Gen}.
\textbf{39}, L423--L437 (2006).

\bibitem {HolevoBOOKa}Holevo, A. Quantum Systems, Channels, Information: A
Mathematical Introduction (De Gruyter, Berlin-Boston, 2012).

\bibitem {AlickFANN}Alicki, R. \& Fannes, M. Continuity of conditional quantum
mutual information. \textit{J. Phys. A: Math. Gen.} \textbf{37},
L55--L57 (2004).

\bibitem {arvind_real_1995a}Arvind, B. Dutta, N. Mukunda, \& R. Simon. The
real symplectic groups in quantum mechanics and optics.
\textit{Pramana} \textbf{45}, 471--497 (1995).

\bibitem {Banchi}Banchi, L., Braunstein, S.L. \& Pirandola, S. Quantum
fidelity for arbitrary Gaussian states. \textit{Phys. Rev. Lett.}
\textbf{115}, 260501 (2015).

\bibitem {bonafidea}Pirandola, S., Serafini, A. \& Lloyd, S. Correlation
matrices of two-mode bosonic systems. \textit{Phys. Rev. A}
\textbf{79}, 052327 (2009).

\bibitem {bonafide2a}Pirandola, S. Entanglement Reactivation in Separable
Environments. \textit{New J. Phys.} \textbf{15}, 113046 (2013).

\bibitem {OptimalDISa}Pirandola, S., Spedalieri, G., Braunstein, S. L., Cerf,
N. J. \& Lloyd, S. Optimality of Gaussian discord. \textit{Phys.
Rev. Lett}. \textbf{113}, 140405 (2014).

\bibitem {HolevoWernera}Holevo, A. S. \& Werner, R. F. Evaluating capacities
of bosonic Gaussian channels.\ \textit{Phys. Rev. A} \textbf{63},
032312 (2001).

\bibitem {TGWa}Takeoka, M., Guha, S. \& Wilde, M. M. Fundamental rate-loss
tradeoff for optical quantum key distribution. \textit{Nat.
Commun.} \textbf{5}, 5235 (2015).

\bibitem {GEWa}Goodenough, K., Elkouss, D. \& Wehner, S. Assessing the
performance of quantum repeaters for all phase-insensitive
Gaussian bosonic channels. Preprint at
https://arxiv.org/abs/1511.08710v1 (2015).

\bibitem {DiscordQKDa}Pirandola, S. Quantum discord as a resource for quantum
cryptography. \textit{Sci. Rep.} \textbf{4}, 6956 (2014).

\bibitem {RMPdiscorda}Modi, K. \textit{et al.} The classical-quantum boundary
for correlations: Discord and related measures.\ \textit{Rev. Mod.
Phys.} \textbf{84}, 1655--1707 (2012).

\bibitem {GerryDa}Adesso, G. \& Datta, A. Quantum versus classical
correlations in Gaussian states. \textit{Phys. Rev. Lett.}
\textbf{105}, 030501 (2010).

\bibitem {ParisDa}Giorda, P. \& Paris, M.G.A. Gaussian Quantum Discord.
\textit{Phys. Rev. Lett.} \textbf{105}, 020503 (2010).

\bibitem {Wolfa}Wolf, M. M., P\'{e}rez-Garc\'{\i}a, D. \& Giedke, G. Quantum
Capacities of Bosonic Channels. \textit{Phys. Rev. Lett.}
\textbf{98}, 130501 (2007).

\bibitem {ErasureChannel}Bennett, C. H., DiVincenzo, D. P., \& Smolin, J. A.
Capacities of Quantum Erasure Channels. \textit{Phys. Rev. Lett.}
\textbf{78}, 3217--3220 (1997).

\bibitem {VedFORM}Vedral, V., Plenio, M. B., Rippin, M. A. \& Knight, P. L.
Quantifying Entanglement. \textit{Phys. Rev. Lett.} \textbf{78},
2275--2279 (1997).

\bibitem {degradablea}Devetak, I. \& Shor, P.W. The Capacity of a Quantum
Channel for Simultaneous Transmission of Classical and Quantum
Information,\ \textit{Commun. Math. Phys.} \textbf{256}, 287--303
(2005).

\bibitem {Synak}Synak, B., Horodecki, K \& Horodecki, M. Bounds on localisable
information via semidefinite programming.\textit{ J. Math. Phys.
}\textbf{46}, 082107 (2005).

\bibitem {depha1}Fukuda, M. \& Holevo, A. S. On Weyl covariant channels.
Preprint at https://arxiv.org/abs/quant-ph/0510148 (2005).

\bibitem {depha2}Pirandola, S., Mancini, S., Braunstein, S. L. \& Vitali, D.
Minimal qudit code for a qubit in the phase-damping channel.
\textit{Phys. Rev. A} \textbf{77}, 032309 (2008).

\bibitem {SquashChannela}Takeoka, M., Guha, S. \& Wilde, M. M. The Squashed
Entanglement of a Quantum Channel. \textit{IEEE Transactions on
Information Theory} \textbf{60}, 4987--4998 (2014).

\bibitem {Weea}Weedbrook, C. \textit{et al.} Quantum Cryptography Without
Switching. \textit{Phys. Rev. Lett.} \textbf{93}, 170504 (2004).

\bibitem {GG02a}Grosshans, F. \& Grangier, P. Continuous Variable Quantum
Cryptography Using Coherent States. \textit{Phys. Rev. Lett.}
\textbf{88}, 057902 (2002).

\bibitem {Freda}Grosshans, F. \textit{et al.}\ Quantum key distribution using
gaussian-modulated coherent states. \textit{Nature }\textbf{421},
238--241 (2003).

\bibitem {CVMDIQKDa}Pirandola, S. \textit{et al.} High-rate
measurement-device-independent quantum cryptography.
\textit{Nature Photon.} \textbf{9}, 397--402 (2015).

\bibitem {Correspondencea}Pirandola, S. \textit{et al.} Reply to `Discrete and
continuous variables for measurement-device-independent quantum
cryptography'. \textit{Nature Photon.} \textbf{9}, 773--775
(2015).

\bibitem {Carloa}Ottaviani, C., Spedalieri, G., Braunstein, S.L. \& Pirandola,
S. Continuous-variable quantum cryptography with an untrusted
relay: Detailed security analysis of the symmetric configuration.
\textit{Phys. Rev. A} \textbf{91}, 022320 (2015).

\bibitem {Gaea}Spedalieri, G. \textit{et al.} Quantum cryptography with an
ideal local relay. \textit{Proceedings of the SPIE Security +
Defence 2015 conference on Quantum Information Science and
Technology}, Toulouse, France (21-24 September 2015). Paper
9648-47. See also Preprint at https://arxiv.org/abs/1509.01113
(2015).

\bibitem {Twowaya}Pirandola, S., Mancini, S., Lloyd, S. \& Braunstein, S. L.
Continuous-variable quantum cryptography using two-way quantum
communication. \textit{Nature Phys.} \textbf{4}, 726--730 (2008).

\bibitem {Carlo2ww}Ottaviani, C. \& Pirandola S. General immunity and
superadditivity of two-way Gaussian quantum cryptography.
\textit{Sci. Rep}. \textbf{6}, 22225 (2016).

\bibitem {BB84a}Bennett, C. H. \& Brassard, G. Quantum cryptography: Public
key distribution and coin tossing.\ \textit{Proc. IEEE
International Conf. on Computers, Systems, and Signal Processing},
Bangalore, pp. 175--179 (1984).

\bibitem {Scarania}Scarani, V. \textit{et al. }The security of practical
quantum key distribution. \textit{Rev. Mod. Phys.} \textbf{81},
1301--1350 (2009).

\bibitem {Decoya}Hwang, W.-Y. Quantum Key Distribution with High Loss: Toward
Global Secure Communication. \textit{Phys. Rev. Lett.}
\textbf{91}, 057901 (2003).

\bibitem {MDI1a}Braunstein, S. L. \& Pirandola, S. Side-channel-free quantum
key distribution. \textit{Phys. Rev. Lett.} \textbf{108}, 130502
(2012).

\bibitem {MDI2a}Lo, H.-K., Curty, M. \& Qi, B. Measurement-Device-Independent
Quantum Key Distribution. \textit{Phys. Rev. Lett.} \textbf{108},
130503 (2012).

\bibitem {B2}Bennett, C. H., DiVincenzo, D. P., Smolin, J. A. \& Wootters, W.
K. Mixed-state entanglement and quantum error correction.
\textit{Phys. Rev. A} \textbf{54}, 3824--3851 (1996).

\bibitem {WernerTELE}Werner, R. F. All teleportation and dense coding schemes.
\textit{J. Phys. A} \textbf{34}, 7081--7094 (2001).

\bibitem {HoroTEL}Horodecki, M., Horodecki, P. \& Horodecki, R. General
teleportation channel, singlet fraction, and quasidistillation.
\textit{Phys. Rev. A} \textbf{99}, 1888--1898 (1999).

\bibitem {SougatoBowen}Bowen, G. \& Bose, S. Teleportation as a Depolarizing
Quantum Channel, Relative Entropy and Classical Capacity.
\textit{Phys. Rev. Lett.} \textbf{87}, 267901 (2001).

\bibitem {Albeverio}Albeverio, S., Fei, S.-M. \& Yang, W.-L. Optimal
teleportation based on Bell measurements. \textit{Phys. Rev. A}
\textbf{66}, 012301 (2002).

\bibitem {Leung}Leung, D. \& Matthews, W. On the power of ppt-preserving and
nonsignalling codes. \textit{IEEE Transactions on Information
Theory} \textbf{61}, 4486--4499 (2015).

\bibitem{AlexAPP} M\"{u}ller-Hermes, A. Transposition in Quantum Information
Theory. \textit{Master Thesis, Technische Universit\"{a}t
M\"{u}nchen} (2012).

\bibitem {Niset}J. Niset, J., Fiurasek, J. \& Cerf, N. J. No-go theorem for
Gaussian quantum error correction. \textit{Phys. Rev. Lett.}
\textbf{102}, 120501 (2009).

\bibitem {SamKimble}Braunstein, S. L. \& Kimble, H. J. Teleportation of
continuous quantum variables. \textit{Phys. Rev. Lett.}
\textbf{80}, 869--872 (1998).

\bibitem {qSIMa}Ji, Z., Wang, G., Duan, R., Feng, Y. \& Ying, M. Parameter
estimation of quantum channels.\textit{ IEEE Trans. Inform.
Theory} \textbf{54}, 5172--5185 (2008).

\bibitem {Processors}Nielsen, M. A. \& Chuang, I. L. Programmable Quantum Gate
Arrays. \textit{Phys. Rev. Lett.} \textbf{79}, 321--324 (1997).

\bibitem {Port-based}Ishizaka, S. \& Hiroshima, T. Asymptotic Teleportation
Scheme as a Universal Programmable Quantum Processor.
\textit{Phys. Rev. Lett.} \textbf{101}, 240501 (2008).

\bibitem {RafalJanek}Kolodynski J. \& Demkowicz-Dobrzanski, R. Efficient tools
for quantum metrology with uncorrelated noise. \textit{New J.
Phys.} \textbf{15}, 073043 (2013).

\bibitem {LupoPIRa}Pirandola, S. \& Lupo, C. Ultimate precision of adaptive
noise estimation. Preprint at https://arxiv.org/abs/1609.02160
(2016).

\bibitem {networkPIRSa}Pirandola, S. Capacities of repeater-assisted quantum
communications. Preprint at https://arxiv.org/abs/1601.00966
(2016).

\bibitem {Lorenzoa}Demkowicz-Dobrzanski, R. \& Maccone, L. Using Entanglement
Against Noise in Quantum Metrology. \textit{Phys. Rev. Lett.}
\textbf{113}, 250801 (2014).

\bibitem {Devetaka}Devetak, I. The private classical capacity and quantum
capacity of a quantum channel. \textit{IEEE Trans. Info. Theory
}\textbf{51}, 44--55 (2005).

\bibitem {Rainsa}Rains, E. M. A semidefinite program for distillable
entanglement. \textit{IEEE Trans. Info. Theory} \textbf{47},
2921--2933 (2001).

\bibitem {Rains2a}Audenaert, K., De Moor, B., Vollbrecht, K. G. H. \& Werner,
R. F. Asymptotic relative entropy of entanglement for orthogonally
invariant states. \textit{Physical Review A} \textbf{66}, 032310
(2002).

\bibitem {Multipointa}Laurenza, R. \& Pirandola, S. General bounds for
sender-receiver capacities in multipoint quantum communications.
Preprint at https://arxiv.org/abs/1603.07262 (2016).

\bibitem {Takeo}Takeoka, M., Seshadreesan, K. P. \& Wilde, M. M. Unconstrained
distillation capacities of a pure-loss bosonic broadcast channel.
Preprint at https://arxiv.org/abs/1601.05563v3 (2016).

\bibitem {CarloSPIE}Ottaviani, C., Laurenza, R., Cope, T. P. W., Spedalieri,
G., Braunstein, S. L. \& Pirandola, S. Secret key capacity of the
thermal-loss channel: Improving the lower bound. \textit{Proc.
SPIE 9996, Quantum Information Science and Technology II, 999609}
(2016); doi:10.1117/12.2244899. Preprint at
https://arxiv.org/abs/1609.02169 (2016).

\bibitem {firstP}Pirandola, S., Laurenza, R., Ottaviani, C., Banchi, L. The
Ultimate Rate of Quantum Cryptography. Preprint at
https://arxiv.org/abs/1510.08863v1 (Version 1, 29 October 2015).

\bibitem {secondP}Pirandola, S. \& Laurenza, R. General Benchmarks for Quantum
Repeaters. Preprint at https://arxiv.org/abs/1512.04945v1 (Version
1, 15 December 2015).

\bibitem {WildeFollowup}Wilde, M. M., Tomamichel, M. \& Berta, M. Converse
bounds for private communication over quantum channels. Preprint
at https://arxiv.org/abs/1602.08898v2 (Version 1, 29 Feb 2016;
Version 2, 7 Sep 2016).

\bibitem {Namiki}Namiki, R. Teleportation stretching for lossy Gaussian
channels. Preprint at https://arxiv.org/abs/1603.05292v1 (2016).

\bibitem {firstS}Pirandola, S., Laurenza, R., Ottaviani, C., Banchi, L. The
Ultimate Rate of Quantum Communications. Preprint at
https://arxiv.org/abs/1510.08863v2 (Version 2, 8 December 2015).

\bibitem {Vaidman}Vaidman, L. Teleportation of quantum states.\textit{ Phys.
Rev. A} \textbf{49}, 1473--1476 (1994).

\end{thebibliography}
\end{document}